\documentclass[onecolumn,amsmath,amssymb,nofootinbib,12pt]{article}
\usepackage{jheppub}
\usepackage{ifpdf}

\usepackage{graphicx,subcaption}
\usepackage{amsfonts}
\usepackage{amsmath}
\usepackage{amssymb}
\usepackage{epsfig}
\usepackage{pdfpages}
\usepackage{graphicx,epstopdf}
\usepackage[makeroom]{cancel}
\usepackage{hyperref}
\hypersetup{pdftex,colorlinks=true,allcolors=blue}
\usepackage{array}
\usepackage{ulem}

\newcommand{\RN}[1]{%
  \textup{\uppercase\expandafter{\romannumeral#1}}%
}

\newcommand{\del}{\partial}
\newcommand{\vac}{\mt{vac}}

\newcommand{\mx}{\mt{max}}
\newcommand{\mn}{\mt{min}}
\newcommand{\past}{\mt{past}}

\newcommand{\be}{\begin{equation}}
\newcommand{\ee}{\end{equation}}
\newcommand{\bea}{\begin{eqnarray}}
\newcommand{\eea}{\end{eqnarray}}
\newcommand{\beq}{\begin{equation}}
\newcommand{\eeq}{\end{equation}}
\newcommand{\beqa}{\begin{eqnarray}}
\newcommand{\eeqa}{\end{eqnarray}}
\newcommand{\beqar}{\begin{eqnarray*}}
\newcommand{\eeqar}{\end{eqnarray*}}

\newcommand{\labell}[1]{\label{#1}} 

\newcommand{\eg}{{\it e.g.,}\ }
\newcommand{\ie}{{\it i.e.,}\ }
\newcommand{\reef}[1]{(\ref{#1})}

\newcommand{\mt}[1]{\textrm{\tiny #1}}

\newcommand{\mC}{\mathcal{C}}

\newcommand{\eps}{\epsilon}

\newcommand{\veps}{\varepsilon}
\def\S{\Sigma}
\newcommand{\cv}{{\cal C}_\mt{V}}
\newcommand{\ca}{{\cal C}_\mt{A}}
\newcommand{\csv}{{\cal C}_\mt{STV}}
\newcommand{\tca}{\tilde{\cal C}_\mt{A}}
\newcommand{\E}{P}

\newcommand{\ctL}{\ell_\mt{ct}}
\newcommand{\ttL}{\tilde\ell_\mt{ct}}

\usepackage{xspace}

\newcommand{\BHO}{BH$_1$\xspace}
\newcommand{\BHT}{BH$_2$\xspace}

\newcommand{\cOR}{{\cal O}_\mt{R}}
\newcommand{\UR}{U_\mt{R}}
\newcommand{\UL}{U_\mt{L}}
\newcommand{\tR}{t_\mt{R}}
\newcommand{\tL}{t_\mt{L}}
\newcommand{\vR}{v_\mt{R}}
\newcommand{\vL}{v_\mt{L}}
\newcommand{\uR}{u_\mt{R}}
\newcommand{\uL}{u_\mt{L}}
\newcommand{\tLc}[1]{t_{\mt{L},c{#1}}}
\newcommand{\tRc}[1]{t_{\mt{R},c{#1}}}
\newcommand{\hatt}{{\hat t}^{*}}

\begin{document}

\preprint{arXiv:1805.nnnnn [hep-th]}
\title{\Large Holographic Complexity in Vaidya Spacetimes\ \  II}

\author[a]{Shira Chapman,}
\author[a, b]{Hugo Marrochio,}
\author[a]{Robert C. Myers}
\affiliation[a]{Perimeter Institute for Theoretical Physics, Waterloo, ON N2L 2Y5, Canada}
\affiliation[b]{Department of Physics $\&$ Astronomy,
University of Waterloo,\\ Waterloo, ON N2L 3G1, Canada}

\emailAdd{schapman@perimeterinstitute.ca}
\emailAdd{hmarrochio@perimeterinstitute.ca}
\emailAdd{rmyers@perimeterinstitute.ca}

\date{\today}

\abstract{In this second part of the study initiated in \cite{Vad1}, we investigate holographic complexity for eternal black hole backgrounds perturbed by shock waves, with both the complexity$=$action (CA) and complexity$=$volume (CV) proposals. In particular, we consider Vaidya geometries describing a thin shell of null fluid with arbitrary energy falling in from one of the boundaries of a two-sided AdS-Schwarzschild spacetime. We demonstrate how known properties of complexity, such as the switchback effect for light shocks, as well as analogous properties for heavy ones, are imprinted in the complexity of formation and in the full time evolution of complexity. Following our discussion in \cite{Vad1}, we find that in order to obtain the expected properties of the complexity, the inclusion of a particular counterterm on the null boundaries of the Wheeler-DeWitt patch is required for the CA proposal.}

\maketitle


\section{Introduction}\label{sec:Intro}

Using perspectives and techniques from quantum information has brought many surprising insights into properties of the AdS/CFT correspondence \cite{AdSCFT,AdSCFTbook} in recent years. The best studied example has been the role of holographic entanglement entropy \cite{Ryu:2006bv,Ryu:2006ef,Hubeny:2007xt, HoloEntEntropy} in understanding the emergence of a semi-classical bulk spacetime \cite{VanRaamsdonk:2009ar, VanRaamsdonk:2010pw, Cool}. A new suggestion in this research program is that the quantum complexity of states in the boundary theory is encoded in a set of gravitational observables, which probe spacetime properties which are inaccessible from the perspective of holographic entanglement entropy \cite{EntNotEnough}.

In the holographic setting, circuit complexity provides a measure of how difficult it is to construct a particular target state from a (simple) reference state by applying a set of (simple) elementary gates, for a review see, \eg \cite{johnw,AaronsonRev}. There are two complementary proposals for the holographic description of the complexity of the boundary states: the complexity=volume (CV) \cite{Susskind:2014rva,Stanford:2014jda} and the complexity=action (CA) \cite{Brown1,Brown2} conjectures. The CV conjecture states that the complexity is dual to the volume of an extremal codimension-one bulk surface anchored at the time slice $\S$ in the boundary on which the state is defined,
\begin{equation}
\cv(\S) =\ \mathrel{\mathop {\rm
max}_{\scriptscriptstyle{\S=\partial \mathcal{B}}} {}\!\!}\left[\frac{\mathcal{V(B)}}{G_N \, L}\right] \, ,\labell{defineCV}
\end{equation}
with $\mathcal B$ corresponding to the bulk surface of interest, and $G_N$ and $L$ denoting Newton's constant and the AdS curvature scale, respectively, in the gravitational theory. On the other hand, the CA proposal states that the complexity is given by evaluating the gravitational action on a region of spacetime, known as the Wheeler-DeWitt (WDW) patch, which can be regarded as the causal development of a space-like bulk surface, \ie a Cauchy surface, anchored on the boundary time slice $\S$. The CA proposal then suggests
\bea
\ca(\S) =  \frac{I_\mt{WDW}}{\pi\, \hbar}\,. \labell{defineCA}
\eea

These conjectures have stimulated a wide variety of recent research efforts investigating the properties of both holographic complexity and circuit complexity in quantum field theories,  \eg  \cite{Roberts:2014isa, subreg1, Cai:2016xho,RobLuis, Format, Growth, subreg2, diverg,2LawComp, EuclideanComplexity1,EuclideanComplexity2,EuclideanComplexity3,Reynolds:2017lwq, koji, qft1, qft2,qft3, fish, brian, Alishahiha:2018tep, Chen:2018mcc, Zhao:2017isy, Fu:2018kcp,Bridges,Moosa,Agon:2018zso}.  In \cite{Vad1}, we investigated the CA and CV proposals for Vaidya spacetimes \cite{Vaid0,OriginalVaidya,VaidyaAdS} describing a thin shell of null fluid collapsing into the AdS vacuum to form a (one-sided) black hole. A surprising result we found was that the standard definition of the WDW action (\eg \cite{RobLuis,Format,Growth}) was inappropriate for these dynamical spacetimes, in that eq.~\reef{defineCA} did not reproduce the desired properties of complexity. However, the situation was rectified by adding an additional surface term \cite{RobLuis} on the null boundaries, which also ensured that the action was independent of the parametrization of the null generators for these boundaries.

The present paper is a direct continuation of \cite{Vad1} where we examine the holographic complexity for Vaidya geometries in which a thin null shell collapses into an eternal black hole --- see figure \ref{EternalShocktEvol}. Such shock wave geometries have already been extensively studied in the context of holographic complexity, \eg \cite{Stanford:2014jda,Bridges,Roberts:2014isa,Brown2}, however, these studies focused on the case where the energy in the shock was small. Using the formalism developed in \cite{Vad1}, we will not need to restrict our attention to this regime of light shocks here. Further, we will investigate the full time evolution of the holographic complexity, \ie including the transient regime, and this will allow us to identify several critical times that arise as the WDW patch (or the maximal volume surface) evolves forward in the background geometry. As well as the time evolution, we will investigate the complexity of formation in these shock wave geometries. As we found in \cite{Vad1}, we will argue that the inclusion of the null surface counterterm is crucial in these dynamical spacetimes in order for the CA proposal \reef{defineCA} to properly produce the expected properties of  complexity, such as the `switchback' effect \cite{Stanford:2014jda,multiple,Bridges}. In the CV calculations, we develop a geometric understanding of certain limits of the rate of change in complexity and relate them to having the extremal surfaces wrapping and/or unwrapping certain critical surfaces behind the past and/or future horizons of the black holes. We summarize the main results in some detail at the beginning of section \ref{sec:Discussion}.

The rest of the paper is organized as follows: In section \ref{bkgd}, we review the Vaidya background geometries in the context of two-sided black holes. We restrict our attention to thin shells of null fluid for which the action vanishes when the thickness shrinks to zero, as we showed in \cite{Vad1}. Next, we investigate the holographic complexity in these background geometries using the CA  proposal in \ref{sec:Shocks}. We evaluate the time evolution and complexity of formation in the presence of light and heavy shock waves, and also examine the consequences of not including the null surface counterterm. In section \ref{sec:VolShocks}, we evaluate the time evolution and complexity of formation using the CV proposal, for both light and heavy shock waves. We review our main results and discuss their physical implications in section \ref{sec:Discussion}, where we also present some future directions. We leave some technical details to the appendices: In appendix \ref{app:CounterTerm}, we evaluate the counterterm contributions to the WDW patch and review its implications for the UV structure of complexity. In appendix \ref{app:AppEternalAdS5}, we present some numerical results for the holographic complexity in higher dimensions using the CA conjecture. Finally, appendix \ref{app:CVShocksDetails} presents some of the relevant details for various intermediate results used in applying the CV conjecture in section \ref{sec:VolShocks}.

\section{Background Geometry}\label{bkgd}

Recall that the (unperturbed) eternal black hole geometry is dual to a thermofield double (TFD) state \cite{Maldacena:2001kr}, which is a pure state in which the degrees of freedom of two identical copies of the boundary CFT are entangled,
\beq
| TFD \rangle \equiv Z^{-1/2} \sum_{n = 0}^{\infty}  e^{-\frac12\,\beta E_n}\, | E_n \rangle_\mt{L} \, | E_n \rangle_\mt{R} \, ,
\label{TFDState}
\eeq
where the two copies are denoted as left (L) and right (R), in analogy to the left and right boundaries of the eternal geometry.  Tracing out either the left or right CFT leaves a thermal density matrix with inverse temperature $\beta$. While this density matrix is invariant under time translations, we can time evolve the two sets of degrees of freedom in the TFD state independently to produce
\beqa
| TFD (\tL, \tR) \rangle &=& \UL (\tL) \, \UR (\tR) \, | TFD \rangle  \, , \nonumber \\
&=& Z^{-1/2} \sum_{n = 0}^{\infty}  e^{-\frac12\,\beta E_n-iE_n(\tL+\tR)}\, | E_n \rangle_\mt{L} \, | E_n \rangle_\mt{R} \, ,
\labell{TFDState2}
\eeqa
where  $U_\mt{L,R}$ are the usual time evolution operators for the corresponding CFTs, \ie $\UL(\tL)=e^{- i H_\mt{L} \tL}$
and $\UR(\tR)=e^{- i H_\mt{R} \tR}$. One immediate observation is that the state is invariant when we shift
\begin{equation}\label{tshift1}
\tL \rightarrow \tL + \Delta t\,, \qquad
\tR \rightarrow \tR - \Delta t\,,
\end{equation}
 \ie the TFD state \reef{TFDState} is invariant if we time evolve with the combined Hamiltonian $H_\mt{L}-H_\mt{R}$.
Of course, this invariance is reflected in the `boost symmetry' of the dual black hole geometry. As a result, the holographic complexity remains unchanged by the above shifts \reef{tshift1}, \ie it only depends on the combination $\tL+\tR$  (\eg see \cite{Stanford:2014jda,Brown2,Growth}).

In the following, we study Vaidya geometries describing a thin shell of null fluid (or shock wave) injected into an eternal black hole background. Following \cite{ShenkerStanfordScrambling,multiple}, these Vaidya geometries describe\footnote{Our geometries are more properly interpreted in terms of a thermal quench, \eg \cite{quench1,quench2}, where some boundary coupling is rapidly varied at $\tR=-t_w$. Instead, eq.~\reef{TFDPertState} corresponds to an excited state in which the excitation becomes coherent at $\tR=-t_w$ (but with no variations of the couplings). The corresponding bulk geometry involves a null shell which emerges from the white hole singularity and reflects off of the right asymptotic boundary at $\tR=-t_w$ to become a collapsing shell, \eg see \cite{ShenkerStanfordScrambling,multiple,Stanford:2014jda}. Since our evaluations of the holographic complexity always involve $\tR>-t_w$, our results would be the same for either geometry. \label{footyXYZ}} a perturbation of the TFD state \reef{TFDState},
\beq
| TFD  \rangle_{pert} =  \cOR(-t_w)\, | TFD  \rangle = \UR (t_w)\, \cOR\, U^{\dag}_\mt{R} (t_w)\, | TFD  \rangle \, , \label{TFDPertState}
\eeq
where $\cOR(-t_w)$ is operator inserted in the right CFT at a time $-t_w$.\footnote{The details of the operator will not be important for our analysis, however, for the special case of $d=2$,  \cite{Anous:2016kss} provides a detailed description of the dual of the Vaidya-AdS$_3$ geometry.} In the second expression, we are describing this precusor as $\cOR(-t_w)=\UR (t_w)\, \cOR\, U^{\dag}_\mt{R} (t_w)$,  \ie $U^{\dag}_\mt{R} (t_w)$ evolves the right degrees of freedom backwards by a time $t_w$, $\cOR$ is inserted and then the right CFT is evolved forward by $t_w$. In the following, we will use the complexity of formation \cite{Format} (in the Vaidya geometry) to evaluate the complexity of the precursor, \ie to compare the complexities of $|TFD\rangle_{pert}$ and $|TFD\rangle$. The nontrivial cancellations in the complexity of the precursor are connected to the switchback effect \cite{Stanford:2014jda,multiple, Bridges}.

We will also examine the complexity of the time evolved state
\beqa
| TFD (\tL, \tR) \rangle_{pert} &=& \UL (\tL) \, \UR (\tR) \, | TFD \rangle_{pert}
\nonumber\\
&=&\UL (\tL) \, \UR (\tR+t_w)\, \cOR\, U^{\dag}_\mt{R} (t_w)\, | TFD  \rangle \labell{TFDPertState2}\\
&=& \UR (\tR+t_w)\, \cOR\,\UR (\tL-t_w) \,  | TFD  \rangle
\, ,  \nonumber
\eeqa
where in the last line, we use the boost symmetry of the TFD state, \ie  $\UL (\tL)\, | TFD  \rangle =\UR(\tR)\, | TFD  \rangle$,  and that $\UL$ commutes with all operators in the right CFT \cite{Stanford:2014jda}. In this case, inserting $\cOR$ at a fixed time $-t_w$ breaks the shift symmetry \reef{tshift1}. However, from the above expression, it is clear that if we combine the previous translations of the left and right times with a shift the insertion time,
\begin{equation}\label{tshift2}
t_w \rightarrow t_w + \Delta t\,,
\end{equation}
then the time-evolved state in eq.~\reef{TFDPertState2} is invariant. We will refer to the combination of eqs.~\eqref{tshift1} and \eqref{tshift2} together as the {\it time-shift symmetry}  of the problem. Of course, this will also produce a symmetry for the holographic complexity and as a result, we will find that the holographic complexity only depends on two combinations of the boundary times, $\tR + t_w$ and $\tL - t_w$, which appear in eq.~\reef{TFDPertState2}.

Let us now turn to the dual geometry in the bulk. As noted above, we consider the AdS-Vaidya spacetimes \cite{VaidyaAdS}, sourced by the collapse of a spherically symmetric shell of null fluid,
\begin{align}\label{MetricV}
&d s^2 = - F(r,v) d v^2 + 2 dr dv + r^2 d \Sigma^2_{k, d-1} \nonumber \, , \\
& \, \text{with} \qquad F(r, v) = \frac{r^2}{L^2} + k - \frac{f_p(v)}{r^{d-2}} \, ,
\end{align}
where $d$ is the dimension of the boundary spacetime and $k$ denotes the geometry of the horizon --- see \cite{Vad1} for further details.\footnote{The parameter $k$ can be $\lbrace +1,0,-1\rbrace$, corresponding to spherical, planar, and hyperbolic horizon geometries. We denote by $\Omega_{k,d-1}$ the dimensionless volume, see for instance \cite{Format,Growth}. Of course, for spherical geometries it becomes simply $\Omega_{1,d-1} = 2\pi^{d/2}/\Gamma(d/2)$, and for hyperbolic and planar black holes, it should be understood as a dimensionless infrared regulated quantity.   \label{footy22}} In particular, we consider the profile,
\beq
f_{\text{p}}(v)=\omega_1^{d-2} \left(1- {\cal H}(v-v_s)\right)
+\omega_2^{d-2} \, {\cal H}(v-v_s)\,,
\label{heavyX}
\eeq
where ${\cal H}(v)$ is the Heaviside step function.\footnote{It is a simple task to generalize the present discussion to BTZ black holes with $d=2$, and we will treat this case separately in the following.} With this profile, the Vaidya geometry  describes the collapse of an infinitely thin shell of null fluid, which raises the mass of the black hole from $M_{1}$ to $M_{2}$ where\footnote{Again, $\Omega_{k,d-1}$ denotes the (dimensionless) volume of the spatial geometry --- see footnote \ref{footy22}.}
\beq
M_i = \frac{(d-1) \, \Omega_{k,d-1}}{16 \pi \, G_N}\,\omega_i^{d-2} \,.
\label{energy}
\eeq

\begin{figure}
\centering
\includegraphics[scale=0.4]{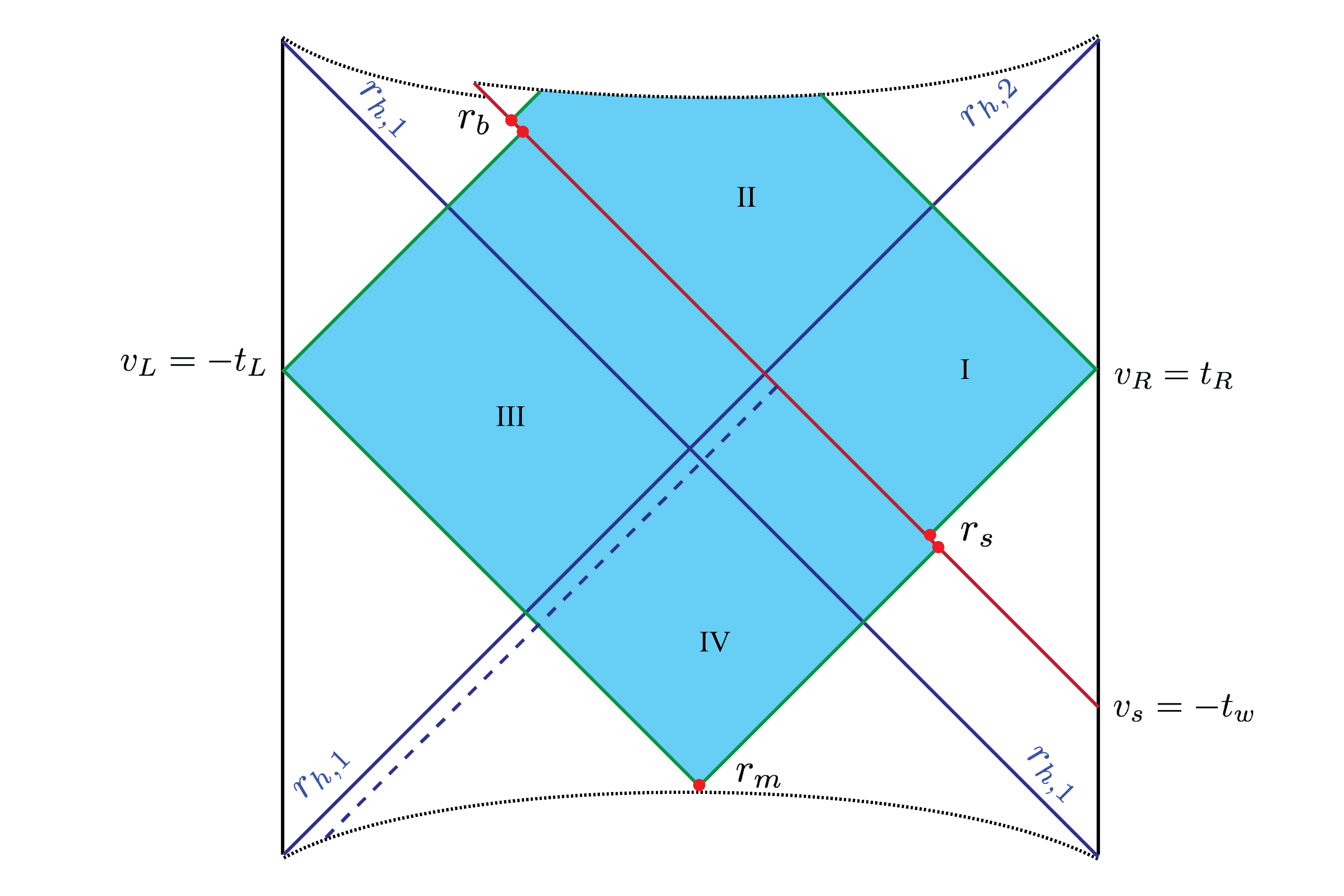}
\caption{Penrose-like diagram for one shock wave on an eternal black hole geometry. At $v_s = - t_w$ a thin shock is injected at the right boundary which raises the mass of the black hole from $M_1$ to $M_2$. We identify three points in the geometry that depend on time, $r_m$, where the boundaries of the WDW patch cross behind the past horizon, $r_s$, where the boundary of the WDW patch crosses the collapsing shell in the right exterior, and $r_b$ where the boundary of the WDW patch crosses the shock wave inside the future black hole.}
\label{EternalShocktEvol}
\end{figure}
Further, in eq.~\reef{heavyX}, we implicitly consider the shock wave as coming from the right boundary at some early time $v_s = - t_w$ (with $t_w>0$), in accord with our description of the boundary state \reef{TFDPertState}.
Hence we have
\beqa
v_\mt{R}< - t_{w} \ :&& \qquad  F(r, v) = f_{1} (r) =  \frac{r^2}{L^2} + k - \frac{\omega_1^{d-2}}{r^{d-2}}\,,\labell{fBH1}\\
v_\mt{R}> - t_{w}\ :&& \qquad  F(r, v) = f_{2} (r) =  \frac{r^2}{L^2} + k - \frac{\omega_2^{d-2}}{r^{d-2}} \, .\labell{fBH2}
\eeqa
where the coordinates $r$ and $v_\mt{R}$ cover the right exterior region and the future black hole interior, labeled I and II in figure \ref{EternalShocktEvol}. There is a corresponding set of coordinates $r$ and $v_\mt{L}$ covering the left exterior region and the past white hole interior, labeled III and IV in the figure. The shock wave does not enter either of the latter regions and so we have
\beq
{\rm for\ all}\ v_\mt{L} \ : \qquad  F(r, v) = f_{1} (r) =  \frac{r^2}{L^2} + k - \frac{\omega_1^{d-2}}{r^{d-2}}\,.\label{fBH1a}
\eeq
On either side of the shock wave, the geometry \reef{MetricV} corresponds to that of a (static) AdS black hole, whose horizon radius $r_{h,i}$ determined by\footnote{Let us note here that for $k=-1$, we will only consider `large' hyperbolic black holes with $r_h>L$, which have the casual structure illustrated in figure \ref{EternalShocktEvol}.}
\begin{equation}\label{horz}
\omega_i^{d-2} = r_{h,i}^{d-2} \left( \frac{r_{h,i}^2}{L^2}+ k \right) \, .
\end{equation}
As noted above, in each of these regions, the mass is given by eq.~\reef{energy} and further, the temperature and entropy become
\beq
T_i
=\frac{1}{4\pi }\left.\frac{\partial f}{\partial r}\right|_{r=r_{h,i}}=\frac{1}{4\pi \,r_{h,i}}\left(d\,\frac{r_{h,i}^2}{L^2} + (d-2)\,k \right) \, ,\quad\qquad  S_i = \frac{\Omega_{k,d-1}}{4G_N}\,r_{h,i}^{d-1}    \, .
\label{effect}
\eeq

We define the tortoise coordinates with respect to eqs.~(\ref{fBH1})--(\ref{fBH1a}) as
\beqa
{\rm for\ all}\ \vL \ \ \&\ \ {\rm for}\ \vR< - t_{w} \ :&& \quad  r^{*}_{1}(r) = - \int_{r}^{\infty}\frac{d r}{ f_{1} (r)}  \,,\labell{rStar1} \\
\vR> - t_{w}\ :&& \quad  r^{*}_{2}(r) = - \int_{r}^{\infty}\frac{d r}{ f_{2} (r)} \, , \labell{rStar2}
\eeqa
where again we chose the range of integration such that both expressions satisfy $\lim_{r \to \infty}~r^{*}_{1,2}(r)\rightarrow 0$. Using the tortoise coordinates, we can define an outgoing null coordinate $u$ and an auxiliary time coordinate $t$ as
\beq
u_{1,2} \equiv v - 2 r_{1,2}^*(r) \, ,\qquad\quad
t_{1,2} \equiv v - r_{1,2}^*(r)  \, .
\label{time3}
\eeq
Again these coordinates are discontinuous across the shell because $f(r)$ jumps from eq.~\eqref{fBH1} to eq.~ \reef{fBH2} at $\vR=-t_w$. In analogy to the diagrams in \cite{Vad1}, we represent the shock wave geometries with Penrose-like diagrams (\eg see figure \ref{EternalShocktEvol}), which can be smoothly ruled with lines of constant $u$ and $v$. As before, since the coordinate $u$ is discontinuous, this introduces a(n unphysical) jump as the outgoing null rays cross the shock wave. Of course, the spacetime is continuous along this surface and the outgoing null rays are smooth, as can be seen by regulating the thin shell to have a finite thickness --- see section $2$ in \cite{Vad1}.

Before we proceed further,  let us comment on synchronizing the times between the left and right boundaries. In principle, the left boundary time $\tL$ is completely independent of the right boundary time $\tR$ in the eternal black hole geometry. However, implicitly, they are synchronized by considering a geometric construction, \eg where an extremal codimension-one surface that runs from one boundary to the other through the bifurcation surface. This surface  connects  the time slice $\tL$ on the left boundary to the time slice $\tR=-\tL$ on the right boundary. The minus sign arises here because of our convention  that the boundary times increase upwards on both boundaries in figure \ref{EternalShocktEvol}. Of course, such extremal surfaces have a more interesting profile in the shock wave geometry, as we will discuss extensively in section \ref{sec:VolShocks}. Hence one might worry that the two boundary times cannot be synchronized in a natural way. However, we observe that the above geometric construction is unaffected for times $\tL>t_w$, for which the extremal surface will reach the right boundary at times $\tR<-t_w$. That is, for late (early) times on the left (right) boundary, the desired extremal surfaces do not meet the shock wave and remain entirely within the portion of the spacetime where $F(v,r)=f_1(r)$. Once the boundary times are synchronized in these regions, this synchronization is straightforwardly extended to the entire boundaries. Implicitly, this is how we match the left and right boundary times in the following.

It will be useful to define some dimensionless quantities in order to express the evolution of complexity, as well as the complexity of formation, in the following. We define,\footnote{Note that here, $z$ and $x$ are defined as the ratio of the AdS scale $L$ or the radius $r$ with the final horizon size $r_{h, 2}$. Using the ratio $w$ one can easily construct quantities normalized by $r_{h, 1}$ instead. }
\begin{equation}\label{eq:wz}
w \equiv \frac{r_{h,2}}{r_{h,1}} \, , \qquad z \equiv \frac{L}{r_{h,2}}\,,\qquad x \equiv \frac{r}{r_{h, 2}} \, ,
\end{equation}
which for positive-energy shock waves, yields $w>1$. Also note that for planar (and BTZ) black holes, $w$ is proportional to the ratio of temperatures, \ie $w=T_2/T_1$, for $k=0$ (or $d=2$) which can be seen by using eq.~\reef{effect}.
The ratio between the masses and the entropies reads
\begin{equation}
\frac{M_2}{M_1} = w^{d} \frac{(1 + k z^2)}{(1 + k z^2 w^2)}
\qquad{\rm and}\qquad \frac{S_2}{S_1} = w^{d-1}\, .
\label{rocker}
\end{equation}
It is also useful to rescale the blackening factor such that
\begin{equation}\label{Dimlessf1f2}
f_{2} (r, L) = \frac{1}{z^2} \tilde f (x, z) \, , \qquad \, \qquad \, f_{1} (r, L) = \frac{1}{w^2 z^2} \tilde f (w x, w z)  \, ,
\end{equation}
where
\begin{equation}
\tilde f (x, z)  \equiv x^2 + k z^2 - \frac{1}{x^{d-2}} \left( 1+ k z^2 \right) \, .
\end{equation}
Note that $\tilde f (x, z)$ is not a function of $z$ for planar holes (\ie $k=0$). Finally, if a physical quantity depends on $z$, we can use eq.~\eqref{effect} to express $z$ as a function of $L T_2$,
\begin{equation}\label{eq:zfuncT}
z = \frac{d}{\sqrt{4 \pi ^2 (L T_2)^2-(d-2) \, d \, k} + 2 \pi L T_2} \, .
\end{equation}

\section{Complexity = Action}\label{sec:Shocks}

In order to evaluate the holographic complexity using the  complexity=action proposal \reef{defineCA}, we begin by writing the total action as
\begin{equation}\label{sumAct}
I = I_\mt{grav} +I_\mt{ct}+ I_\mt{fluid} \, ,
\end{equation}
where $I_\mt{fluid}$ is the null fluid action, as constructed in section 2 of \cite{Vad1}. There we showed that $I_\mt{fluid}$ vanishes on shell, and so the imprint of the shock wave in the CA calculations comes only from its backreaction on the metric. Further,  we showed in \cite{Vad1} that as the width of the null shell shrinks to zero, the action of the spacetime region occupied by the shell itself vanishes. Hence for an infinitely thin shell, as introduced in eq.~\reef{heavyX}, the action can be evaluated by separately evaluating the action on the portion of the WDW patch above the shell and that below the shell.

The gravitational action can be written as \cite{RobLuis}
\begin{equation}\label{THEEACTION}
\begin{split}
I_\mt{grav} = & \frac{1}{16 \pi G_N} \int_\mathcal{M} d^{d+1} x \sqrt{-g} \left(\mathcal R + \frac{d(d-1)}{L^2}\right) \\
&\quad+ \frac{1}{8\pi G_N} \int_{\mathcal{B}} d^d x \sqrt{|h|} K + \frac{1}{8\pi G_N} \int_\Sigma d^{d-1}x \sqrt{\sigma} \eta
\\
&\quad  + \frac{1}{8\pi G_N} \int_{\mathcal{B}'}
d\lambda\, d^{d-1} \theta \sqrt{\gamma} \kappa
+\frac{1}{8\pi G_N} \int_{\Sigma'} d^{d-1} x \sqrt{\sigma} a \, .
\end{split}
\end{equation}
The bulk integral in the manifold $\mathcal{M}$ is proportional to the Einstein-Hilbert action with the Ricci scalar $\mathcal{R}$ and the negative cosmological constant $\Lambda=-d(d-1)/(2L^2)$. In addition, the integral along the boundary $\mathcal{B}$ is the usual Gibbons-Hawking-York (GHY) boundary term for smooth spacelike and timelike boundaries, proportional to the trace of the extrinsic curvature $K$. When the boundary includes null hypersurfaces, there is a similar surface term which integrates $\kappa$ along these portions of the boundary $\mathcal{B}^{'}$. The latter  indicates how the null coordinate $\lambda$ along the boundary deviates from affine parametrization. Finally, there are the codimension-two joint terms: if the intersection involves timelike or spacelike boundaries, the Hayward joint terms \cite{Hay1,Hay2} at $\Sigma$ are defined by the ``boost angle" $\eta$ between the normal vectors of these hypersurfaces. If either of the boundaries forming the joint is null, there is an equivalent contribution at $\Sigma^{'}$ given in terms of an analogous ``angle'' $a$, which depends on the null normals involved, as constructed in \cite{RobLuis}.

In eq.~\eqref{sumAct}, we have also included an additional surface term for the null boundaries \cite{RobLuis},
\begin{equation}\label{counter}
I_\mt{ct} = \frac{1}{8 \pi G_N} \int_{\mathcal{B}'} d \lambda \, d^{d-1} \theta \sqrt{\gamma}\, \Theta \,\log \left(\ctL \Theta \right) \, ,
\end{equation}
where $\Theta = \partial_{\lambda} \log \sqrt{\gamma}$ is the expansion scalar, and $\ctL$ is an arbitrary constant length scale.
This counterterm is not needed to produce a well defined variational principle for the gravitational action, \ie $I_\mt{ct}$ only depends on intrinsic boundary data. Instead it was introduced in \cite{RobLuis} to ensure that the action is reparametrization invariant. In \cite{Vad1}, we found that including the counterterm was essential if the WDW action was to reproduce certain properties expected of the complexity in dynamical spacetimes and hence it is included in (most of) our CA calculations here.

However, we will also expand our arguments indicating that the counterterm \reef{counter} is an essential part of the gravitational action by considering our results after we remove the counterterm contributions. In this case, we must come to grips with the various ambiguities arising from  the surface and joint terms associated with the null boundaries, \eg see discussion in \cite{RobLuis}. We follow the prescription in \cite{RobLuis} where we set $\kappa=0$ by choosing an affine parametrization for the null normals. Further, we fix the overall normalization of these null vectors by their inner product with the asymptotic timelike Killing vector at the boundary, $\hat t = \partial_t$, \ie we set $\hat t  \cdot k = \pm \alpha$.

In section \ref{CAshock}, we evaluate the time evolution (including the null counterterm) for the Vaidya spacetimes described in section \ref{bkgd}, and then we examine their complexity of formation in section \ref{CoF1}. In section \ref{CANoCT}, we (mostly) focus on the BTZ black hole (\ie $d=2$) and demonstrate that the CA calculations without the inclusion of the counterterm fail to produce the expected behaviour of holographic complexity. In appendix \ref{app:CounterTerm}, we discuss some further details on the influence of the null counterterm in Vaidya spacetimes. In appendix \ref{app:AppEternalAdS5} we discuss the complexity evolution for higher ($d>2$) dimensional black holes, focusing mostly on $d=4$.

\subsection{Time Evolution} \label{CAshock}

Consider the shock wave spacetime represented in figure \ref{EternalShocktEvol}, with the Penrose-like diagram describing the geometry in eqs.~(\ref{fBH1}--\ref{fBH1a}). The null shell is injected at the right boundary at $v_s = - t_w$ (with $t_w>0$), raising the mass of the black hole from $M_1$ to $M_2$. We can study the time evolution of the holographic complexity in many different ways. However, for simplicity, we will focus on a symmetric time evolution with $\tL=t/2=\tR$ starting at $\tL= \tR =0$, in analogy to the analysis in \cite{Growth}.

We identify the three positions which are important in defining the WDW patch, depending on the time: $r_{b}$ is where the boundary of the WDW patch originating from the left boundary meets the shock wave inside the future black hole; $r_{s}$ is the surface where the WDW boundary in the right exterior meets the shock wave; and $r_{m}$ is where the past null boundary segments of the WDW patch meet inside the white hole region --- see figure \ref{EternalShocktEvol}. Of course, depending on the parameters of the problem, $r_{b}$ and $r_{m}$ could be behind the singularities. In particular, if $r_m<0$, the WDW patch has a spacelike boundary segment running along the past singularity. In this section, we carefully evaluate all these possibilities in the shock wave black hole geometry, and show how the critical times where $r_b$ and $r_m$ cross $r=0$ produce transitions between different behaviours of the holographic complexity.

We calculate the bulk action given by eq.~\eqref{THEEACTION} by using the same prescription discussed in \cite{Vad1}, \ie implicitly we evaluate the total gravitational action as the sum of the action evaluated on the regions comprising the WDW patch to the future and the past of the shock wave.  We start here by identifying the three positions introduced above (\ie $r_b$, $r_s$ and $r_m$) as functions of the times\footnote{The minus sign arises for $\tL$ because our convention is that the left boundary time increases as we run upwards in figure \ref{EternalShocktEvol}, while $\vL$ increases as we move down diagonally towards the bottom left corner.} $\vL  = - \tL$ and $\vR = \tR$ at which the WDW patch is anchored on the left and right boundaries,
\begin{align}
&\tL - t_w =  2 r^{*}_{1} (r_{b}) \, , \nonumber \\
&\tR + t_w = - 2 r^{*}_{2} (r_{s}) \, ,  \label{eq:rwrsrm} \\
&\tL - t_w = 2 r_{1}^{*} (r_s) - 2 r_{1}^{*}(r_m)  \nonumber \, .
\end{align}
Recall that $t_w$ is defined to be positive. Given eqs.~\reef{rStar1} and \reef{rStar2}, the time evolution of these position is relatively simple. For example, if $\tL$ is held fixed, eq.~\eqref{eq:rwrsrm} implies that
\begin{equation}\label{eq:DertR}
\frac{d r_b}{d \tR} = 0 \, , \, \qquad \, \qquad \, \frac{d r_s}{d \tR} = - \frac{f_{2}(r_s)}{2} \, , \, \qquad \, \qquad \,  \frac{d r_m}{d \tR} = - \frac{f_{1}(r_m)}{2} \frac{f_{2}(r_s)}{f_{1}(r_s)} \, .
\end{equation}
Similarly, when $\tR$ is held constant, the evolution with the left boundary time becomes
\begin{equation}\label{eq:DertL}
\frac{d r_b}{d \tL} = \frac{f_{1}(r_b)}{2} \, , \, \qquad \, \qquad \, \frac{d r_s}{d \tL} = 0 \, , \, \qquad \, \qquad \,  \frac{d r_m}{d \tL} = - \frac{f_{1}(r_m)}{2}  \, .
\end{equation}
Recall that we will be interested in $\tL=t/2=\tR$ in the following, and so when required we can combine the above results in the appropriate linear combination.

\subsubsection*{Bulk contribution}

We start by evaluating the bulk action for the WDW patch represented in figure \ref{EternalShocktEvol}. As before, the Einstein-Hilbert contribution to the action is $\mathcal{R} - 2 \Lambda = - \frac{2 d}{L^2}$ in $d+1$ bulk dimensions. The total bulk action reads
\begin{align}
& I_{\text{bulk}} = \left( \frac{\Omega_{k, d-1}}{16 \pi \, G_N} \right) \left( - \frac{2 d}{L^2}\right) \bigg[ \int_{r_s}^{r_{max}} \, d r \, r^{d-1} \, \left( - 2 r^{*}_{2}(r)  \right)  + \int_{r_b}^{r_s} \, d r \, r^{d-1}  \, \left( \tR + t_w \right) \nonumber \\
&  \int_{r_{h, 1}}^{r_s} \, d r \, r^{d-1}  \, \left(2 r^{*}_{1}(r_s)  - 2 r^{*}_{1}(r) \right) +   \int_{r_{m}}^{r_{h, 1}} \, d r \, r^{d-1}  \, \left(- \tL + t_w - 2 r^{*}_{1}(r )  + 2 r^{*}_{1}(r_s) \right) + \nonumber  \\
& + \int_{r_{h, 1}}^{r_{max}} \, d r \, r^{d-1}  \, \left( - 2 r^{*}_{1}(r) \right) + \int_{r_b}^{r_{h, 1}} \, d r \, r^{d-1}  \, \left( -t_w  + \tL  - 2 r^{*}_{1}(r) \right) + \nonumber \\
&+  \int_{0}^{r_{b}} \, d r \, r^{d-1}  \, \left( \tR + t_w   - 2 r^{*}_{2}(r) +  2 r^{*}_{2}(r_b) \right)   \bigg] \, . \label{BulkEternalSTotal}
\end{align}
Now, we fix the left boundary time $\tL$ and vary $\tR$ in the right boundary, as in eq.~\eqref{eq:DertR}. The time derivative of the bulk action with respect to the right boundary reads
\begin{equation}\label{eq:BulkActEtRight}
\frac{d I_{\text{bulk}}}{d \tR} = -  \frac{\Omega_{k, d-1}}{8 \pi \, G_N \, L^2}     \left[ r_s^d \left(1 - \frac{f_2 (r_s)}{f_1 (r_s)} \right) + r_m^d \, \frac{f_2 (r_s)}{f_1 (r_s)}  \right] \, .
\end{equation}
In addition, we can write the time derivative with respect to the left boundary evolution $\tL$ in eq.~\eqref{eq:DertL},
\begin{equation}\label{eq:BulkActEtLeft}
\frac{d I_{\text{bulk}}}{d \tL} = -  \frac{\Omega_{k, d-1}}{8 \pi \, G_N \, L^2}     \left[ r_m^d   - r_b^d \left(1 - \frac{f_1 (r_b)}{f_2 (r_b)} \right)   \right] \, .
\end{equation}
With respect to a symmetric time evolution $\tL = \tR = t/2$, we sum the linear combination of eqs.~\eqref{eq:BulkActEtRight} and \eqref{eq:BulkActEtLeft},
\begin{equation}\label{eq:BulkActEtSym}
\frac{d I_{\text{bulk}}}{d t } = -  \frac{\Omega_{k, d-1}}{16 \pi \, G_N \, L^2}     \left[  r_m^d \left( 1 + \frac{f_2 (r_s)}{f_1 (r_s)} \right)   - r_b^d \left(1 - \frac{f_1 (r_b)}{f_2 (r_b)} \right) + r_s^d \left(1 - \frac{f_2 (r_s)}{f_1 (r_s)} \right)   \right] \, .
\end{equation}

\subsubsection*{Boundary surface contributions}

We now turn our attention to the boundary surface contributions in the action in eq.~\eqref{THEEACTION}. As suggested in \cite{RobLuis}, we choose the normals to the null boundaries to be  affinely parametrized (before and after the shock wave --- see discussion in section 2 of \cite{Vad1}). Therefore the parameter $\kappa$ and the corresponding boundary term vanishes for all of these null boundary segments. There are, however, two possible boundary contributions to the action, namely, evaluating the Gibbons-Hawking-York (GHY) term on a spacelike (regulator) surface right before the future singularity, and also in the regime that $r_m$ is behind the singularity, a similar contribution arises from the past singularity.\footnote{We do not consider possible surface terms at the UV regulator surfaces because these contributions will be independent of time, \eg see \cite{diverg,Growth}.}

We will denote the critical times at which $r_m$ leaves the past singularity (\ie $r_m$ becomes positive) as $\tLc1$ and $\tRc1$. From eq.~\eqref{eq:rwrsrm}, we have
\begin{equation}\label{tRc1}
\tRc1  = -t_w - 2 r^{*}_{2} (r_s) \, , \, \qquad \, \qquad \, \tLc1=t_w + 2 r^{*}_{1} (r_s)- 2 r^{*}_{1} (0)  \,.
\end{equation}
We can apply this result as follows: If we choose a value for $\tRc1$, then the first equation determines a particular value of $r_s$ and the second equation determines the value $\tLc1$ at which the WDW patch lifts off the past singularity. Similarly, if instead we choose a value for $\tLc1$, we can apply the equations in the opposite order to determine the value of $\tRc1$ at which the WDW patch lifts off the past singularity.
There is also a second critical time for the left boundary, which we will denote $\tLc2$ and it is the time at which the crossing point $r_b$ touches the singularity, \ie
\begin{equation}\label{tLc2}
\tLc2   = t_w+ 2 r^{*}_{1} (0)  \, .
 \end{equation}
Of course, whether there are critical times in the range of time evolution that we are studying depends on how early and how energetic the shock wave was. However, once the latter parameters are chosen, one can determine with eqs.~\eqref{tRc1}  and \eqref{tLc2} whether there are critical times and find their respective values.

Now, we first investigate the GHY term at the future singularity. As usual \eg \cite{Brown2,Format}, we introduce a regulator surface at $r=\veps$ and after evaluating the GHY term on this surface, we take the limit $\veps\to0$. Since $\tL > 0$ and $v_s = - t_w < 0$, there are two possibilities: $\tL < \tLc2$ for which the crossing point $r_b$ arises in the black hole interior region; and $\tL> \tLc2$ for which the future null boundary of WDW patch from the left boundary reaches the singularity without crossing the shock wave.
For $\tL < \tLc2 $, we have
\begin{equation}\label{GHFut_Tot}
I^{(f)}_{GHY} = - \frac{\Omega_{k,d-1}}{8 \pi G_N} \left( \frac{- d \omega_{2}^{d-2}}{2} \right) \left( \tR + t_w + 2 r_{2}^{*}(r_b) -  2 r_{2}^{*}(0)  \right) \, ,
\end{equation}
which leads to
\begin{equation}\label{GHFut}
\frac{d I^{(f)}_{GHY}}{d \tR} = \frac{\Omega_{k,d-1}}{8 \pi G_N}  \,\frac{d \omega_{2}^{d-2}}{2}
\qquad {\rm and} \qquad
\frac{d I^{(f)}_{GHY}}{d \tL} = \frac{\Omega_{k,d-1}}{8 \pi G_N} \, \frac{d \omega_{2}^{d-2}}{2}
 \, \frac{f_1 (r_b)}{f_2 (r_b)} \, .
\end{equation}
In contrast, for $\tL > \tLc2 $, we have
\begin{equation}\label{GHFut_Tot_Norway}
I^{(f)}_{GHY} = - \frac{\Omega_{k,d-1}}{8 \pi G_N} \left( \frac{- d \omega_{2}^{d-2}}{2} \right) \left( \tR + t_w \right)  - \frac{\Omega_{k,d-1}}{8 \pi G_N} \left( \frac{- d \omega_{1}^{d-2}}{2} \right) \left(- t_w + \tL - 2 r^{*}_{1}(0) \right) \, ,
\end{equation}
which results in
\begin{equation}\label{GHFut_Norway}
\frac{d I^{(f)}_{GHY}}{d \tR} =  \frac{\Omega_{k,d-1}}{8 \pi G_N}\,   \frac{d \omega_{2}^{d-2}}{2}
\qquad {\rm and}
\qquad
\frac{d I^{(f)}_{GHY}}{d \tL} =  \frac{\Omega_{k,d-1}}{8 \pi G_N}  \, \frac{d \omega_{1}^{d-2}}{2} \, .
\end{equation}

Now, the GHY contribution from the past singularity follows a similar analysis. Whenever $\tR < \tRc1 $ or $\tL <\tLc1 $,
the WDW patch intersects the past singularity  and one finds the following GHY contribution
\begin{equation}\label{GHPast_Tot}
I^{(p)}_{GHY} =  \frac{\Omega_{k,d-1}}{8 \pi G_N} \left( \frac{ d\, \omega_{1}^{d-2}}{2} \right) \left( - \tL + t_w + 2 r_1^{*} (r_s) - 2 r_{1}^{*}(0) \right)   \, .
\end{equation}
The time derivatives of this result then become
\begin{equation}\label{GHPast}
\frac{d I^{(p)}_{GHY}}{d \tR} = - \frac{\Omega_{k,d-1}}{8 \pi G_N} \, \frac{d \omega_{1}^{d-2}}{2} \, \frac{f_{2}(r_s) }{f_{1}(r_s)}
\qquad{\rm and}\qquad
\frac{d I^{(p)}_{GHY}}{d \tL} = - \frac{\Omega_{k,d-1}}{8 \pi G_N} \, \frac{d \omega_{1}^{d-2}}{2} \, .
\end{equation}

\subsubsection*{Joint contributions}

We now focus on the joint contributions to the action \eqref{THEEACTION} evaluated on the WDW patch. In principle, such contributions arise where the null boundaries intersect the UV regulator surfaces near the asymptotic boundary. However, these contributions are time independent and so we ignore them in the following. Similarly, there are joint contributions where the null boundaries intersect the regulator surfaces $r=\veps$ near the singularities but these vanish in the limit $\veps\to0$. This leaves  three possible different contributions coming from joints at $r=r_b$, $r_s$ and $r_m$, as shown in figure \ref{EternalShocktEvol}.\footnote{In the case where $r_b<0$, there would be additional joints where the shock wave hits the future singularity, \ie on the regulator surface $r=\eps$ at $v=-t_w$. However, these again yield a vanishing contribution in the limit $\veps\to0$.}  The joint contributions at $r_s$ and $r_b$ are analogous to the ones discussed in the one sided geometry in \cite{Vad1}, while the contribution from $r_m$ is similar to the joint action found in unperturbed eternal black holes \cite{Growth}.

We start by evaluating the sum of joint contributions where the past null boundary of the WDW patch crosses the shock wave, \ie at $r=r_s$. The relevant null normals on the past boundary are
\beq
k^{p}_{\mu} d x^{\mu}  =\left\lbrace \begin{matrix}
&\alpha \left( - d v + \frac{2}{f_{2}(r)} d r \right)&&
{\rm for}&& r>r_s\,,\\
&\tilde \alpha \left( - d v + \frac{2}{f_{1}(r)} d r \right)&&
{\rm for}&& r<r_s\,.
\end{matrix}\right.
\label{pastrs}
\eeq
Further we introduce the two normals along the collapsing shock wave,
\beqa
v> - t_w \ :&&\qquad k^\mt{s+}_{\mu} d x^{\mu}  =  - \beta d v \, ,
\nonumber\\
v< - t_w \ :&&\qquad k^\mt{s--}_{\mu} d x^{\mu}  =  \ \  \beta d v \, .
\labell{shellnAg}
\eeqa
The sum of the two joint contributions then reads
\begin{equation}\label{Joint_rs}
I^{(\RN{1})}_{\text{joint}} = \frac{\Omega_{k, d-1} \, r_s^{d-1}}{8 \pi G_N} \, \log \left( \frac{\alpha f_{1}(r_s)}{\tilde \alpha f_2(r_s)} \right) \, .
\end{equation}

We note that in eq.~\reef{pastrs}, the normalization constant $\alpha$ was fixed with the usual asymptotic condition $k^p\cdot \hat t=-\alpha$ \cite{RobLuis}. However, to fix the normalization constant $\tilde\alpha$  below the shell, we demand that the null boundary is affinely parametrized across the shock wave, \ie $\kappa=0$, following \cite{Vad1}. The latter constraint imposes
\begin{equation}\label{AffinePast}
\frac{\tilde \alpha}{\alpha} = \frac{f_1 (r_s)}{f_2 (r_s)} \, .
\end{equation}
As a consequence, the corner contributions at $r_s$ vanish, as was discussed for the one-sided collapse in \cite{Vad1}, \ie
\begin{equation}\label{JointrsVan}
I^{(\RN{1})}_{\text{joint}} = 0 \, .
\end{equation}

Next, we evaluate the sum of joint terms where the future null boundary of the WDW patch crosses the shock wave, \ie at $r=r_b$.  The (outward-directed) null normal to this future boundary is
\beq
k^{f}_{\mu} d x^{\mu}  =\left\lbrace \begin{matrix}
&\alpha \left( - d v + \frac{2}{f_{1}(r)} d r \right)&&
{\rm for}&& r>r_b\,,\\
&\hat \alpha \left( - d v + \frac{2}{f_{2}(r)} d r \right)&&
{\rm for}&& r<r_b\,.
\end{matrix}\right.
\label{futurerb}
\eeq
Using the null normals along the shock wave in eq.~\eqref{shellnAg}, the total contribution to the action is
\begin{equation}\label{Joint_rb}
I^{(\RN{2})}_{\text{joint}} = \frac{\Omega_{k, d-1} \, r_b^{d-1}}{8 \pi G_N} \, \log \left( \frac{\alpha f_{2}(r_b)}{\hat \alpha f_1(r_b)} \right) \, .
\end{equation}
Once again, the condition of affine parametrization across the shock wave fixes the ratio between the normalization constants in eq.~\reef{futurerb} with
\begin{equation}\label{AffineFut}
\frac{\hat \alpha}{\alpha} = \frac{f_2 (r_b)}{f_1 (r_b)} \, .
\end{equation}
Therefore, the joint contributions at $r_b$ also vanishes ,
\begin{equation}\label{JointrsVan}
I^{(\RN{2})}_{\text{joint}} = 0 \, .
\end{equation}

Finally, we turn to the possible contribution from the joint where the two past null boundaries of the WDW patch meet inside the white hole region, \ie at $r_m$. This joint contribution is evaluated with $k^{p}$ in eq.~\reef{pastrs} on the right boundary (with $r<r_s$), and
\beq
k^{L}_{\mu} d x^{\mu} = \alpha\, d v
\label{hoppy}
\eeq
for the normal to the left null boundary.\footnote{Without loss of generality, we are assuming here that the null normals are normalized at both the left and right boundaries with the same constant. In fact, when we add the counterterm \reef{counter} which ensures reparametrization invariance to the null boundaries, the total action will be independent of $\alpha$.} The resulting joint contribution then reads, with the affine parametrization condition \eqref{AffinePast},
\begin{equation}\label{Joint_rm}
I^{(\RN{3})}_{\text{joint}} = -  \frac{\Omega_{k, d-1} \, r_m^{d-1}}{8 \pi G_N} \, \log \left( \frac{ | f_{1}(r_m) |}{\alpha \, \tilde \alpha} \right) = -  \frac{\Omega_{k, d-1} \, r_m^{d-1}}{8 \pi G_N} \, \log \left( \frac{ | f_{1}(r_m) | f_2(r_s) }{\alpha^2 \, f_1(r_s) } \right)  \, .
\end{equation}

The time derivatives of this joint contribution then become
\begin{align}
&\frac{d I^{(\RN{3})}_{\text{joint}}}{d \tR} =  \frac{(d-1) \Omega_{k, d-1}}{16 \pi G_N} \, r_m^{d-2} f_1(r_m) \frac{f_2(r_s)}{f_1(r_s)} \log \left[   \frac{|f_1(r_m) | f_2 (r_s)}{\alpha^2 f_1 (r_s)} \right]  \nonumber    \\
&\qquad\qquad + \frac{\Omega_{k, d-1}}{16 \pi G_N}  \frac{f_2(r_s)}{f_1(r_s)} \left[ \frac{2 r_m^{d}}{L^2}  + (d-2) \omega_1^{d-2} \right] + \frac{\Omega_{k, d-1} \, r_m^{d-1}}{16 \pi G_N}  \left[ f_2^{'}(r_s) -  f_1^{'}(r_s) \frac{f_2 (r_s)}{f_1 (r_s)} \right] \, , \nonumber \\
&\frac{d I^{(\RN{3})}_{\text{joint}}}{d \tL} =  \frac{(d-1) \Omega_{k, d-1}}{16 \pi G_N} \, r_m^{d-2} f_1(r_m) \log \left[   \frac{|f_1(r_m)| f_2(r_s)}{\alpha^2 f_1(r_s)} \right]  \nonumber    \\
& \qquad\qquad+ \frac{\Omega_{k, d-1}}{16 \pi G_N}  \left[ \frac{2 r_m^{d}}{L^2}  + (d-2) \omega_1^{d-2} \right]   \, . \label{ICorner3}
\end{align}

\subsubsection*{Counterterm contributions}

Next we examine the contributions of the  counterterm \reef{counter} to the time derivative of the holographic complexity. The counterterm is evaluated on each of the four null boundaries of the WDW patch in appendix \ref{app:CounterTerm}, but only three of these contribute to the growth rate. First, for the right past boundary, we have
\beqa
 I^{(\RN{1})}_{\mt{ct}} &=& ``UV\ terms"
 - \frac{\Omega_{k, d-1}}{8 \pi G_N} \, r_m^{d-1} \left[ \log\!{\left( \frac{(d-1) \alpha\ctL}{r_m} \right)} +\frac1{d-1}\right]\labell{count1}\\
 &&\qquad\qquad
+ \frac{\Omega_{k, d-1} }{8 \pi G_N}\,\left(  r_s^{d-1}-  r_m^{d-1} \right) \, \log\!\left( \frac{f_1(r_s)}{f_2(r_s)} \right)\,.
\nonumber
\eeqa
The above expression corresponds to eq.~\reef{count02} after we have substituted the affine parametrization condition \reef{AffinePast}. We have also left implicit the terms coming from the UV regulator surface at $r=r_\mx$, since they are time independent and so do not contribute to the growth rate. Of course, if we are considering early times (\ie $\tR<\tRc1$ or $\tL<\tLc1$) when this boundary ends on the past singularity, we simply set $r_m=0$ in the above expression leaving only the contribution for the crossing point $r=r_s$.

For the left future boundary, we find
\begin{equation}
I^{(\RN{2})}_{\mt{ct}} = ``UV\ terms"  + \frac{\Omega_{k, d-1} }{8 \pi G_N}\, r_b^{d-1} \, \log\!\left( \frac{f_2(r_b)}{f_1(r_b)} \right)\, ,
\label{count2}
\end{equation}
by substituting eq.~\reef{AffineFut} for $\hat\alpha$ into eq.~\reef{count03}. Here we are implicitly assuming that this boundary always terminates on the future singularity and for late times (\ie $\tL>\tLc2$) when this boundary does not cross the shock wave, we simply set $r_b=0$ above. We also consider the left past boundary, for which eq.~\reef{count04} yields
\begin{equation}
I^{(\RN{3})}_{\mt{ct}} =``UV\ terms"  - \frac{\Omega_{k, d-1}}{8 \pi G_N} \, r_m^{d-1} \left[ \log\!{\left( \frac{(d-1) \alpha\ctL}{r_m} \right)}
 +\frac1{d-1}\right]  \, .
\label{count3}
\end{equation}
We might note that this contribution is identical to the first line in eq.~\reef{count1}. Again,  we set $r_m=0$ above for early times (\ie $\tR<\tRc1$ or $\tL<\tLc1$) when this boundary ends on the past singularity. Finally, we also have the counterterm contribution for the right future boundary in eq.~\reef{count05} but as noted above, it will not contribute to the complexity growth rate, since we only consider the regime when this surface terminates at the future singularity at $r=0$.

We now evaluate the time derivative of these three contributions in turn by using eqs.~\reef{eq:DertR} and \reef{eq:DertL} to evaluate the time derivatives of $r_s$, $r_b$ and $r_m$. Let us begin with eq.~\reef{count2} and consider the regime
$\tL<\tLc2$, which yields
\begin{align}
&\frac{d I^{(\RN{2})}_{\mt{ct}} }{d \tR}  = 0  \, , \nonumber \\
&\frac{d I^{(\RN{2})}_{\mt{ct}} }{d \tL}  = - \frac{\Omega_{k, d-1}}{16 \pi G_N} \left[   \frac{2 r_b^{d}}{L^2} \left(1 - \frac{f_1 (r_b)}{f_2 (r_b)} \right) + (d-2) \left(\omega_1^{d-2} - \omega_{2}^{d-2} \frac{f_1 (r_b)}{f_2 (r_b)} \right)\right]
 \label{ICount2}     \\
& \qquad\qquad\qquad + \frac{(d-1) \Omega_{k, d-1}}{16 \pi G_N} \, r_b^{d-2} f_1(r_b) \log\! \left( \frac{f_2(r_b) }{f_1(r_b)} \right) \, .\nonumber
\end{align}
Of course, for later times $\tL>\tLc2$, both of these time derivatives vanish since $r_b=0$.

Given the similarities between eqs.~\reef{count1} and \reef{count3}, we combine $I^{(\RN{1})}_{\mt{ct}}$ and $I^{(\RN{3})}_{\mt{ct}}$ in evaluating the time derivatives.
For early times (\ie $\tR<\tRc1$ or $\tL<\tLc1$), we set $r_m=0$, and the time derivatives only act on $r_s$ producing
\begin{align}
&\frac{d\ }{d \tR} \left(  I^{(\RN{1})}_{\mt{ct}} +  I^{(\RN{3})}_{\mt{ct}} \right)  \bigg{|}_{r_m = 0} =\frac{\Omega_{k, d-1}}{16 \pi G_N} \left[ \frac{2 r_s^{d}}{L^2} \left(1 - \frac{f_2 (r_s)}{f_1 (r_s)} \right) + (d-2)\left( \omega_2^{d-2} - \omega_{1}^{d-2} \frac{f_2 (r_s)}{f_1 (r_s)}\right) \right]   \nonumber    \\
& \qquad\qquad\qquad\qquad\qquad\qquad\qquad
- \frac{(d-1) \Omega_{k, d-1}}{16 \pi G_N} \, r_s^{d-2} f_2(r_s) \log \left[ \frac{f_1(r_s) }{f_2(r_s)} \right] \, , \label{ICount1norm} \\
&\frac{d\ }{d \tR} \left(  I^{(\RN{1})}_{\mt{ct}} +  I^{(\RN{3})}_{\mt{ct}} \right)  \bigg{|}_{r_m = 0} = 0  \nonumber \, .
\end{align}
At later times (\ie $\tR<\tRc1$ or $\tL<\tLc1$),  $r_m$ becomes a dynamical variable and so there are additional contributions to the above time derivatives
\begin{align}
\frac{d\ }{d \tR} \left(  I^{(\RN{1})}_{\mt{ct}} +  I^{(\RN{3})}_{\mt{ct}} \right) =& \frac{d\ }{d \tR} \left(  I^{(\RN{1})}_{\mt{ct}} +  I^{(\RN{3})}_{\mt{ct}} \right)  \bigg{|}_{r_m = 0}  - \frac{\Omega_{k,d-1} (d-1)}{16 \pi G_N} \, r_m^{d-2}  \left[ f_2^{'} (r_s) - \frac{f_2 (r_s)}{f_1 (r_2)} f_1^{'} (r_s) \right] \nonumber \\
&+ \frac{\Omega_{k,d-1} (d-1)}{8 \pi G_N} \, r_m^{d-2} f_{1}(r_m) \frac{f_2 (r_s)}{f_1 (r_s)} \log \left[ \frac{(d-1)}{r_m} \frac{f_1 (r_s)}{f_2 (r_s)} \ctL \, \alpha \right] \, \nonumber \\
\frac{d\ }{d \tL} \left(  I^{(\RN{1})}_{\mt{ct}} +  I^{(\RN{3})}_{\mt{ct}} \right) =& \frac{\Omega_{k,d-1} (d-1)}{8 \pi G_N} \, r_m^{d-2} f_{1}(r_m) \log \left[ \frac{(d-1)}{r_m} \frac{f_1 (r_s)}{f_2 (r_s)} \ctL \, \alpha \right]   \, . \label{ICount13}
\end{align}
Of course, we will see below that when these counterterm contributions are combined with those from the rest of the gravitational action, the dependence on the normalization constant $\alpha$ is completely eliminated --- see also eq.~\reef{finale}.

\subsubsection{Time Dependence of Complexity}

We can now evaluate the time dependence of the holographic complexity by summing the various  expressions above. However, we can consider many different forms for the time evolution, \eg varying $\tL$ alone or $\tR$ alone. For simplicity, we will focus on the symmetric case where we vary $t=\tL+\tR$ while fixing $\tL-\tR=0$. This approach is closely related to the time evolution studied in \cite{Growth} for an unperturbed eternal black hole (without any shock waves). In principle though, the results above would easily allow one to study the evolution of the holographic complexity according to any other linear combination $t'=a\,\tL+b\,\tR$.

Further, in analogy with \cite{Growth}, we will focus on the evolution for $t>0$.\footnote{Although we also consider some times slightly before $t=0$.} However, there remains two important factors in determining how the holographic complexity grows, namely, the time at which the shock wave is sent in from the right boundary and its mass, \ie the values of $t_w$ and $M_2-M_1$. In particular, these will determine the geometry of the WDW patch as discussed around eqs.~\reef{tRc1} and \reef{tLc2}. That is, as seen in studying the CA conjecture in the unperturbed eternal black hole, \eg \cite{Brown2,Growth}, we are generally in a regime where $r_m<0$ at $t=0$ and the WDW patch touches some interval on the past singularity. Hence in the present situation, the shock wave parameters effect the critical time $t_{c1}$ when $r_m$ becomes positive and the WDW patch terminates above the past singularity. This critical time is determined by setting $\tLc1=t_{c1}/2=\tRc1$ in eq.~\reef{tRc1}, which then yields
\begin{equation}\label{tc1}
t_{c1}  = 2t_w - 4 r^{*}_{1} (0) +4 r^{*}_{1} (r_s)\, , \, \qquad \, \qquad \,  r^{*}_{1} (r_s)+r^{*}_{2} (r_s)=  r^{*}_{1} (0)-t_w  \,.
\end{equation}
Here the second equation determines the value of $r_s$ which should be substituted into the first to
determine $t_{c1}$. Generally, increasing $t_w$ or the mass of the shock, \ie increasing $M_2-M_1$, increases the value of $t_{c1}$.\footnote{These statements can be confirmed as follows: For a fixed ratio $w=r_{h,2}/r_{h, 1}$, the following identity holds
\begin{equation}
\frac{d t_{c1}}{d t_w} =  \frac{2(\omega_{2}^{d-2} -\omega_{1}^{d-2})}{r_s^{d-2} (f_1 (r_s) + f_2 (r_s))} \, .
\end{equation}
Assuming $M_2>M_1$, the above derivative is positive and so the effect of increasing $t_w$ here is to increase the value of $t_{c1}$. Similarly, we have
\begin{equation}
\frac{d t_{c1}}{d w} = - \frac{4 f_2 (r_2)}{f_2 (r_2) + f_1 (r_2)} \, \frac{d r^{*}_{2}}{d w} \bigg|_{r_s} \, ,
\end{equation}
and increasing $M_2$ for a fixed $r_s$ decreases $ r^{*}_{2}(r_s)$. Hence increasing the mass of the second black hole also increases $t_{c1}$.}

Now similarly, we are generally in a regime at $t=0$ where the WDW patch touches some interval on the future singularity. If we evolve forward in time, this interval simply expands but there is another critical time $t_{c2}$ where the interval includes the point where the shock wave hits the singularity. That is, $t_{c2}$ is the time when $r_b$ vanishes (and then becomes negative for larger values of $t$). Substituting $\tLc2=t_{c2}/2$ into eq.~\eqref{eq:rwrsrm}, we find this critical time to be
\beq\label{tc2}
t_{c2} = 2t_w+ 4 r^{*}_{1} (0) \, .
\end{equation}
Here again, the effect of increasing $t_w$ is to increase $t_{c2}$, while varying $M_2-M_1$ has no effect on the value of $t_{c2}$ (if we assume that $M_1$ is fixed).

We would like to add one more critical time to this list, in analogy with the evolution of the holographic complexity for the unperturbed eternal black hole in \cite{Growth}. In that instance, there was actually an interval $-t_{c1}\le t\le t_{c1}$ in which the complexity did not change. We will find a similar plateau in the case of the shock wave geometries where the rate of change is small but since shock wave breaks the time-shift symmetry, we introduce a new critical time $t_{c0}$ to denote the beginning of this period, $-t_{c0}\le t\le t_{c1}$. Geometrically, this time is the time at which the WDW patch lifts off of the future singularity if we push $t$ to sufficiently negative values. This critical time can be determined in a similar way to finding $t_{c1}$ and the result is
\begin{equation}\label{tc0}
t_{c0}  = 2t_w - 4 r^{*}_{2} (0) +4 r^{*}_{2} (r_b)\, , \, \qquad \, \qquad \,  r^{*}_{1} (r_b)+r^{*}_{2} (r_b)=  r^{*}_{2} (0)-t_w  \,.
\end{equation}
Here again, we determine $r_b$ from the second equation and  then substitute this value into the first equation to determine $t_{c0}$. We may note that $r_b <r_{h,1}$, \ie the (future) null boundary of the WDW (on the left) crosses the shock wave behind the black hole horizon, and so one can easily show that $t_{c0}<2t_w$. In some sense, $t=-2t_w$ is the next critical time since at this point the right boundary time slice coincides with the point on the boundary surface where the shock wave originates.

Comparing eqs.~\reef{tc1} and \reef{tc2}, we find
\beq\label{diffc}
t_{c2}-t_{c1} =8 r^{*}_{1} (0) -4 r^{*}_{1} (r_s(t_{c1}))\,.
\eeq
Now with our conventions $r^{*}_{1} (0)$ and $r^*_1(r_s)$ will be negative quantities --- see eq.~\reef{rStar1} --- and so there is a competition to determine the sign of this difference.\footnote{Again, $d=2$ is a special case with $r^{*}_{1} (0)=0$  --- see eq.~\reef{loppit} below.}
However, at least if $t_w$ and/or ($M_2$--$M_1$) are sufficiently large, we expect that $t_{c2}-t_{c1} >0$. In this scenario, there are three regimes of the WDW patch geometry to be considered,
\begin{align}
&\ \ \RN{1}\ : -t_{c0}< t < t_{c 1} \, \qquad       \text{$r_b$, $r_s$ exist; $r_m<0$}    \nonumber \\
&\ \RN{2}\ :\ \  t_{c 1} < t < t_{c 2}  \, \qquad       \text{$r_b$, $r_s$, $r_m$, exist}        \label{ActionRegimes} \\
&\RN{3}\ :\qquad t > t_{c 2}  \, \qquad  \quad \text{$r_s$, $r_m$ exist; $r_b<0$ }. \nonumber
\end{align}

For the regime $\RN{1}$ in eq.~\eqref{ActionRegimes}, the total rate of change of complexity consists of the bulk contribution in eq.~\eqref{eq:BulkActEtSym} (with $r_m=0$), eqs.~\eqref{GHFut} and \eqref{GHPast} for the GHY contributions from the past and future singularities, respectively, and the two counterterm contributions from eqs.~\eqref{ICount1norm} and \eqref{ICount2}. The final result then becomes
\begin{align}
\frac{d \mathcal{C}_A^{(\RN{1})} }{d t} =&  \frac{M_2}{\pi} \left( 1 + \frac{f_1(r_b)}{f_2(r_b)} \right) - \frac{M_1}{\pi} \left( 1 + \frac{f_2(r_s)}{f_1(r_s)} \right)   \, \nonumber \\
& + \frac{M_1}{2 \pi} \frac{r_b^{d-2}}{\omega_{1}^{d-2}} f_1 (r_b) \log \frac{f_2 (r_b)}{f_1 (r_b)} - \frac{M_2}{2 \pi} \frac{r_s^{d-2}}{\omega_{2}^{d-2}} f_2 (r_s) \log \frac{f_1 (r_s)}{f_2 (r_s)} \, . \label{RateSymt1}
\end{align}

In regime $\RN{2}$, we only need the GHY contribution in eq.~\eqref{GHFut} from the future singularity, the counterterm in the left future boundary of the WDW patch in eq.~\eqref{ICount2} and the sum of the joint contribution at $r_m$ in eq.~\eqref{ICorner3} with the two past counterterm contributions, given by eq.~\eqref{ICount13}. In this case, the total reads
\begin{align}
\frac{d \mathcal{C}_A^{(\RN{2})} }{d t} =&  \frac{M_2}{\pi} \left( 1 + \frac{f_1(r_b)}{f_2(r_b)} \right)  + \frac{M_1}{2 \pi} \frac{r_m^{d-2}}{\omega_{1}^{d-2}} f_1 (r_m) \left(1 +  \frac{f_2 (r_s)}{f_1 (r_s)} \right) \log \left(  \frac{|f_1 (r_m)| (d-1)^2 \ctL^2}{r_m^2} \right)   \, \nonumber \\
& + \frac{M_1}{2 \pi} \frac{r_b^{d-2}}{\omega_{1}^{d-2}} f_1 (r_b) \log \frac{f_2 (r_b)}{f_1 (r_b)} - \frac{M_2}{2 \pi} \frac{r_s^{d-2}}{\omega_{2}^{d-2}} f_2 (r_s) \log \frac{f_1 (r_s)}{f_2 (r_s)} \, . \label{RateSymt2}
\end{align}

In the last regime, the GHY contribution from the future singularity is given by eq.~\eqref{GHFut_Norway}  and the sum of the joint at $r_m$ in eq.~\eqref{ICorner3} with the counterterm contributions from the past boundaries in eq.~\eqref{ICount13}. Hence the rate of growth in this regime is
\begin{align}
\frac{d \mathcal{C}_A^{(\RN{3})} }{d t} =&  \frac{M_1 + M_2}{\pi} + \frac{M_1}{2 \pi} \frac{r_m^{d-2}}{\omega_{1}^{d-2}} f_1 (r_m) \left(1 +  \frac{f_2 (r_s)}{f_1 (r_s)} \right) \log \left( \frac{|f_1 (r_m)|(d-1)^2 \ctL^2 }{r_m^2} \right)  \, \nonumber \\
&  - \frac{M_2}{2 \pi} \frac{r_s^{d-2}}{\omega_{2}^{d-2}} f_2 (r_s) \log \frac{f_1 (r_s)}{f_2 (r_s)} \, . \label{RateSymt3}
\end{align}

\noindent{\textbf{Dimensionless variables}}

In addition to the dimensionless variables in eq.~\reef{eq:wz}, it is also useful to define dimensionless coordinates corresponding to the three positions $r_b$, $r_s$ and $r_m$,
\begin{equation}\label{dimx3}
x_s \equiv \frac{r_s}{r_{h, 2}}   \, , \,  \, \qquad \,   x_m \equiv \frac{r_m}{r_{h, 1}}   \, , \,  \, \qquad \,    x_b \equiv \frac{r_b}{r_{h, 1}}  \, .
\end{equation}
Note that these definitions are not completely harmonious with the definition $x=r/r_{h,2}$ in eq.~\reef{eq:wz}. In the definition above, we have chosen to normalize $r_s$ with $r_{h, 2}$ such that the minimum value of $x_s$ is one, but $r_b$ and $r_m$ are normalized with $r_{h, 1}$ such that the maximum value of $x_b$ and $x_m$ is $1$ as well. The addition of the counterterm \reef{counter} introduces one more length scale $\ctL$ and so it will be useful to define the dimensionless quantity
\begin{equation}\label{dimlct}
 \ttL \equiv \frac{\ctL}{L} \, .
\end{equation}

Using the rescaled blackening factors in eq.~\eqref{Dimlessf1f2}, we can rewrite the rates of change of complexity in eqs.~\eqref{RateSymt1}, \eqref{RateSymt2}, \eqref{RateSymt3} and \eqref{dimlct} as
\small
\begin{align}
&\frac{d \mathcal{C}_A^{(\RN{1})} }{d t} =  \frac{M_2}{\pi} \left( 1 + \frac{\tilde f(x_b, w z)}{w^2 \tilde f(x_b/w, z)} \right) - \frac{M_1}{\pi} \left( 1 + \frac{w^2 \tilde f(x_s, z)}{\tilde f(w x_s, w z)} \right)   \, \label{RateSymtDimless1} \\
&\qquad + \frac{M_1}{2 \pi} \frac{x_b^{d-2}}{1 + k z^2 w^2} \tilde f (x_b, w z) \log \frac{w^2 \tilde f(x_b/w,z)}{\tilde f (x_b, w z)} - \frac{M_2}{2 \pi} \frac{x_s^{d-2}}{1+ k z^2} \tilde f (x_s, z) \log \frac{ \tilde f (w x_s, w z)}{w^2 \tilde f (x_s, z)} \nonumber \, , \\
&\frac{d \mathcal{C}_A^{(\RN{2})} }{d t} =    \frac{M_2}{\pi} \left( 1 + \frac{\tilde f(x_b, w z)}{w^2 \tilde f(x_b/w, z)}  \right)
\label{RateSymtDimless2} \\
&\qquad+ \frac{M_1}{2 \pi} \frac{x_m^{d-2}}{1+ k z^2 w^2} \left( 1 + \frac{w^2 \tilde f (x_s, z)}{ \tilde f (w x_s, w z)}\right) \tilde f(x_m, w z) \log  \frac{|\tilde f (x_m, w z)| (d-1)^2  \ttL^2}{x_m^2}  \nonumber \\
&\qquad  +  \frac{M_1}{2 \pi} \frac{x_b^{d-2}}{1 + k z^2 w^2} \tilde f (x_b, w z) \log \frac{w^2 \tilde f(x_b/w,z)}{\tilde f (x_b, w z)} -  \frac{M_2}{2 \pi} \frac{x_s^{d-2}}{1+ k z^2} \tilde f (x_s, z) \log \frac{ \tilde f (w x_s, w z)}{w^2 \tilde f (x_s, z)} \nonumber \, , \\
&\frac{d \mathcal{C}_A^{(\RN{3})} }{d t} =  \frac{M_1 + M_2}{\pi} - \frac{M_2}{2 \pi} \frac{x_s^{d-2}}{1+ k z^2} \tilde f (x_s, z) \log \frac{ \tilde f (w x_s, w z)}{w^2 \tilde f (x_s, z)}\label{RateSymtDimless3} \\
&
\qquad+  \frac{M_1}{2 \pi} \frac{x_m^{d-2}}{1+ k z^2 w^2} \left( 1 + \frac{w^2 \tilde f (x_s, z)}{ \tilde f (w x_s, w z)}\right)  \tilde f(x_m, w z) \log  \frac{|\tilde f (x_m, w z)| (d-1)^2 \ttL^2}{x_m^2}   \,
\nonumber \, .
\end{align}

\normalsize

\noindent{\textbf{Early and late time behaviours}}

We now turn our attention to two simple limits for the rate of change of complexity. First, let us consider early times which means that we should consider the growth rate given in eq.~\reef{RateSymt1}. Now, if $t_w$ is sufficiently large, then $r_s$ approaches $r_{h,2}$ and $r_b$ approaches $r_{h,1}$, \ie $f_2(r_s),\ f_1(r_b)\to0$. In this limit, the growth rate in eq.~\eqref{RateSymtDimless1} simplifies to
\begin{equation}\label{EarlyTimesRate}
\frac{d \mathcal{C}^{(\RN{1})}_{A}}{d t} \bigg|_{t_w \rightarrow \infty} = \frac{M_2 - M_1}{\pi}  + \mathcal{O} \left( T_1(2 t_w - t) e^{- \pi T_1 ( 2 t_w - t) }  \right) \, ,
\end{equation}
\ie it is simply proportional to the difference of masses.

Another simple limit occurs at late times, when the growth rate is given by  eq.~\eqref{RateSymtDimless3}. In this case, irrespective of the value of $t_w$, $r_m$ and $r_s$ approach $r_{h,1}$ and $r_{h,2}$, respectively, \ie $f_1(r_m),\ f_2(r_s)\to0$. In this case, the growth rate of the holographic complexity is given by the sum of the black hole masses, \ie
\begin{equation}\label{LateTimeEtShock}
\frac{d \mathcal{C}^{(\RN{3})}_{A}}{d t} \bigg|_{t \rightarrow \infty} = \frac{M_1 + M_2}{\pi} + \frac{M_1}{2}\left( \frac{d + (d-2) k z^2 }{1 + k z^2} \right) T_1 t  \,  e^{-  \pi T_1 (t - 2 t_w - 4 r^{*}_{1}(r_{h,2}))} + \mathcal{O} \left( e^{-  \pi T_1 t }
\right) \, ,
\end{equation}
as was previously argued in \cite{Brown2}. Further, we note that the second term in eq.~\eqref{LateTimeEtShock} is always positive, and therefore $d\ca/dt$ approaches the previous late time limit from above. Similar behaviour was found for the unperturbed eternal black holes in \cite{Growth}. In analogy to these earlier results, we also note that the next correction term, of order $e^{-  \pi T_1 t }$, depends on the normalization factor $\ttL$. We add that more generally, the dependence on $\ttL$ is more pronounced at early times.

\subsubsection{A case study: BTZ black holes}

It is instructive to analyze the particular case of BTZ black holes (\ie $d=2$), since the positions $r_s$, $ r_m$ and $r_b$ can be determined analytically. First, the two blackening factors are given by
\begin{equation}\label{locker}
f_1(r) = \frac{r^2 - r_{h,1 }^2}{L^2}  \, \qquad{\rm and}\qquad \, f_2(r) = \frac{r^2 - r_{h,2}^2}{L^2} \, .
\end{equation}
For each black hole (so $r_h$ can be either $r_{h,1}$ and $r_{h,2}$), the physical quantities characterizing the black hole solutions are
\begin{equation}\label{BTZquantities}
M = \frac{\Omega_{k, 1}\, r_h^2}{16 \pi G_N L^2 }\, ,  \qquad T = \frac{r_h}{2 \pi L^2} \,,\qquad S=\frac{\Omega_{k, 1} r_h}{4 G_N} = \frac{\pi}{6}\,c\, \Omega_{k, 1} L T\, ,
\end{equation}
and $c=3L/(2G_N)$ is the central charge of the boundary CFT. In the above formulas, $k=0$ and 1 correspond to the Ramond and Neveu-Schwarz vacuum, respectively, of the boundary theory \cite{couscous}.

With the blackening factors in eq.~\eqref{locker}, we can evaluate the tortoise coordinates in eqs.~\reef{rStar1} and \reef{rStar2} as
\begin{equation}\label{loppit}
r_1^{*}(r) = \frac{L^2 }{2 r_{h,1}} \log \left(\frac{|r-r_{h,1}|}{r+r_{h,1}}\right) \, , \, \, \qquad \, r_2^{*}(r) = \frac{L^2}{2 r_{h,2}}  \log \left(\frac{|r-r_{h,2}|}{r+r_{h,2}}\right) \, .
\end{equation}
We choose to normalize the time coordinates by the temperature of the final (more massive) black hole, which reads
\begin{equation}
T_2 =  \frac{r_{h, 2}}{2 \pi L^2} \, .
\end{equation}
For the BTZ geometry, $w$ which is the ratio of the horizon sizes in eq.~\reef{eq:wz} is just the ratio of the temperatures,
\begin{equation}\label{ratt}
w = {T_2}/{T_1} \, .
\end{equation}
Further, we note that $M_2/M_1=w^2$ and $S_2/S_1=w$.

Now combining eqs.~\eqref{eq:rwrsrm}, \reef{dimx3} and \reef{loppit}, as well as the above ratio \reef{ratt}, yields the following:
\begin{align}
&x_s = \frac{1 + e^{- 2 \pi T_2 (\tR + t_w)} }{1 - e^{- 2 \pi T_2 (\tR + t_w)}} \, , \, \qquad \,  x_b = \frac{ e^{- 2 \pi T_1 (\tL - t_w)} - 1}{e^{- 2 \pi T_1 (\tL - t_w)}+1} \, , \nonumber \\
&\qquad  x_m = \frac{w x_s+1 -\left(w x_s-1\right)e^{-2 \pi  T_1 \left(\tL- t_w\right)}}{w x_s+1+\left(w x_s-1\right)e^{-2 \pi  T_1 \left(\tL- t_w\right)}} \, . \label{BTZ_xsxmxw}
\end{align}
With these three expressions, the growth rates in eqs.~(\ref{RateSymtDimless1}--\ref{RateSymtDimless3}) are implicitly expressed entirely in terms of boundary quantities. Further, from eq.~\reef{locker}, we see that $r^*_1(0)=0$ and hence the critical times in eqs.~(\ref{tc1}--\ref{tc0}) simplify to
\beq
t_{c2}=2t_w\,,\qquad
t_{c1}=2t_w- 4|r_1^*(r_s)|
\qquad {\rm and} \qquad
t_{c0}=2t_w- 4|r_2^*(r_b)|\,.
\label{tcrit}
\eeq
While we do not have an analytic expression for $r_s$, it is easily determined numerically by combining the expressions in eqs.~\reef{tc1} and \reef{locker}, and similarly for $r_b$. We return to examine the critical times $t_{c1}$ and $t_{c0}$ in more detail in a moment. In any event, we see that we are in the situation with $t_{c1}<t_{c2}$ and so the evolution of the holographic complexity is described by the scenario in eq.~\reef{ActionRegimes} and so let us explicitly examine $d\ca/dt$ in a few examples.

\begin{figure}
\centering
\includegraphics[scale=0.6]{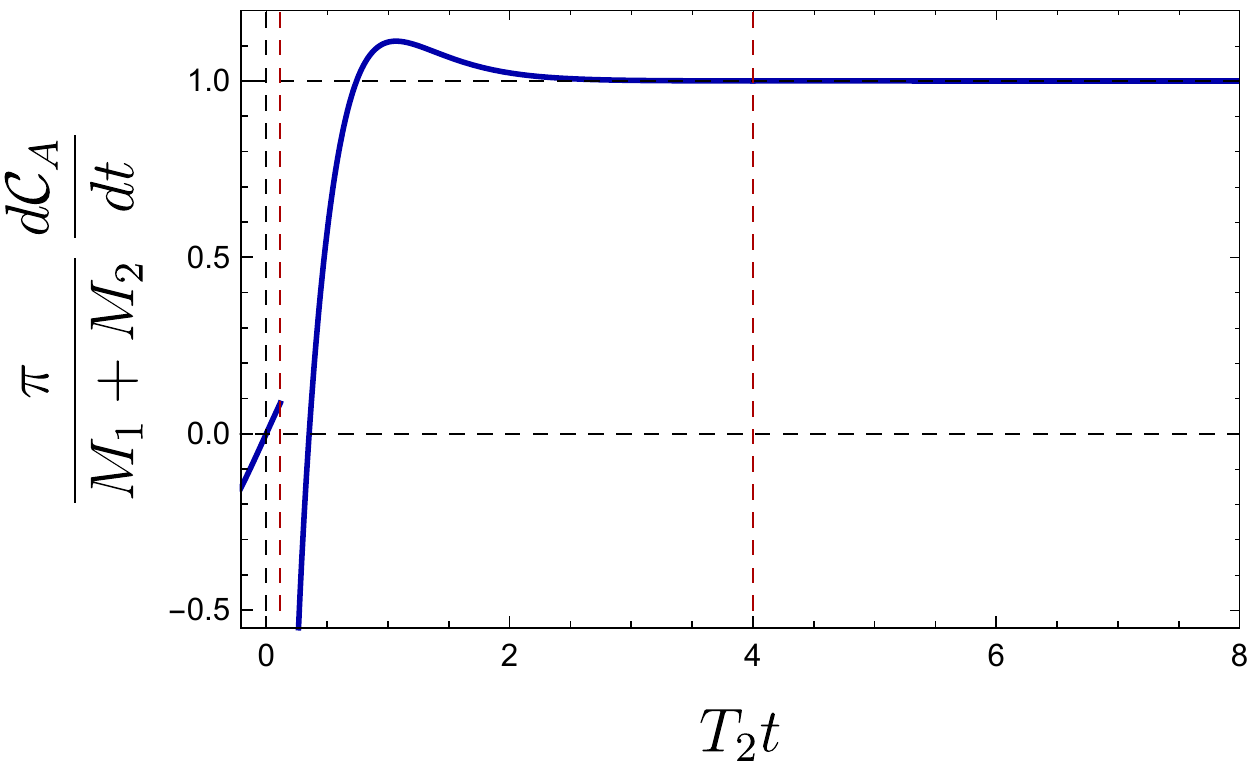}
\includegraphics[scale=0.6]{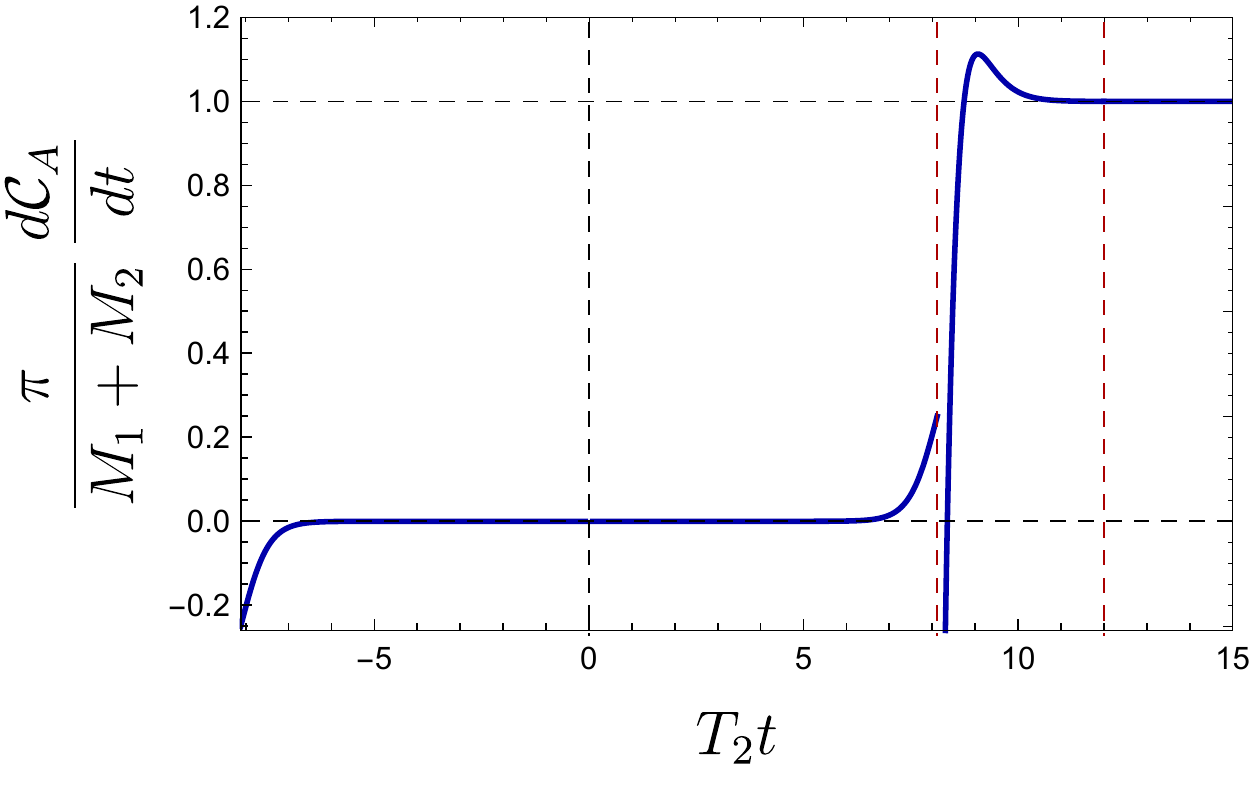}
\caption{Time derivative of complexity, evolving both boundaries as $\tL=\tR=\frac{t}{2}$ with $T_2 t_w =  2$ (left) and  $T_2 t_w = 6$ (right). We have set $w=1+10^{-5}$ and $\ttL=1$. The condition on $z$ implies that the smaller black hole is at the Hawking-Page transition, for both cases. The lower horizontal dashed line (near zero) corresponds to $(M_2-M_1)/(M_2+M_1)$, and by construction, the late time limit approaches $1$ at the higher horizontal line. The horizontal axis in both figures starts from the respective $t_{c0}$ in eq.~\reef{tcrit}. The first vertical black dashed line appears at $t=0$, while the vertical red dashed lines appear at $t_{c1}$ (left) and $t_{c2}=2t_w$ (right), see also eq.~\reef{tcrit}. There is a negative spike right after $t_{c1}$, where $x_m$ is close to the past singularity. For the earlier shock wave in the right figure, there is a long regime where the rate of change is close to zero. In both cases, the late time limit is approached from above. }
\label{TimeDepBTZShockBotht}
\end{figure}

In figure \ref{TimeDepBTZShockBotht}, we show $d\ca/dt$ for a very light shock wave where $w=1+10^{-5}$. In the left panel, we show the results for $T_2 t_w = 2$, and in the right, for $T_2 t_w = 6$. For both cases, the growth rate is essentially zero over the period $-t_{c0}\lesssim t\lesssim t_{c1}$, however, this is a longer time period for a larger value of $t_w$. Immediately after $t_{c1}$, there is a negative spike in the rate of growth, which is similar to the one found for the eternal BTZ black hole with the inclusion of the counterterm in appendix A of \cite{Growth}. Note that this very small initial growth rate is consistent with eq.~\reef{EarlyTimesRate} since the difference $(M_2-M_1)= (w^2-1)M_1$. Further, the separation $t_{c2}-t_{c1}$ (as well as $2t_w-t_{c0}$) appears to be independent of $t_w$. We will examine these observations further in the following.

\begin{figure}
\centering
\includegraphics[scale=0.6]{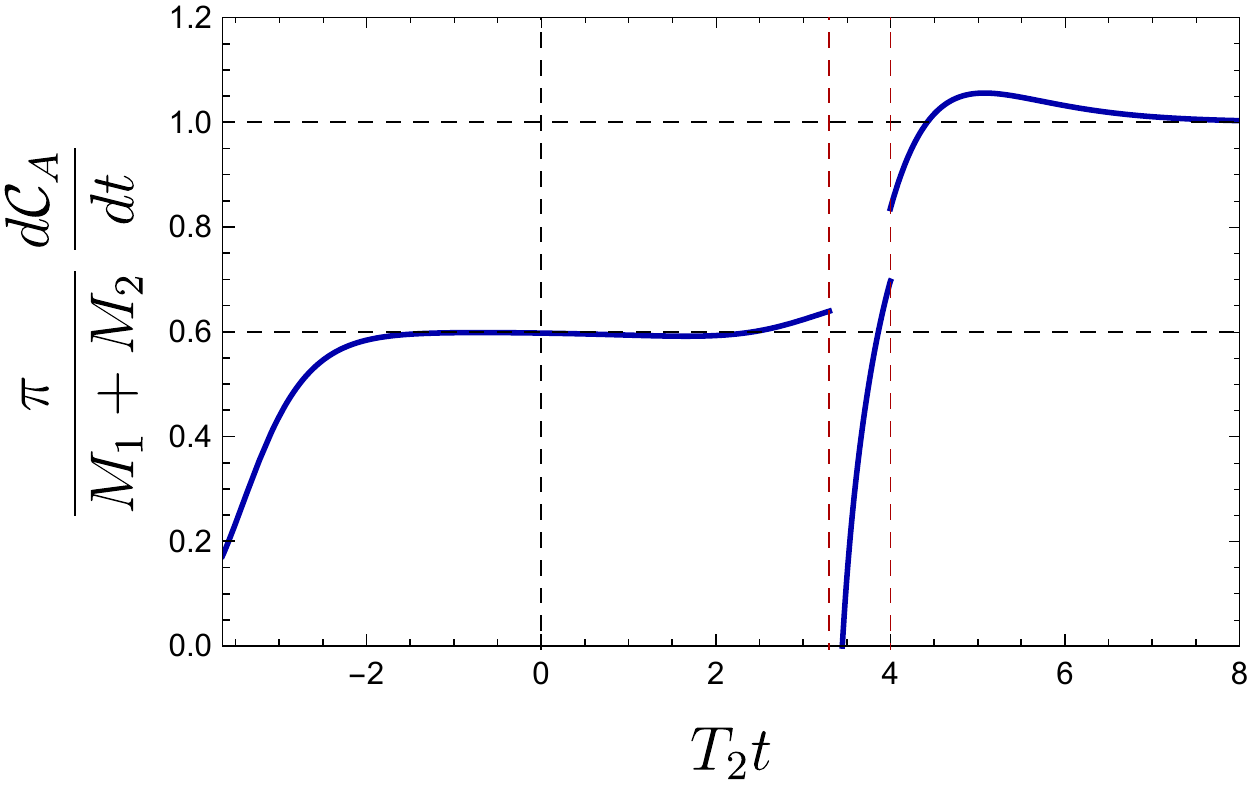}
\includegraphics[scale=0.6]{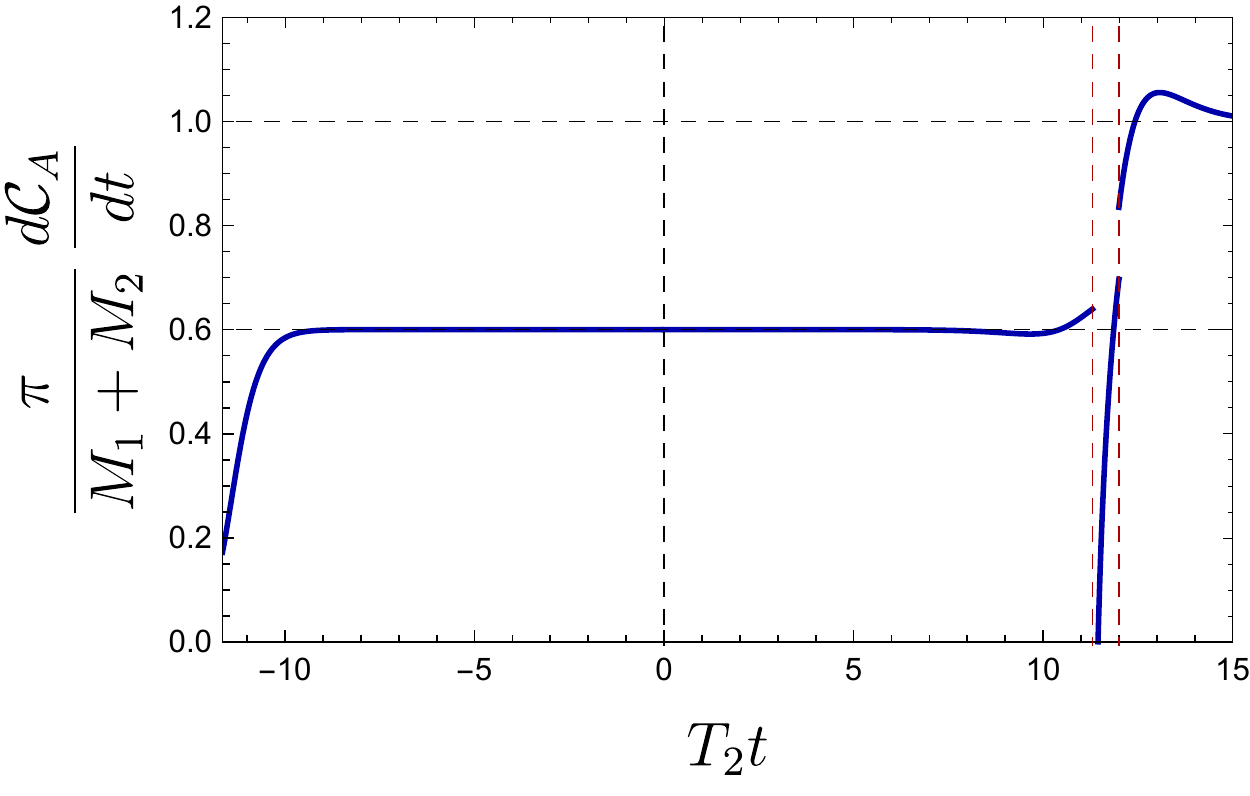}
\caption{Time derivative of complexity, evolving both boundaries as $\tL=\tR=\frac{t}{2}$, with $T_2 t_w =  2$  (left) and  $T_2 t_w = 6$ (right). In both cases, we have set $w=2$ and $ \ttL =1$. The lower horizontal black dashed line corresponds to the time derivative at early times, \ie $(M_2 - M_1)/\pi$ in eq.~\reef{EarlyTimesRate}, and the higher line to the late time limit, \ie $(M_2 + M_1)/\pi$ in eq.~\reef{LateTimeEtShock}. The horizontal axis starts at $t_{c0}$, and the critical times $t_{c1}$ and $t_{c2}$ are shown by the left and right vertical dashed red lines, respectively. There is a negative spike right after $t_{c1}$, where $x_m$ is close to the past singularity. Pushing the shock wave to the past increases the plateau where the time derivative is given roughly by the difference of the masses. }
\label{TimeDepBTZShockBotht_w2}
\end{figure}

In figure \ref{TimeDepBTZShockBotht_w2}, we show $d\ca/dt$ for a heavier shock wave where $w=2$ (\ie the temperature doubles or the black hole mass increases by a factor of four) and $z=1/w$, such that the smaller black hole is at the Hawking-Page transition. In the left panel, we show the results for $T_2 t_w = 2$, and in the right, for $T_2 t_w = 6$. For both cases, the growth rate is significantly lower (than  the final rate) in the period $-t_{c0}\lesssim t\lesssim t_{c1}$. This plateau is more evident in the case with a larger value of $t_w$. Rather than vanishing in this period, $d\ca/dt$ is given by the difference $M_2$--$M_1$, as in eq.~\reef{EarlyTimesRate}. Note that the separation $t_{c2}-t_{c1}$ (as well as $2t_w-t_{c0}$) again appears to be independent of $t_w$, but is a smaller interval (when normalized by $T_2$) than with a very light shock wave, as in figure \ref{TimeDepBTZShockBotht}.

\subsubsection*{Critical times in BTZ}

Here, we examine the critical times for the special case $d=2$ in more detail. Recall that in this case, eq.~\reef{locker} yields $r^*_1(0)=0$ and hence the critical times in eqs.~\eqref{tc1}-\eqref{tc0} simplify to the expressions given in \reef{tcrit}. Hence the critical time where the endpoint of the shock wave on the future singularity enters the WDW patch is simply given by $t_{c2}=2t_w$.\footnote{Note that this corresponds to $\tL=t_w$ and this simple result arises from the special property that the singularity is a straight horizontal line in a Penrose diagram describing the BTZ black hole \cite{SingBow}.} However, the critical times $t_{c1}$ and $-t_{c0}$ where the WDW patch lifts off of the past singularity and first impinges on the future singularity, respectively, have a more interesting structure. From eq.~\reef{EarlyTimesRate}, we found that during the period $-t_{c0}\lesssim t\lesssim t_{c1}$,\footnote{This was regime I in eq.~\reef{ActionRegimes}.} the growth rate of the holographic complexity is roughly proportional to the difference of the masses, at least when $t_w$ is sufficiently large. This plateau with $d\ca/dt\simeq (M_2-M_1)/\pi$ is clearly shown in figures \ref{TimeDepBTZShockBotht} and \ref{TimeDepBTZShockBotht_w2}.\footnote{In these figures, we can see that there are significant tails on the plateau in the interval $-t_{c0}\le t\le t_{c1}$. A better estimate of the length of the plateau can be determined from the analytic expressions of $x_b$ and $x_s$ in eq.~\eqref{BTZ_xsxmxw}, as follows: The plateau is in the regime where both $x_s$ and $x_b$ are close to $1$. Let us define the ``boundaries'' of the plateau with the conditions that $x_s \approx 1 + 2 e^{- \gamma \pi} + \mathcal{O} (e^{- 2 \gamma \pi} ) $ and $x_b \approx 1 - 2 e^{-\gamma \pi} + \mathcal{O} (e^{- 2 \gamma \pi} ) $, where $\gamma$ is a number of order $1$. It then follows that the length of the plateau is approximately $T_2 \Delta t_{\text{pl}} \approx 4 T_2 t_w - \gamma (w+1)$.}

From eq.~\reef{tcrit}, we have $t_{c1}=2t_w- 4|r_1^*(r_s)|$ for the $d=2$ shock wave geometries. We would like to understand this result in terms of boundary quantities and this is most simply done by considering various limits. First, suppose that the shock wave is very heavy, \ie $w$ in eq.~\reef{ratt} is a large parameter. Recalling that $t_{c1}$ is the critical time when $x_m$ becomes positive, we may use eq.~\eqref{BTZ_xsxmxw} to find
\begin{equation}\label{tc1Largew}
t_{c1} = 2 t_w - \frac{2}{ \pi T_{2}} + \mathcal{O} \left( \frac{1}{w^2 \, T_2} \right) \, ,
\end{equation}
for large $w$.
For very high temperatures, the above expression implies that this critical time approaches $t_{c2}$, \ie $t_{c1} \rightarrow 2 t_w=t_{c2}$.

We also consider the case of a very light shock for which $w$ can be parametrized as $w = 1 + \epsilon$. Using eqs.~\reef{BTZquantities} and \reef{ratt}, the ratio of the masses  is given by
\begin{equation}
\frac{M_2}{M_1} = w^2 = 1 + 2 \epsilon + \epsilon^2 \, ,
\end{equation}
and hence the energy of the shock $E$ is given by
\beq \frac{E}{M_1}=2 \epsilon + \epsilon^2 \simeq 2 \epsilon\,.
\label{eshok}\eeq
Now again using eq.~\eqref{BTZ_xsxmxw}, we have in the limit $e^{-2 \pi T_2 t_w} \ll\epsilon\ll1$
\begin{equation}
t_{c1} =  2 t_w + \frac{1}{\pi T_1} \log \frac{\epsilon}{2}   -\frac{\epsilon}{2 \pi T_1} + \mathcal{O} \left( \frac{\epsilon^2}{T_1} \right) \, .
\label{lavla}\end{equation}
Following \cite{ShenkerStanfordScrambling}, we can relate the first correction to the scrambling time \cite{Sekino:2008he}. If one considers $E$ to be of the order of the energy of a few thermal quanta of energy, then we may use eqs.~\reef{BTZquantities} and \reef{eshok} to write\footnote{Note that we chose $E\simeq 2T_1$ to simplify the subsequent expressions.}
\beq
\frac2\epsilon\simeq\frac{4M_1}{E}\simeq\frac{4M_1}{2T_1}\simeq S_1\,.
\label{hatch}
\eeq
Hence eq.~\reef{lavla} becomes
\begin{equation}
t_{c1} = 2 ( t_w  - t^{*}_{scr}) + \mathcal{O} \left( \epsilon  \right)\label{lavla2}
\end{equation}
where
\begin{equation}\label{Scramblingt}
 t^{*}_\text{scr} \equiv \frac{1}{2 \pi T_1} \log{\frac{2}{ \epsilon}}=\frac{1}{2 \pi T_1} \log S_1\, .
\end{equation}

\begin{figure}
\centering
\includegraphics[scale=0.6]{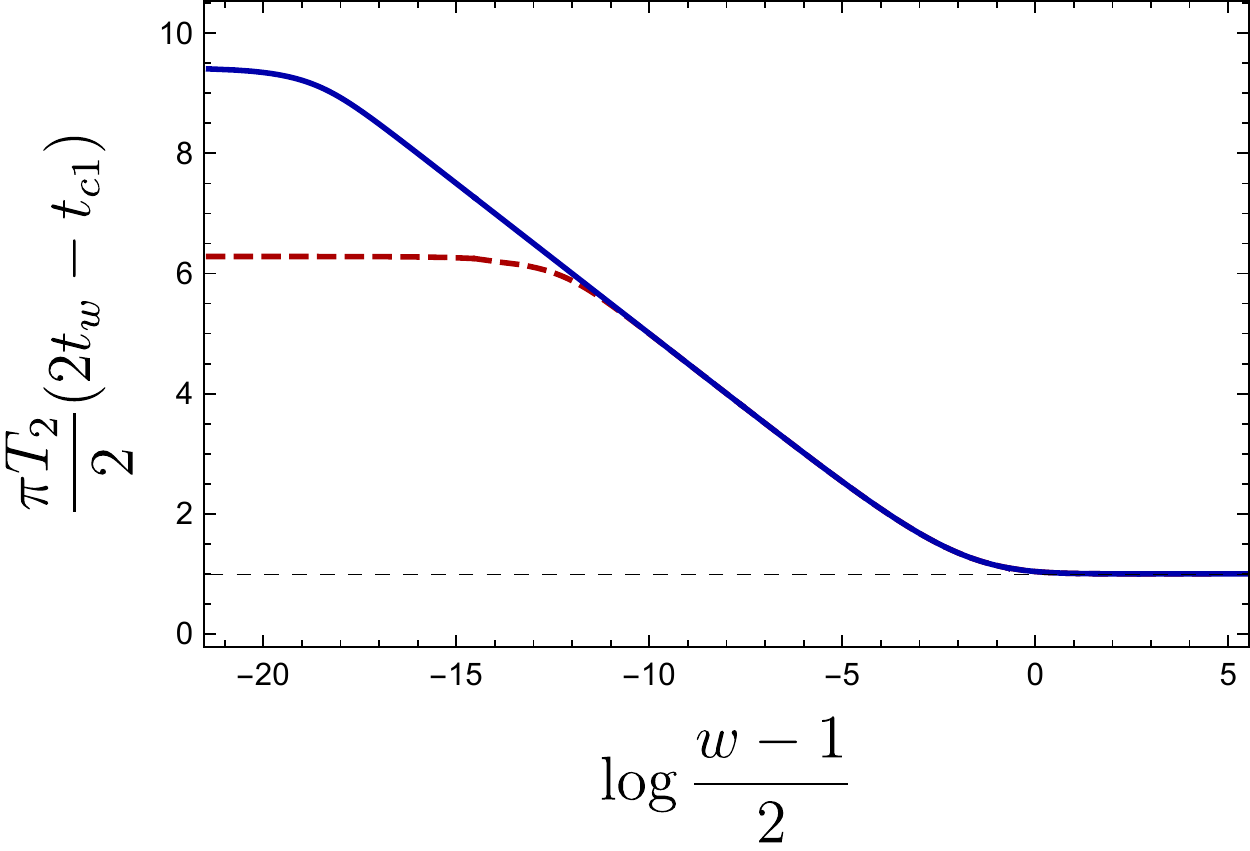}
\includegraphics[scale=0.6]{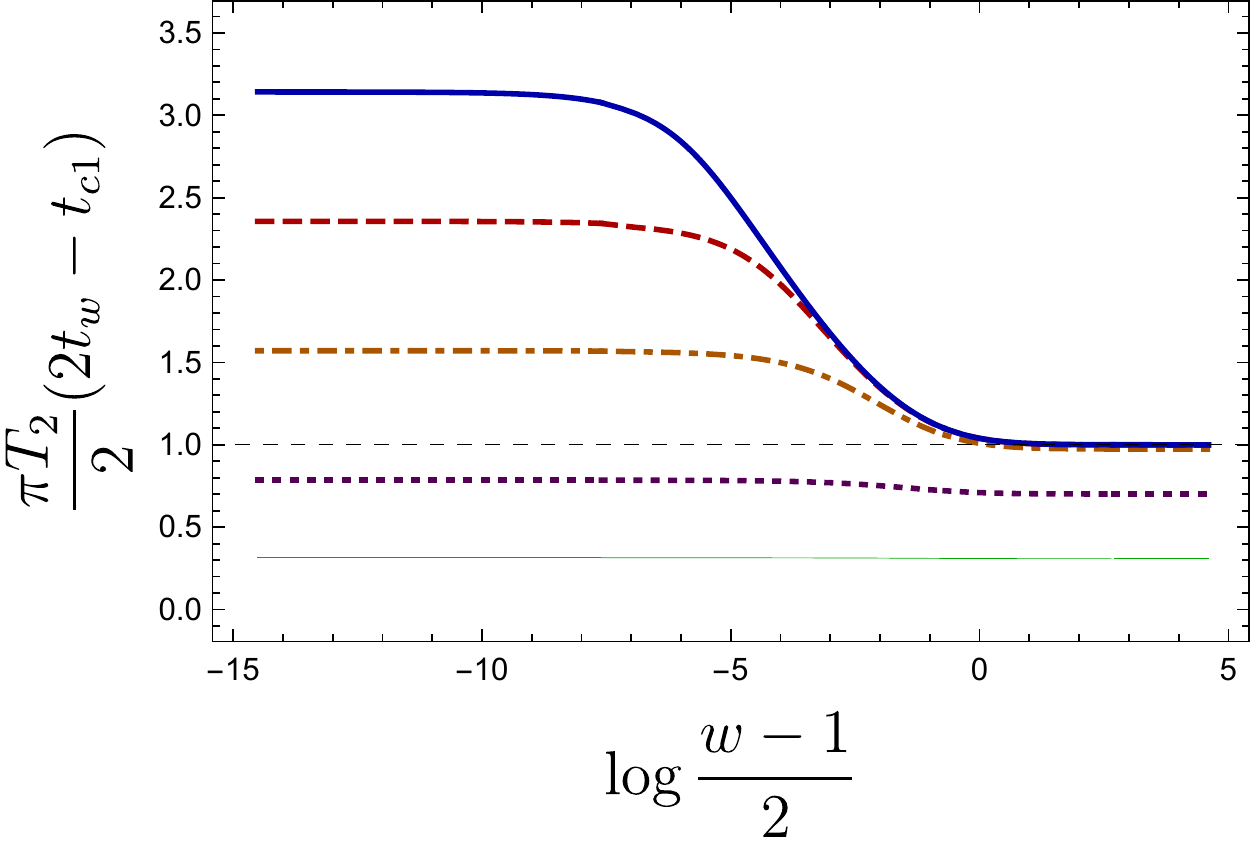}
\caption{Dependence of the critical time $t_{c1}$ on $\log\frac{w-1}2=\log(\epsilon/2)$, which parametrizes the energy in the shock wave. In the left panel, we show the behaviour of $t_{c1}$ for early shock waves: $T_2 t_w = 3$ in solid blue, and $T_2 t_w = 2$ in dashed red. In this case, we see the transition between the light shock behaviour \reef{lavla} and the heavy shock behavior \reef{tc1Largew}. In the right panel, we show the behaviour of $t_{c1}$ when the shock wave is not sent very early, \ie with $T_2 t_w = 0.1, 0.25, 0.5, 0.75, 1$ from bottom to top.  For these parameters, the range of $w$ that has approximately a log dependence starts appearing as the shock wave is sent earlier (larger $t_w$). The horizontal thin dashed black line is just $1$ (for both panels). }
\label{tc1Plots}
\end{figure}

Having evaluated the behaviour of the critical time $t_{c1}$ for  heavy and light shocks in eqs.~\reef{tc1Largew} and \reef{lavla}, respectively, we plot the numerical solution from eq.~\reef{tcrit} in figure \ref{tc1Plots}. In the left panel, we show the behaviour of $t_{c1}$ as a function of $\log(\epsilon/2)$ for early shock waves, \ie for which $x_s-1 \simeq  2 e^{-2 \pi T_2 t_w} \ll 1$.  In the figure, we clearly see the transition between the light shock behaviour (where $2t_w-t_{c1}$ depends linearly on $\log(\epsilon/2)$) and the heavy shock behaviour (where $2t_w-t_{c1}$ is constant) and that the value of $w$ where the transition occurs is independent of $t_w$. Recall that an essential assumption in deriving eq.~\eqref{lavla} was that the order of limits $e^{- 2 \pi T_2 t_w} \ll \epsilon \ll 1$ held. The geometrical interpretation of this limit is that $x_s$ is exponentially close to $1$, and therefore corrections of order $e^{-2 \pi T_2 t_w}$ are much smaller than corrections to the energy from the shock wave, which are of order $\epsilon$.  Therefore in figure \ref{tc1Plots} where the value of $t_w$ is fixed for each curve, we see there is a regime of very small $\epsilon$ where $\epsilon\lesssim e^{- 2 \pi T_2 t_w}$ where the
difference $2t_w-t_{c1}$ again saturates at some constant value.
In the right panel of figure \ref{tc1Plots}, we show the behaviour when the shock waves are not sent very early. In this regime, $t_{c1}$ just transitions between two different constant regimes without much of a linear regime in between. Further, increasing the mass ratio decreases the critical time.  For large $w$, notice that most curves (with big enough $t_w$) saturate to $1$, which is consistent with the large $w$ expansion in eq.~\eqref{tc1Largew}. Note that the plots produced here in figure \ref{tc1Plots} also represent the difference $t_{c2} - t_{c1}$, since for $d=2$ from eq.~\eqref{tcrit} we have $t_{c2} =  2 t_{w}$. Hence in the early shock regime, the right panel shows that the separation between these two critical times is independent of $t_w$ (except for very small $\epsilon$), as observed in figures \ref{TimeDepBTZShockBotht} and \ref{TimeDepBTZShockBotht_w2}.

\begin{figure}
\centering
\includegraphics[scale=0.6]{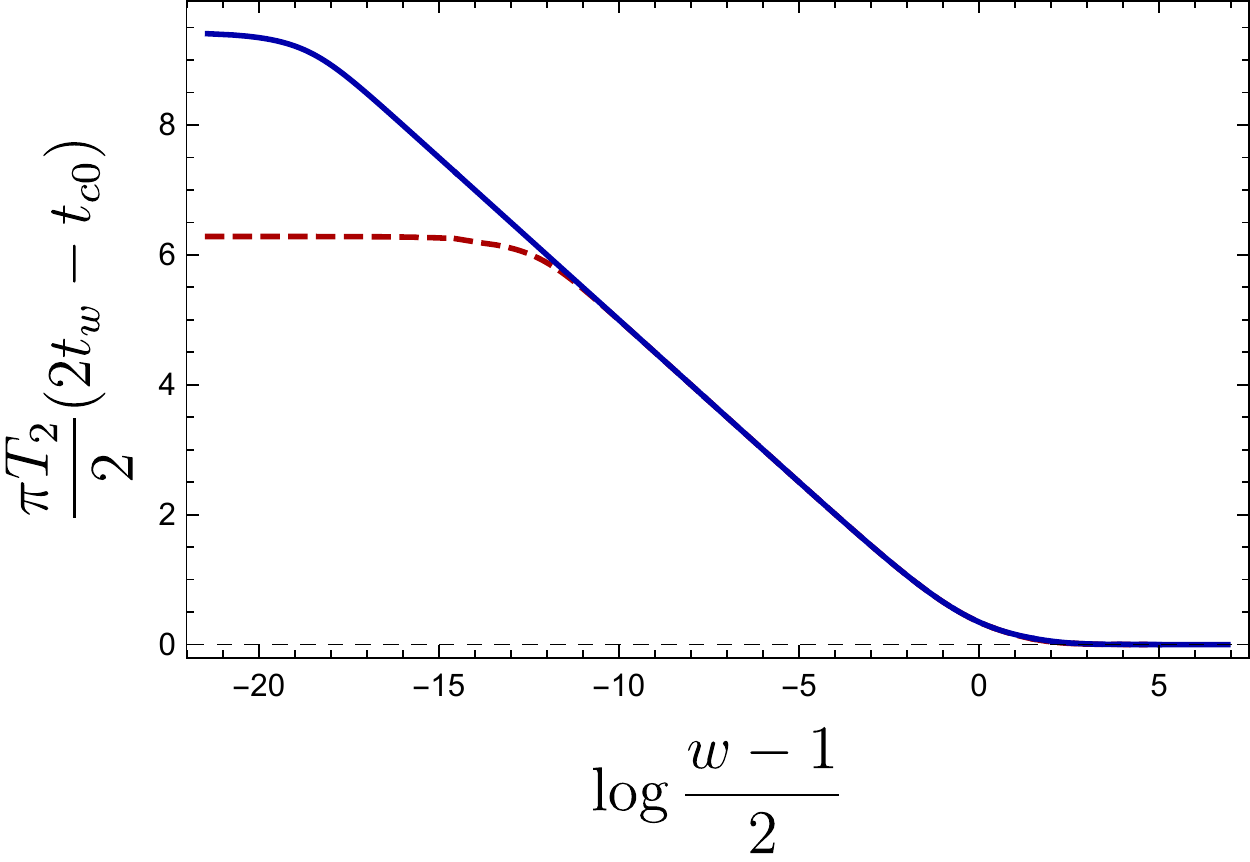}
\includegraphics[scale=0.6]{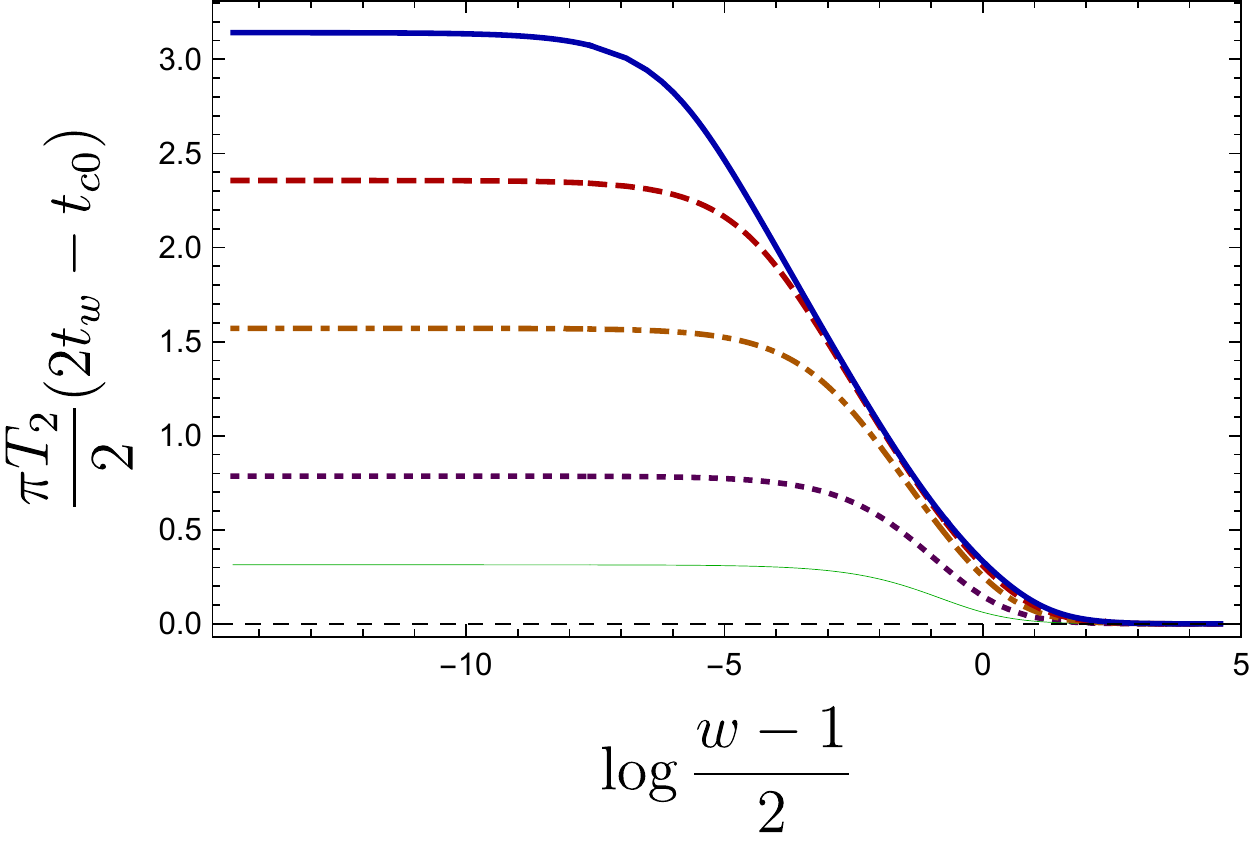}
\caption{Dependence of the critical time $t_{c0}$ on the energy of the shock wave, parametrized by the temperature ratio $w$ for BTZ. In the left, we show the behaviour of $t_{c0}$ with respect to early shock waves with $T_2 t_w = 3$ in solid blue, $T_2 t_w = 2$ in dashed red. Similarly to $t_{c1}$ in figure \ref{tc1Plots}, as $w$ approaches one, there is a stretched range of $w$ such that $t_{c1}$ grows as a logarithm, and the earlier the shock wave the longer this log regime. Also, we see that it approaches $2 t_{w}$ in the large $w$ regime. In the right, we show the behaviour of $t_{c0}$ when the shock wave is not sent early enough, with $T_2 t_w = 0.1, 0.25, 0.5, 0.75, 1$ from bottom to top. As the shock wave is sent earlier, the region with log dependence becomes more pronounced.}
\label{tc0Plots}
\end{figure}

Next, we turn our attention to $t_{c0}=2t_w- 4|r_2^*(r_b)|$ from eq.~\eqref{tcrit}, which for early shock waves (\ie $t_w$ large) represents the beginning (\ie at $t=-t_{c0}$) of the plateau where the rate of growth is approximately $(M_{2} - M_{1})/\pi$.
We begin by considering light shock waves in the limit with $e^{-2 \pi T_1 t_w}\ll\epsilon\ll1$. In this scenario, $r_b \rightarrow r_{h,1}$ as the ratio of temperatures approaches one (\ie $w\to1$), and $r_b \rightarrow 0$ as $w$ increases. Therefore, we can expand eq.~\eqref{tc0} for $w=1+\epsilon$ with $\epsilon$ small,\footnote{Note that in this case, the expression would be simpler if we had defined the scrambling time with $T_2$. That is, using $t^{*}_{\text{scr}}=\frac1{\pi T_2}\log(2/\epsilon)$, rather than the definition in eq.~\reef{Scramblingt}, would remove the $\epsilon\log\epsilon$ correction in eq.~\reef{what}.}
\begin{equation}\label{what}
t_{c0} = 2 \left( t_{w} - t^{*}_{\text{scr}} \right)  + \frac{\epsilon}{ \pi T_1} \left( \log \frac{2}{\epsilon} - \frac12\right) +  \mathcal{O}(\epsilon^2 \log \epsilon) \, .
\end{equation}
For heavy shock waves (\ie large $w$), $t_{c0}$ scales as
\begin{equation}
t_{c0} = t_{w} \left( 2 - \frac{4}{w^2} + \mathcal{O} (w^{-4}) \right) \, .
\end{equation}
In figure \ref{tc0Plots}, we show the numerical solution of eq.~\eqref{tc0} for $T_2 t_{w} = 2$ in the left panel, and $T_2 t_{w} = 0.25$ in the right panel. For the early shock wave and small $\epsilon$, we see that $t_{c0}$ depends linearly on $\log\epsilon$. As a result, the plateau (where the derivative is close to zero) will extend far into the past. If the shock wave is not sent early enough the range with this log dependence is much shorter, similar to the behaviour found for $t_{c1}$.

We focused our analysis of the critical times here on the special case of $d=2$ because many features, such as the dependence of $x_s$ and $x_b$ on $t_w$, were analytic. In addition, since eq.~\reef{loppit} yields $r^{*}(0) = 0$ for $d=2$, $t_{c2} - t_{c1}$ was always positive and $t_{c2}$ was simply given by $2 t_w$. We investigate higher dimensions (in particular $d=4$) in appendix \ref{app:AppEternalAdS5}. There, the fact that $r^{*}(0) \neq 0$ leads to some modifications for shock waves not inserted early enough, \ie for small $t_w$, we may find that $t_{c2} - t_{c1}$ is negative. On the other hand, if the shock wave is sent early enough, it is also true that in higher dimensions, there is a plateau of rate of change $(M_2-M_1)/\pi$ that extends for a length of time of approximately $4 t_w$.

\subsection{Complexity of Formation} \label{CoF1}

In this section, we consider the complexity of formation, as previously discussed in \cite{Format}. There, we compared the complexity of preparing two copies of the boundary CFT in the thermofield double state (TFD) at $\tL = \tR = 0$ to the complexity of preparing each of the CFT's in its vacuum state,
\beq\label{CoFdef}
\Delta\mC=\mC(|TFD\rangle)-\mC(|0\rangle\otimes|0\rangle)\,.
\eeq
Using the CA conjecture \reef{defineCA}, the holographic calculation was to evaluate the WDW action for $\tL = \tR = 0$ in an eternal black hole background and subtract that for two copies of the AdS vacuum geometry. This difference removed the UV divergences leaving a finite quantity. For neutral black holes,\footnote{These calculations were extended to charged black holes in \cite{Growth}.} we found that at high temperatures generally $\Delta\mC$ was proportional to the entropy of the black hole or alternatively, the entanglement entropy in the TFD state,  plus small curvature corrections. However, $d=2$ was a special case where for the BTZ black hole, $\Delta\mC$ was a constant proportional to the central charge.
\begin{figure}
\centering
\includegraphics[scale=0.4]{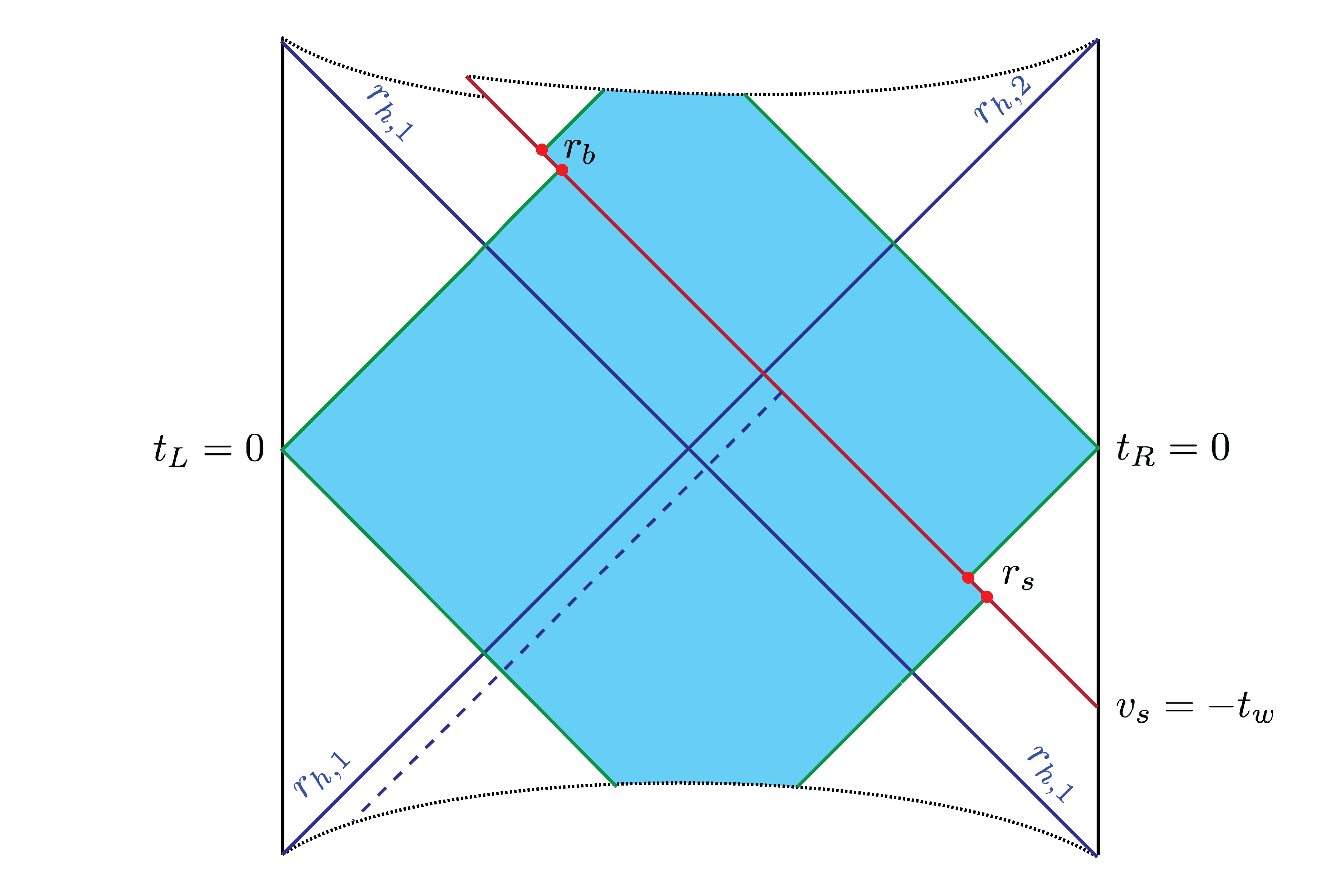}
\caption{Penrose-like diagram for one shock wave sent from the right boundary at $v_s = - t_w$ on an eternal black hole geometry, with the Wheeler-DeWitt patch anchored at $\tL=\tR=0$, which we will call the complexity of formation, in analogy to the case studied in \cite{Format} of the unperturbed eternal black hole geometry. There are two coordinates $r_b$ and $r_s$ that are usually given by a transcendental equation as functions of $w$ and $t_w$, as shown in eq.~\eqref{eq:rwrsrm}. The effect of crossing the collapsing shell from the right boundary is to increase the surface of the WDW patch above the past singularity.  }
\label{PenroseShockForm}
\end{figure}

In the following, we aim to evaluate the complexity of formation for the perturbed state dual to the shock wave geometry, again at $\tL = \tR = 0$.\footnote{The boundary times are synchronized according to the procedure outlined at the beginning of this section --- see the discussion above eq.~\reef{eq:wz}.} The resulting $\Delta\ca$ can be studied as a function of $t_w$ and $M_2$--$M_1$. We illustrate this setup with the Penrose diagram of figure \ref{PenroseShockForm}. The calculation follows straightforwardly from the considerations of the previous subsection. For instance, the bulk integral is obtained with $r_m = 0$ and also by setting $\tL = \tR =0$ in eq.~\eqref{BulkEternalSTotal}. Also, we have to subtract two copies of the vacuum, which was discussed in detail in \cite{Format}. We have then
\begin{align}\label{FormEternalSBulk}
&\Delta I_{\text{bulk}} =  \left( \frac{\Omega_{k, d-1}}{16 \pi \, G_N} \right) \left( - \frac{2 d}{L^2}\right) \bigg[   \int_{r_s}^{r_{max}} d r \, r^{d-1} (- 2 r^{*}_{2}(r))  +  \int_{r_b}^{r_{max}} d r \, r^{d-1} (- 2 r^{*}_{1}(r))   \\
&+ \int_{0}^{r_s} d r \, r^{d-1} \left(t_w + 2 r^{*}_{1}(r_s) - 2 r^{*}_{1}(r) \right) + \int_{0}^{r_b} d r \, r^{d-1} \left( t_w+ 2 r^{*}_{2}(r_b) - 2 r^{*}_{2}(r) \right)  \bigg] - 2 I_{\text{bulk,vac}} \nonumber \, ,
\end{align}
where $r_b$ and $r_s$ are given by eq.~\eqref{eq:rwrsrm} with $\vL=\vR=0$, and $I_{\text{bulk,vac}}$ is the appropriate vacuum bulk integral given by
\begin{equation}
2 I_{\text{bulk,vac}} =  \left( \frac{\Omega_{k, d-1}}{16 \pi \, G_N} \right) \left( - \frac{2 d}{L^2}\right) \int_{0}^{r_{max}} d r \,  r^{d-1} (- 4 r^{*}_{\vac} (r) ) \, ,
\end{equation}
with
\begin{equation}
r^{*}_{\vac} (r) = -\int_{r}^{\infty} \frac{d r}{k+ r^2/L^2} \, , \, \qquad \, \qquad \lim_{r\to \infty} r^{*}_{\vac} (r)  = 0 \, .
\end{equation}

The only nonvanishing contributions from the boundary surfaces are the two GHY contributions at the past and future singularities, given by eqs.~\eqref{GHFut_Tot} and \eqref{GHPast_Tot} with $\vL=\vR=0$, which results in
\begin{equation}\label{FormEternalSGH}
\Delta I_{GHY} =  \frac{d \, \Omega_{k,d-1}}{16 \pi G_N} \left[ \omega_{1}^{d-2} \left( t_w+ 2 r_1^{*} (r_s) - 2 r_{1}^{*}(0) \right) + \omega_{2}^{d-2} \left(  t_w+ 2 r_{2}^{*}(r_b) -  2 r_{2}^{*}(0)  \right) \right]    \, .
\end{equation}
Finally, we need to add the contribution of the two counterterms in eqs.~\eqref{count1} and \eqref{count2} with $r_m = 0$. The UV contributions cancel when subtracting the vacuum, so as a result we have
\begin{equation}\label{FormEternalSCt}
\Delta I_{\mt{ct}} = \frac{\Omega_{k, d-1} }{8 \pi G_N} \left[ r_s^{d-1} \log \left( \frac{f_{1}(r_s)}{f_2(r_s)} \right) +  \, r_b^{d-1} \log \left( \frac{f_{2}(r_b)}{f_1(r_b)} \right)   \right] \, .
\end{equation}
As argued in \cite{Format}, the joint contributions at the UV regulator surface precisely cancel the same contributions from the vacuum geometries.

Combining all of these contributions then yields the desired complexity of formation,
\begin{equation}\label{CoFSum0}
\Delta \ca = \frac{ \Delta I_{\text{bulk}} + \Delta I_{GHY} + \Delta I_{\mt{ct}} }{\pi} \, .
\end{equation}
This result is more complicated than the complexity of formation for the unperturbed BH geometry, since the points $r_s$ and $r_b$ are determined by inverting a transcendental equation (for higher dimensional black holes). However, it can be studied analytically for $d=2$ and we consider this special case in the following. We will also consider $\Delta\ca$ for planar AdS$_5$ black holes in appendix \ref{app:AppEternalAdS5}.

It is straightforward to evaluate eq.~\eqref{CoFSum0} for $d=2$, and it is instructive to compare the result to the complexity of formation for the unperturbed BTZ black hole. The latter was evaluated in \cite{Format}, where we found $\Delta\mC_\mt{NS}=-c/3$ when subtracting the complexity of the Neveu-Schwarz vacuum (\ie $k=+1$).\footnote{With $c$ being the central charge of the boundary CFT, which is given by $c= 3 L /(2 G_N)$.} Comparing the result for the perturbed state to $\Delta\mC_\mt{NS}$ then yields
\begin{align}\label{FormationShockBTZ}
&\frac{\Delta \ca-\Delta\mC_\mt{NS}}{|\Delta\mC_\mt{NS}|} =  LT_1 \bigg[ \frac{w^2-1}{w}\,\log \left(\frac{x_s-1}{x_s+1}\right)+w x_s \log \left(\frac{w^2 x_s^2-1}{w^2 \left(x_s^2-1\right)}\right) \nonumber \\
&\qquad\quad\qquad\qquad\qquad+ \left(w^2-x_b^2\right) \log \left(\frac{1+x_b}{1-x_b}\right)-\frac{w^2-x_b^2}{w}\,\log \left(\frac{w+x_b}{w-x_b}\right) \nonumber \\
&\qquad\quad\qquad\qquad\qquad+ x_b\, \log \left(\frac{w^2-x_b^2}{1-x_b^2}\right) \bigg] \, .
\end{align}
Here the coordinates $x_s$ and $x_b$ are given by eq.~\eqref{BTZ_xsxmxw} with $\tL=\tR=0$,
\begin{equation}\label{BTZ_Form_xsxw}
x_s = \frac{1 + e^{- 2 \pi T_2  t_w} }{1 - e^{ - 2 \pi T_2  t_w}} \, , \, \qquad \, x_b = \frac{ 1-e^{ -2 \pi T_1 t_w} }{1+ e^{-2 \pi T_1  t_w}} \, .
\end{equation}

\begin{figure}
\centering
\includegraphics[scale=0.6]{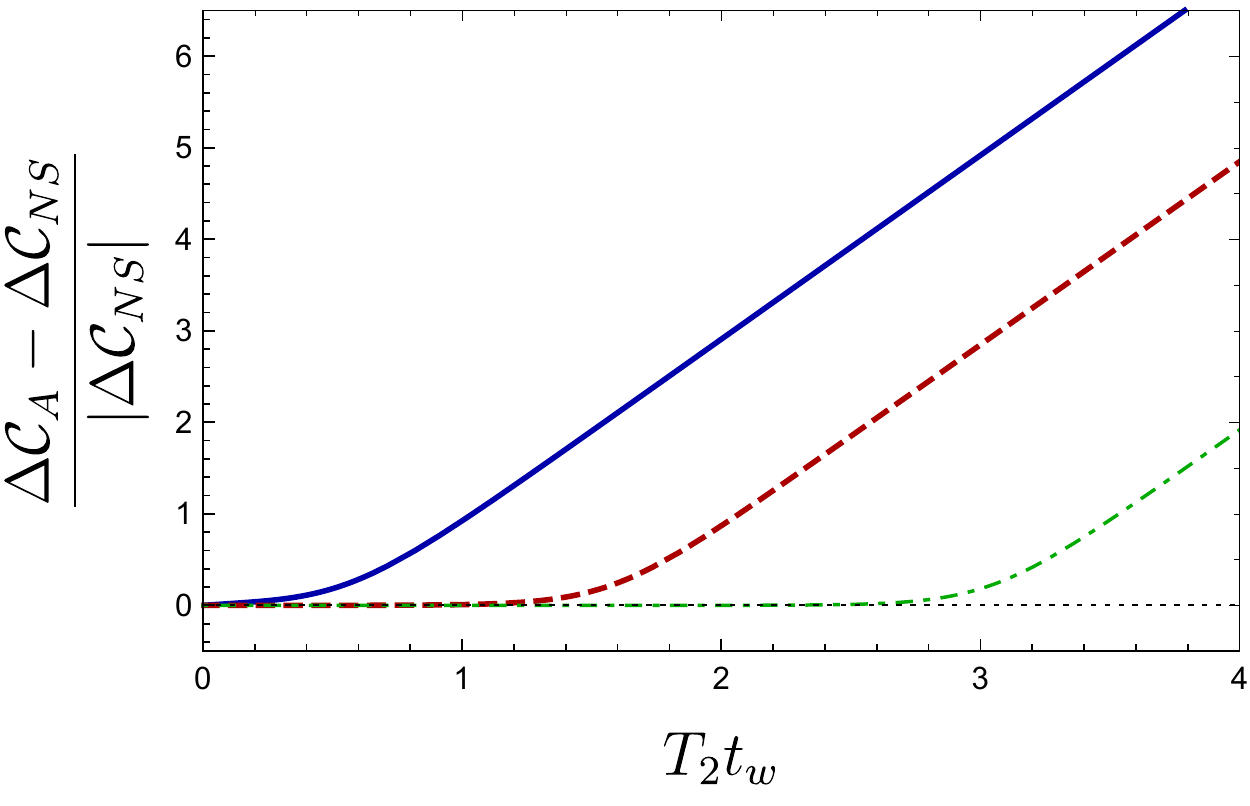}
\includegraphics[scale=0.6]{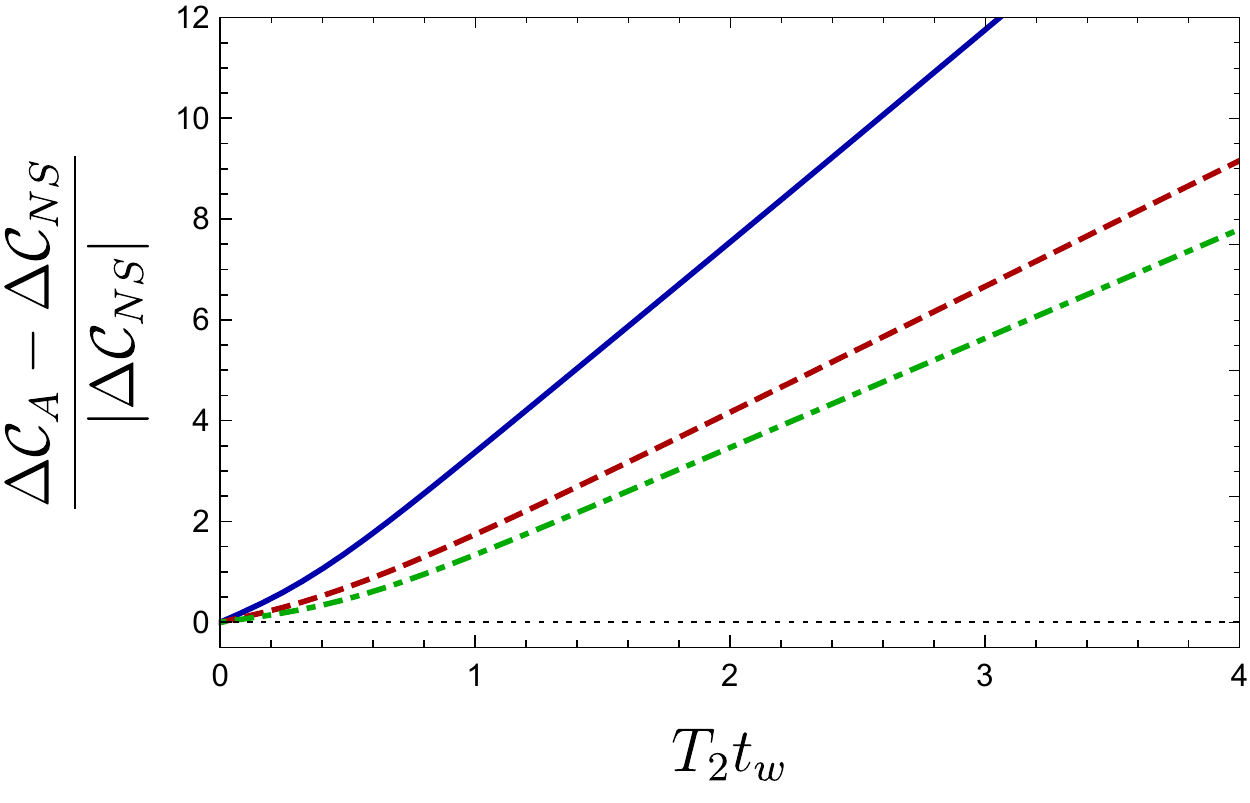}
\caption{The complexity of formation for BTZ black holes for $z=1/w$, such that the smaller black hole is at the Hawking-Page transition, or alternatively we could normalize the complexity of formation by the entropy, which would remove the overall multiplicative factor of $z$, cf. eq.~\eqref{FormationShockBTZ}. In the left panel, we evaluate the complexity of formation for light shock waves as a function of $T_2 t_w$. The energies of the shock waves are parametrized by the temperature ratio $w$, with $w=1+10^{-1}$ (solid blue), $w=1+10^{-4}$ (dashed red) and $w=1+10^{-8}$ (dot-dashed green). For a period of time of the order of the scrambling time \reef{Scramblingt}, the complexity of formation is approximately the same as the unperturbed state. For early shocks (\ie larger $t_w$), the complexity of formation grows linearly with $t_w$. In the right panel, we show the complexity of formation for heavier shocks, $w=4$ (solid blue), $w=2.5$ (dashed red) and $w=1.5$ (dot-dashed green).  For these parameters, we see that the complexity of formation starts changing immediately and rapidly approaches a regime of linear growth with increasing $t_w$. }
\label{FormationBTZShock}
\end{figure}

In the left panel of figure \ref{FormationBTZShock}, we show the effect of a light shock wave on the complexity of formation as a function of $t_w$. Initially there is a period where $\Delta\ca=\Delta\mC_\mt{NS}$ after which $\Delta\ca$ begins to grow linearly. As
the shock is made lighter (\ie as $w$ is brought closer to one), this period over which the complexity of formation is essentially unchanged grows longer. In the period of linear growth, the slope seems more or less the same independent of $w$. In the right panel, we show the effect of heavier shock waves. In this regime, the complexity of formation starts changing immediately, even for small $t_w$, and $\Delta\ca$ rapidly enters a regime of linear growth with increasing $t_w$. In appendix \ref{app:AppEternalAdS5}, similar features are found with shock wave geometries which are inserted into a  planar AdS$_5$ black hole spacetime.

We want to investigate the behaviour of figure \ref{FormationBTZShock} analytically in the case of a very light shock wave with $w=1+\epsilon$. We start by analyzing eq.~\eqref{FormationShockBTZ} in the limit where the shock wave enters at a very early time, \ie $T_2 t_w\gg1$. In eq.~\eqref{BTZ_Form_xsxw}, the coordinates $x_s$ and $x_b$  become
\begin{equation}
x_{s} = 1 + 2 e^{- 2 \pi T_1 t_w} + \mathcal{O}(\eps e^{- 2 \pi T_1 t_w}, e^{- 4 \pi T_1 t_w} ) \, ,  \qquad x_b = 1 - 2 e^{- 2  \pi T_1 t_w} + \mathcal{O}(e^{- 4 \pi T_1 t_w } ) \, .
\end{equation}
In this limit, the leading order behaviour of eq.~\reef{FormationShockBTZ} reduces to
\begin{equation}\label{CoFBTZApprox}
\frac{\Delta \ca-\Delta\mC_\mt{NS}}{|\Delta\mC_\mt{NS}|} = L T_1 \log \left( \frac{(w x_s -1)(w - x_b)}{(x_s -1)(1 -  x_b)} \right) + \mathcal{O} \left( \epsilon,\, e^{- 2 \pi T_2 t_w} \right) \, .
\end{equation}
Now there are two interesting regimes to consider: $\eps\ll 2 e^{- 2 \pi T_2 t_w} $ and $\epsilon\gg 2 e^{- 2 \pi T_2 t_w} $. Of course, the transition between these two regimes occurs when $\epsilon \approx 2 e^{-2 \pi T_1 t_w}$, \ie when
$t_w\approx \frac{1}{2\pi T_1}\log(2/\eps)=t^{*}_{\mt{scr}}$ using eq.~\eqref{Scramblingt}.
That is, the transition occurs when the perturbation of the thermofield double state is made approximately one scrambling time before the complexity of formation is evaluated!

 In the first regime, we can simply approximate $w \approx 1$ in the argument of the $\log$ in eq.~\eqref{CoFBTZApprox}, and as a consequence, the latter becomes
\begin{equation}
\frac{\Delta \ca-\Delta\mC_\mt{NS}}{|\Delta\mC_\mt{NS}|}  =  \mathcal{O} \left( e^{- 2 \pi T_2 t_w} \right) \,
\end{equation}
where  we have omitted order $\epsilon$ corrections because by assumption they were smaller than the exponential. This is the regime where the complexity of formation is essentially the same as the unperturbed geometry in figure \ref{FormationBTZShock}.

In the second regime with $\epsilon\gg e^{- 2 \pi T_2 t_w} $, the denominator of the $\log$ in eq.~\eqref{CoFBTZApprox} becomes the dominant part, with
\begin{equation}
\frac{\Delta \ca-\Delta\mC_\mt{NS}}{|\Delta\mC_\mt{NS}|} = 2L T_1 \left[2\pi T_1t_w + \log \left( \frac\epsilon2 \right)  \right]+ \mathcal{O} (\epsilon) \, .
\end{equation}
Hence this second regime is where $\Delta\ca$ grows linearly with $t_w$ in figure \ref{FormationBTZShock}. Using the expressions in eq.~\reef{BTZquantities} (with $\Omega_{1,1}=2\pi $ and $2M_1=S_1 \,T_1$) and the scrambling time in eq.~\eqref{Scramblingt}, as well as
$|\Delta\mC_\mt{NS}|=c/3$, we can rewrite the last result as
\beq
\Delta \ca=\Delta\mC_\mt{NS} + \frac{4M_1}\pi \left( t_w - t^*_\text{scr}\right) + \mathcal{O} (\epsilon) \, .
\label{latelate2}
\eeq

Hence we can approximate the complexity of formation in both regimes with the following simple expression:
\begin{equation}\label{FormBTZLightSh}
\Delta \ca \simeq \Delta\mC_\mt{NS} + \Theta \left( t_w - t^{*}_{\text{scr}}\right)  \, \frac{4 M_1}{\pi} \left( t_w- t^{*}_{\text{scr}}\right)     \, .
\end{equation}

\begin{figure}
\centering
\includegraphics[scale=0.6]{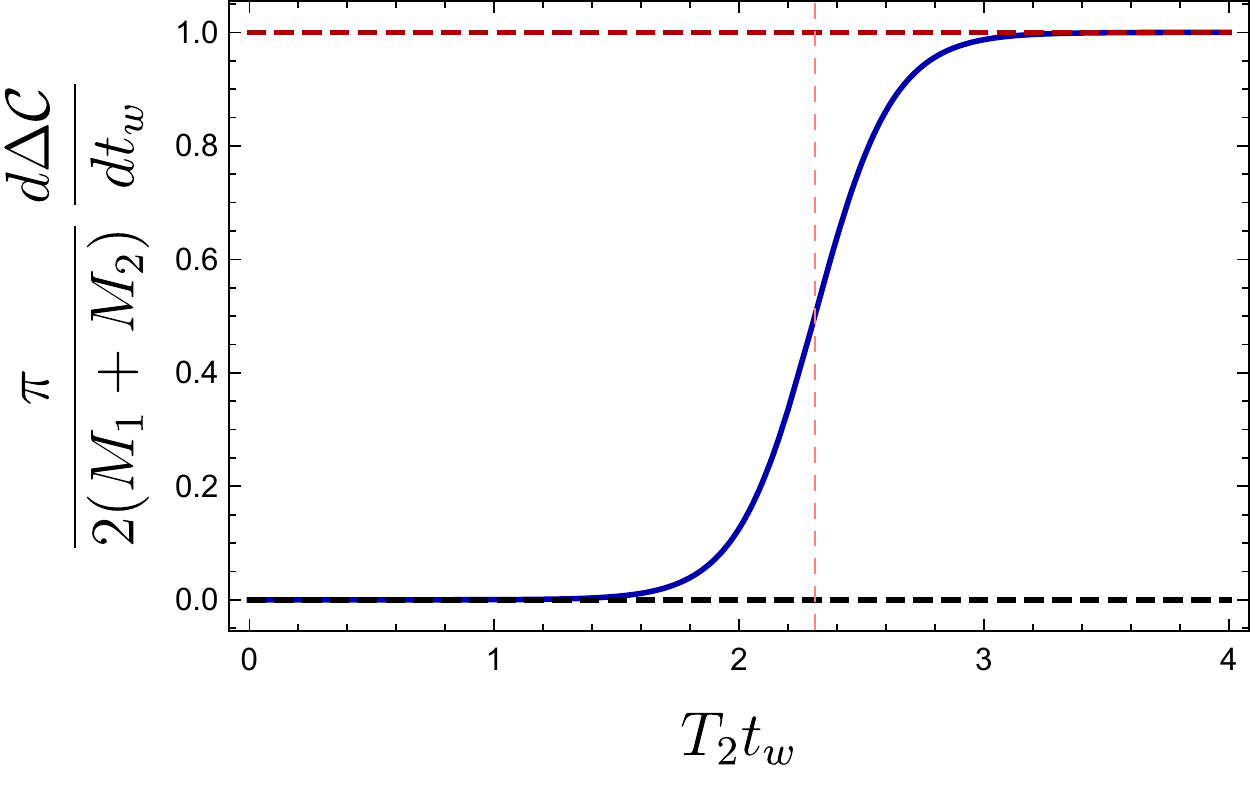}
\includegraphics[scale=0.6]{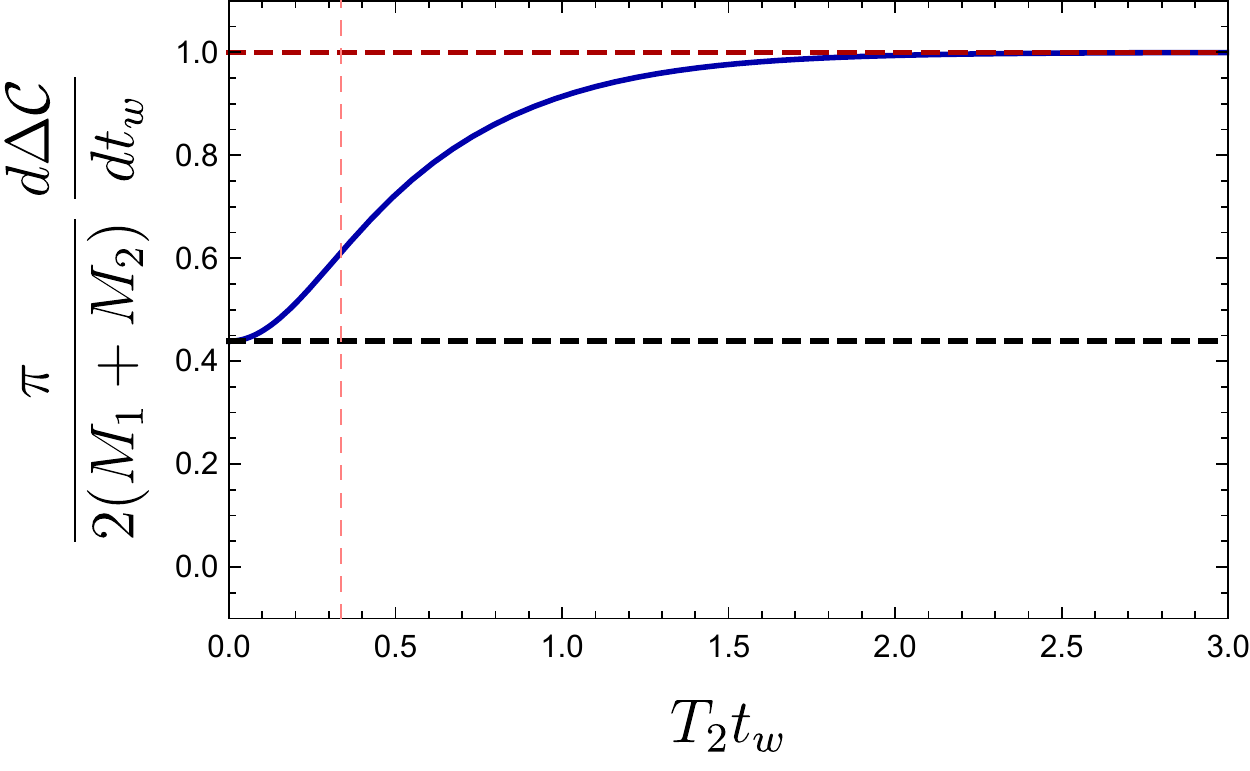}
\caption{The derivative of the complexity of formation with respect to $t_w$. The black dashed line is the expected slope at $t_w=0$, and the dashed red line is the slope at large $t_w$ (normalized in the plot to approach $1$). For both panels, we adopt $z=1/w$. The left panel illustrates the behaviour for a light shock wave, with $w=1 +10^{-6}$. In this regime, described by eq.~\eqref{FormBTZLightSh}, the slope is approximately zero until $t\simeq t^{*}_{\text{scr}}$ (vertical dashed line), at which point it rapidly rises to the final constant value, $4M_1/\pi$. The right panel illustrates the behaviour for a heavy shock wave, with $w=2$. In this regime, the slope starts at $\gamma_0$ in eq.~\reef{SlopeFormEarly} and rapidly rises to the final constant value $2(M_1+M_2)/\pi$. In this case, the vertical dashed line indicates $t_w=t_\text{del}$ from eq.~\reef{laterlate}.}
\label{FormationBTZShockDer}
\end{figure}
\begin{figure}
\centering
\includegraphics[scale=0.6]{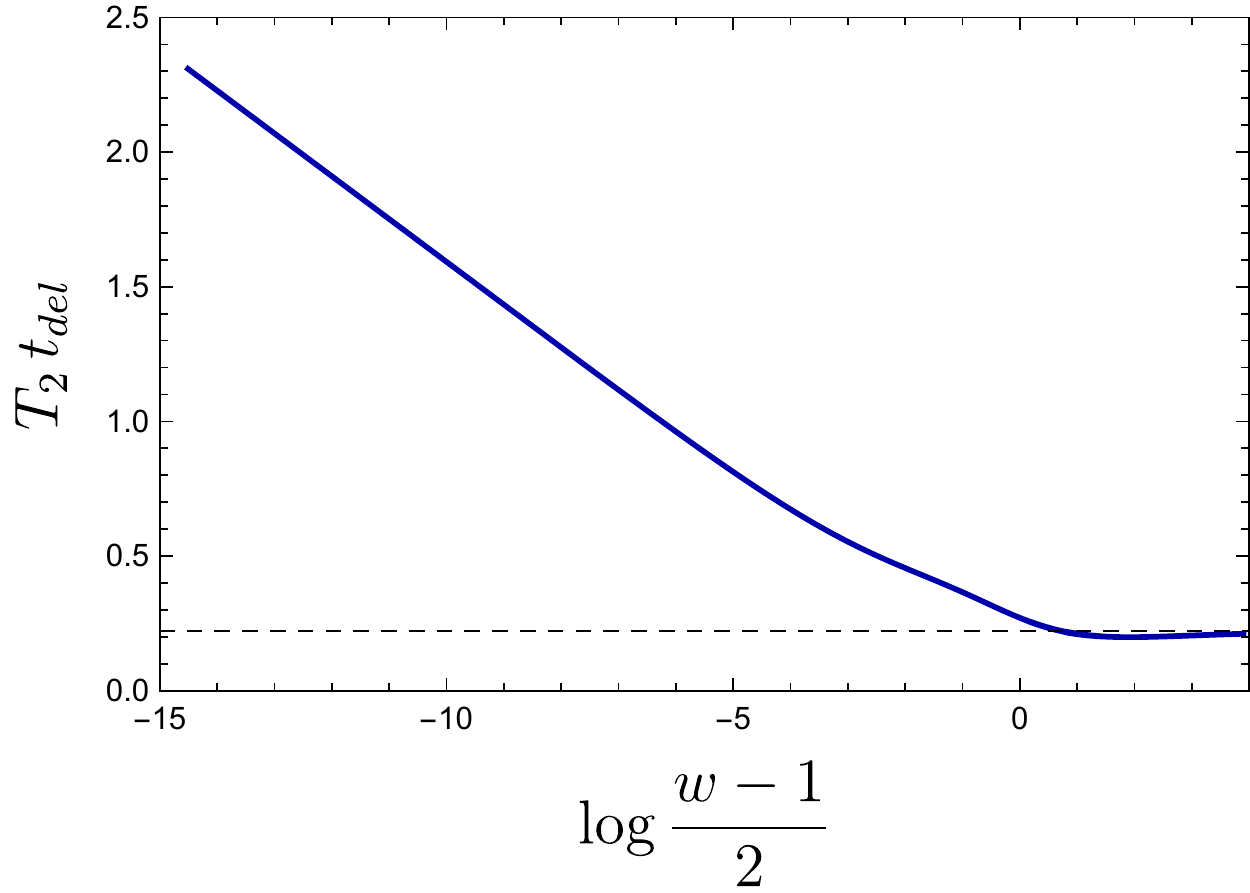}
\caption{The `delay' time \eqref{laterlate} as a function of the energy in the shock wave. For $w\sim1$, we have a line of slope --1, which is characteristic of the scrambling time in eq.~\eqref{Scramblingt}, as expected from eq.~\reef{newark2}. For heavy shock waves, $t_\text{del}$ approaches a constant proportional to $1/T_2$, as shown in eq.~\reef{newark}. }
\label{twInterceptBTZ}
\end{figure}

For the heavy shock waves, \ie (relatively) large $w$, we see from the right panel of figure \ref{FormationBTZShock} that $\Delta\ca$ begins increasing immediately as $t_w$ increases from zero. From  eq.~\eqref{FormationShockBTZ}, we can evaluate the slope of this increase,
\beq
\frac{d \Delta \ca}{d t_w} \bigg|_{t_w \rightarrow 0^{+}} = \frac{1}{\pi} \left( M_2 -M_1 + M_1 \log \left[ \frac{M_2}{M_1} \right] \right) \equiv \gamma_0 \, . \label{SlopeFormEarly} \eeq
Of course, this result vanishes for light shock waves, with $M_2 \approx M_1$, and we recover the result in eq.~\eqref{FormBTZLightSh}.
We can also use  eq.~\eqref{FormationShockBTZ} to determine the linear growth for larger values of $t_w$,
\beq
\Delta \ca  \simeq\Delta\mC_\mt{NS} + \frac{2}{\pi} \left( M_2 +M_1 \right) (t_w - t_\text{del}) +{\cal O}\left(t_w e^{- t_w} \right) \label{SlopeFormLate}
\eeq
where the generalized `delay' time can be written as
\begin{equation}
t_{\text{del}}=  \frac{w+1}{2 \pi T_2\, (w^2+1)} \left[  (2 w-1) \log\!\left(\frac2{w-1}\right)+\frac{2 w^2}{w+1}\,\log w+\log\!\left(\frac2{w+1}\right)\right]\, .
\label{laterlate}
\end{equation}
It is straightforward to see that this expression reduces to the scrambling time \reef{Scramblingt} when $w\to 1$. More precisely, with $w=1+\epsilon$ and $\epsilon\ll1$,
\beq
t_{\text{del}}= t^*_{\text{scr}} + \frac{1}{4 \pi T_1} \left( 1 + 5 \log \frac{2}{\epsilon} \right) \epsilon + \mathcal{O} ( \epsilon^2 \log \epsilon) \, .\label{newark2}
\eeq
Eq.~\reef{SlopeFormLate} provides an extension to general values of $w$ of eq.~\reef{latelate2}, which applies only for light shock waves (\ie $w\simeq 1$), and in this sense, $t_\text{del}$ replaces $t^*_{\text{scr}}$ for general shocks. Roughly, we can think of this time as characterizing when there is a transition between the early time behaviour given in eq.~\reef{SlopeFormEarly} and the late time behaviour given in eq.~\reef{SlopeFormLate}. As shown in figure \ref{FormationBTZShock}, this is a sharp transition between two distinct regimes for light shock waves, but not for the heavy shock waves. In the latter case, there is not an extended period of time where eq.~\reef{SlopeFormEarly} applies. In any event, considering $t_{\text{del}}$ for large values of $w$, we find to leading order
\begin{equation}
t_{\text{del}} = \frac{1}{\pi T_2}\left[\log2 - \frac{1}w\,\log w+\frac{1 + \log 2}{w}+ \mathcal{O} \left( w^{-2} \right)
 \right]\, ,\label{newark}
\end{equation}
that is, the delay time is simply a constant proportional to the inverse temperature of the final black hole, which is then a small time in the limit of large $w$.

Figure \ref{FormationBTZShockDer} shows the variation of $d \Delta \ca/d t_w$ for a light and a heavy shock wave. We see that for the light shock wave, the slope vanishes initially but then rapidly rises to the final constant value at $t\simeq t^{*}_{\text{scr}}$, corresponding to the two regimes shown in the left panel of figure \ref{FormationBTZShock} --- see also eq.~\eqref{FormBTZLightSh}. Instead, for the heavy shock wave, the slope is initially nonvanishing and proportional to $\gamma_0$ in equation \eqref{SlopeFormEarly} and rises quickly to the final constant value, again in agreement with the results shown in the right panel of figure \ref{FormationBTZShock}. Figure \ref{twInterceptBTZ} shows $t_\text{del}$ as a function of $\log\left(\frac{w-1}2\right)$. For $w\sim1$, a line of slope --1 appears since $t_\text{del}\simeq t^*_\text{scr}$, as shown in eq.~\reef{newark2}. For heavy shock waves, $t_\text{del}$ approaches a constant proportional to $1/T_2$, as shown in eq.~\reef{newark}.

\subsection{`Complexity' without the Counterterm} \label{CANoCT}

In this section, we turn our attention to the effects of dropping the counterterm \reef{counter} from the bulk action \reef{sumAct}. For stationary spacetimes, the WDW action does not seem to be effected in an important way if this surface term is not included.  However, in studying holographic complexity for the formation of a black hole \cite{Vad1}, we recently found that the counterterm is an essential ingredient for the CA proposal. The most dramatic effect of  dropping the counterterm was found for $d=2$ (and $k=+1$) where, without the counterterm, the holographic complexity actually  decreased throughout the black hole formation process and the rate of change only approached zero for asymptotically late times.
In the following, we show that without the counterterm, the holographic calculations fail to reproduce the expected late time growth rate and that the complexity of formation in $d=2$ does not exhibit the behaviour that is characteristic of the switchback effect.

Note that without the counterterm \reef{counter}, we must deal with the ambiguities associated with the surface and joint terms on the null boundaries of the WDW patch. We follow the standard prescription proposed in \cite{RobLuis} where we set $\kappa=0$ by choosing affine parametrization for the null normals. Further, we fix the overall normalization of these null vectors with $\hat t  \cdot k = \pm \alpha$ (where $\hat t$ is the asymptotic Killing vector producing time flow in the boundary). Of course, we have already adopted these conventions in the previous sections and so it is straightforward to simply drop the counterterm contributions in eqs.~\eqref{count1}-\eqref{count3} (and implicitly, also eq.~\reef{count05}) from the previous analysis. We tentatively denote the resulting quantity as `complexity' \ie
\beq
\tca=\frac{I_\mt{WDW}-I_\mt{ct}}{\pi}\,,\label{laughter}
\eeq
but as in \cite{Vad1}, we will find  that  this gravitational observable fails to behave in the manner expected of complexity.

\subsubsection{Time Evolution}

Here, we evaluate the growth rate of $\tca$ for  the three different regimes described in eq.~\eqref{ActionRegimes}. First, in regime $\RN{1}$ (\ie $-t_{c0}< t < t_{c 1}$), the total rate of change of complexity only receives contributions from the bulk term in eq.~\eqref{eq:BulkActEtSym} (with $r_m=0$), and from the GHY surface terms in eqs.~\eqref{GHFut} and \eqref{GHPast} at the past and future singularities, respectively. The growth rate then becomes
\begin{align}
\frac{d  \mathcal{\tilde C}_A^{(\RN{1})} }{d t} =& - \frac{M_2}{(d-1) \pi} \frac{r_s^d}{\omega_2^{d-2} L^2}  \left( 1- \frac{f_2 (r_s)}{f_1 (r_s)} \right) +\frac{M_1}{(d-1) \pi} \frac{r_b^d}{\omega_1^{d-2} L^2}   \left( 1- \frac{f_1 (r_b)}{f_2 (r_b)} \right)   \, \nonumber \\
& + \frac{d M_2}{2 (d-1) \pi} \left( 1+ \frac{f_1 (r_b)}{f_2 (r_b)} \right) - \frac{d M_1}{2(d-1) \pi} \left( 1 + \frac{f_2 (r_s)}{f_1 (r_s)} \right) \, . \label{RateSymt1Noct}
\end{align}
In regime $\RN{2}$ (\ie $ t_{c 1} < t < t_{c 2}$), the WDW patch has lifted off of the past singularity and so in addition to the bulk contribution \eqref{eq:BulkActEtSym} and the GHY contribution \eqref{GHFut} from the future singularity, we also have the joint contribution at $r_m$ in eq.~\eqref{ICorner3}. Combining these then yields
\begin{align}
\frac{d  \mathcal{\tilde C}_A^{(\RN{2})} }{d t} =& - \frac{M_2}{(d-1) \pi} \frac{r_s^d}{\omega_2^{d-2} L^2}  \left( 1- \frac{f_2 (r_s)}{f_1 (r_s)} \right) +\frac{M_1}{(d-1) \pi} \frac{r_b^d}{\omega_1^{d-2} L^2}   \left( 1- \frac{f_1 (r_b)}{f_2 (r_b)} \right)   \, \nonumber \\
& \qquad + \frac{d M_2}{2 (d-1) \pi} \left( 1+ \frac{f_1 (r_b)}{f_2 (r_b)} \right) + \frac{(d-2) M_1}{2 (d-1) \pi} \left( 1 + \frac{f_2 (r_s)}{f_1 (r_s)} \right) \nonumber \\
&+ \frac{M_1}{\pi (d-1) } \frac{r_m^{d-1} r_s}{\omega_{1}^{d-2} L^2} \left[   \left( 1 - \frac{f_2 (r_s)}{f_1 (r_s)} \right) + \frac{(d-1) L^2}{2 r_s^d}  \left( \omega_2^{d-2} - \omega_1^{d-2} \, \frac{f_2 (r_s)}{f_1 (r_s)} \right)  \right]  \nonumber \\
&+ \frac{M_1}{2 \pi} \frac{r_m^{d-2}}{\omega_1^{d-2}} f_1 (r_m) \left( 1 + \frac{f_2 (r_s)}{f_1 (r_s)}  \right) \, \log \left[ \frac{|f_1 (r_m)|}{\alpha^2} \, \frac{f_2 (r_s)}{f_1 (r_s)} \right]  \, . \label{RateSymt2Noct}
\end{align}
In the final regime $\RN{3}$ (\ie $t > t_{c 2}$), the relevant contributions are: the bulk term given by  eq.~\eqref{eq:BulkActEtSym} (with $r_b =0$), the GHY contribution from the future singularity given by eq.~\eqref{GHFut_Norway} and the joint term at $r_m$ given by eq.~\eqref{ICorner3}. The rate of change of the complexity in this regime is
\begin{align}
\frac{d  \mathcal{\tilde C}_A^{(\RN{3})} }{d t} =& - \frac{M_2}{(d-1) \pi} \frac{r_s^d}{\omega_2^{d-2} L^2}  \left( 1- \frac{f_2 (r_s)}{f_1 (r_s)} \right) + \frac{d(M_2 + M_1)}{2 (d-1) \pi}  + \frac{(d-2) M_1}{2 (d-1) \pi} \left( 1 + \frac{f_2 (r_s)}{f_1 (r_s)} \right)   \, \nonumber \\
&+ \frac{M_1}{\pi (d-1) } \frac{r_m^{d-1} r_s}{\omega_{1}^{d-2} L^2} \left[   \left( 1 - \frac{f_2 (r_s)}{f_1 (r_s)} \right) + \frac{(d-1) L^2}{2 r_s^d}  \left( \omega_2^{d-2} - \omega_1^{d-2} \, \frac{f_2 (r_s)}{f_1 (r_s)} \right)  \right]  \nonumber \\
&+ \frac{M_1}{2 \pi} \frac{r_m^{d-2}}{\omega_1^{d-2}} f_1 (r_m) \left( 1 + \frac{f_2 (r_s)}{f_1 (r_s)}  \right) \, \log \left[ \frac{|f_1 (r_m)|}{\alpha^2} \, \frac{f_2 (r_s)}{f_1 (r_s)} \right]  \, . \label{RateSymt3Noct}
\end{align}

\noindent{\textbf{Early and late time behaviours}}

Of course, the critical times in the time evolution depend only on the background geometry and so these basic features in the time evolution remain unchanged if we choose to study $\tca$, without the counterterm contributions. However, if the shock wave was injected early enough, there were two clear plateaus in $d\ca/dt$ (which included the counterterm contributions), given by eqs.~\eqref{EarlyTimesRate} and \eqref{LateTimeEtShock}. So we examine to see to what extent these plateaus arise for $d\tca/dt$.

The first plateau is found in the regime of large $t_w$, such that $r_s$ is very close to $r_{h,2}$ and $r_b$ to $r_{h,1}$. In this limit, the growth rate in eq.~\eqref{RateSymt1Noct} becomes
\begin{align}
\frac{d  \tca^{(\RN{1})} }{d t}   \bigg{|}_{t_w \rightarrow \infty} &= \frac{d-2}{2 (d-1) \pi} \left( M_2 - M_1 \right)  + \frac{k z^2}{(d-1) \pi}\left( \frac{M_2}{1+ k z^2} - \frac{w^2 \,M_1}{1+ k w^2 z^2} \right) \nonumber \\
&\qquad\qquad   + \mathcal{O} \left( e^{-\pi T_1 (2 t_w -t)} \right) \, . \label{EarlyLimitNoCT}
\end{align}
Comparing to eq.~\eqref{EarlyTimesRate}, we see that here we also have a similar plateau with the rate being proportional to $(M_2-M_1)$, at least for $k=0$, but in general there are curvature corrections to this result. Further note that for the BTZ black hole (\ie $d=2$), the time derivative is always zero, irrespective of the shock wave energy
\beq
\frac{d  \tca^{(\RN{1})} }{d t}   \bigg{|}_{d=2} = 0 \, ,   \label{EarlyLimitBTZNoCT}
\eeq
because $k$ does not play a role in the BTZ geometries.

The late time limit, analogously to that in eq.~\eqref{LateTimeEtShock}, is approached as $r_s$ is close to $r_{h,2}$ and $r_m$ close to $r_{h, 1}$. However, without the inclusion of the counterterm, there are further considerations if the shock wave is light. If we denote the ratio of horizons as $w = 1 + \epsilon$, with $\epsilon$ small, there are two regimes to consider, that depend on a time scale related to the scrambling time in eq.~\eqref{Scramblingt}, defined as
\begin{equation}
\hatt = \frac{1}{\pi T_1} \log \frac{2}{\epsilon} - 2  t_w =  2 t^{*}_{\text{scr}} - 2 t_w \, .
\label{ScramblingDiff}
\end{equation}
If the late time regime is such that $t > \hatt$, then we can evaluate eq.~\eqref{RateSymt3Noct} for $x_m$ and $x_s$ approaching $1$, which yields
\begin{align}
\frac{d  \tca^{(\RN{3})} }{d t}   \bigg{|}_{t \rightarrow \infty,\,  t > \hatt }&= \frac{M_1}{\pi}\left(1+\frac{w(d+1)}{2(d-1)}\right)   + \frac{d-2}{2  (d-1)}\, \frac{M_2}\pi  + \frac{kz^2}{(d-1)  (1 + k z^2)}\,\frac{ M_2}\pi    \nonumber \\
& - \frac{kz^2\,w}{2} \,\frac{(d+1) w^2  - (d-1) }{(d-1)  (1 + k w^2 z^2)}
\,\frac{M_1}\pi + \mathcal{O} \left( T_2 t e^{- \pi T_1 (t - 2 t_w)} \right)  \, . \label{LateLimitNoCT}
\end{align}
In contrast to eq.~\eqref{LateTimeEtShock}, the late time rate here is not proportional to the expected sum of the masses, even for the planar horizons ($k=0$) or in the limit of light but still non-zero shocks ($w\sim1$). For simplicity, let's rewrite eq.~\eqref{LateLimitNoCT} for planar black holes ($k=0$) and light shocks, such that $M_2 \approx M_1$. The late time limit then reads
\beq
\frac{d  \tca^{(\RN{3})} }{d t}   \bigg{|}_{t \rightarrow \infty, \,  t > \hatt} = \frac{2 M_1}{\pi}\left(1+\frac{1}{4(d-1)}  \right)  + \mathcal{O} \left(\frac{T_2 t}{w-1} \, e^{- \pi T_1 (t - 2 t_w)} \right)  \, . \label{LateLimitNoCTW1}
\eeq

If one wants to consider a shock wave with exactly zero energy, such that $w=1$, then $\hatt$ given by eq.~\eqref{ScramblingDiff} goes to infinity, which is equivalent to a regime where $t< \hatt$. This is equivalent to setting $w =1$ in eq.~\eqref{RateSymt3Noct}, which simplifies to
\begin{equation}
\frac{d  \mathcal{\tilde C}_A^{(\RN{3})} }{d t} \bigg{|}_{w =1} = \frac{2 M_1}{\pi} + \frac{M_1}{2 \pi} \frac{r_m^{d-2}}{\omega_1^{d-2}} f_1 (r_m)  \, \log \left[ \frac{|f_1 (r_m)|}{\alpha^2} \, \right]    \, ,
\end{equation}
and is simply the rate of change of the eternal black hole discussed in \cite{Growth}. This demonstrates that the order of limits does not commute.

In addition, the heavy shock wave regime of the rate of change given by eq.~\eqref{LateLimitNoCT} can be calculated by considering the limit $w \rightarrow \infty$. The rate of change becomes then, for $k=0$ for simplicity,
\begin{equation}
\frac{d  \mathcal{\tilde C}_A^{(\RN{3})} }{d t} \bigg{|}_{w \rightarrow \infty, t \rightarrow \infty} =  \frac{M_2}{2 \pi} \left( \frac{(d-2)}{ (d-1)} + \mathcal{O} \left( \frac{T_2 t }{w^{d-1}} \, e^{- \pi T_1 (t - 2 t_w)} \right) \right) \, ,
\end{equation}
which as expected is half of the one sided collapse value without the inclusion of the counterterm in eq.~($3.29$) of \cite{Vad1}, and it is vanishing for BTZ (d=2).

Consider as an example the BTZ black hole, with $k=0$ and $d=2$. The late time regime for $t> \hatt$ reads
\begin{equation}
\frac{d  \tca^{(\RN{3})} }{d t}   \bigg{|}_{d  = 2,\, t \rightarrow \infty} = \frac{1}{ \pi} \left( M_1 + \frac32 \sqrt{{M_2}{M_1}} \right) +  \mathcal{O} \left( \frac{T_2 t}{w-1} \, e^{- \pi T_1 (t - 2 t_w)} \right) \, ,\label{LateLimitBTZNoCT}
\end{equation}
where again we substituted $w^2=M_2/M_1$ for $d=2$. This expression again fails to produce the expected late time limit but we also see an unusual nonlinear dependence on the masses, \ie $\sqrt{M_2M_1}$.

\begin{figure}
\centering
\includegraphics[scale=0.8]{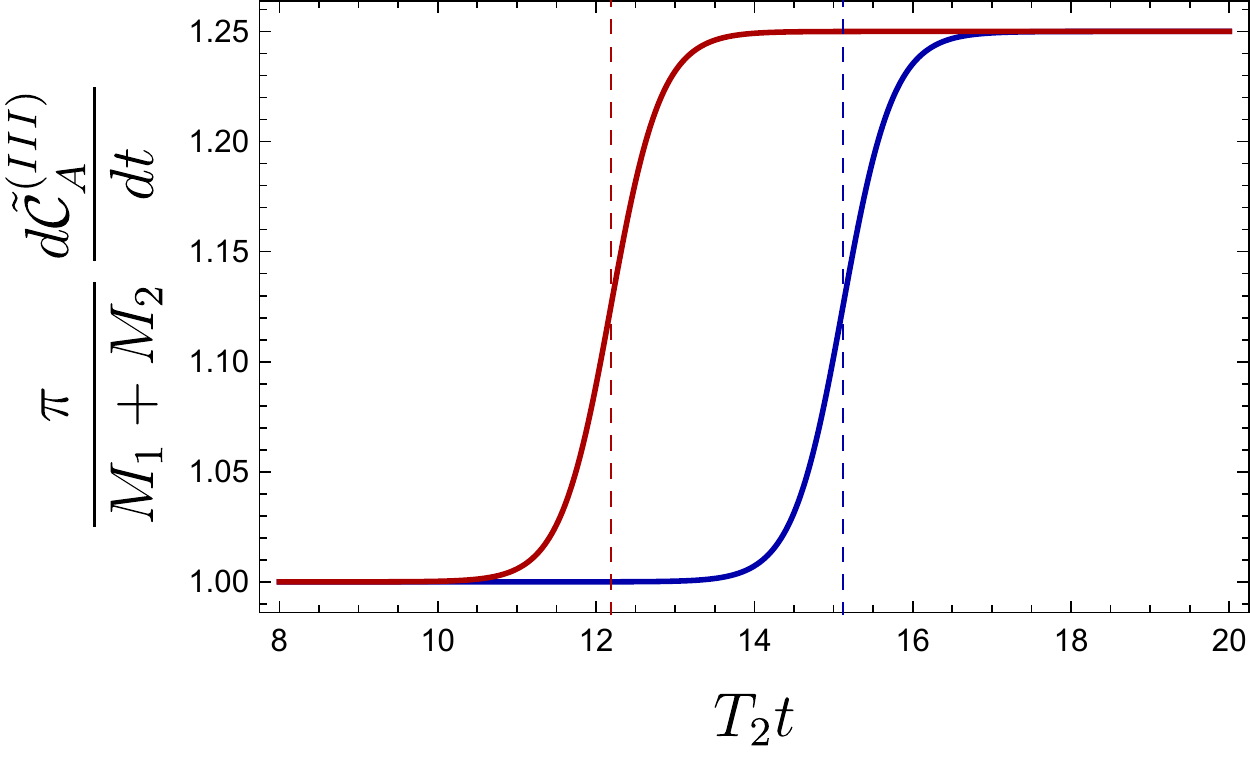}
\caption{Rate of change of complexity for BTZ ($d=2$) black hole without the addition of the counterterm, given by eq.~\eqref{RateSymt3Noct}. We study the transition between late time rates for very light shock waves, with $w = 1 + 10^{-33}$ (red) and $w = 1 + 10^{-37}$ (blue). The vertical lines represent the characteristic transition times $\hatt$, given by eq.~\eqref{ScramblingDiff}. For light but non-zero shock waves, the late time limit is similar to the eternal black hole for $t <   \hatt$, but for $ t> \hatt$, it becomes the rate in eq.~\eqref{LateLimitBTZNoCT} (with $M_1 \approx M_2$).}
\label{LateTimeNoCTT2tw6}
\end{figure}
In figure \ref{LateTimeNoCTT2tw6}, we investigate these limits for a very light but non-zero shock wave. We show the growth rate for BTZ black holes without the addition of the counterterm given by eq.~\eqref{RateSymt3Noct}, with $w = 1 + 10^{-33}$ (red) and $w = 1 + 10^{-37}$ (blue) and the vertical line representing $ \hatt $ as in eq.~\eqref{ScramblingDiff}. For times which are large but smaller than $\hatt$, the effective late time limit is the same as the eternal black hole, but at $ t \simeq\hatt $, there is a sharp transition to the late time limit of eq. \eqref{LateLimitBTZNoCT} (with $M_1 \approx M_2$).

\begin{figure}[b]
\centering
\includegraphics[scale=0.6]{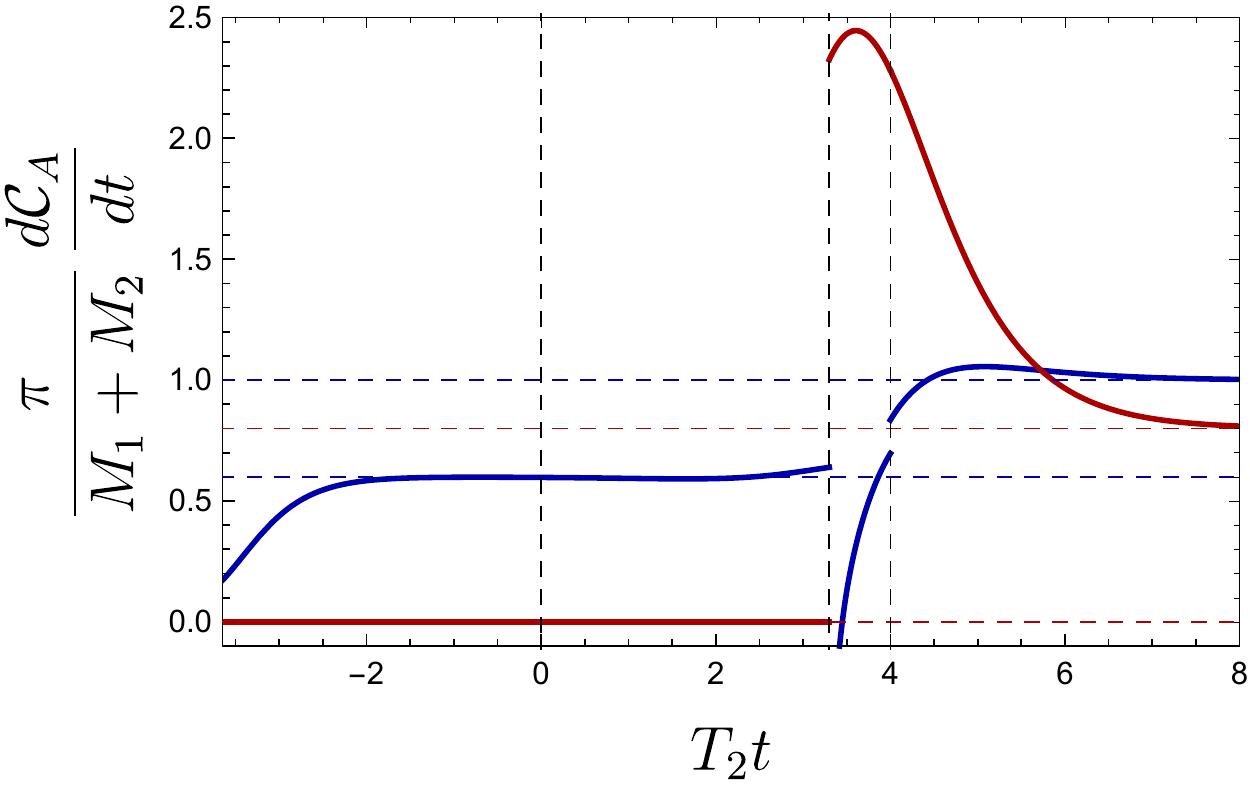}
\includegraphics[scale=0.6]{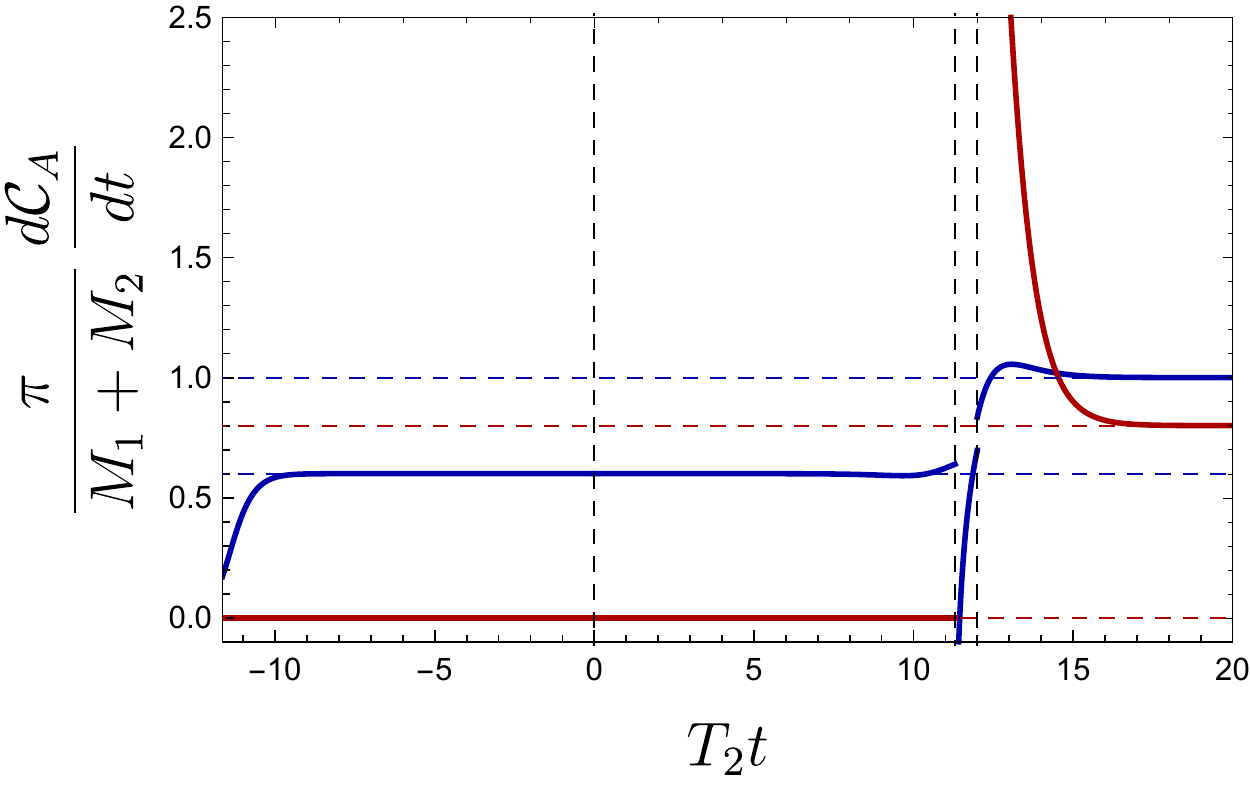}
\caption{ Rate of change of complexity evolving both boundaries as $\tL=\tR=\frac{t}{2}$,  $w=2$, $z=1/w$, $\ttL= 1 $ and $ \alpha =1$. In the left we evaluate  $T_2 t_w = 2$ and in the right  $T_2 t_w = 6$, and the blue curve is the rate of change with the inclusion of counterterm, while the red line is the rate of change without it.  Despite being a shock wave that doubles the temperature, the rate of change is exactly zero without the inclusion of the counterterm for $t_{c0}< t< t_{c1}$, as opposed to being proportional to the difference of masses. In addition, there is a large positive peak after $t_{c1}$ for the red curve, in contrast to the (short) negative spike of the blue curve. The peak in the red curve in the right figure is similar to the one in the left, but is sharper and reaches higher values the earlier the shock wave is sent.
Finally, the late time limit is given by eq.~\eqref{LateLimitBTZNoCT}, in contrast to $(M_1 + M_2)/\pi$, as discussed in section \ref{CAshock}.
}
\label{NewBTZw2BOTH}
\end{figure}
We compare the rate of change for BTZ with and without the inclusion of the counterterm in figure \ref{NewBTZw2BOTH}. We focus on heavy shock waves ($w=2$), since for early times (before $t_{c1}$) there is a bigger discrepancy of rates, \ie vanishing without the counterterm or proportional to $M_2 - M_1$ with the counterterm. In addition, immediately after $t_{c1}$ there is a large positive peak in the rate without counterterm, due to a factor of $f_{2}(r_s)$ in eq.~\eqref{RateSymt2Noct}, which approaches zero much faster than $f_1 (r_m)$ for early shocks, since the exponent of $r_s$ approaching $r_{h,2}$ at late times is proportional to $T_2 (t + 2 t_w)$, while $r_m$ approaching $r_{h,1}$ is controlled by an exponent proportional to $T_1 (t - 2 t_w)$ for late times. Finally, the late time limit of the rate of change without the inclusion of the counterterm, as discussed previously, approaches the late time limit given by eq.~\eqref{LateLimitBTZNoCT}, which does not reduce to the eternal black hole result for light shock waves.

\newpage

\subsubsection{Complexity of Formation}

Turning now to the complexity of formation \reef{CoFdef}  but evaluated with eq.~\reef{laughter}, \ie without  the counterterm contribution \reef{FormEternalSCt}. We consider the BTZ black hole (\ie $d=2$) as a simple example. In this case, our previous result \reef{FormationShockBTZ} is replaced with
\small
\begin{align}
\frac{\Delta \tca-\Delta\mC_\mt{NS}}{|\Delta\mC_\mt{NS}|}  &= L T_1 \bigg[ 2 \coth ^{-1}\!\left(w x_s\right)-\frac{2}{w}\left( w^2-1\right) \coth ^{-1}\!\left(x_s\right) +(w^2 - x_b^2) \log \!\left(\frac{1+x_b}{1 - x_b}\right) \nonumber \\
&\qquad\qquad-w \log\! \left(\frac{w+x_b}{w - x_b} \right)+\frac{2 x_b^2 }{w}\tanh ^{-1}\!\left(\frac{x_b}{w}\right) -\log\! \left( \frac{w x_s+1}{w x_s -1}\right) \bigg ] \, . \label{CoFNoCT}
\end{align}
\normalsize

\begin{figure}
\centering
\includegraphics[scale=0.8]{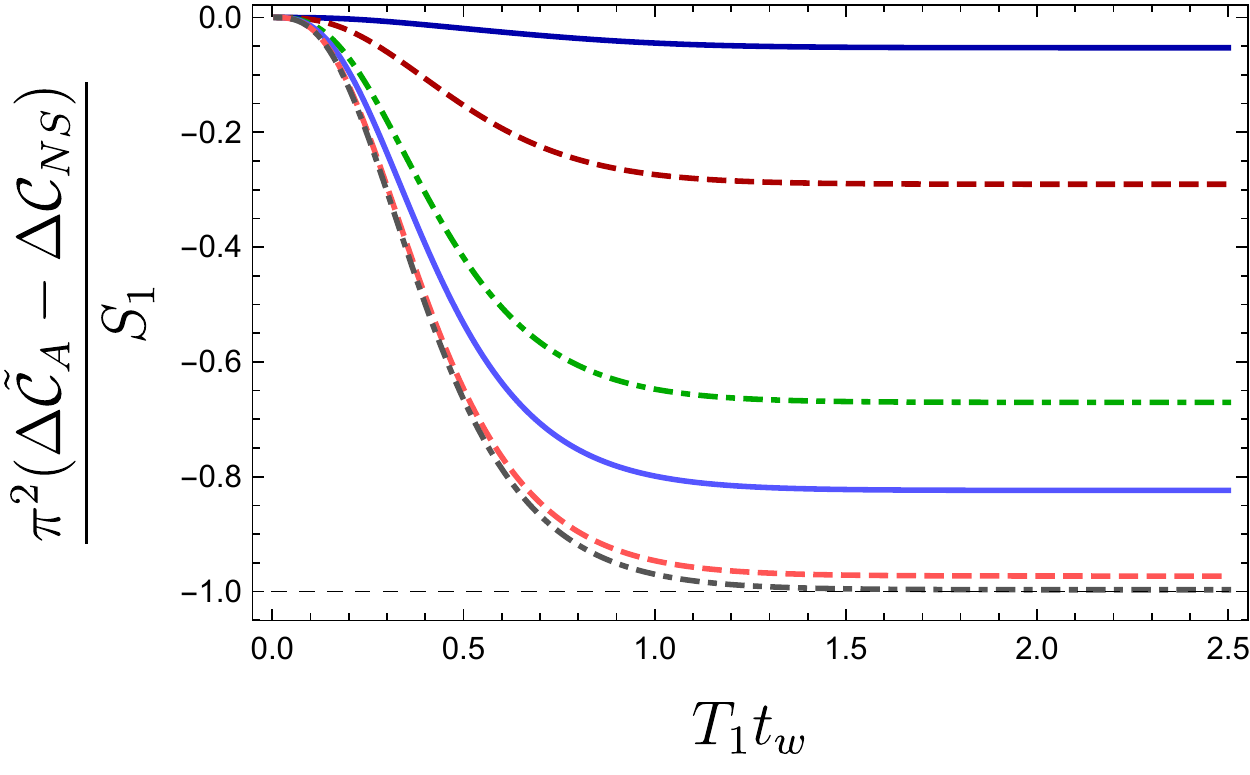}
\caption{ Complexity of Formation for BTZ black holes with no counterterm added to the null boundaries (and $\kappa=0$). From top to bottom, we consider light to heavy shock waves, with $w=1+10^{-2}$, $w=1+ 10^{-1}$, $w=1.5$, $w = 2$, $w = 5$ and $w=15$. For $T_1 t_w$ of order $1$, the complexity of formation with respect to the unperturbed one saturates to a constant, being close to zero for light shocks and $-S_1/\pi^2$ for heavy ones, represented by the horizontal black dashed line, as given by eq.~\eqref{gamam}. }
\label{FormationBTZKappaZero}
\end{figure}
We numerically evaluate eq.~\eqref{CoFNoCT} and plot the complexity of formation as a function of the insertion time for light and heavy shock waves in figure \ref{FormationBTZKappaZero}. For both light and heavy shock waves, the complexity of formation just approaches a constant value for large $t_w$. Further, the latter value is less than the original $\Delta\mC_\mt{NS}=-c/3$ found with $t_w=0$ (or alternatively, with no shock wave). Note that the transition to the final $\Delta\tca$ essentially saturates after $T_1 t_w\sim 1$. We can evaluate the large $t_w$ limit of eq.~\eqref{CoFNoCT} analytically to find
\begin{equation}
\frac{\Delta \tca-\Delta\mC_\mt{NS}}{|\Delta\mC_\mt{NS}|}   \bigg{|}_{t_w \rightarrow \infty} =  - L T_1 \, \frac{w^2 -1}{w} \, \log \left[ \frac{w + 1}{w -1} \right]   + \mathcal{O}\! \left( T_2 t_w e^{- 2 \pi T_1 t_w }\right) \, . \label{CoFNoCTLargetw}
\end{equation}
For light shock waves, the above difference vanishes as $w\to1$, \ie $\Delta \tca\to\Delta\mC_\mt{NS}$. On the other hand, for very heavy shocks (\ie $w \to\infty$), we find
\beq
\Delta \tca\big|_{w\to\infty}=\Delta\mC_\mt{NS}-S_1 /\pi^2\,.
\label{gamam}
\eeq
That is, for heavy shocks injected at early times, the complexity of formation decreases and it does so in a way that only depends on the initial black hole and not on the final black hole.

Of course, these results for $\Delta\tca$ contrast with the previous results (including the counterterm) in figure \ref{FormationBTZShock}, where the complexity of formation began to grow linearly for large $t_w$ for both light and heavy shock waves. Therefore, if one studies the complexity of formation without the inclusion of the counterterm, there is no dependence on the scrambling time or on how early the shock wave was inserted. As we discuss in section \ref{sec:Discussion}, this means that $\tca$ fails to exhibit the switchback effect for $d=2$. Of course, this only strengthens the argument that this gravitational observable cannot be interpreted in term of complexity in the boundary theory.

\section{Complexity = Volume}\label{sec:VolShocks}

\begin{figure}
\centering
\includegraphics[scale=1.5]{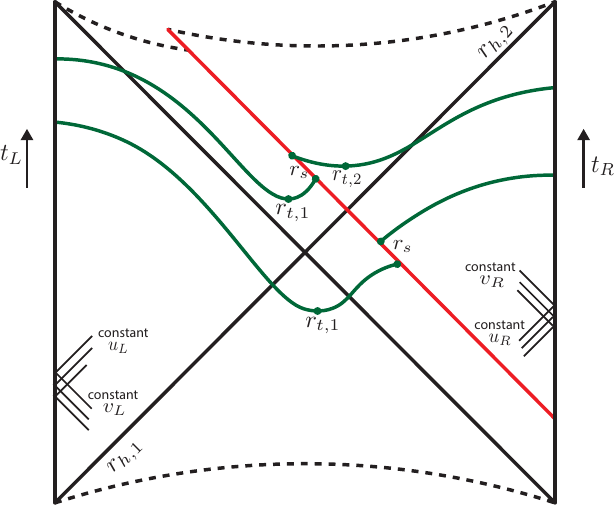}
\caption{Illustration of maximal volume surfaces in the shock wave geometry. Some special points are indicated in the drawing: $r_{t,1}$ the turning point in \BHO (if it exists), $r_{t,2}$ the turning point in \BHT (if it exists) and $r_s$ the point in which the maximal volume surface crosses the shell. We have also indicated in the figure lines of constant $u$ and $v$ induced from either boundary as well as the direction of the time coordinates on the left and right boundaries.}
\label{fig:drawing2}
\end{figure}

In this section, we apply the complexity=volume conjecture \reef{defineCV} to evaluate the rate of change of complexity and also the complexity of formation for the shock wave backgrounds described in section \ref{bkgd}. These geometries describe a transition between a black hole of mass $M_1$ and radius $r_{h,1}$ and a black hole of mass $M_2$ and radius $r_{h,2}$ caused by an incoming shock wave from the right boundary at $\vR=v_s=-t_w$. The relevant setup is illustrated in figure \ref{fig:drawing2}, which is again a `Penrose-like' diagram built in such a way that lines of constant $v$, $u$ and $t$ look continuous in the diagram while lines of constant $r$ do not. To simplify the discussion, we will denote the (black hole) region before the transition by \BHO and the one after the transition by \BHT in what follows. Note that in this section, we use $r_s$ to denote the point in which the maximal surfaces cross the shell. We will be interested in surfaces that are anchored on the asymptotic boundaries at times $\tL$ and $\tR$. The synchronization of the two boundary times was already explained in the paragraph above eq.~\eqref{eq:wz}. Much of the following discussion borrows from our analysis of CV complexity in \cite{Vad1} and we refer the interested reader there for further details.

To describe the full geometry, it will be sometimes useful to replace the coordinate $v$  with $u=v-2 r^*(r)$ (as in eq.~\reef{time3}).\footnote{It will also be convenient to distinguish the Eddington-Finkelstein coordinates, $\vR$ and $\uR$, specifying null rays ending on the right AdS boundary from $\vL$ and $\uL$, anchored to the left AdS boundary.} In each of the two spacetime regions, the metric \reef{MetricV} reads
\begin{equation}
ds^2
=  -f_i(r) dv^2 + 2 dr dv + r^2 d \Sigma_{k,d-1}^2
=
-f_i(r) du^2 - 2 dr du + r^2 d \Sigma_{k,d-1}^2
\label{jumper}
\end{equation}
with $f_i(r)$, the appropriate blackening factor given in eqs.~\eqref{fBH1} and \eqref{fBH2}. The maximal volume surfaces then extremize the following functional
\begin{equation}\label{actLL1}
\mathcal{V} = \Omega_{k,d-1} \int d \lambda\,r^{d-1} \sqrt{-f \dot{v}^2 +2 \dot{v}  \dot{r}}
= \Omega_{k,d-1} \int d \lambda\,r^{d-1} \sqrt{-f \dot{u}^2 -2 \dot{u}  \dot{r}}\, ,
\end{equation}
where taking advantage of the symmetry of boundary, these directions have been integrated out. Further, $\lambda$ is the  radial coordinate intrinsic to the surface, which will increase  along the surface moving from the left asymptotic AdS boundary to the right boundary, \ie from left to right in figure \ref{fig:drawing2}. We fix the reparametrization symmetry of the volume functional with the convenient gauge choice \cite{Vad1}:
\begin{equation}\label{ParamLL1}
\sqrt{-f \dot{v}^2 +2 \dot{v} \dot{r}}=\sqrt{-f \dot{u}^2 -2 \dot{u} \dot{r}}=r^{d-1}\,.
\end{equation}
In addition, to cover the full spacetime, we will need to work with the coordinates $\vR$, $\uR$ and $\vL$, $\uL$ induced from the right/left boundaries, respectively (see figure \ref{fig:drawing2}).
Recall from eqs.~\reef{rStar1} and \reef{rStar2}, we choose the constant of integration in the definition of the tortoise coordinate such that $r^*_\infty=0$. The boundary times $\tL$ and $\tR$ run upwards on their respective boundaries according to our conventions and hence on the left boundary, we have $\vL=\uL=-\tL$ while on the right boundary, $\vR=\uR=\tR$.

Of course, the metric profile \reef{heavyX} makes a sharp transition at $\vR=v_s$ but otherwise the geometry is independent of $v$. Hence in the \BHO and  \BHT regions separately, there is a conserved `$v$-momentum' $P_i$,
\begin{equation}\label{eq:consEVaid22}
\begin{split}
P_{i} =  \frac{\del \mathcal{L}}{\del \dot v}=\frac{r^{d-1}(\dot r- f_i(r)\,  \dot v)}{\sqrt{-f_i(r)\, \dot v^2 + 2 \dot v \dot r}} = \frac{r^{d-1}(-\dot r- f_i(r)\, \dot u)}{\sqrt{-f_i(r)\, \dot u^2 - 2 \dot u \dot r}},
\end{split}
\end{equation}
which with the above gauge choice takes the form
\begin{equation}\label{eq:consEVaid}
P_i = \dot r-f_i(r)\, \dot v= - \dot r - f_i(r)\, \dot u \,.
\end{equation}
Using  eqs.~\eqref{ParamLL1} and \eqref{eq:consEVaid} to solve for $\dot r$ and $\dot v$ (as well as $\dot u$), we find
\begin{equation}
\begin{split}
\dot r_{\pm}\left[P_i,r\right]  = &\, \pm {\sqrt{f_i(r)\,	 r^{2(d-1)} + \E_i^2}}, \\
\dot v_{\pm}\left[P_i,r\right]  = &\, \frac{\dot r-\E_i}{f_i(r) } = \frac{1}{f_i(r) } \left( -\E_i  \pm \sqrt{f_i(r)\,	 r^{2(d-1)} + \E_i^2} \right), \\
\dot u_{\pm}\left[P_i,r\right]  = &\, -\frac{\dot r+\E_i}{f_i(r) } = \frac{1}{f_i(r) } \left( -\E_i  \mp \sqrt{f_i(r)\,	 r^{2(d-1)} + \E_i^2} \right)\, .
\end{split}\label{onedot}
\end{equation}
Here, we have added the $\pm$ subscripts to indicate whether $r$ is increasing/decreasing as we move along the surface from left to right.

An intuitive picture of the dynamics of the extremal surfaces is given by recasting the $\dot r$ equation as a Hamiltonian constraint in each black hole region,
\begin{equation}
\dot  r^2 + U_{i}(r)   =  \E_{i}^2\,,
\label{yarnSho}
\end{equation}
where the effective potential is given by
\begin{equation}\label{PotentialShocks}
U_{i}(r)=-f_{i}(r)\, r^{2(d-1)}\, .
\end{equation}
In this framework, $P_{i}^2$ plays the role of the conserved energy. We will come back to this classical Hamiltonian picture when studying specific examples below.

Now the maximal surface is determined by fixing two boundary conditions, namely, the times $\tL$ and $\tR$ at which it is anchored on the two asymptotic AdS boundaries. However, it will be more convenient to trade these for fixing $P_1$, the conserved momentum in the \BHO region, and $r_s$, the radius where the surface crosses the shock wave.\footnote{To be precise, we also have to specify the sign of $\dot r_1(r_s)$ when the surface reaches the shell.} For given values of $P_1$ and $r_s$, we can obtain $\dot v(r_s)$ and $\dot r_1(r_s)$ by evaluating eq.~\eqref{onedot} at $r=r_s$ and with $i=1$:
\begin{equation}\label{eq:onedotSho}
\begin{split}
\dot  r_1(r_s) = &\, \pm {\sqrt{f_1(r_s)	 r_s^{2(d-1)} + \E_1^2}},
\\
\dot v_1(r_s)  =&\, \frac{1}{f_1(r_s) } \left( -\E_1  \pm \sqrt{f_1(r_s)	 r_s^{2(d-1)} + \E_1^2} \right).
\end{split}
\end{equation}
The second order equations of motion derived from eq.~\reef{actLL1} can be integrated in the vicinity of the shock wave to conclude that $\dot v$ is continuous across the shell (\ie $\dot v_2(r_s)=\dot v_1(r_s)$) while $\dot r$ undergoes a jump given by
\begin{equation}\label{eq:rdotJumpShock}
\dot r_2(r_s) = \dot r_1(r_s) + \frac{\dot v_1(r_s)}{2} \left(f_2(r_s)-f_1(r_s)\right)  \, .
\end{equation}
Of course, from eq.~\eqref{eq:consEVaid}, the conserved momentum must also jump when crossing the shell and we find
\begin{equation}\label{eq:P2JumpShock}
P_2 = \dot r_2(r_s)-f_2(r_s) \,\dot v_2(r_s) \, .
\end{equation}

There are a number of different scenarios to be considered for the shape of the surface, namely the surface may pass into the black/white hole region of \BHO and it may or may not admit a {\it turning point} (a point where $\dot r$ changes sign) in \BHO and/or in \BHT. The conserved momentum $P_1$ in \BHO is positive when the surfaces pass into the black hole part of spacetime and negative when the surfaces pass into the white hole region. The reason for this becomes obvious when expressing the conserved momentum in terms of an auxiliary time coordinate behind the horizon since in the black/white hole part this coordinate has to increase/decrease along the surface respectively. Of course, when $P_1=0$ the maximal surface in the \BHO side is a line of constant time  going through the bifurcation surface.
When turning points exist (where $\dot r=0$), their positions can be obtained from eq.~\reef{yarnSho} by solving
\begin{equation}\label{eq_shocks:rturn}
P_i^2 + f_i(r_{t,i}) r_{t,i}^{2(d-1)}=0\,,
\end{equation}
where $r_{t,1}$ denotes the turning point in the \BHO region (if it exists) and  $r_{t,2}$, the turning point in \BHT (if it exists).
Of course when passing in the white hole region of \BHO, we must at least admit a turning point in \BHO.

Given our boundary conditions $r_s$ and $P_1$, we are able to integrate out to the left boundary to determine $\tL$, while  with $P_2$ fixed by eq.~\reef{eq:P2JumpShock}, we can similarly determine $\tR$. Here we will simply present the expressions for the left and right boundary times, as well as the volume and its time derivative, for the  various cases described above. The interested reader is referred to appendix \ref{app:CVShocksDetails} for the full derivation. To simplify these expressions, we also define
\begin{equation}\label{tauR}
\tau \left[P,r\right] \equiv \frac{1}{f(r)}-\frac{P}{f(r)\sqrt{f(r) r^{2(d-1)} + P^2}}, \qquad
R \left[P,r\right] \equiv \frac{r^{2(d-1)}}{\sqrt{f(r) r^{2(d-1)} + P^2}}.
\end{equation}
and we will add subscripts ($\tau_{1,2}$, $R_{1,2}$) to specify which blackening factor (alternatively, which horizon radius) is being used.
For the right boundary time $\tR$, we obtain
\begin{equation}\label{tRshocksF}
\tR +t_w =
\begin{cases}
\int_{r_{t,2}}^\infty \tau_2 \left[P_2,r\right] dr - \int_{r_{t,2}}^{r_s} \tau_2\left[-P_2,r\right] dr & \qquad \text{turning point in \BHT}
\\
\int_{r_s}^\infty \tau_2 \left[P_2,r\right] dr & \qquad \text{otherwise.}
\end{cases}
\end{equation}
For the left boundary time $\tL$, we obtain
\footnotesize
\begin{equation}\label{tLshocksF}
\hspace{-15pt} \tL -t_w =
\begin{cases}
2 r_1^*(r_{t,1}) + \int_{r_{t,1}}^\infty \tau_1 \left[P_1,r\right] dr + \int_{r_{t,1}}^{r_s} \tau_1\left[P_1,r\right] dr
& {P_1>0 \text{ (black hole)},\atop \text{turning point in \BHO}}
\\
2 r_1^*(r_s)+\int_{r_s}^\infty \tau_1 \left[P_1,r\right] dr &
    {P_1>0 \text{ (black hole)},\atop \text{no turning point in \BHO}}
\\
 2 r_1^*(r_s) - 2 r_1^*(r_{t,1})
-\int_{r_{t,1}}^\infty \tau_1 \left[-P_1,r\right] dr - \int_{r_{t,1}}^{r_s} \tau_1\left[-P_1,r\right] dr
&  P_1<0 \text{ (white hole)}
\\
r_1^*(r_s) &  P_1=0.
\end{cases}
\end{equation}
\normalsize
We note that the integrals are well behaved near the black hole horizons due to a cancellation of logarithmic divergences coming from integrating separately the two parts of the expression for $\tau[P,r]$. We also note that  the only combinations of times which appear here are $\tR  +t_w$ and $\tL -t_w$. This is a consequence of the time-shift symmetry that was discussed in the introductory remarks (\ie see discussion after eq.~\reef{TFDPertState2}) and hence the same behaviour occurred in the CA subsection (\eg see  eq.~\eqref{eq:rwrsrm}).

The volume of the maximal surface will in general be given by the sum of the relevant volumes in \BHO and \BHT, which we denote by $\mathcal{V}_1$ and $\mathcal{V}_2$
\begin{equation}\label{VTshocksF}
\mathcal{V}=\mathcal{V}_1+\mathcal{V}_2
\end{equation}
where
\begin{equation}\label{VT2shocksF}
\begin{split}
\mathcal{V}_1= & \, \Omega_{k,d-1}
\begin{cases}
\int_{r_{t,1}}^{r_{\text{max}}} R_1 \left[P_1,r\right] dr + \int_{r_{t,1}}^{r_s} R_1 \left[P_1,r\right] dr & \qquad \text{turning point in \BHO}
\\
\int_{r_s}^{r_{\text{max}}} R_1 \left[P_1,r\right] dr & \qquad \text{otherwise}
\end{cases}
\\
\mathcal{V}_2 = & \,
\Omega_{k,d-1}
\begin{cases}
\int_{r_{t,2}}^{r_{\text{max}}} R_2 \left[P_2,r\right] dr + \int_{r_{t,2}}^{r_s} R_2 \left[P_2,r\right] dr & \qquad \text{turning point in \BHT}
\\
\int_{r_s}^{r_{\text{max}}} R_2 \left[P_2,r\right] dr & \qquad \text{otherwise}
\end{cases}
\end{split}
\end{equation}
and where $r_{\text{max}}$ indicates the position of the UV regulator surface.

\subsection{Time Evolution} \label{CVshock}

The time derivative of the volume admits a very simple form, common to all the cases above (see appendix \ref{app:CVShocksDetails} for the derivation)
\begin{equation}\label{VtshocksF}
\frac{1}{\Omega_{k,d-1}} \frac{d \mathcal{V}}{d t} = P_1 \frac{d \tL}{d t} + P_2 \frac{d \tR}{d t}
\end{equation}
where $t$ is a time parameter specifying the time evolution of the two boundary times according to $\tR(t)$, $\tL(t)$. In this way we can study general patterns for the time evolution, \eg symmetric $\tL=\tR$ and antisymmetric $\tL=-\tR$.  We will focus mainly on the case of symmetric boundary times with $\tR=\tL=t/2$ in which case we obtain for the complexity using eq. \eqref{defineCV} with $\ell=L$
\begin{equation}\label{CtshocksF}
\frac{d \cv}{d t} = \frac{\Omega_{k,d-1}}{2 G_N L} \left(P_1+P_2\right).
\end{equation}
Before we proceed, let us note that all the above results continue to be valid for $d=2$ when substituting the blackening factors with those of BTZ black holes.

\begin{figure*}[t!]
    \centering
      \begin{subfigure}[t]{0.31\textwidth}
        \centering
        \includegraphics[scale=0.8]{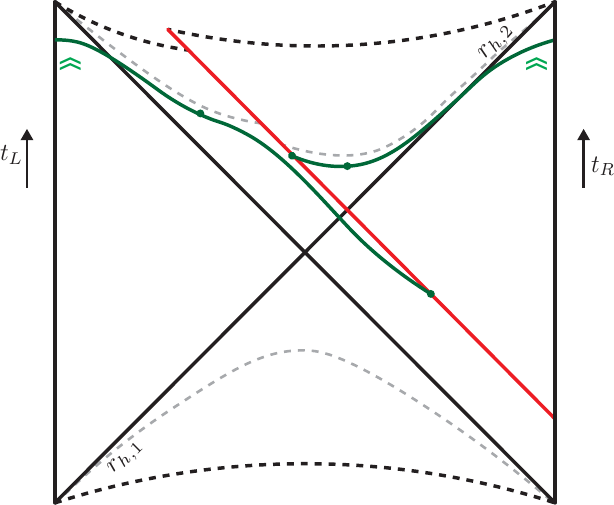}
        \caption{Late times}\label{s1}
    \end{subfigure}
    ~
    \begin{subfigure}[t]{0.31\textwidth}
        \centering
        \includegraphics[scale=0.8]{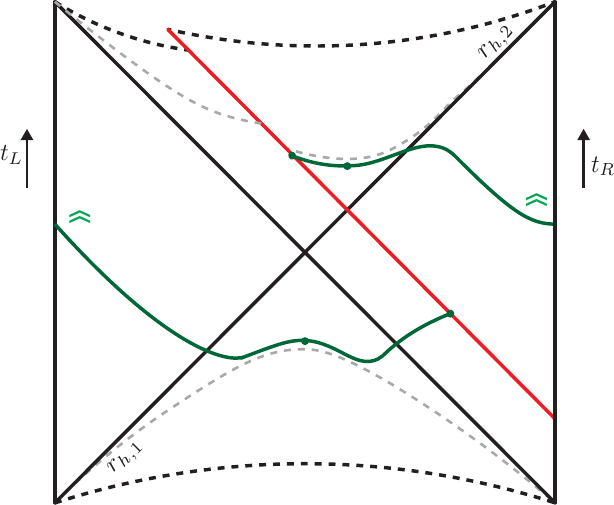}
        \caption{$t=0$ ledge}\label{s2}
    \end{subfigure}
    ~
    \begin{subfigure}[t]{0.31\textwidth}
        \centering
        \includegraphics[scale=0.8]{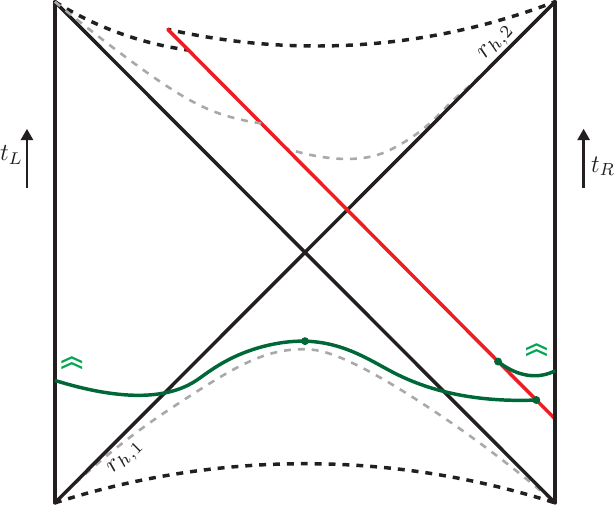}
        \caption{Early times}\label{s3}
    \end{subfigure}%
    \caption{The shape of the surface in the various limits discussed in eqs.~\eqref{lim1}, \eqref{lim2}, \eqref{lim3} for symmetric time evolution $\tL=\tR=t/2$ after an early shock.}
    \label{surfShapes}
\end{figure*}

{\bf Early and late time limits:}
The relation between the boundary times and the momentum parameters $P_1$ and $P_2$ is not given by a closed analytic expression and it requires some numerical treatment as we are about to describe below. However, we were able to make the  following general observations for  planar black holes, as well as BTZ black holes in $d=2$.
In the late time limit, the momenta $P_1$ and $P_2$ correspond to the maximal values of the potentials \eqref{PotentialShocks} according to
\begin{equation}
\begin{split}
&\del_r \left[f_1(r_{1,m}) r_{1,m}^{2(d-1)} \right] = 0, \qquad
f_1( r_{1,m})  r_{1,m}^{2(d-1)} + P_{1,m}^2 = 0,
\\
&\del_r \left[f_2(r_{2,m}) r_{2,m}^{2(d-1)} \right] = 0, \qquad
f_2( r_{2,m})  r_{2,m}^{2(d-1)} + P_{2,m}^2 = 0,
\end{split}
\end{equation}
and this means that the surface extends along the critical surfaces of constant $r=r_{1,m}$ in \BHO and $r=r_{2,m}$ in \BHT (see figure \ref{s1}). The reason is that the expressions in eqs.~\eqref{tRshocksF}-\eqref{tLshocksF} for the boundary times become divergent in this limit since the limit of integration $r_{t,i}$ also approaches $r_{i,m}$ and the expression in the denominator of eq.~\eqref{tauR} becomes approximately linear in $r-r_{i,m}$. Hence the integrals become logarithmically divergent at this point, which leads to very large times $t_\mt{R}+t_w$, $t_\mt{L}-t_w$. This argument holds both for planar and for spherical geometries. For the planar geometry these equations can be solved for analytically and yield
\begin{equation}\label{maxfaceplanar}
r_{1,m} = \frac{r_{h,1}}{2^{1/d}}, \qquad
r_{2,m} = \frac{r_{h,2}}{2^{1/d}}, \qquad
P_{1,m} = \frac{r_{h,1}^d}{2 L} , \qquad
P_{2,m} = \frac{r_{h,2}^d}{2 L} .
\end{equation}
Substituting this back into eq.~\eqref{CtshocksF} and using the definition of the mass in eq.~\eqref{energy} we obtain
\begin{equation}\label{lim1}
\frac{d \mathcal{C}_V}{dt} = \frac{4 \pi (M_1+M_2)}{d-1}.
\end{equation}
And of course, for light shocks this reduces to the known result for eternal black holes upon setting $M_1=M_2=M$.
In our numerical studies, we have always observed that this limit is approached from below.

If the perturbation is sent early enough we observe a period of time during which the growth rate of the complexity, centered around $t=0$, is constant, cf. figure \ref{fig:tlight}. The width of this ledge is approximately $4 t_w$ and it appears due to the fact that in this region we have approximately $P_1\approx -P_{1,m}$ and $P_2\approx P_{2,m}$. The geometric picture indicates that now the surface is unwrapping the critical surface $r=r_{1,m}$ behind the past horizon of \BHO while wrapping onto the critical surface $r=r_{2,m}$ in \BHT (see figure \ref{s2}). The reason is again that for $t_w$ very large and for $t_\mt{L}=t_\mt{R}=0$, the time integrals in eqs.~\eqref{tRshocksF}-\eqref{tLshocksF} should be very large and of opposite signs. This is indeed the case for the momenta above since the lower limit of integration approaches $r_{i,m}$ and the denominator is approximately linear in $r-r_{i,m}$. Substituting these values for the momenta into eq.~\eqref{CtshocksF} and using again the definition of the mass in eq.~\eqref{energy} leads to
\begin{equation}\label{lim2}
\frac{d \mathcal{C}_V}{dt} = \frac{4 \pi (M_2-M_1)}{d-1}.
\end{equation}
We note that for light shock this indicates a rate of computation which is close to zero.

One last interesting limit is the one where the boundary times are only slightly above the shock, \ie $\tR\approx - t_w$ (and hence $t\approx -2 t_w$). This scenario is demonstrated in figure \ref{s3}. In this case we still have $P_1\approx -P_{1,m}$, and in addition we know that $x_s \rightarrow \infty$. We can now use the relations \eqref{eq:onedotSho}-\eqref{eq:P2JumpShock} to obtain the value of $P_2$ in this limit
\begin{equation}
P_2 \approx P_{2,m} - 2 P_{1,m}\, ,
\end{equation}
which using eqs. \eqref{CtshocksF}, \eqref{maxfaceplanar} and \eqref{energy} yields
\begin{equation}\label{lim3}
\frac{d \mathcal{C}_V}{dt} = \frac{4 \pi (M_1-3M_2)}{d-1}.
\end{equation}
This limit satisfies the consistency check that when $M_1=M_2=M$ (\ie a light shock) we recover a negative rate of
\begin{equation}
\frac{d \mathcal{C}_V}{dt} = -\frac{8 \pi M}{d-1},
\end{equation}
as expected for eternal black holes of mass $M$ at early (negative) times.

For BTZ black holes, the same conclusions hold upon replacing the blackening factor with that in eq.~\eqref{locker} and the masses with those given by eq.~\eqref{BTZquantities} and of course setting $d=2$ in all the above equations. In particular the maximal points of the potential still correspond to those given in eq.~\eqref{maxfaceplanar} with $d=2$ and the three limits \eqref{lim1}, \eqref{lim2} and \eqref{lim3} are still valid.

{\bf Critical time:}  As mentioned above, around $t=0$ we observe a long regime in which the rate of computation is approximately constant and proportional to the difference of the masses $M_2-M_1$. We would like to estimate the duration of this regime. In our numerical solutions, we observed that this regime extends more or less equally in positive and negative times and that one can approximately identify the end of it with the transition of the surfaces from the white hole part to the black hole part of \BHO (cf. figure \ref{fig:tlight}). In other words the symmetric time configuration for which $P_1=0$ defines a critical time $t_{c,\text{v}}$ in which this regime ends.\footnote{Recall that with $P_1=0$, the extremal surface passes through the bifurcation surface, separating the black hole and the white hole regions.} This critical time can be evaluated for planar ($d\geq3$) and BTZ geometries in closed form in the limit of early shocks by using the fact that it is associated with the following values of the surface momenta -- $P_1\approx0$ and $P_2\approx P_{2,m}$. The geometric picture lies in between those illustrated in figures \ref{s1} and \ref{s2}. The surface inside \BHO is a straight line going through the bifurcation surface while it still wraps around the surface of constant $r=r_{m,2}$ inside \BHT. Solving the matching condition \eqref{eq:P2JumpShock} with those values of the surface momenta, we are able to extract the value of $r_s$ at the critical time
\begin{equation}
r_{s,c} = \frac{r_{h,2}^{2 }}{ (2 r_{h,2}^d-r_{h,1}^d)^{1/d} }.
\end{equation}
Substituting this into the expression for $\tL$ in eq.~\eqref{tLshocksF} with $P_1=0$, we obtain
\begin{equation}\label{cvShotc}
t_{\mt{L},c} = t_{c,\text{v}}/2 = -v_s + r_1^*(r_{s,c}).
\end{equation}
For instance, in BTZ, using the relevant tortoise coordinate in eq. \eqref{loppit}, as well as $v_s=-t_w$ and  $w=T_2/T_1=\sqrt{M_2/M_1}$ (see eqs.~\eqref{eq:wz}-\eqref{rocker}), we obtain
\begin{equation}\label{cvShotcBTZ}
\begin{split}
\hspace{-17pt} t_{c,\text{v}} =  2 t_w +
\frac{L^2}{r_{h,1}} \log \left(\frac{r_{h,2}^2-r_{h,1} \sqrt{2 r_{h,2}^2-r_{h,1}^2}}{r_{h,2}^2+r_{h,1} \sqrt{2 r_{h,2}^2-r_{h,1}^2}}\right)
 = 2 t_w +
\frac{1}{2 \pi T_1} \log \left(\frac{\frac{w^2}{\sqrt{2 w^2-1}}-1}{\frac{w^2}{\sqrt{2 w^2-1}}+1}\right).
\end{split}
\end{equation}

For light shocks, this critical time can be expanded as
\begin{equation}\label{cvShotcBTZcrit}
t_{c,\text{v}} =
 2 t_w +
\frac{1}{\pi T_1} \log \left(w-1\right) + O(w-1) = 2(t_w-t_{\text{scr}}^*) + \frac{\log 2}{ \pi T_1}  + O(w-1)
\end{equation}
where the scrambling time \eqref{Scramblingt} (with $\epsilon\equiv w-1$) appears in a manner similar to eq.~\reef{lavla} for the CA approach.   For heavy shocks, we can expand eq.~\reef{cvShotcBTZ} for large $w$, which yields
\begin{equation}
t_{c,\text{v}}
 =  2 t_w -
\frac{\sqrt{2}}{ \pi T_2}  + \mathcal{O}\left(\frac{1}{w^2T_2}\right)
\label{cvheavydel}
\end{equation}
where we have used the relation $w T_1=T_2$. Again this expression is similar to the critical time in eq.~\reef{tc1Largew} found for the CA calculations.

Since this regime in which the rate of computation is proportional to the difference of the masses extends more or less equally in negative times we expect it to last approximately for $\Delta t= 2 t_c$. For instance, if we consider very early shocks of a fixed non-vanishing energy ($t_w$ very large and $w$ fixed and finite) we expect that $\Delta t \approx 4 t_w$. This duration will be (significantly) shortened for light shocks by the scrambling time. We will confirm these predictions as well as the results of eqs.~\eqref{cvShotc}-\eqref{cvShotcBTZ} with numerical examples below.

{\bf A comment on dimensionless quantities:} It is possible to express all our results in eqs.~\eqref{eq:onedotSho}-\eqref{eq_shocks:rturn}, \eqref{tauR}-\eqref{tLshocksF} and \eqref{CtshocksF}, in terms of the dimensionless quantities in eq.~\eqref{eq:wz}, as well as the dimensionless surface momenta
\begin{equation}\label{dimlescv}
p_{1} \equiv \frac{L P_1}{r_{h,1}^d}, \qquad p_{2} \equiv \frac{L P_2}{r_{h,2}^d}.
\end{equation}
We have found this representation particularly useful for the case of planar black holes in $d\geq3$ and for BTZ black holes for which the dependence on $z$ cancels and one is able to demonstrate that the profile of the rate of change in complexity normalized by the sum of masses as a function of $T_2 \, t$, depends on $w$ and $T_2 \, t_w$, but not on $z$.

\subsection{Complexity of Formation} \label{CoF2}
As discussed in eq.~\eqref{CoFdef}, the complexity of formation was originally defined as the difference between the complexity of the TFD state and that of two copies of the vacuum. Here we evaluate the complexity of formation for the shock wave geometries using the complexity=volume proposal, by evaluating the complexity at $\tR=\tL=0$ and subtracting two copies of the equivalent result for empty AdS. For this purpose, it will be convenient to define
\begin{equation}
R_0[r] = \frac{r^{(d-1)}}{\sqrt{f_{\vac}(r)}}
\end{equation}
where $f_{\vac}(r)=\frac{r^2}{L^2}+k$ is the corresponding `blackening' factor for the AdS vacuum. As we have mentioned before, there are different scenarios for the shape of the surface. In particular, it can have turning points in \BHO and/or in \BHT (see figure \ref{surfShapes2} in appendix \ref{app:CVShocksDetails}). These different cases will have different expressions for the complexity of formation as we detail below. We have reorganized the integrals in such a way that they are all finite in the limit $r_{\max}\rightarrow \infty$ and so we may replace the relevant integration limits by $\infty$. The complexity of formation is a sum of two contributions
\begin{equation}\label{DVTshocksF}
\Delta \cv =\Delta \cv{}_1+\Delta \cv{}_2
\end{equation}
where
\footnotesize
\begin{equation}\label{DVT2shocksF}
\begin{split}
\hspace{-10pt} \Delta \cv{}_1= & \, \frac{\Omega_{k,d-1}}{GL}
\begin{cases}
\int_{r_{t,1}}^{\infty} \left(R_1 \left[P_1,r\right]-R_0[r] \right) dr + \int_{r_{t,1}}^{r_s} R_1 \left[P_1,r\right] dr
 -\int_{0}^{r_{t,1}} R_0[r]
 & ~ \text{turning point in \BHO}
\\
\int_{r_s}^{\infty} \left( R_1 \left[P_1,r\right]-R_0[r] \right) dr -\int_{0}^{r_{s}} R_0[r] & ~ \text{otherwise}
\end{cases}
\\
\hspace{-10pt} \Delta \cv{}_2 = & \, \frac{\Omega_{k,d-1}}{GL}
\begin{cases}
\int_{r_{t,2}}^{\infty} \left(R_2 \left[P_2,r\right]-R_0[r]\right) dr + \int_{r_{t,2}}^{r_s} R_2 \left[P_2,r\right] dr
 -\int_{0}^{r_{t,2}} R_0[r] & ~\text{turning point in \BHT}
\\
\int_{r_s}^{\infty} \left(R_2 \left[P_2,r\right]-R_0[r] \right) -\int_{0}^{r_{s}} R_0[r] dr & ~  \text{otherwise.}
\end{cases}
\end{split}
\end{equation}
\normalsize
We will be able to study the complexity of formation numerically as a function of $t_w$ and the masses $M_1$, $M_2$. When studying BTZ black holes, we will take $f_{\vac}(r)$ to correspond to the Neveu-Schwarz vacuum, \ie $k=+1$.

When plotting our numerical results, we found it convenient to subtract from the complexity of formation in the shock wave geometry the equivalent complexity for forming an eternal AdS black hole of radius $r_{h,1}$. In this way, we expect that near $t_w=0$ the result will vanish. We have used the following values for the complexity of formation of eternal black holes with  $d=2$ starting from the Neveu-Schwarz vacuum
\begin{equation}\label{eq:dcVNS}
\Delta \mathcal{C}_{V,NS} = \frac{4 \pi L}{G_N},
\end{equation}
and for planar black holes in $d\geq3$
\begin{equation}
\Delta \mathcal{C}^{\text{eternal}}_{V,1} = \frac{\sqrt{\pi} \Omega_{0,d-1}}{G_N} \frac{(d-2)\Gamma\left(1+\frac{1}{d}\right)}{(d-1)\Gamma\left(\frac{1}{2}+\frac{1}{d}\right)} r_{h,1}^{d-1},
\end{equation}
which were previously evaluated in \cite{Format}. Equivalently, after the subtraction we are evaluating the difference of complexities between the shock wave geometry and an eternal black hole of radius $r_{h,1}$ and this amounts to substituting in eq.~\eqref{VT2shocksF} $R_0[r]$ by $R_1[0,r]$ and replacing the $0$ in the integration limit by $r_{h,1}$ (perhaps with some reorganization of the integrals).

{\bf Early and late time limits of the complexity of formation:}
The symmetry of the problem under time shifts (recall eqs.~\eqref{tshift1}-\eqref{tshift2}), which includes shifting the time of the shock wave, implies that the effect on the complexity of pushing the shock wave to the past by an amount of time $d t_w$ is the same as that obtained by studying the time evolution of the complexity under anti-symmetric time evolution with $d \tR=-d \tL=d t_w$. We can use this understanding to make some general statements about the slope of the complexity of formation with respect to the time of the shock $t_w$ in certain limits for planar and BTZ black holes.

The first limit we consider is that of shocks injected at times $t_w \approx 0$. In this case we expect that $x_s \rightarrow \infty$. We also expect that the configuration is extremely similar to studying the complexity of formation in \BHO without the shock, which implies that $P_1 \approx 0$. Using the matching condition \eqref{eq:P2JumpShock}, this leads to $P_2\approx P_{2,m}-P_{1,m}$ and using eq.~\eqref{VtshocksF} with $d \tR=-d \tL=d t_w$, we obtain
\begin{equation}\label{lim4}
\frac{d \Delta \mathcal{ C}_V}{dt_w} = \frac{8 \pi (M_2-M_1)}{d-1}.
\end{equation}
In fact, we will see that this approximation of the slope holds for a range of $t_w$ which increases logarithmically with the inverse of the energy of the shock. This range is another indicator of the scrambling time of the system and we will come back to this point in our numerical analysis.

For very early shocks (large $t_w$), the $\tL=\tR=0$ configuration of the complexity of formation is very similar to the one that appears in figure \ref{s2}. This implies that $P_1 \approx -P_{1,m}$ and $P_2 \approx P_{2,m}$. Substituting  this together with $d \tR=-d \tL=d t_w$ into eq.~\eqref{VtshocksF}, we obtain
\begin{equation}\label{lim5}
\frac{d \Delta \mathcal{C}_V}{dt_w} = \frac{8 \pi (M_1+M_2)}{d-1}.
\end{equation}

One comment is in order before we proceed: The linear growth with the slope in eq.~\eqref{lim4} turns out to be superimposed with a tiny exponential growth, which governs the transition to the regime in \eqref{lim5}. The coefficient in this exponential growth is none other than the Lyapunov exponent, which characterizes the scrambling of information in the system. We will come back to this point in detail in section \ref{sec:Discussion}.

{\bf A comment on the complexity of formation and dimensionless variables:}
When expressing our results in terms of the dimensionless coordinates \eqref{dimlescv} for planar geometries (with $d\geq 3$), it is possible to demonstrate that the combination
\begin{equation}\label{cvShoshift1}
\frac{\Delta \cv - \Delta \mathcal{C}_{V,1}^{\text{eternal}}}{S_1}
\end{equation}
where the entropy $S_1$ of \BHO was defined in eq.~\eqref{effect}, is independent of $z$. For BTZ black holes in $d=2$, the same statement holds when replacing $\Delta \mathcal{C}_{V,1}^{\text{eternal}}$ by $\Delta \mathcal{C}_{V,NS}$ and the entropy by that for BTZ, see eq.~\eqref{BTZquantities}.
In this case, the $z$ independence implies that the extra complexity generated by the shock wave grows linearly with the temperature (just as $S_1$ does). This contrasts with the original complexity of formation $\Delta \mathcal{C}_{V,NS}$ in eq.~\eqref{eq:dcVNS}, which was a fixed constant.
We will plot the combination \eqref{cvShoshift1} in our numerical analysis below.

\subsection{Numerical Analysis}

\subsubsection{$d=2$}

\begin{figure}
\centering
\includegraphics[scale=0.8]{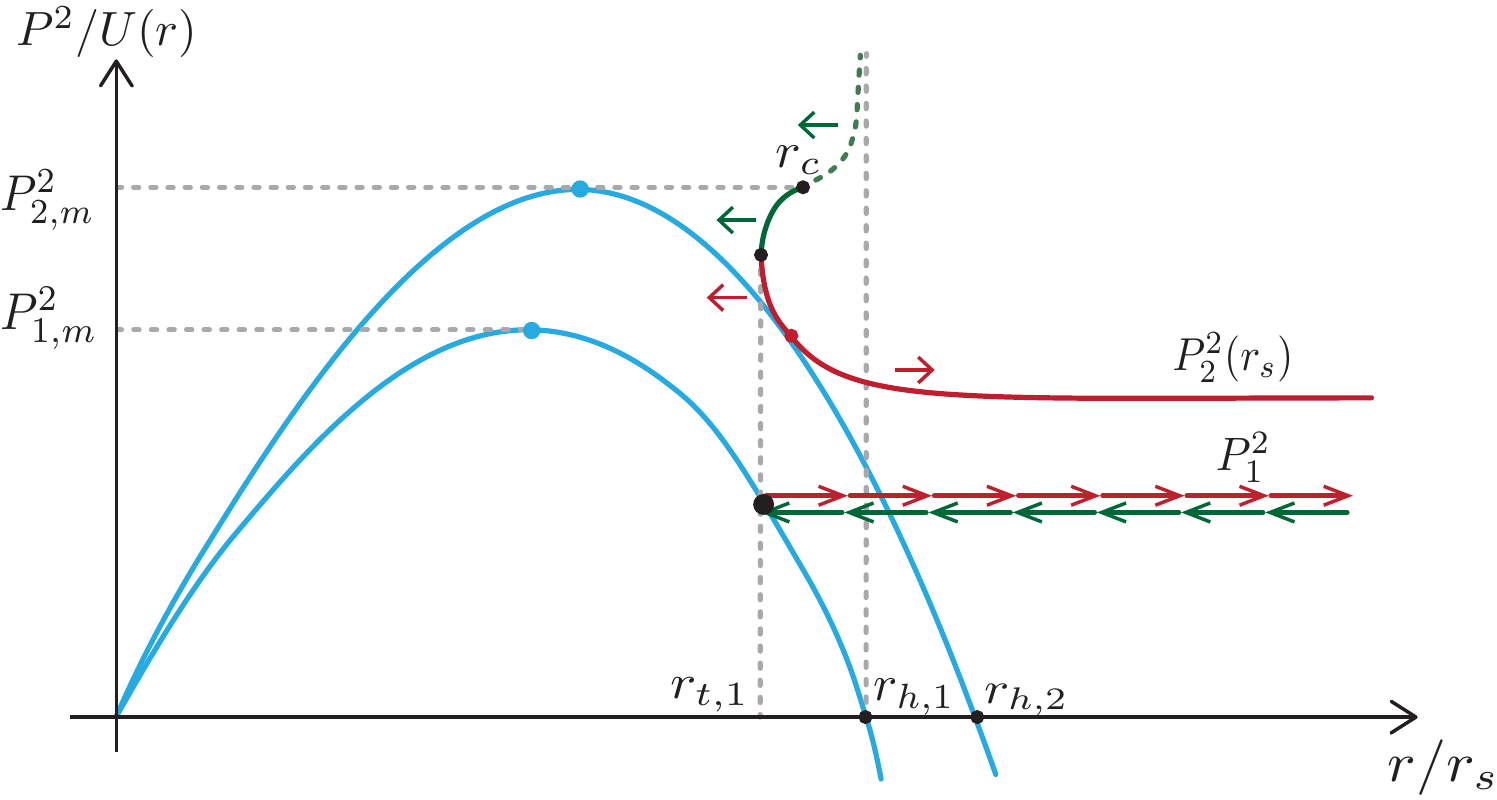}
\caption{Illustration of the potentials $U_{1,2}(r)$ (see eq.~\eqref{PotentialShocks}), superimposed by a plot of $P_2^2(r_s,P_1)$  (see eq.~\eqref{eq:P2JumpShock}) for a specific value of $P_1^2$. We have indicated by arrows the sign of $\dot r$ before hitting the shell (on the $P_1^2$ lines) and right after hitting the shell (next to the $P_2^2(r_s)$ lines). Red lines are associated with surfaces that have a turning point in \BHO and green lines with those that do not. The plot was made for $P_1>0$ but a similar plot can be made for $P_1<0$. The special point $r_c$ where $P_2$ becomes equal to the maximal value $P_{2,m}^2$ was very important in our analysis since in many of the cases studied (early shocks, symmetric boundary times) the solution for $r_s$ was found to be extremely close to it. The dashed part of the green curve $P_2^2(r_s)$ is associated with surfaces that will end on the singularity after crossing the shell and are therefore irrelevant  for our analysis.}
\label{fig:PotentialShocks}
\end{figure}

In this section, we collect some numerical results for shock waves in BTZ black holes, \ie with the boundary dimension $d=2$.
In obtaining these results, we found it very useful to draw plots of the potentials in eq.~\eqref{PotentialShocks}, superimposed by the value of $(P_2)^2$ predicted from the jump condition \eqref{eq:P2JumpShock} as a function of $r_s$ --- see figure \ref{fig:PotentialShocks}. These plots for a given value of $P_1$ helped us develop an intuition for the numerical range in which we expect to find the crossing point $r_s$ for the solution of symmetric boundary times for early shocks.

\begin{figure}
\centering
\includegraphics[scale=0.434]{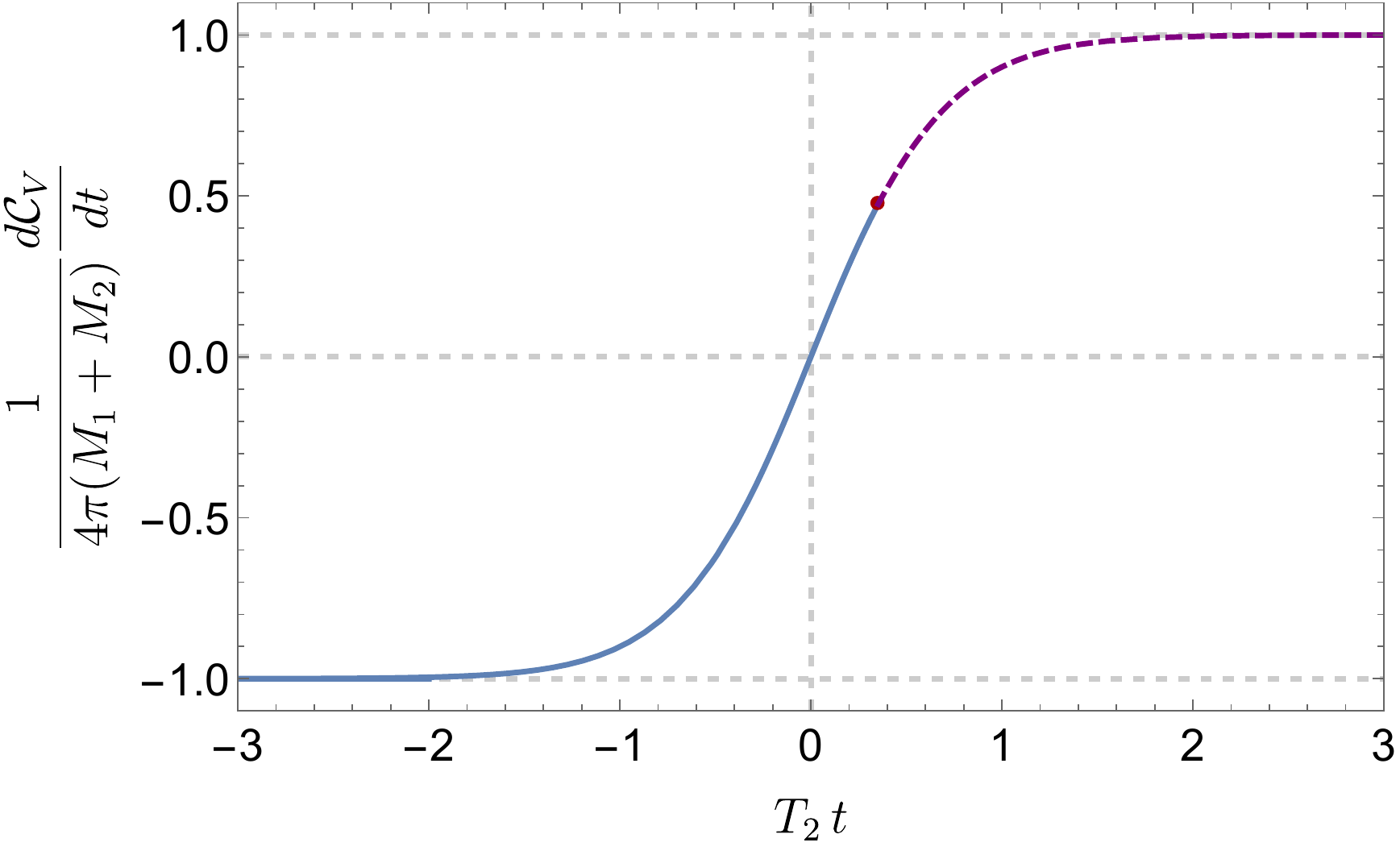}
\includegraphics[scale=0.438]{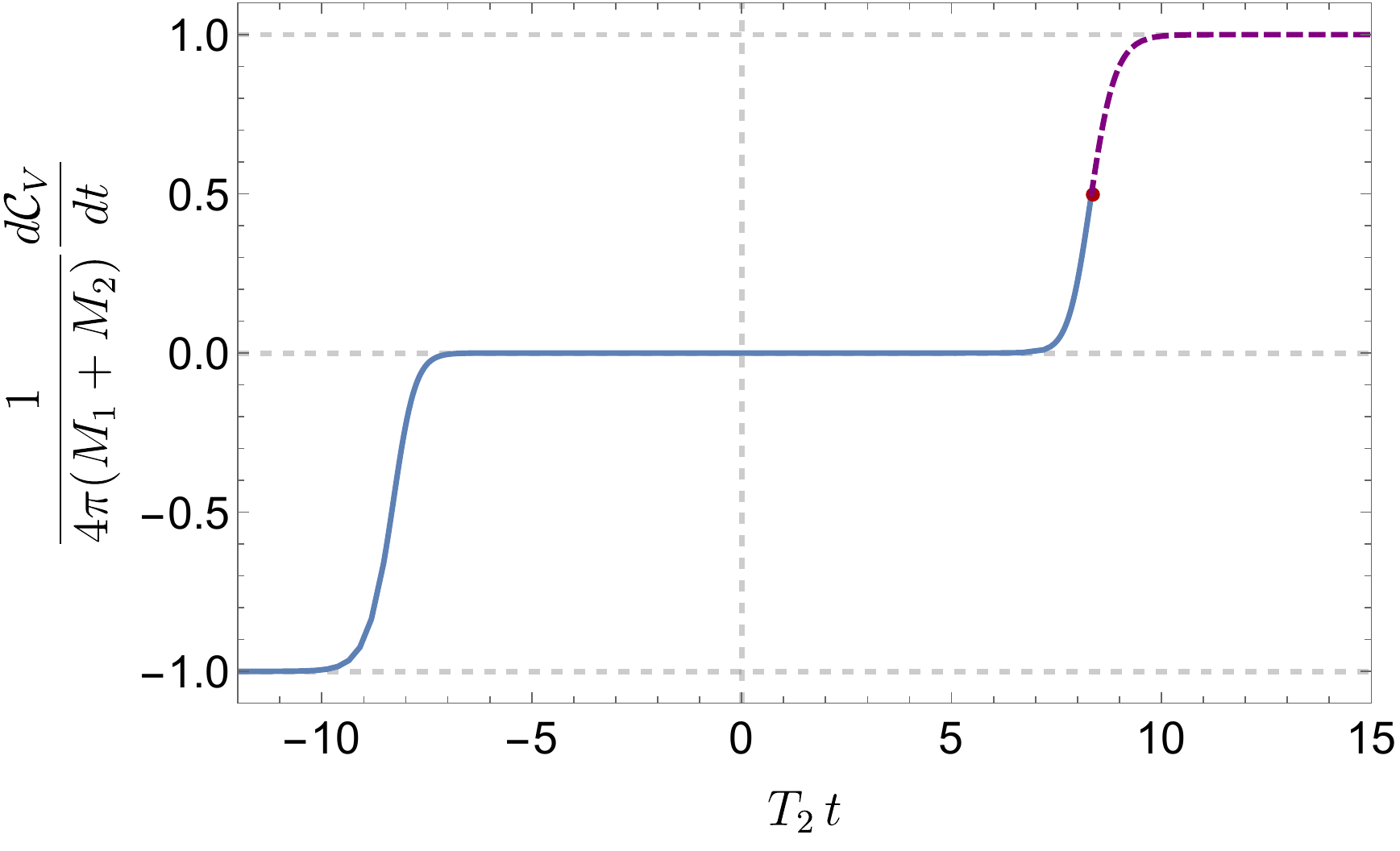}
\caption{Time derivative of the complexity evolved with symmetric boundary times $\tL=\tR=t/2$ as a function of $T_2 t$ normalized by the sum of masses for light shocks with $w=1+10^{-5}$ for BTZ black holes (\ie $d=2$)  for $T_2 t_w=2$ (left) and $T_2 t_w=6$ (right). The horizontal dashed lines indicate the limits \eqref{lim1} and \eqref{lim3} which are very close to those of the configuration without the shock, \ie $\pm 1$ in our normalization. We see that the late time limit is approached from below. For early shock times, a plateau develops near $t=0$ in which the rate of computation is proportional to the difference of the masses and hence is nearly vanishing (see eq.~\eqref{lim2}).}
\label{fig:tlight}
\end{figure}

\begin{figure}
\centering
\includegraphics[scale=0.43]{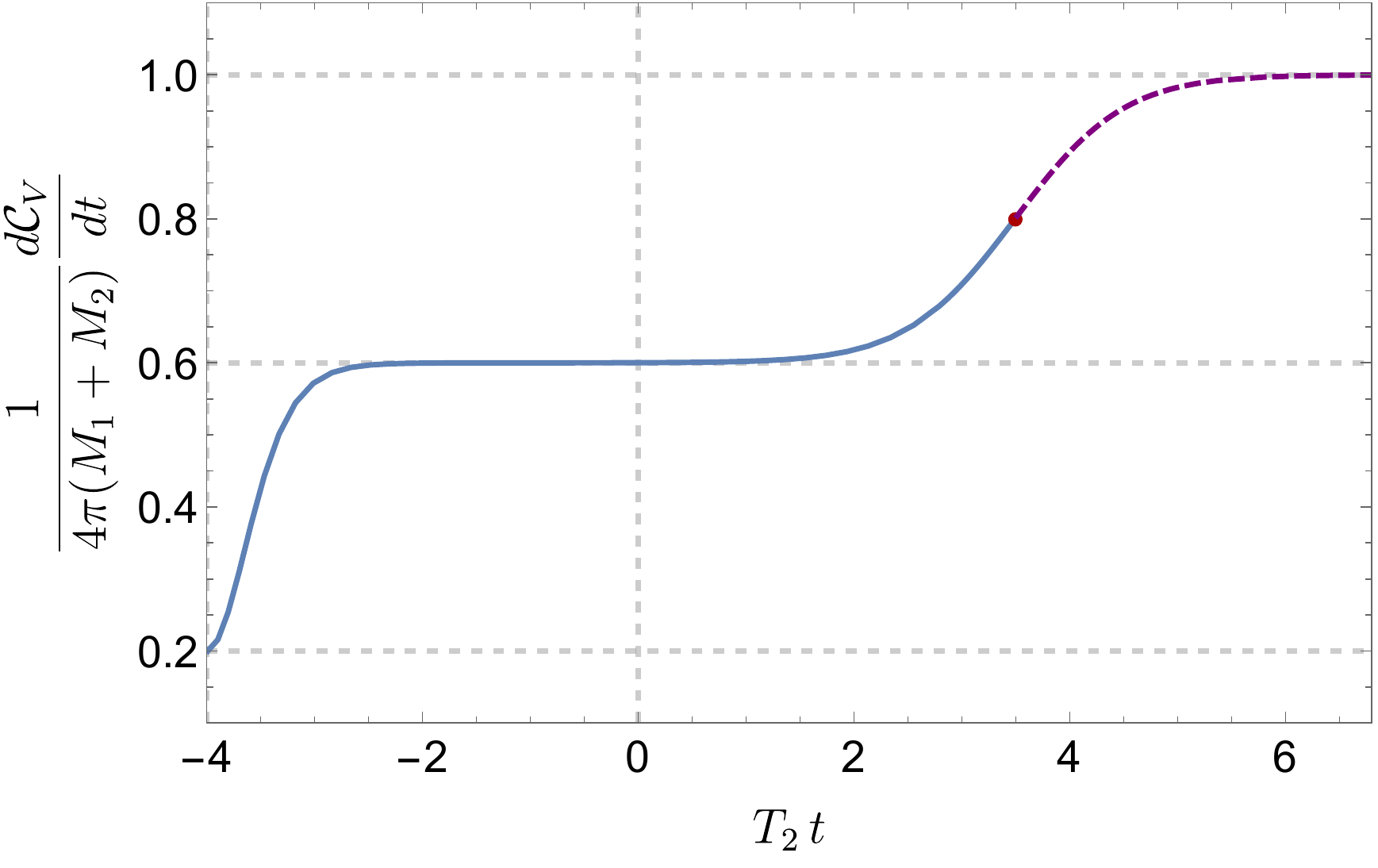}
\includegraphics[scale=0.45]{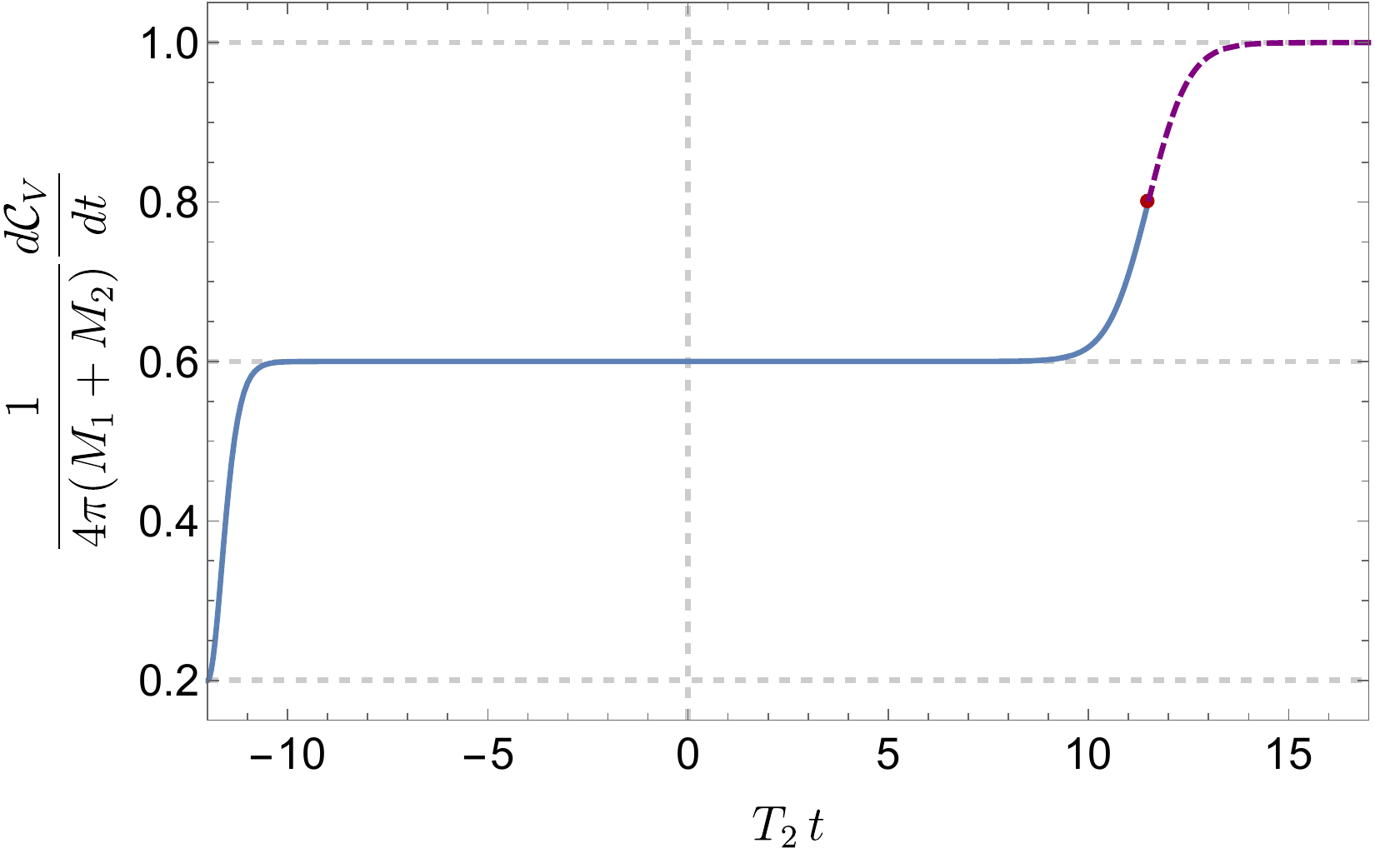}
\caption{Time derivative of the complexity evolved with symmetric boundary times $\tL=\tR=t/2$ as a function of $T_2 t$ normalized by the sum of masses for finite size shocks with $w=2$ for BTZ black holes (\ie $d=2$) for $T_2 t_w=2$ (left) and $T_2 t_w=6$ (right). The horizontal dashed lines indicate the limits \eqref{lim1}, \eqref{lim2} and \eqref{lim3}. We see that the late time limit is approached from below. For early shock times, a plateau develops near $t=0$ in which the rate of computation is proportional to the difference of the masses (see eq.~\eqref{lim2}).}
\label{fig:theavy}
\end{figure}

\begin{figure}
\centering
\includegraphics[scale=0.47]{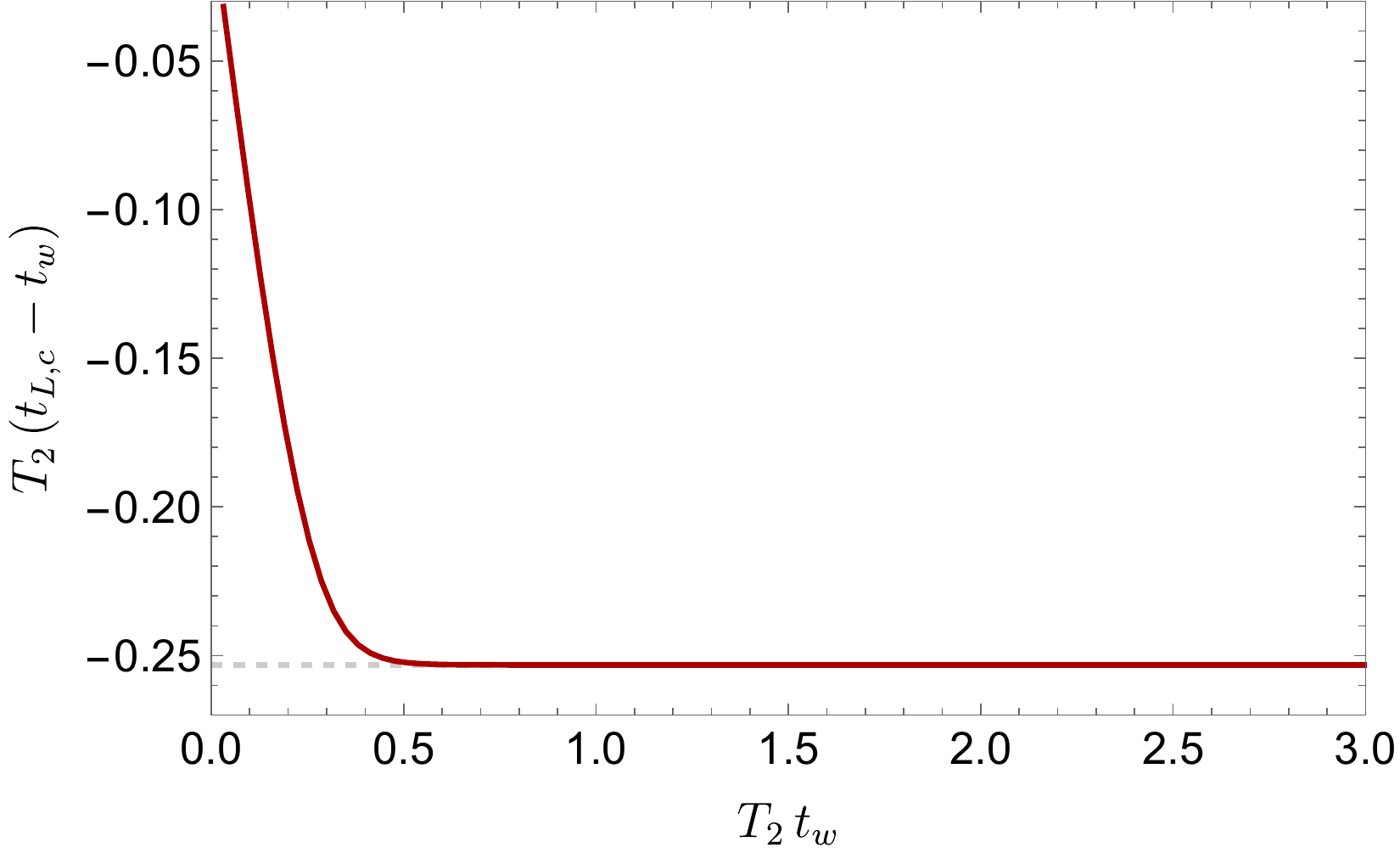}
\includegraphics[scale=0.44]{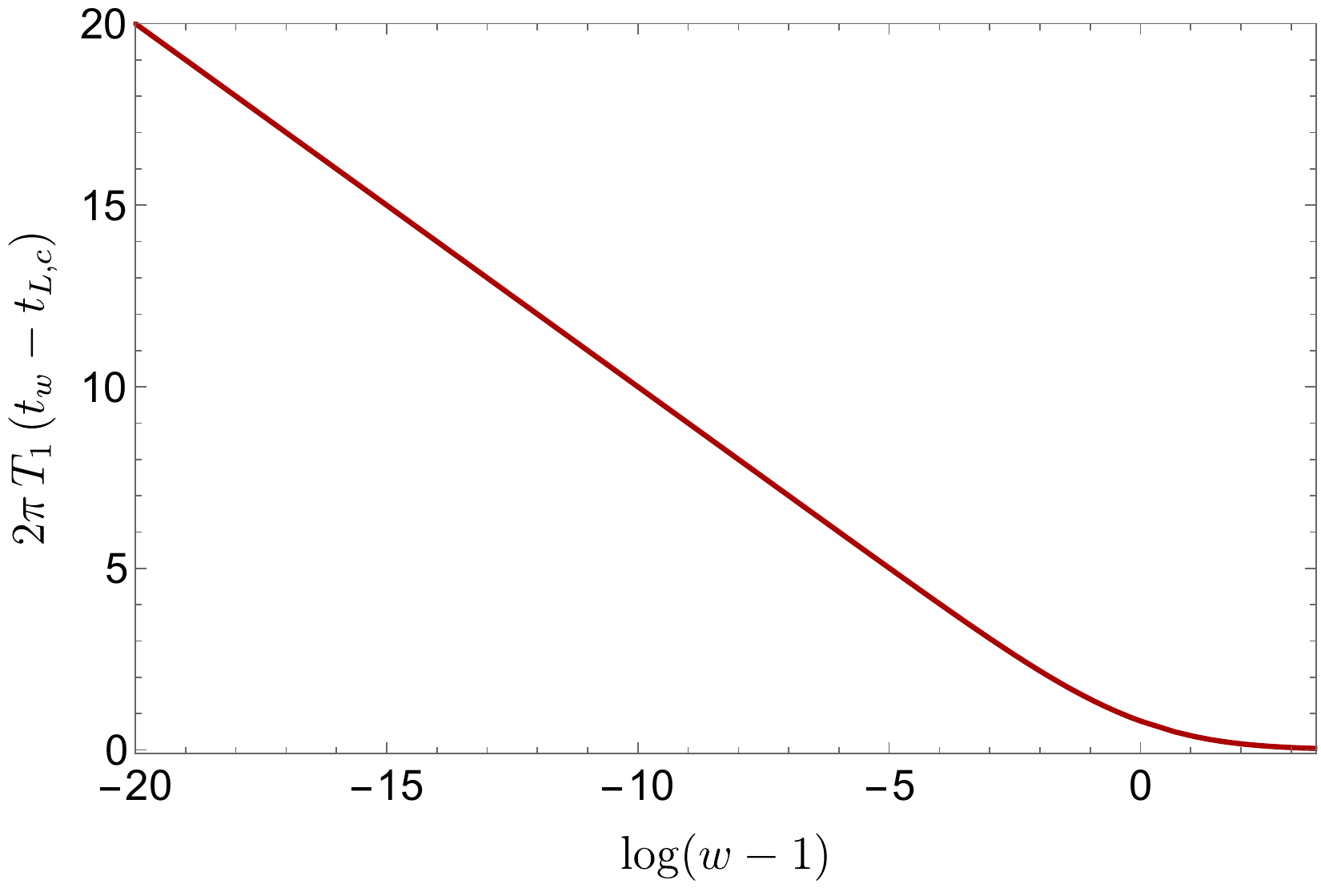}
\caption{Left: critical time in the symmetric boundary time evolution $\tR=\tL=t/2$ of the extremal surface passing through the bifurcation surface in \BHO ($P_1=0$) in shocks of $d=2$ BTZ black holes with $w=2$ as a function of the time of the shock $t_w$. The dashed gray line indicates the prediction of eq.~\eqref{cvShotcBTZ} for early shocks. Right: critical time obtained for early shocks with $ T_2 \, t_w = 6$ for various values of $w$. As indicated by eq.~\eqref{cvShotcBTZcrit}, the logarithmic dependence on $w-1$ is an indicator of the scrambling time for light shocks. The critical times in these two plots are related to the ending of the plateau in which the rate of computation is proportional to the difference of the masses, see figures \ref{fig:tlight}-\ref{fig:theavy}, where the relevant point was indicated by a red dot.}
\label{fig:trnstime}
\end{figure}

Figures \ref{fig:tlight} and \ref{fig:theavy} contain the time derivatives of the complexity for a symmetric boundary time evolution $\tL=\tR=t/2$ as a function of $t$ for light and heavy shocks, respectively. The figures demonstrate that for early shocks, a plateau develops in which the rate of computation is proportional to the difference of the masses, \ie vanishes for light shocks. As explained earlier, we may want to define a critical time as the point in the symmetric time evolution for which $P_1=0$ and which indicates the end of this plateau. We have plotted this critical time as a function of the time of the shock for a heavy shock $w=2$ and checked its agreement with the (early shock) prediction in eq.~\eqref{cvShotcBTZ} in figure \ref{fig:trnstime}. We have also pointed out in eq.~\eqref{cvShotcBTZcrit} that the behavior of this critical time for light shocks is an indicator of the scrambling time. We verify this statement by plotting the critical time as a function of $w$ in figure \ref{fig:trnstime}.

We find it insightful to describe the evolution of the extremal surface that came into play in these numerical solutions. Evolving from past to future for times $\tL=\tR>-t_w$, we have: 1. The surface passes behind the past horizon in \BHO and has a  turning point only in \BHO; 2. Part of the surface moves into the future horizon in \BHT and a turning point in \BHT develops; 3. The surface crosses the bifurcation surface and moves behind the future horizon in \BHO, we still have two turning points; 4. If the shock is light, the turning point in \BHO disappears for a certain range of surface momenta, then reappears in the late time limit. This regime only takes place when there exists a range of energies for which the crossing point between the red and green curves in figure \ref{fig:PotentialShocks} moves below $P_2^m$, which happens if $w<\sqrt{\frac{1}{2}+\frac{1}{\sqrt{2}}}$ for energies $|4 p_1+1-2 w^2|<\sqrt{1+4 w^2-4 w^4}$ where $p_1$ was defined in eq.~\eqref{dimlescv}.

\begin{figure}
\centering
\includegraphics[scale=0.46]{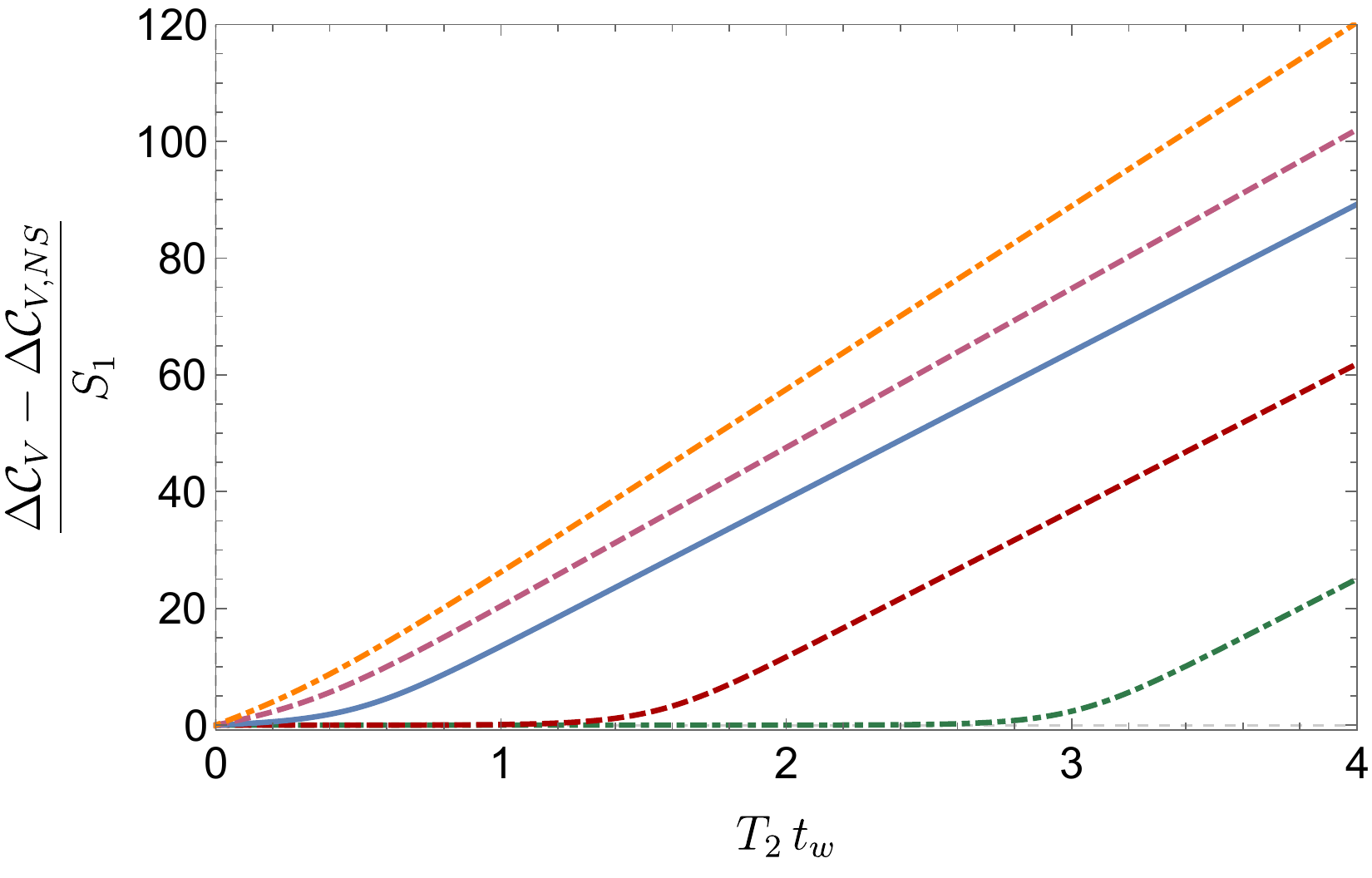}
\includegraphics[scale=0.46]{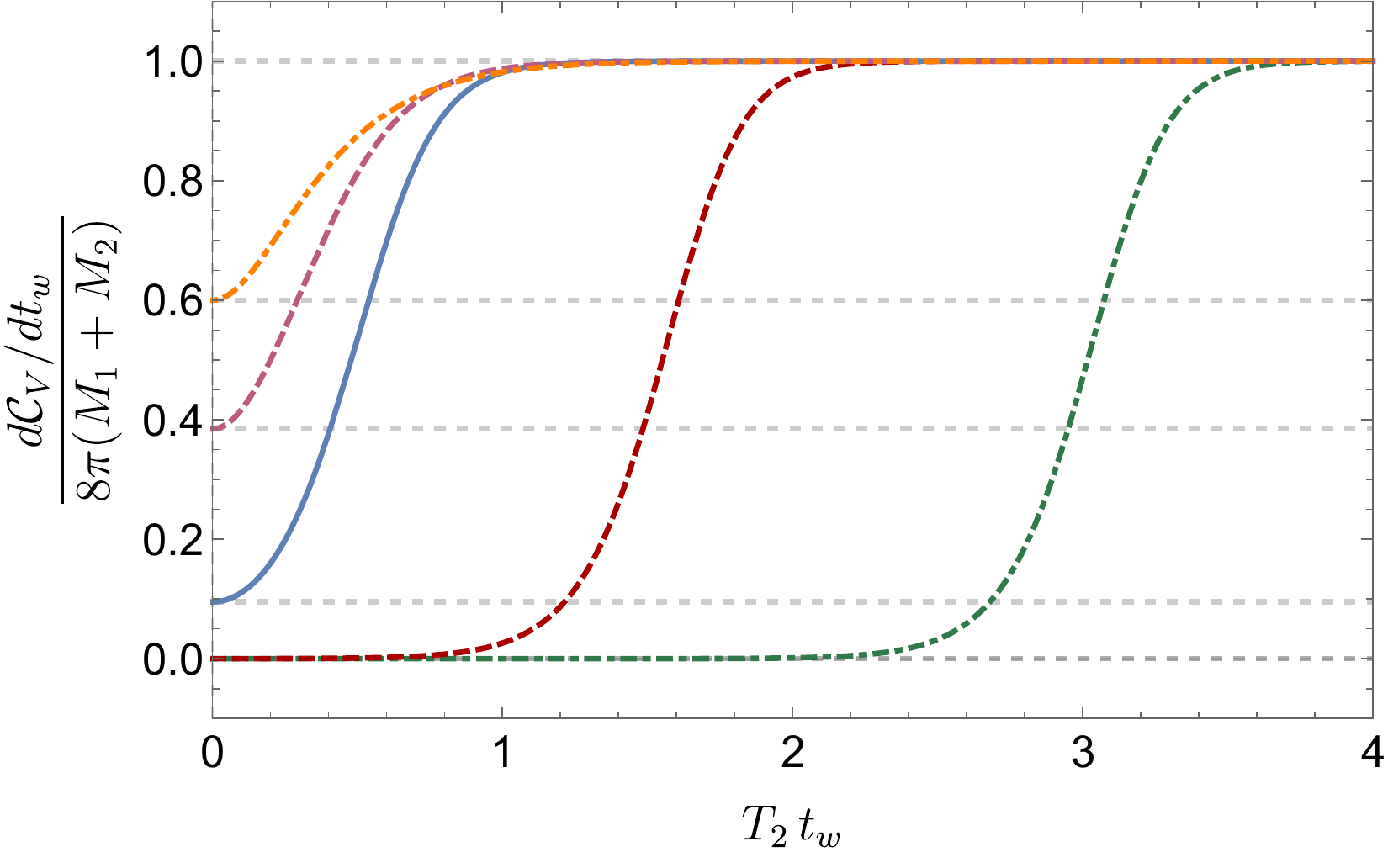}
\caption{Complexity of Formation as a function of the shock time $t_w$ (left) and its derivative (right) for various energies of the shocks -- $w=2$ (orange, dot-dashed), $w=1.5$ (pink, dashed), $w=1.1$ (blue, solid), $w=1+10^{-4}$ (red, dashed) and $w=1+10^{-8}$ (green, dot-dashed). The slope of this plot is proportional to the difference of the masses for small $t_w$ (not so early shocks) and to the sum of the masses for large $t_w$ (early shocks) and matches the predictions in eqs.~\eqref{lim4}-\eqref{lim5}, indicated by dashed gray lines on the right plot.}
\label{fig:formationcvcv}
\end{figure}

\begin{figure}
\centering
\includegraphics[scale=0.46]{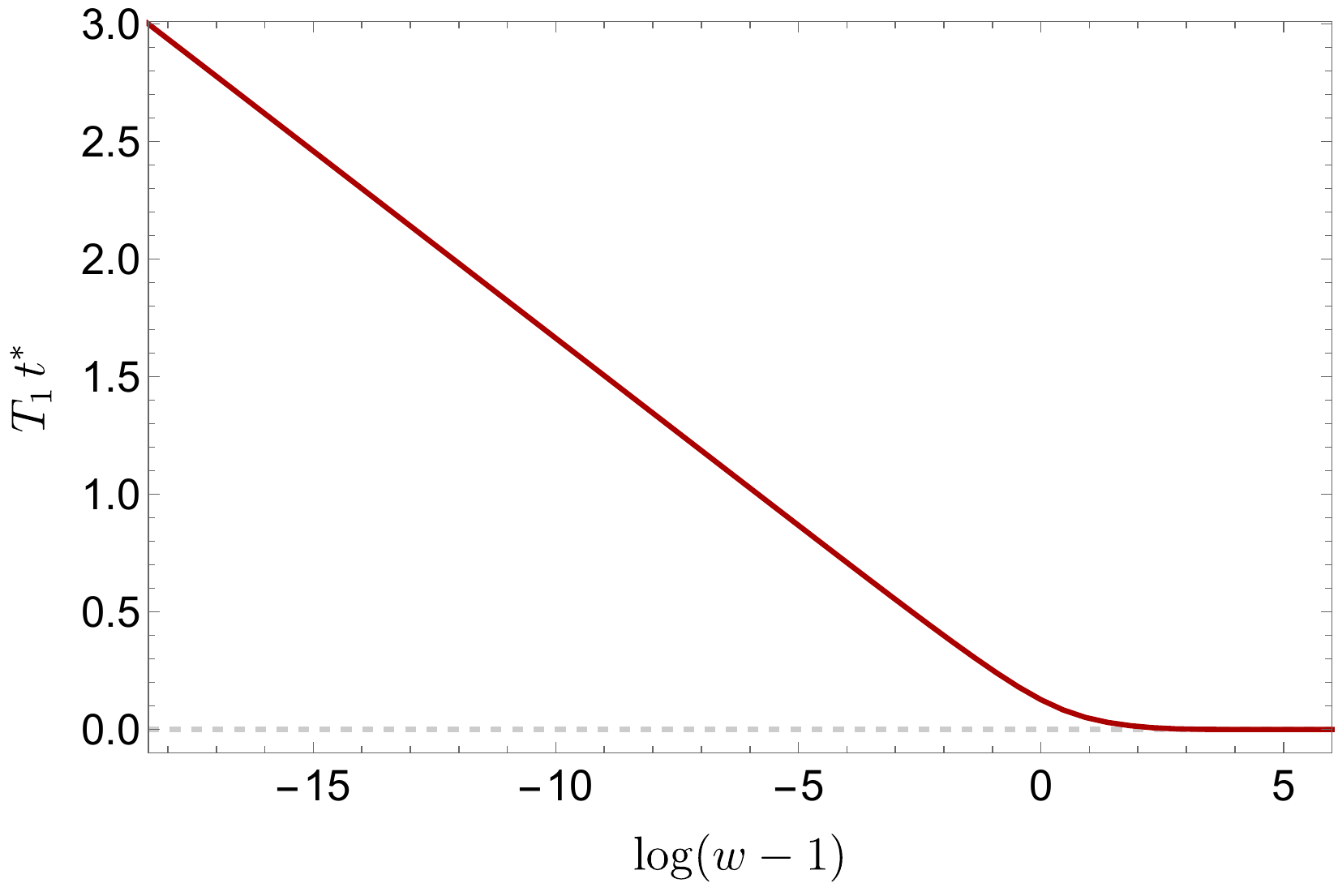}
\caption{Characteristic time of crossing between the two regimes in the complexity of formation in which the slopes are proportional to the difference of the masses, see eq.~\eqref{lim4}, and the sum of the masses, see eq.~\eqref{lim5}. The logarithmic dependence on the energy of the shock in this plot is another diagnostic of the scrambling time (cf. figure \ref{fig:trnstime}, right panel) and matches closely the fit in eq.~\eqref{tscrformcv}.}
\label{fig:sctimecv}
\end{figure}

We have also studied numerically the complexity of formation for various energies of the shock as a function of the shock time $t_w$. The results are extremely similar to those obtained using the complexity=action prescription and can be found in figure \ref{fig:formationcvcv} (c.f. figure \ref{FormationBTZShock} in the CA subsection). We see that for light shocks, a long plateau develops in which the rate of computation is proportional to the difference of the masses. The length of this plateau is logarithmic in the inverse energy of the shock. To confirm that we have plotted the length of this plateau as a function of $w$ for light shocks and checked that it agrees with the scrambling time in eq.~\eqref{Scramblingt}, up to a small constant shift. We have defined the length of the plateau as the time for which the two straight line asymptotic regimes corresponding to the limits in eqs.~\eqref{lim4} and \eqref{lim5} intersect. The small constant shift is a consequence of the fact that the constant in eq.~\eqref{Scramblingt} was fixed arbitrarily there in order to simplify the expressions in the CA subsection. If we fit the plot in figure \ref{fig:sctimecv}, we find a very close agreement with
\begin{equation}\label{tscrformcv}
t^*_{\text{form}}=\frac{1}{2\pi T_1} \log \left(\frac{\gamma_2}{2 (w-1)} \right)\,, \qquad \gamma_2=3.1412\pm0.0008\,,
\end{equation}
\ie the factor 2 in the argument of the logarithm in eq.~\reef{Scramblingt} is replaced here by $\pi/2$.

\subsubsection{$d=4$}
\begin{figure}
\centering
\includegraphics[scale=0.43]{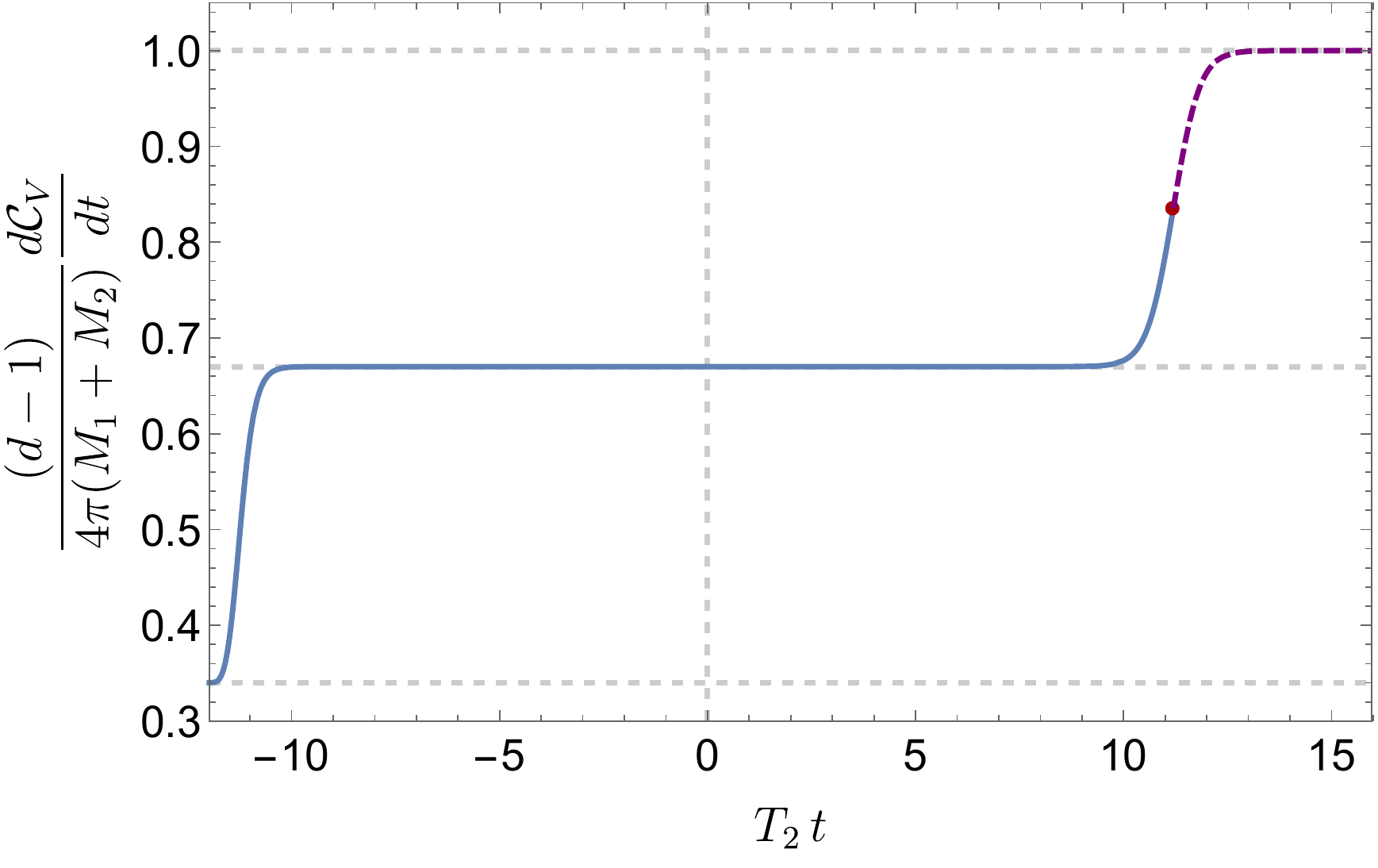}
\includegraphics[scale=0.435]{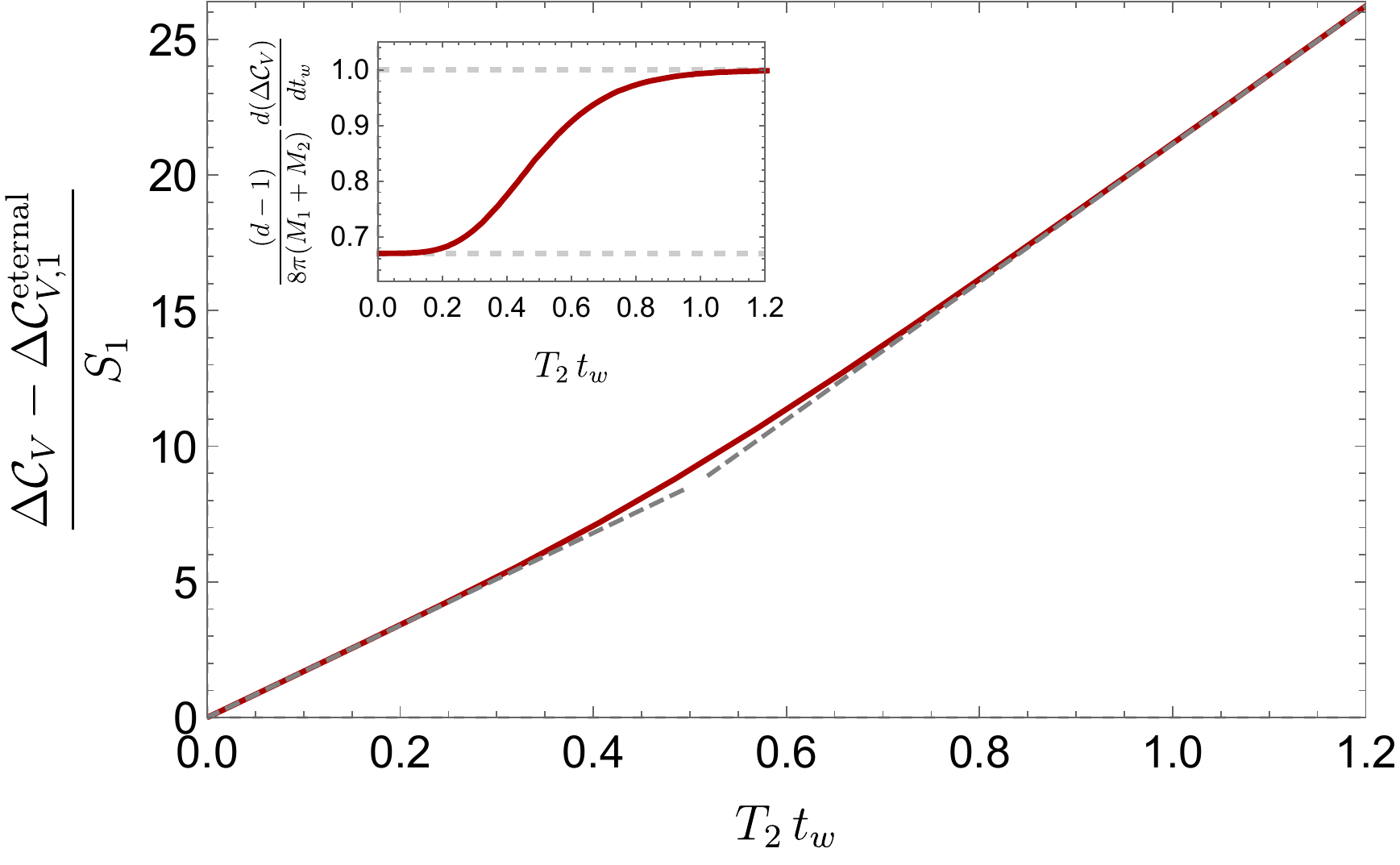}
\caption{Numerical results for the complexity in shock wave backgrounds in planar ($k=0$) black holes in $d=4$ with
$w\equiv r_{h,2}/r_{h,1}=1.5$. Left: rate of change of complexity as a function of time for symmetric evolution with $\tL=\tR=t/2$ with an early shock sent at $T_2 t_w = 6$. We have started the plot at $t=-2 t_w$ and indicated by horizontal dashed lines three different limits (in the relevant normalization as indicated on the axis) -- the limit of times close to $\tL=\tR=-t_w$ (see eq.~\eqref{lim3}), the plateau near $t=0$  (see eq.~\eqref{lim2}), and the late time limit (see eq.~\eqref{lim1}). The red dot indicates the point in which $P_1=0$ which signals the transition of the surface from the white hole to the black hole region in \BHO. Right: Complexity of formation as a function of the shock time $t_w$ shifted and normalized as in eq.~\eqref{cvShoshift1}. We have indicated by dashed lines the predicted slopes (see eqs.~\eqref{lim4}-\eqref{lim5}). The inset shows the derivative and its agreement with the predicted slopes.}
\label{fig:cvd4Shocks}
\end{figure}

We have repeated the analysis above for the case of planar black holes in $d=4$. The results for the time derivative of complexity under symmetric time evolution and for the complexity of formation can be found in figure \ref{fig:cvd4Shocks} and they are in agreement with the predicted limits in eqs.~\eqref{lim1}-\eqref{lim2}, \eqref{lim3} and \eqref{lim4}-\eqref{lim5}.
We have not studied the case of spherical black holes, however, we would like to suggest that the various limits in this case are still  related to the maximal points of the potentials. Of course, this will not give an exact proportionality to the masses but rather the late time limit will be modified by curvature corrections proportional to $1/(LT)^2$. There are interesting questions to be asked, \eg is the late time limit for the rate of computation approached from below like in the time dependent eternal BH or from above like in Vaidya? Is the limit above or below the planar one? We leave these questions for future study.

\section{Discussion}\label{sec:Discussion}

Using the framework established in \cite{Vad1}, we studied  holographic complexity in Vaidya geometries \reef{MetricV} describing a shock wave propagating into an eternal black hole. Of course, this situation has already been well studied not only for a single shock wave but also for many shock waves as well, \eg \cite{ShenkerStanfordScrambling, multiple, Stanford:2014jda,Brown1,Brown2,Roberts:2014isa}.  New directions investigated here were to study the full time dependence of the holographic complexity for both light and heavy shocks in sections \ref{CAshock} and \ref{CVshock}, and to evaluate the complexity of formation in sections \ref{CoF1} and \ref{CoF2}. In both cases, we examined both the CA and CV approaches. In the following, we review our results in these calculations, and also consider their physical implications.

\subsection*{Complexity of Formation}

The complexity of formation for the shock wave geometries was examined in sections \ref{CoF1} and \ref{CoF2} using the CA and CV approaches, respectively. Recall that as originally studied in \cite{Format},\footnote{These calculations were extended to charged AdS black holes in \cite{Growth}.}  the complexity of formation  was defined as the difference between the complexity of preparing two copies of the boundary CFT in the thermofield double state (TFD) at $\tL =\tR = 0$ and the complexity of preparing each of the CFT's in its vacuum state, as shown in eq.~\reef{CoFdef}. A key feature of this quantity is that the difference of the complexities is UV finite.

In extending these calculations to the perturbed black holes, we first considered light shocks,
\ie $w=r_{h,2}/r_{h,1}\simeq1$, sent from the right boundary at some early time $\tR=-t_w$
--- see left panel in figures \ref{FormationBTZShock} and \ref{FormationAdS5} for $d=2$ and $d=4$, respectively, with the CA approach and the
 green and red curves in figure \ref{fig:formationcvcv} for $d=2$, with the CV approach.
In this case, $\Delta{\cal C}$ was essentially unchanged for a wide range of $t_w$. But then beyond some critical $t_w$, \ie for earlier shock waves, $\Delta{\cal C}$ grew linearly as\footnote{The results for $\Delta\cv$ in the present equation, as well as in eqs.~\reef{boil1} and \reef{boil2}, only apply for planar black holes. For spherical geometries, one can still express the various limits of ${d\Delta\cv}/{d t_w}$ and ${d\cv}/{d t}$ using the extremal points of the potentials \eqref{PotentialShocks}. However, in the spherical case these values are no longer simply related to the masses. \label{gamma9}}
\beq\label{rock}
\frac{d\Delta{\cal C}_A}{d\,t_w}= \frac{4M_1}\pi
\qquad{\rm and}\qquad
\frac{d\Delta{\cal C}_V}{d\,t_w}= \frac{16 \pi M_1}{d-1}\,.
\eeq
The critical injection time where this transition occurred was precisely given by the scrambling time, as defined in eqs.~\reef{Scramblingt} and \reef{tscrformcv} for the CA and CV approaches, respectively. Hence as a rough approximation, the complexity of formation can be described by two linear regimes, as shown in eq.~\reef{FormBTZLightSh} for $\Delta{\cal C}_A$. Of course, this is a manifestation of the switchback effect \cite{Stanford:2014jda}. That is, we can think that we begin with the unperturbed TFD state at $\tL=0=\tR$. We then evolve the state backwards in time to $\tR=-t_w$, where we make the perturbation dual to the insertion of the shock wave, and then we evolve forward in time again to the initial time, \ie $\tR=0$. The perturbation has a minor effect on the final state for $t_w<t^*_{\rm scr}$, with the backward and forward time evolution essentially canceling out, and hence the complexity of formation remains unchanged for these perturbations. However, the perturbation begins to have a dramatic effect in modifying the final state for $t_w>t^*_{\rm scr}$ and hence we see that the complexity of formation begins to grow at this point --- see further discussion below.

For higher energy shock waves on the other hand, the perturbation brings in an appreciable energy and accesses new degrees of freedom. Hence the state is modified even for small $t_w$ and the regime in which the $\Delta{\cal C}$ is unchanged is absent. Instead, the complexity of formation as a function of $t_w$ starts increasing right away with initial rate given by
\begin{equation}\label{boil1}
\frac{d\Delta\ca}{d\,t_w}\bigg|_{t_w=0}= \frac{(M_2-M_1)}\pi
\qquad{\rm and}\qquad
\frac{d\Delta\cv}{d\,t_w}\bigg|_{t_w=0}= \frac{8 \pi (M_2-M_1)}{d-1}\,,
\end{equation}
as in eqs.~\eqref{SlopeAdS5Small} and \eqref{lim4}, respectively.\footnote{Note that we have ignored the logarithmic term for $\Delta\ca$ in eq.~\eqref{SlopeFormEarly}, which only appears for $d=2$ but not higher boundary dimensions.} That is, this initial growth rate is driven by the energy in the shock wave, \ie it is proportional to the difference of the masses of the two black holes. As the injection time $t_w$ continues to increase, $d\Delta{\cal C}/dt_w$ increases  and soon saturates to a constant rate proportional to the sum of the masses
\begin{equation}\label{boil2}
\frac{d\Delta\ca}{d\,t_w}= \frac{2(M_1+M_2)}\pi
\qquad{\rm and}\qquad
\frac{d\Delta\cv}{d\,t_w}= \frac{8 \pi (M_1+M_2)}{d-1}\,.
\end{equation}
Of course, the latter matches eq.~\reef{rock} in the limit where $M_2\to M_1$.

Let us observe that we can connect these results to the time evolution of holographic complexity for one-sided black holes studied in \cite{Vad1}. As discussed above eq.~\eqref{lim4}, the derivative of the complexity of formation with respect to $t_w$ can be related via the time shift symmetry to the antisymmetric time evolution of the complexity, \ie using eqs.~\eqref{tshift1} and \eqref{tshift2},
\beq
\frac{d\Delta\mC}{d\,t_w}\ =\ \frac{d \,\mC}{d\,\tR}\ - \ \frac{d \,\mC}{d\,\tL}\,,
\label{assym2}
\eeq
where we have used the fact that the complexity of preparing each CFT in its vacuum state is independent of the time and so we could replace $\Delta  \mC$ with $\mC$ on the right hand side.
The limit of one-sided black holes is obtained by setting $M_1 = 0$, $M_2=M$ and studying the dependence on the time $t_0=\tR+t_w$, \ie the time after the perturbation is inserted. The second time interval $\tL-t_w$ appearing in our expressions will not play a role after setting $M_1=0$. Hence in eqs.~\eqref{boil1} and \eqref{boil2}, we set  $M_1 = 0$ and $M_2=M$  and trade the time evolution for an evolution in $t_0$. Then taking the limit $t_w\to0$ corresponds to the early time limit $t_0\rightarrow 0$, for which eq.~\reef{boil1} yields  ${d\ca}/{d t_0} = M/\pi$ and ${d\cv}/{d t_0}= {8 \pi M}/{(d-1)}$. Similarly taking $t_w\to\infty$ corresponds to the late time limit $t_0\rightarrow \infty$, for which eq.~\reef{boil1} yields  ${d\ca}/{d t_0}= {2M}/\pi$ and ${d\cv}/{d t_0}= {8 \pi M}/{(d-1)}$. Indeed these limits precisely match eqs.~(3.47), (3.48) and (3.77) in \cite{Vad1}. In particular, the relative factor of 2 between the early and late time limits of the CA results in eqs.~\eqref{boil1} and \eqref{boil2} is the same ratio observed between the early and late time limits of one sided black holes, see \eg figure $4$ in \cite{Vad1}. Similarly, the equal early and late time limits from the CV results match the constant rate of change in complexity obtained for planar one sided black holes using the CV conjecture. Note that in taking this limit, the details of the left boundary time become unimportant and so one can also extract the same limit from the symmetric time evolution, as we explain below.

\subsection*{Time Evolution}

We have also extended the previous studies of holographic complexity in shock wave geometries \cite{Stanford:2014jda,Roberts:2014isa} by studying the full time evolution of the holographic complexity for both light and finite energy shocks and using both the CA and the CV approaches. For simplicity, we focused on the symmetric time evolution $\tL=\tR=t/2$, which is readily compared with the time evolution in the unperturbed black hole backgrounds studied in \cite{Growth}.

Let us begin by discussing the light shocks. Using both conjectures, we observed that if the shock was sent earlier than the scrambling time, the rate of change in complexity was approximately vanishing for a long period of time, centered around $t=0$. At later times, the rate of growth of the complexity rapidly approaches the growth rate found in the unperturbed geometry. For the CA conjecture, we defined a number of critical times which characterized the transitions between the different regimes of the complexity growth. These were: $-t_{c,0}$, the time in the past where the WDW patch enters the future singularity; $t_{c1}$, the time in which the point $r_m$ leaves the past singularity; and $t_{c,2}$, the time in which the crossing point $r_b$ enters  the singularity, see figure \ref{EternalShocktEvol}. The plateau we have mentioned where the rate of computation vanishes appears for $-t_{c0}\lesssim t\lesssim t_{c1}$, see the right panel in figure \ref{TimeDepBTZShockBotht}. Using the CV conjecture, we observed a similar behaviour of the time derivative of the complexity, as shown in the right panel of figure \ref{fig:tlight}. We could characterize the period over which the rate of change in complexity was nearly vanishing by another critical time $t_{c,\text{v}}$ defined as the characteristic time in which the extremal surfaces pass through the bifurcation surface in \BHO. In the limit of early and light shocks in BTZ black holes, we were able to derive analytic expressions for the various critical times
\begin{equation}\label{crititi}
t_{c1} = t_{c0} = 2 t_w - \frac{1}{ \pi T_1}\log \frac{2}{\epsilon} + O(\epsilon \log \epsilon)\,,
\eeq
and $t_{c2}=2t_w$ in the CA approach, while
\beq\label{crititi2}
t_{c,\text{v}} = 2 t_w - \frac{1}{ \pi T_1}\log \frac{1}{\epsilon} + O(\epsilon )
\end{equation}
in the CV approach. Hence, we see the appearance of the scrambling time
\begin{equation}\label{crititi3}
 t^*_{\text{scr}} = \frac{1}{2 \pi T_1}\log \frac{2}{\epsilon}
\end{equation}
in shortening of the plateau of constant complexity from $2t_w$.  This is another manifestation of the switchback effect \cite{Stanford:2014jda}, as we explain below.

For heavier shocks, the regime of vanishing computation rate was replaced by a regime in which the rate of computation was approximately constant and proportional to the difference of the masses, \ie the energy carried by the shock wave,\footnote{Again, for $\cv$, the result here and in the following equation only apply for planar black holes --- see footnote \ref{gamma9}.}
\begin{equation}\label{rat4}
\frac{d\ca}{dt}= \frac{(M_2-M_1)}\pi
\qquad{\rm and}\qquad
\frac{d\cv}{dt}= \frac{4 \pi (M_2-M_1)}{d-1}\,.
\end{equation}
There is then a rapid transition where the rate of computation approaches the late-time limit, which is proportional to the sum of the masses,
\begin{equation}\label{ratesrates}
\frac{d\ca}{dt}= \frac{(M_1+M_2)}\pi
\qquad{\rm and}\qquad
\frac{d\cv}{dt}= \frac{4 \pi (M_1+M_2)}{d-1}\,.
\end{equation}
This late time limit was approached from above using the CA conjecture and from below using the CV conjecture in all the cases analyzed.

As before, we can relate these results to analogous rates found for one-sided black holes in \cite{Vad1}. In particular, in eq.~\reef{ratesrates}, we set $M_1=0$  and replace $t_\mt{R}+t_w=t/2+t_w=t_0$, $M_2=M$, which yields ${d\ca}/{dt_0}= {2M}/\pi$ and ${d\cv}/{dt_0}= \frac{8 \pi M}{d-1}$. These then match the late time limits in eqs.~(3.48) and (3.77) of \cite{Vad1}. In all these cases, we have assumed that the value of $t_w$ was large and therefore they correspond to the $t_0\rightarrow \infty$ limit of the one-sided black holes.

The CV conjecture suggests a simple geometric picture to explain these rates in terms of surfaces wrapping around constant $r$ surfaces behind the future and past horizons, see figure \ref{surfShapes}. A similar interpretation was suggested in \cite{ying1} starting from boost symmetry principles in terms of two `tapes' storing the forward and backward Hamiltonian evolution in the past and future interiors of the black hole. Our geometric picture makes it clear that evolving only $\tL$ starting at $\tL=\tR=0$ will result in decreasing complexity.
In fact, this situation is very similar to the one described in figure \ref{s2}, where evolving the right boundary time upwards caused the surface to wrap around the critical surface in \BHT, while evolving the left boundary time upwards caused the surface to unwrap the critical surface in \BHO, resulting in a rate of change in complexity proportional to $M_2-M_1$ at early times. Of course, if we only evolve $\tL$ while holding $\tR=0$ fixed, we expect to be left with a negative rate of change of the complexity proportional to $-M_1$. We will return to this point below.

\subsection*{Null Surface Counterterm}

Our calculations using the complexity=action proposal in section \ref{sec:Shocks} included the counterterm \reef{counter} on the null boundaries of the WDW patch. Adding this surface term does not modify many key results for the CA proposal for eternal black holes, \eg the complexity of formation \cite{Format} or the late-time rate of growth \cite{Growth}. But it does modify the details of the transient behaviour in the time evolution \cite{Growth}. However, these comments are limited to the behaviour of holographic complexity in stationary spacetimes.
In studying holographic complexity in Vaidya spacetimes \cite{Vad1}, we found that the counterterm is an essential ingredient for the CA proposal. In particular, we showed that for geometries describing black hole formation, one does not recover the expected late time growth for general $d$. This effect was most dramatic for $d=2$ (and $k=+1$) where the growth rate was actually negative throughout the process, \ie without the counterterm, the complexity appeared to decrease. In section \ref{CANoCT}, we also considered dropping the counterterm in our present calculations.  There we found that without the counterterm, the holographic calculations again fail to reproduce the expected late time growth rate and that the complexity of formation does not exhibit the behaviour that is characteristic of the switchback effect. Hence the gravitational observable associated with $I_\mt{WDW}-I_\mt{ct}$ simply does not behave like complexity of the boundary state, and the results in section \ref{CANoCT} reinforce our previous arguments that the counterterm should be regarded as an essential ingredient for the CA proposal.

One interesting aspect of the counterterm is that the structure of the UV divergences in the holographic complexity is  modified, as was first noted in \cite{Simon2}, and as is discussed in appendix \ref{app:CounterTerm}. Without the counterterm, the leading UV divergence takes the form (see eq.~\reef{totUV1})
\beq
\tca^\mt{UV}\sim  \frac{{\cal V}(\Sigma)}{\delta^{d-1}}\, \log\!{\left( \frac{L}{\alpha\,\delta } \right)}\,, \label{totUV1a}
\eeq
where ${\cal V}(\Sigma)$ is the (total) volume of the time slice $\Sigma$ on which the boundary state resides. To remove the AdS scale from $\tca^\mt{UV}$, \cite{diverg} suggested  that one should choose
\beq
\alpha=L/\ell\,,
\label{chooice2}
\eeq
where $\ell$ might be some other length scale associated with the microscopic rules used to define the complexity in the boundary theory. This choice then yields
\beq
\mC^\mt{UV}\sim  \frac{{\cal V}(\Sigma)}{\delta^{d-1}}\, \log\!{\left( \frac{\ell}{\delta } \right)}\,. \label{totUV1b}
\eeq
When the counterterm contributions are included, the $\alpha$ dependence in eq.~\reef{totUV1a} is eliminated and the leading UV divergence takes the form (see eq.~\reef{totUV3})
\beq
\ca^\mt{UV}\sim \frac{{\cal V}(\Sigma)}{\delta^{d-1}}\, \log\!{\left( \frac{(d-1)\ctL }{L} \right)} \,. \label{totUV3a}
\eeq
Of course, this expression suffers from the same deficiencies as eq.~\reef{totUV1a}, \ie it contains the AdS scale which has no interpretation in the boundary theory, and it is ambiguous because the counterterm scale is undetermined. However, as before, we can use the latter ambiguity to eliminate the AdS scale. In particular, if we choose
\beq
\ctL=\frac{L}{d-1}\,\frac{\ell}{\delta}\,,
\label{chooice}
\eeq
then the leading UV divergence takes the same form as in eq.~\reef{totUV1b}. Of course, as before, one is left with the ambiguity of fixing the scale $\ell$.

Now in comparing holographic complexity with calculations of complexity in a (free) scalar field theory \cite{qft1,qft2}, it was noted that the leading contribution to the complexity took precisely the form given in eq.~\reef{totUV1b}.\footnote{We note that the logarithmic factor does not seem to arise for a free fermion \cite{qft3,Khan:2018rzm}. That is, the leading singularity takes the form $\mC^\mt{UV}\sim {\cal V}(\Sigma)/\delta^{d-1}$, which is similar to the holographic result found using the complexity=volume.} In this case, the scale $\ell$ corresponds to the width of the unentangled Gaussian reference state appearing in the evaluation of the complexity. Hence it was suggested that the freedom of choosing this scale in the field theory calculations of complexity could be associated with the ambiguity of fixing $\alpha$ in the complexity=action proposal. Since we are now advocating that the latter proposal must include the null surface counterterm \reef{counter}, we must instead associate this freedom with the ambiguity in fixing the counterterm scale, \eg as in eq.~\reef{chooice}.

However, we would like to point out a difference in these two possibilities. This difference is highlighted by first choosing $\ell$ to be a UV scale. For example, with $\ell=e^\sigma\, \delta$, the logarithmic factor in eq.~\reef{totUV1b} simply provides a numerical factor\footnote{In the QFT calculations \cite{qft1,qft2}, this choice could be motivated by the fact that it renders the unitary connecting the (unentangled) reference state and the vacuum state continuous in momentum. That is, the unitary approaches the identity when the momentum approaches the cutoff.} and the leading UV divergence reduces to $\mC^\mt{UV}\sim  \sigma \,{\cal V}(\Sigma)/\delta^{d-1}$.
However, with this choice, eq.~\reef{chooice2} yields $\alpha=e^{-\sigma}\,L/\delta$ while eq.~\reef{chooice} yields
$\ctL=e^{\sigma}\,L/(d-1)$. Hence in previous discussions without the counterterm (\eg \cite{Growth}), the UV cutoff $\delta$ appears in the transient behaviour of $d\tca/dt$, while $d\ca/dt$ is completely independent of $\delta$ after the counterterm is included in the gravitational action. In contrast, if $\ell$ is chosen to be an IR scale, the leading UV divergence \reef{totUV1b} is enhanced by the extra logarithmic factor $\log(\ell/\delta)$, and  $d\tca/dt$ is independent of $\delta$ while this UV cutoff explicitly appears in  $d\ca/dt$.

Of course, an interesting question is if either of these two behaviours is reflected in the QFT calculations of complexity. The effect of the reference scale on the complexity of the thermofield double state in a free scalar field theory was recently studied in \cite{therm0}. In this case, the transient behaviour in the time evolution does exhibit a nontrivial dependence on the reference scale. However, there is no potentially divergent behaviour found either in the case that $\ell\sim\delta$ or that $\ell$ remains an arbitrary IR scale. We might add however that this mismatch may not be very surprising. In particular, we note that the spectrum of the free scalar is not `chaotic' enough to produce the linear growth found for holographic complexity.

Another interesting comparison that one might make is with the results of the covariant regulator used for the BTZ black hole in \cite{Brown2}. In this case, the boundary of the WDW patch is defined by two timelike geodesics that originate at the past joint and reach out to $r=r_\mx$ before falling back to the future singularity. While the leading UV divergent term takes the form ${\cal V}(\Sigma)/\delta^{d-1}$, there is an explicit $\log\delta$ term appearing in the transient contributions to the rate of growth. Hence this regulator produces a result that is similar to the standard action calculations without the null surface term.

\subsection*{Integrated Complexity}
\begin{figure}
\centering
\includegraphics[scale=0.6]{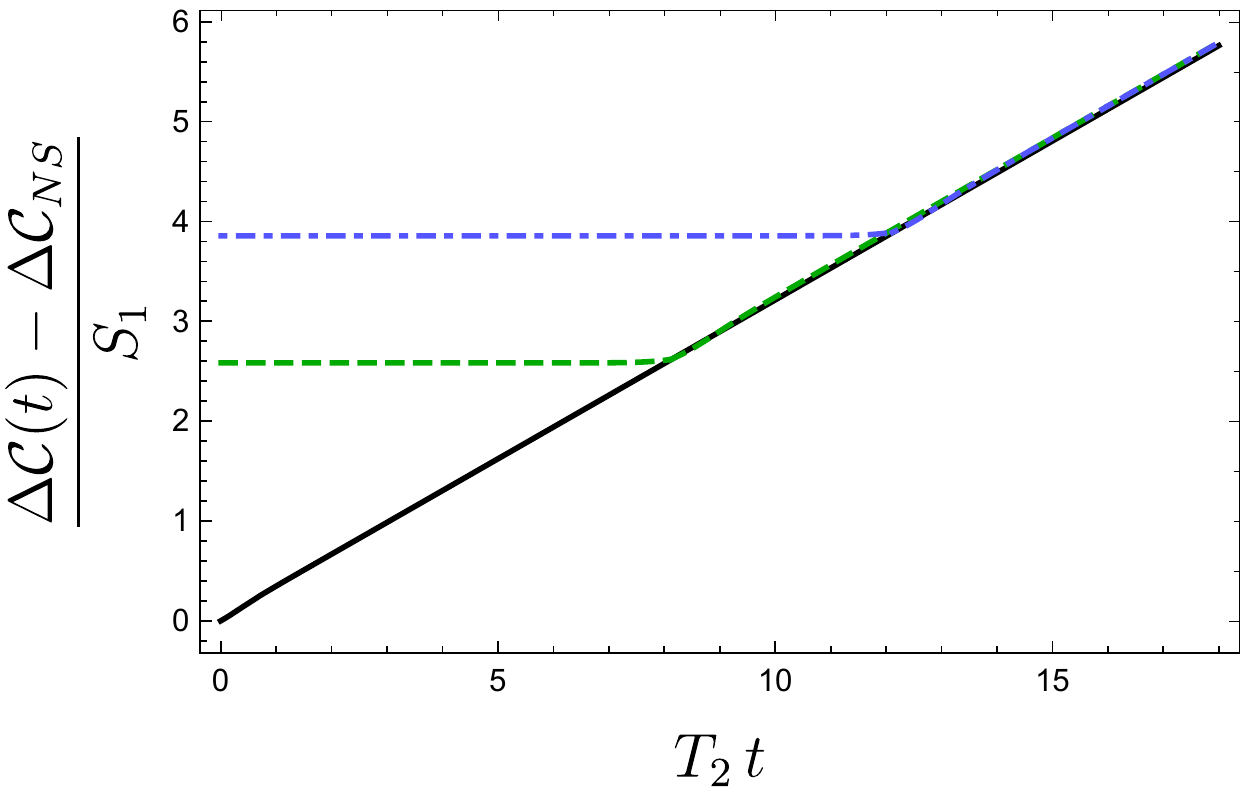}~
\includegraphics[scale=0.45]{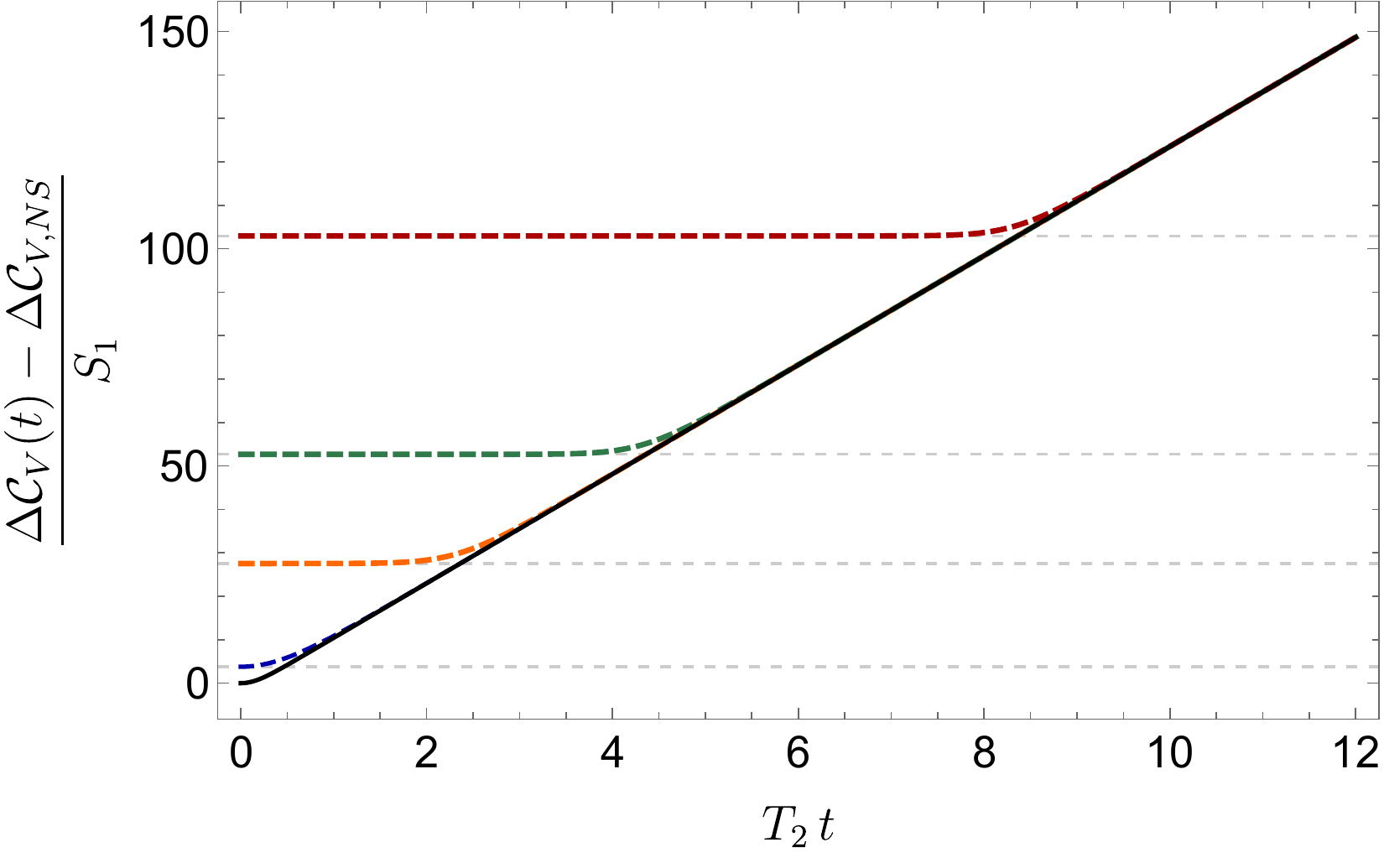}
\caption{Complexity evolution in $d=2$ for light shock waves. Left: Using the CA proposal for $w=1+10^{-5}$ and $ \ttL=1$. The unperturbed evolution is indicated by a solid blue line. We also plot the complexity evolution in the presence of shock waves -- with $T_2 t_w = 6$ (dashed green) and $8$ (dot-dashed light-blue). The initial values for those curves was fixed according to the complexity of formation. We see that the complexity does not change for a long period of time, and at late times, the complexity follows that of the unperturbed evolution. Right: Time evolution of complexity using the volume conjecture for $w=1+10^{-5}$. The unperturbed evolution is plotted in black and the evolution in the presence of the shock is in dashed colored lines --- $T_2 t_w = 6$ (red), $4$ (green), $3$ (orange) and $2$ (blue). The evolution again begins with a plateau at early times and then joins the unperturbed evolution.}
\label{LightShockCompEvol}
\end{figure}

We can also consider the behaviour of the integrated complexity, as shown for early and light shocks in BTZ black holes in figure \ref{LightShockCompEvol}. Comparing to the vacuum complexity, we start with the complexity of formation at $t=0$ but with an early shock wave. This is much larger than the complexity of formation of the unperturbed black hole, \ie we are in the linearly rising regime in, \eg the left panel of figure \ref{FormationBTZShock}. The complexity remains constant up to the critical time given by eq.~\eqref{crititi} or eq.~\reef{crititi2}. At (more or less) this time, the complexity matches that of the unperturbed black hole (of mass $M_1$) and it begins to grow such that the evolution is indistinguishable from the unperturbed evolution of an eternal black hole.  That is, the effect of inserting these early (but light) shock waves is to lift the value of the initial complexity and then it remains fixed for a (long) initial period. For later times, the complexity not only grows in  the same manner as, but is also essentially equal to, that of the unperturbed TFD. We note that this is a feature of the symmetric time evolution (\ie $\tL=\tR=t/2$), and we will discuss further how this is in accordance to quantum circuit models in the spirit of \cite{Stanford:2014jda}.

\begin{figure}
\centering
\includegraphics[scale=0.6]{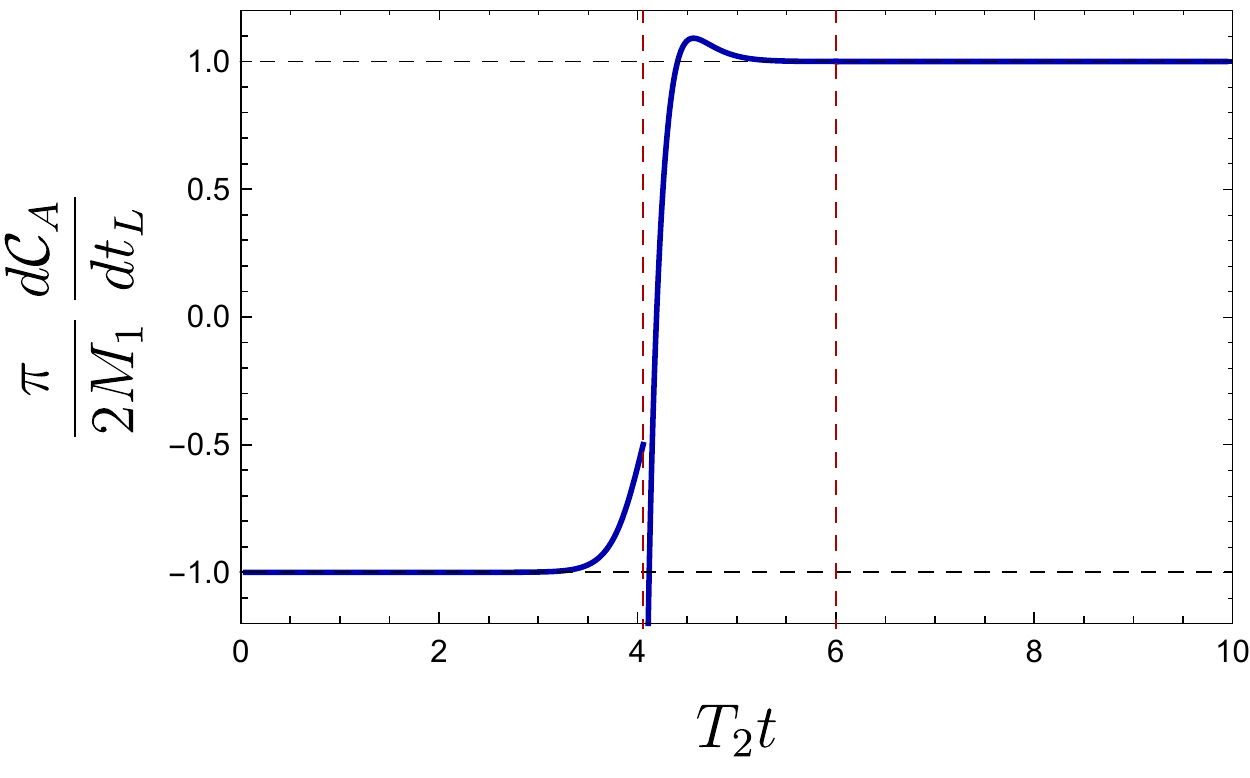}~
\includegraphics[scale=0.6]{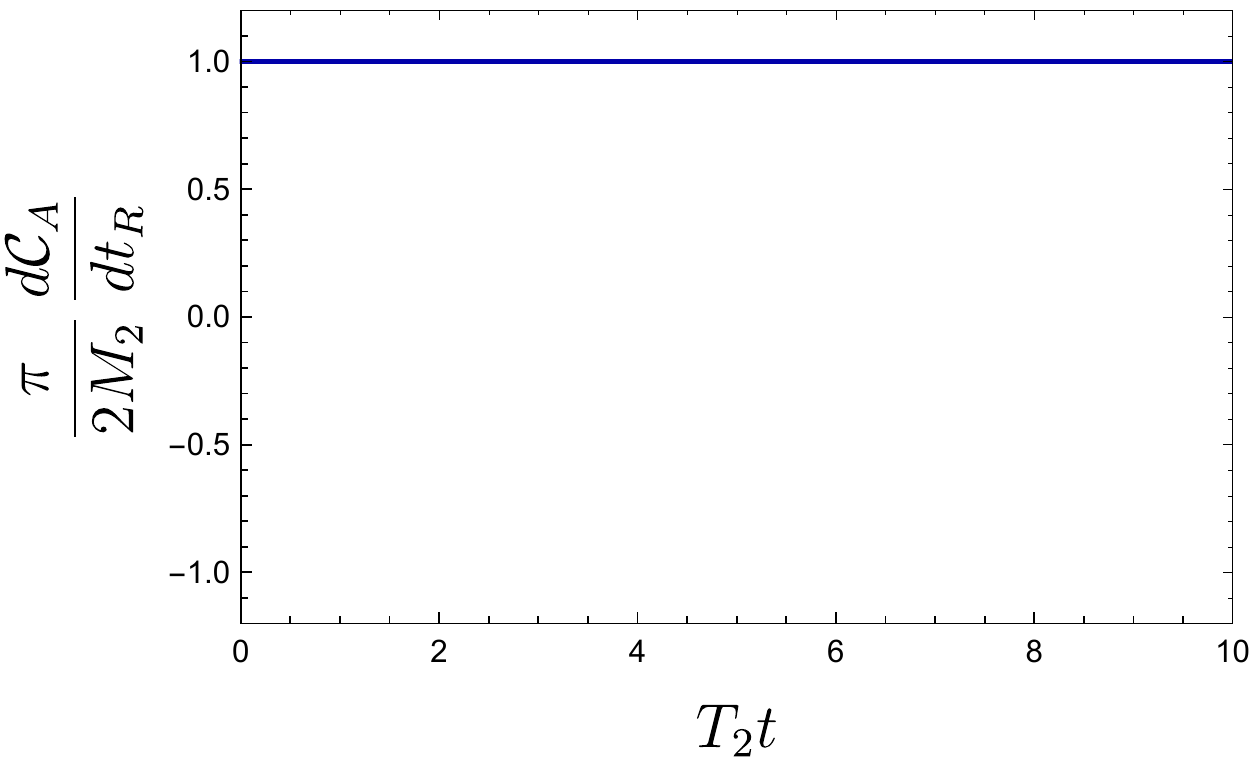}
\caption{The time evolution of complexity using the CA proposal for $d=2$. In the left panel, we fix $\tR=0$ and evolve $\tL$, while in the right panel, we fix $\tL=0$ and evolve $\tR$. In both cases, we have used $w= 1+ 10^{-5}$, $t_w T_2 = 6$ and in the left panel $\ttL=1$. The vertical red lines in the left panel indicate the critical times at which $r_m$ leaves the past singularity and later, when $r_b$ enters the future one. For the right boundary time evolution, since the shock wave is sent early enough, $r_s$ is already very close to $r_{h,2}$ and the flat profile is due to the fact that fixing $\tL= 0$ means that $r_m$ stays behind the past singularity.}
\label{DiffEvoltLtR}
\end{figure}

It is possible to interpret this behaviour in terms of summing two independent evolutions for the left and right boundary times.\footnote{We thank Adam Brown for suggesting this explanation.} For early enough shocks, when evolving the right boundary time while holding the left time fixed, the complexity begins increasing immediately at the late time limit, \ie $d\ca/d\tR\simeq 2 M_2/\pi$. In contrast, when evolving $\tL$ with $\tR$  fixed, the complexity decreases until $\tL\sim t_w-t^*_{\rm scr}$ and then makes a rapid transition to increasing with the same late-time growth rate. However, in both of these periods, the rate is governed by the mass of the past black hole, \ie $d\ca/d\tL\simeq \pm 2 M_1/\pi$.\footnote{Of course, similar behaviour occurs with the CV perspective, and can be explained using the scenario in figure \ref{s2}, where evolving only the right boundary time results in wrapping around the critical surface inside \BHT, while evolving the left boundary time results in unwrapping the critical surface behind the past horizon of \BHO.}  For a light shock wave with $M_2\simeq M_1$, summing these two behaviours together produces the initial period where the complexity is constant in the symmetric time evolution. We show this behaviour for the CA proposal in the BTZ geometry in figure \ref{DiffEvoltLtR}. In this example, we set $w= 1+ 10^{-5}$ and $t_w T_2 = 6$, and we also fix the boundary time which is not evolved at zero. We see in these examples that for times smaller than $2 (t_w -t^*_{\rm scr})$, the left evolution is negative and opposite to the right evolution, such that it has vanishing rate in the symmetric case, as shown in figure \ref{TimeDepBTZShockBotht}. In addition, the behaviour for heavy shock waves follows from this discussion. For times smaller than $2 t_w$, the left evolution contributes with $\approx -2M_1/\pi$, while the right one with $\approx +2M_2/\pi$. Hence the rate is proportional to the difference of masses in this initial phase of the symmetric evolution, as in figure \ref{TimeDepBTZShockBotht_w2}. We will discuss below a quantum circuit model that explains this behaviour.

To close this discussion, we recast various results from section \ref{sec:Shocks} in order to develop  some analytic understanding of the fact that the integrated holographic complexities match in the perturbed and unperturbed black holes (when working with light shocks and symmetric time evolution). We have already noted that for light shocks, the complexity of formation as a function of $t_w$ remains constant until the critical injection time $t_{w,c}$ given by eqs.~\reef{Scramblingt} and \reef{tscrformcv} for the CA and CV approaches, respectively. That is,
\begin{equation}\label{crititiw}
t^{\text{A}}_{w,c}=t^*_{scr}, \qquad{\rm and}\qquad t^{\text{V}}_{w,c}=t^*_{scr}+\frac{1}{2\pi T_1} \log \frac{\gamma_2}{4}\, ,
\end{equation}
where $\gamma_2\simeq \pi$ (and further, we might note that $\log(\pi/4)\simeq-0.24$).
We therefore expect that for $t_{w} > t_{w,c}$ the complexity of formation will be increased compared to the unperturbed complexity of formation by
\begin{equation}\label{DisCompofFormLight}
\Delta \mathcal{C}_A \simeq \Delta \mathcal{C}_{\text{unp}} + \frac{4 M_1}{\pi} \left(  t_w -  t^{\text{A}}_{w,c} \right)
\quad{\rm and}\quad
\Delta \mathcal{C}_V \simeq \Delta \mathcal{C}_{\text{unp}} + \frac{16 \pi M_1}{(d-1)} \left( t_w -  t^{\text{V}}_{w,c} \right)\, .
\end{equation}
Similarly, we can also approximate the time derivative of the complexity as a step function, which changes at the critical time $t_{c1}$ (CA) and $t_{c,\text{v}}$ (CV), from a rate proportional to the difference of the masses (\ie nearly vanishing for light shocks) to a rate proportional to the sum of the masses. Hence for times larger than this critical time, we have
\begin{equation}\label{DisCtShLight}
\mathcal{C}_A (t) \simeq \Delta \mathcal{C}_A +  \frac{2 M_1}{\pi} \left( t - t_{c1} \right) \,
\quad{\rm and}\quad
\mathcal{C}_V (t) \simeq \Delta \mathcal{C}_V +   \frac{8 \pi M_1}{(d-1)} \left( t - t_{c,\text{v}} \right) \,.
\end{equation}
Combining these two equations together yields
\begin{equation}
\mathcal{C}_A (t) \approx \Delta \mathcal{C}_{\text{unp}} +  \frac{2 M_1}{\pi}  t \, ,
\quad{\rm and}\quad
\mathcal{C}_V (t) \approx \Delta \mathcal{C}_{\text{unp}} +   \frac{8 \pi M_1}{(d-1)} \left(  t- \frac{1}{\pi T_1}\log \frac{\gamma_2}{2}\right) \, ,
\end{equation}
again for times larger than the critical time.
Note that this expression does not depend on $t_w$ and as a consequence the time evolutions for different $t_w$ all unify after a certain point. Further, the expression on the right hand side in both instances is approximately the time evolution of the unperturbed thermofield double. Hence the complexity of formation and the rate of growth are modified in such a way that at large times not only the rate of change of complexity matches the unperturbed result, but also the complexity itself. The extra $\gamma_2$ factor in the CV result in fact helps increase the accuracy of this argument due to the initial transient regime of the unperturbed TFD.

\begin{figure}
\centering
\includegraphics[scale=0.58]{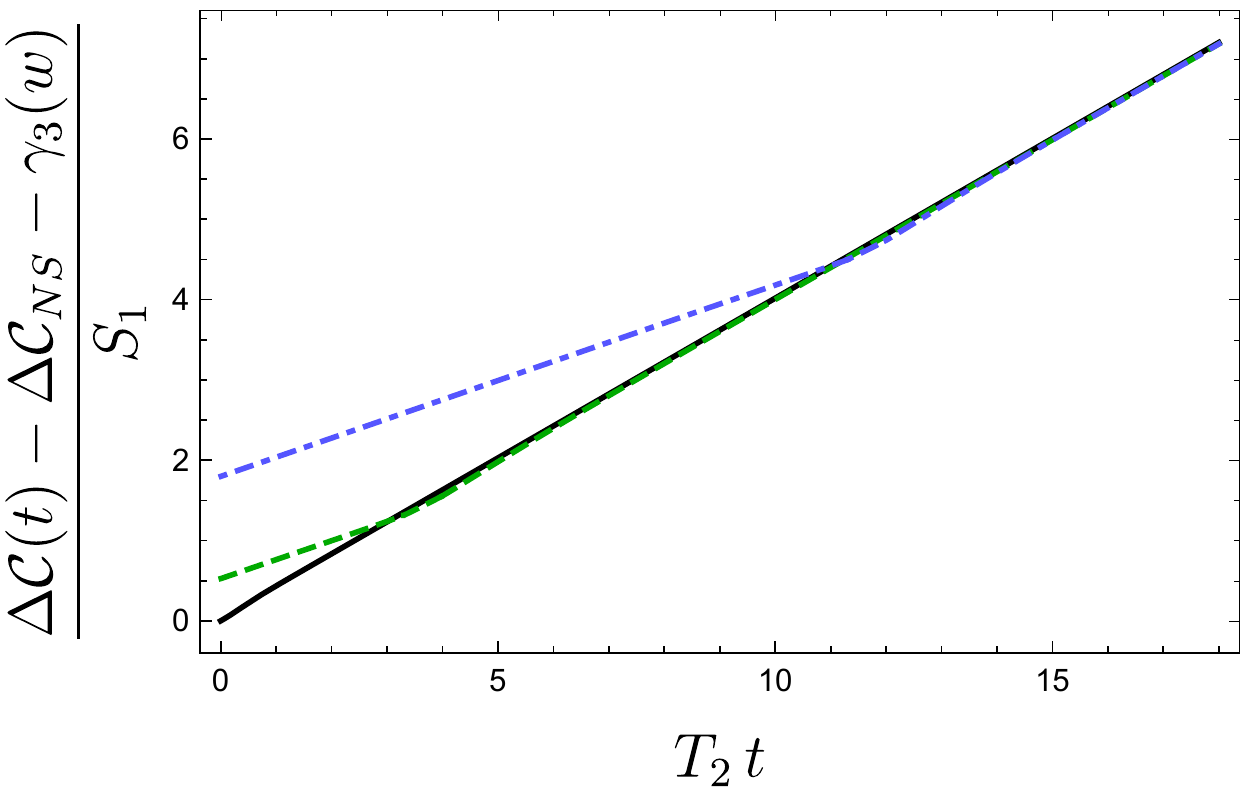}\
\includegraphics[scale=0.45]{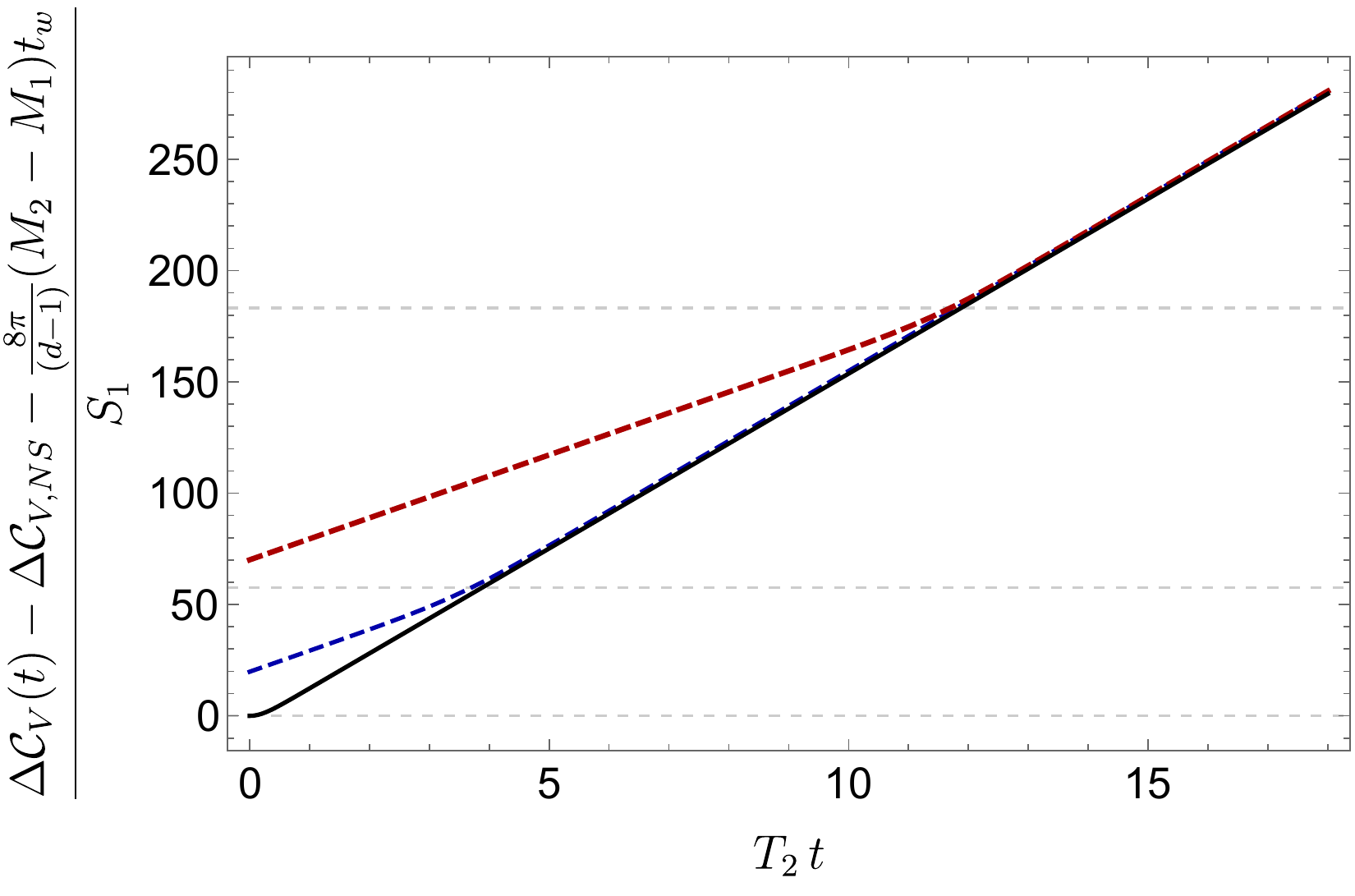}
\caption{Complexity evolution in $d=2$ for heavy shock waves. Left: Using the CA proposal for $w=2$ and $\ttL=1$. We show the full complexity profile for two different times,  $T_2 t_w = 2$ (dashed green), $T_2 t_w = 6$ (dot-dashed light blue). Right: Time evolution of complexity using the volume conjecture for $w=2$. The unperturbed evolution of an eternal BH of mass $\frac{1}{2} (M_1+M_2)$ is plotted in black and the evolution in the presence of the shock is in dashed colored lines for $T_2 t_w = 6$ (red), $T_2 t_w = 2$ (blue). The shift between the curves is proportional to $(M_2-M_1)t_w$ and matches the prediction of eq.~\eqref{heavyShift}.}
\label{HeavyShockCompEvol}
\end{figure}

A similar cancellation does {\it not} occur for heavier shocks (see figure \ref{HeavyShockCompEvol}), but one can use a similar reasoning to that above (neglecting the scrambling time) to show that the relative shift between the holographic complexities at late times is approximately given by
\begin{align}
&\Delta\mathcal{C}_A (t) -\mathcal{C}_{A,NS}  \approx  \frac{M_1 + M_2}{\pi} \, t +  \gamma_3\nonumber \\
&\Delta \mathcal{C}_V(t)-\mathcal{C}_{V,NS} \approx \frac{4 \pi }{(d-1)} (M_1+M_2) t+\frac{8 \pi }{(d-1)} (M_2-M_1) t_w \,, \label{heavyShift}
\end{align}
where the shift in the CA result is more complicated,
\begin{equation}
\gamma_3 \equiv \frac{2 (M_1 + M_2)}{\pi} \left( t_w - t_{\text{del}} \right) - 2 \frac{M_1}{\pi} t_{c1} \, ,
\end{equation}
where $t_{c1}$ and $t_{\text{del}}$ are given by eqs.~\eqref{tc1}  and \reef{laterlate}, respectively. For heavy and early shock waves, using eqs.~\eqref{tc1Largew} and \eqref{newark} for the BTZ black hole, the shift simplifies to
\begin{equation}
\gamma_3  \approx  \frac{2 (M_2 - M_1)}{\pi} t_w + \frac{4 M_1}{\pi^2 T_2} - \frac{2 (M_1 + M_2) \log{2} }{\pi^2 T_2}  + \mathcal{O} \left( \frac{\log{w}}{w} \right) \, .
\end{equation}
In figure \ref{HeavyShockCompEvol}, we explicitly show that the integrated complexities line up very closely after shifting the curves by the term proportional to $(M_2-M_1) t_w$ for both the CA and CV results.

\subsection*{Complexity=(Spacetime Volume)}

Recently it was also suggested that the holographic description of boundary complexity might be simply given by the spacetime volume of the WDW patch \cite{Couch:2016exn}. Hence this CSV proposal defines
\begin{equation}\label{defineCSV}
\csv = \frac{\nu}{16 \pi G_N L^2}\, \int_\mt{WDW} d^{d+1} x \sqrt{-g} \, ,
\end{equation}
where $\nu$ is some numerical constant. It is straightforward to test the behaviour of this proposal in the present situation since the integrand is simply a constant in the bulk integral of the action, \ie $\mathcal{R} - 2 \Lambda=-2d/L^2$. Therefore we can use our previous results for the bulk integrals to evaluate
\begin{equation}
\csv = - \frac{\nu}{2 d}\, I_{bulk}  \, . \label{BulkTotilde}
\end{equation}

For the symmetric time evolution, \ie $\tR=\tL=t/2$, we examine the growth rate for $t < t_{c1}$ in the case of early shock waves, which can be read from eqs.~\eqref{eq:BulkActEtSym} and \eqref{BulkTotilde}, with $r_m = 0$, $r_b \rightarrow r_{h,1} $ and $r_s \rightarrow r_{h,2} $,
\begin{equation}
\frac{d \csv}{d t} \bigg{|}_{t_w \rightarrow \infty} = \frac{\nu}{2 d (d-1) (1+ k w^2 z^2)} \left( M_2 - M_1\right) \, ,
\end{equation}
which we can compare to the results for the CA and CV proposals in eq.~\reef{rat4}. Further, for
the late time limit, when $r_m$ approaches $r_{h,1}$ and  $r_s$ approaches $r_{h,2}$, we find
\begin{equation}
\frac{d \csv}{d t} \bigg{|}_{t \rightarrow \infty} = \frac{\nu}{2 d (d-1)(1+ k w^2 z^2)} \left( M_2 + M_1\right) \, ,
\end{equation}
which we can compare to eq.~\reef{ratesrates}.

It is straightforward to calculate the complexity of formation for planar black holes ($k=0$) in higher dimensions, following the analysis at the end of appendix \ref{app:AppEternalAdS5}. In this case, we simply consider the large $t_w$ regime of eq.~\eqref{FormEternalSBulk} rescaled as in eq.~\eqref{BulkTotilde} and use the tortoise coordinates in eq.~\eqref{TortoisePlanarGend} which results in the following simple expression
\begin{equation}
\frac{d \Delta \csv}{d t_w} \bigg{|}_{t_w \rightarrow \infty} = \frac{\nu}{d (d-1)} \left( M_2 + M_1\right) \, .
\end{equation}
Next, we evaluate the dependence of the complexity of formation on $t_w$ when the latter is close to zero, in which case $r_s$ is close to the boundary. In this regime, we have from eq.~\eqref{FormEternalSBulkNotE}
\begin{equation}
\frac{d  \Delta \csv}{d t_w} \bigg{|}_{t_w \rightarrow 0} = \frac{\nu}{d (d-1)} \left( M_2 - M_1\right) \, .
\end{equation}
In the limit where $M_2 \gg M_1$, both results can be related to the time evolution for the one-sided collapse, as discussed above for the CA and CV results, \ie these rates can be matched with eq.~($3.21$) in \cite{Vad1}, rescaled by the factor in eq.~\eqref{BulkTotilde}.

Despite having a different overall multiplicative constant, the general properties of complexity seem to be satisfied by the CSV proposal \reef{defineCSV}. In particular, with this approach, the holographic complexity exhibits the switchback effect for any boundary dimension (including $d=2$), and the late time rate of change has a smooth limit for light shock waves. Of course, given the simple relation of the spacetime volume to the bulk integral in the CA calculations, one can suggest another simple possibility. Namely, that the holographic complexity is described by the surface and joint terms in the gravitational action alone evaluated on the boundaries of the WDW patch. Our present calculations suggest that if we drop the bulk integral from eq.~\reef{THEEACTION}, the sum of the remaining surface and joint terms obey the expected properties of complexity, up to an overall normalization. Of course, to better understand this possibility and the CSV proposal more generally, it would be interesting to examine the results for background spacetimes in which matter fields deform the geometry in an interesting way. Of course, a simple example would be to compare the results of these new proposals to the results of the CA and CV proposals for charged black holes given in \cite{Growth}.

Let us also add that \cite{Couch:2016exn} suggested a connection between the CSV proposal \reef{defineCSV} and using the `thermodynamic volume' to define the complexity,\footnote{However, we must add that this connection was recently called into question by \cite{miny}. In particular, the simple calculations of \cite{Couch:2016exn} were shown to not apply for black holes with scalar hair.} which may further hint at connections to the black hole chemistry program, \eg see \cite{KubiznakChem} for a review. Since the late time limit of geometries with two horizons, such as Reissner-Nordstrom black holes, reduces to a simple expression of differences of `internal energies',\footnote{Strictly speaking the quantity associated to the inner Cauchy horizon does not have the usual thermodynamic interpretation, but nonetheless it is a useful identification to simplify the formulas. For example, see the early suggestions in \cite{shanming, Huang:2016fks}, and recent results in the context of Lovelock theories in \cite{Cano:2018aqi}. } the authors of \cite{Couch:2016exn} suggested recasting complexity as a function of such extended thermodynamics variables. It would be an interesting future research direction to examine the physical consequences of these proposals for holographic complexity, in particular in the presence of shock waves.

\subsection*{Circuit Model}

Next we would like to consider the connections of the behaviours observed in our holographic results to the switchback effect in more detail. Following the discussion of \cite{Stanford:2014jda}, we can interpret our results with some general considerations about quantum circuit models. As discussed in section \ref{bkgd}, the boundary state of interest is the perturbed thermofield double state \reef{TFDPertState}, in which the precursor
\beq
\cOR(-t_w)=\UR (t_w)\, \cOR\, U^{\dag}_\mt{R} (t_w)
\label{prek}
\eeq
is inserted in the right CFT (where $U_\mt{L,R}=\exp[-iH_\mt{L,R}t]$ are the usual time evolution operators). Of course, if $\cOR$ was the identity operator, nothing would change in the state since the unitaries $\UR(t_w)$ and $\UR^{\dag}(t_w)$ would simply cancel in eq.~\reef{prek}. On the other hand, if $\cOR$ is a localized simple operator, $\UR(t_w)$ and $\UR^{\dag}(t_w)$ would still approximately cancel until times of the order of the scrambling time $t^{*}_{\text{scr}}$, when the effect of the perturbation $\cOR$ has propagated throughout the system. However, the behaviour will be somewhat different for `heavy' operators which inject a finite amount of energy into the system and allow the circuit to access many new degrees of freedom. Therefore we begin with a discussion of the simple operators and return to consider the heavy operators afterwards.

As discussed in section \ref{bkgd}, evolving the perturbed state independently in the left and right times yields the expression in eq.~\reef{TFDPertState2},
\beq
\left| TFD (\tL, \tR)\right\rangle_{pert} = \UR(\tR+t_w)\, \cOR\, \UR(\tL-t_w) \left| TFD  \right\rangle \, . \label{dirc}
\eeq
One immediate observation is that  there are two time scales appearing here: $\tR + t_w$ and $\tL - t_w$, precisely matching the holographic results in section \ref{sec:Shocks}. Of course,  these are the invariant combinations that were naturally picked out by the time-shift symmetry described by eqs.~\reef{tshift1} and \reef{tshift2}. However, we would like to understand whether this perspective of the circuit models provides a deeper explanation of the behaviour of the holographic complexity.

\begin{figure}
\centering
\includegraphics[scale=0.45]{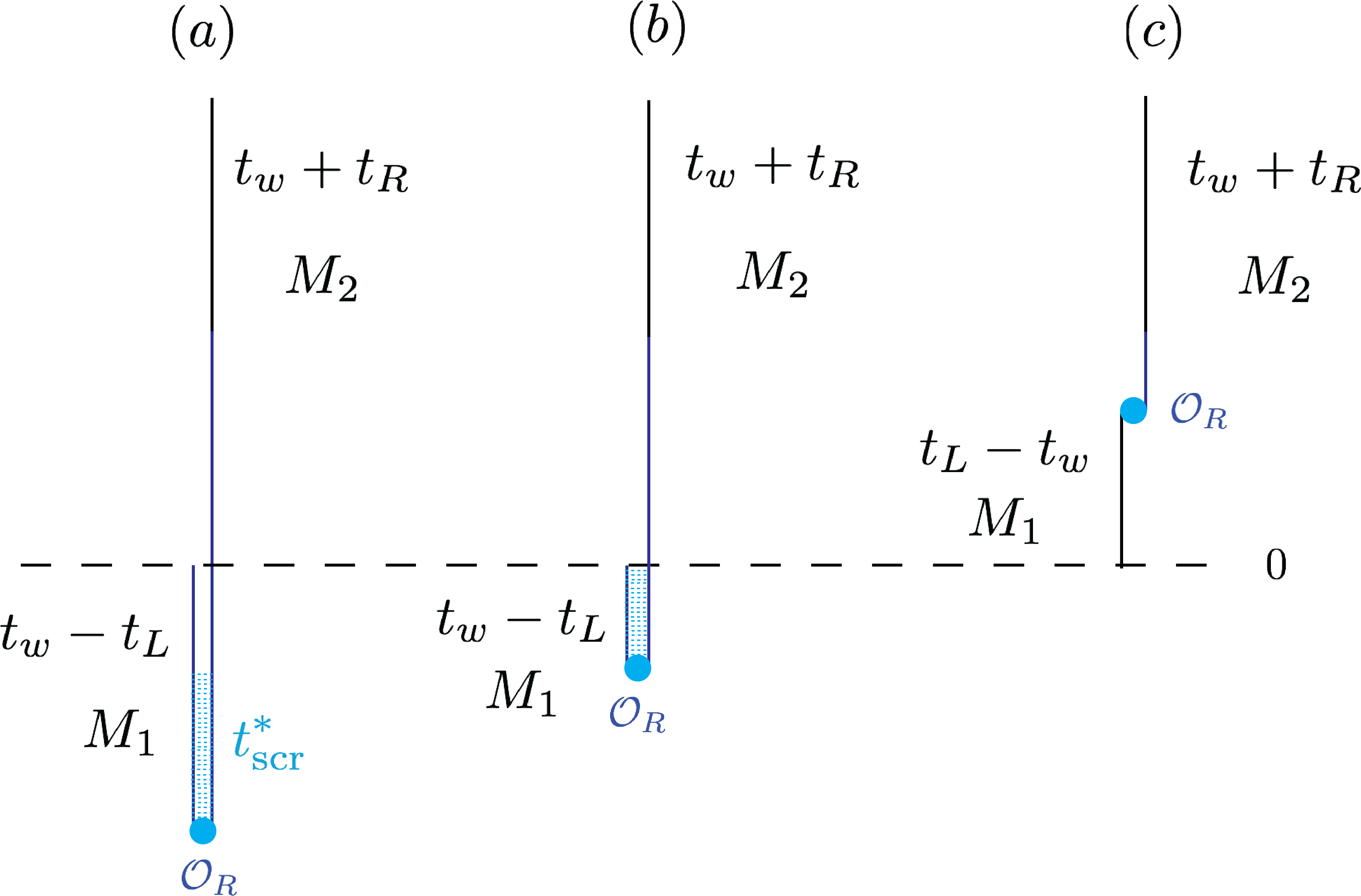}
\caption{A representation of the insertion of a simple perturbation at an early time $-t_w$ for the thermofield double state as in \eqref{dirc}, in analogy to the construction in figure 6 of \cite{Stanford:2014jda}.
(a) The regime where  $\tL  < (t_w - t^{*}_{\text{scr}})$, which corresponds to rate of change under symmetric time evolution proportional to the difference of masses ($M_2-M_1$). There is a cancellation in the time fold only during the scrambling time, which has to be accounted for in both sides of the evolution.  (b) Transient regime that still represents a late time regime for light shock waves, such that $\tL > (t_w - t^{*}_{\text{scr}}) $ but $\tL < t_w$, so there is some folding to an earlier time. However, if such folding is smaller than the scrambling time, there is an effective cancellation of the gates, and because $M_2 \approx M_1$, the complexity matches that in regime (c). (c) The late time behaviour where $\tL > (t_w - t^{*}_{\text{scr}}) $ and $\tL > t_w$, so there is no folding backwards in the insertion of the operator.}
\label{TimeInsertionsShockwave}
\end{figure}

The time evolution of the TFD state perturbed by a simple operator $\cOR$ is schematically portrayed in figure \ref{TimeInsertionsShockwave}. Along each leg of these sketches, we assume that new gates are being laid out at a fixed rate in the circuit preparing the desired state \cite{Stanford:2014jda}. Further, as is evident from the holographic results or can be argued on more general grounds \cite{EntNotEnough,Susskind:2014rva}, the rate is expected to be proportional to the energy of the system.\footnote{More precisely, one would argue that the rate is proportional to $TS$, the product of the temperature and the entropy. However, for a CFT as in the holographic framework, this product is proportional to the energy.  In the following, we set this rate to be $2M$, twice the mass of the dual black hole.} However, after the operator $\cOR$ is inserted, the evolution `folds back' in circuit space and the circuit complexity experiences the switchback effect, as illustrated in the figure. That is, most of the gates laid out in (the final stages of) the initial evolution are canceled out as the second stage of the evolution begins. This cancellation of gates is only effective for  the scrambling time $t^{*}_{\text{scr}}$. Hence as illustrated in the figure, there are three regimes of interest which are distinguished by the value of $\tL-t_w$.

In figure \ref{TimeInsertionsShockwave} (a), if $t_w$ is very large with respect to $\tL$, the initial (\ie furthest-to-the-right) operator evolves the state backwards by $|\tL-t_{w}|$, which we assume is bigger than the scrambling time. Then the second $\UR$ carries the state forward again (assuming $\tR$ is positive). The switchback effect comes into play and while the complexity grows on both legs of the evolution, the two time-evolution operators (at least partially) cancel out by an amount proportional to the scrambling time, as illustrated by the blue shaded region in the first panel of figure \ref{TimeInsertionsShockwave}. That is, the complexity for the perturbation created by a simple operator grows as
\beqa
\tL-t_w<-t^*_{\rm scr}\ :\quad\mathcal{C}_{pert} &\approx&  2M_1 |\tL-t_{w}| +  2M_2 (\tR + t_{w}) - 4 M_1 t^*_{\rm scr}   \labell{boil3} \\
&\approx& 4 M_1 (t_w - t^*_{\rm scr}) + \Delta M ( t + 2 t_w - 2 t^*_{\rm scr} )\,,
\nonumber
\eeqa
where we substituted $\tL=\tR=t/2$ for the symmetric time evolution studied in section \ref{sec:Shocks} in the second line.
Here, we have also kept small corrections of order $\Delta M \equiv M_2 -M_1$. Hence we see that $d{\cal C}_{pert}/dt \sim \Delta M$ is proportional to the difference of masses, as found for the holographic complexity, \eg in eq. \eqref{EarlyTimesRate}. However, since $M_2\approx M_1$, this rate is very close to zero and the complexity effectively remains constant.

The next regime corresponds to $-t^*_{\rm scr}<\tL-t_w<0$, as illustrated in part (b) of figure \ref{TimeInsertionsShockwave}. In this range, $\UR(\tL-t_w)$ evolves the state backward and $\UR(\tR+t_w)$ evolves forward again. However, the switchback effect produces a cancellation for the duration of the first segment because it is less than the scrambling time. Hence the effective growth of the complexity is simply given by
\beqa
-t^*_{\rm scr}<\tL-t_w<0\ : \quad\mathcal{C}_{pert} &\approx & 2M_2 (\tR + t_{w})-   2M_1 |\tL-t_{w}|  \,,
\label{boil5}\\
&\approx &2M_1\, t + 2 \Delta M\, t_w\,, \nonumber
\eeqa
where again we substituted $\tL=\tR=t/2$ and kept the corrections of order $\Delta M$. Therefore in this second regime, the rate of growth already matches that in the unperturbed thermofield double state, \ie $d{\cal C}_\mt{pert}/dt=d{\cal C}/dt $.

Of course, the final regime is when $\tL-t_{w}$ is positive, as sketched in part (c) of figure \ref{TimeInsertionsShockwave}.
In this case, both segments of the evolution move forward in time, and there is no opportunity for the switchback effect to modify the complexity and so the complexity simply grows as
\beqa
0<\tL-t_w\ : \quad\mathcal{C}_{pert}& \approx &  2 M_2(\tR + t_{w}) + 2 M_1 (\tL-t_{w})
\label{boil6}\\
&\approx&  2M_1\,t + \Delta M\, (t+2t_w) \, , \nonumber
\eeqa
where the second line corresponds to the symmetric time evolution.
Again, in this third regime, the growth rate matches that of the unperturbed state.

Hence this simple model identifies two critical times for the symmetric time evolution after a simple perturbation, namely $t_{c1}=2(t_w-t^*_{\rm scr})$ and $t_{c2}=2t_w$. Comparing to eqs.~\eqref{crititi}-\eqref{crititi3}, we see that these times are in good agreement with our holographic results for light shocks in BTZ. Looking at the growth rate suggested by the circuit model, we see that there are essentially two phases. Initially, the growth rate is almost zero and  at  $t=t_{c1}$, the complexity begins to grow with the same rate of the unperturbed state. Of course, this behaviour is in good agreement with the holographic results where we can see a rapid rise from zero to $2M_1$ after $t=t_{c1}$, as shown in the right panels of figures \ref{TimeDepBTZShockBotht} and \ref{fig:tlight}.

We can also compare the circuit model to the holographic results with a light shock for the complexity of formation by simply setting $\tL=\tR=0=t$. In this case, if we increase $t_w$ from zero, we start in the (second) regime described by eq.~\reef{boil5}. The switchback effect (almost) completely cancels the forward and backward evolution and hence the complexity and the complexity of formation are the same as in the unperturbed state. However, upon reaching $t_w\simeq t^*_{\rm scr}$, we enter the (final) regime described by eq.~\reef{boil3}. Hence the complexity of formation grows linearly with $t_w$ for $t_w\gtrsim t^*_{\rm scr}$. Again, this behaviour is in agreement with our holographic results discussed above, and \eg as shown in eq.~\reef{FormBTZLightSh} for $\Delta{\cal C}_A$.

Our holographic calculations also considered heavy shock waves but in these cases, the perturbation is no longer dual to a simple operator.  Rather the dual description would involve `heavy' operators $\cOR$, which inject a finite amount of energy and allow the circuit to access new degrees of freedom. In this case, one does not expect a cancellation of the gates when the time evolution reverses. In particular, we can approximate the number of degrees of freedom before and after the perturbation as $S_1\sim M_1/T_1$ and $S_2\sim M_2/T_2$.  Following \cite{Bridges,Brown:2016wib}, we might analyze the circuit after the time reversal in terms of an epidemic model, however, the size of the initial infection is now of order $S_2-S_1$. Hence if $S_2$ exceeds $S_1$ by some finite factor, we expect that the infection rapidly spreads through all of the degrees of freedom, \ie in a single time step --- see further comments in the next subsection. In other words, the scrambling time in the above discussion is replaced by a much shorter delay time with
\beq
t_{\text{del}}\sim 1/T_2\,,
\label{delate}
\eeq
which matches our holographic results for heavy shocks, \eg as in eqs.~\reef{tc1Largew} and \reef{cvheavydel} for the CA and CV approaches, respectively.

Hence for heavy operators, the transition between regimes essentially occurs when $\tL - t_w$ changes sign, \ie $t=2t_w$. Following the above analysis of the circuit model, initially the complexity grows as
 \beqa
\tL<t_w\ : \quad\mathcal{C}_{pert} &\approx&  2M_1 |\tL-t_{w}| +  2M_2 (\tR + t_{w})   \labell{boil3a} \\
&\approx& 2 (M_2+M_1) t_w  + (M_2-M_1)  t \,,
\nonumber
\eeqa
 where the second line describes the symmetric time evolution,
 $\tL=\tR=t/2$. Similarly in the second regime, the growth is instead described by
\beqa
\tL>t_w\ : \quad\mathcal{C}_{pert} &\approx& 2 M_1 (\tL-t_{w}) + 2 M_2(\tR + t_{w})     \labell{boil6a} \\
&\approx& 2 (M_2-M_1) t_w  + (M_2+M_1)  t \,.
\nonumber
\eeqa
Hence the rate of growth begins with $d{\cal C}_{pert}/dt\propto(M_2-M_1)$, but then makes a transition to
$d{\cal C}_{pert}/dt\propto(M_2+M_1)$ for $t\gtrsim 2t_w$.
Of course, this agrees with the behaviour of the holographic complexity with heavy shocks, \eg as shown in
figures \ref{TimeDepBTZShockBotht_w2} and \ref{fig:theavy}.

Setting $t=0$, we can compare the complexity of formation in our holographic calculations for heavy shocks and in the simple circuit model. In particular, for very small injection times, \ie $t_w\lesssim t_{\text{del}}$, eq.~\reef{boil6a} would apply, yielding $d\Delta\mC/dt_w\propto M_2-M_1$. However, the complexity rapidly transitions to the behaviour in eq.~\reef{boil3a} where $d\Delta\mC/dt_w\propto M_2+M_1$. Of course, holography yields precisely this behaviour, as discussed in  eqs.~\reef{boil1} and \reef{boil2}.

\subsection*{Simple Models}
A simple model was proposed in \cite{Bridges,Brown:2016wib} for the evolution of the complexity in terms of an epidemic spreading of infected qubits when the system is evolved in time. The authors were considering the time evolution of a single qubit operator $W$ given by the precursor $W(t)=U(t) W U(-t)$ and the suggestion was that the number of infected qubits $s(t)$ satisfies the following differential equation
\begin{equation}\label{difep}
\ell \, \frac{ds}{d t} = \frac{K-s}{K-1} \, s\qquad \longrightarrow\qquad
s(t) = \frac{K e^{t/\ell}}{K-1+e^{t/\ell}}
\end{equation}
where $K$ is the number of degrees of freedom and $\ell$ is a characteristic time step in the circuit. The boundary condition for the solution on the right was chosen as $s(t=0)=1$ since originally there was only a single infected qubit. Of course, this solution tends asymptotically to $K$. The complexity is then given by integrating over the infected qubits since these count the number of gates in the circuit which do not cancel out, and this leads to
\begin{equation}\label{csimple32}
\mathcal{C}_{\text{epidemic}}(t)=\frac{1}{\ell} \int_0^t dt \, s(t) = K \log \left(1+e^{(t-t^*)/\ell}\right)
\end{equation}
where $t^* = \ell  \log K$ is the scrambling time. Comparing to the holographic results, one makes the natural identifications that $K\sim S$ and $\ell\sim\beta$.

It is not hard to generalize the above epidemic model to describe the complexity of the precursor  $\UR(\tR+t_w)\, \cOR\, \UR(\tL-t_w)$ appearing in  eq.~\eqref{dirc}, with two independent time evolutions from the left and right sides of the perturbation. One interesting feature of the epidemic toy model compared to the previous subsection is that it yields naturally the scrambling time. In addition as we will see, it gives rise to a regime of suppressed exponential growth which is characteristic of chaotic systems and which can be observed in our holographic results. The time evolution can again be pictured as sketched in figure \ref{TimeInsertionsShockwave}. However, cases $(a)$ and $(b)$ will be treated together here as they both have $\tL-t_w<0$, and we begin by describing the behaviour in this regime.
Throughout the following, we will assume that $t_\mt{L}+t_\mt{R}>0$, and of course, as in the holographic calculations, we assume $\tR+t_w>0$.

\begin{figure}
\centering
~~~~~~~~~~~~~\includegraphics[scale=1.5]{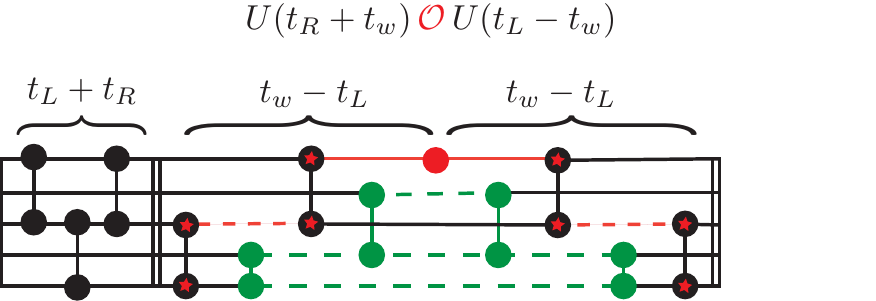}
\caption{Illustration of the spread of infected qubits in the epidemic model when $t_\mt{L}+t_\mt{R}>0$ and $t_w-t_\mt{L}>0$.}
\label{fig:cmep1}
\end{figure}
The circuit relevant for $\tL-t_w<0$ is depicted in figure \ref{fig:cmep1} where the simple operator $\cOR$ perturbing the circuit is indicated by the red dot and infected qubits are indicated by red stars. The two qubit gates that cancel out on the two sides of the unitary evolution are colored in green. In this case, we have to account for the cancellation of gates and this replaces the upper limit of integration in eq.~\eqref{csimple32} by $t_w-t_\mt{L}$. Of course, we should then add a relevant count of the gates in the part of the circuit $U_\mt{R}(t_\mt{R}+t_w)$ that goes beyond $t_w-t_\mt{L}$, \ie in the final period of length $(t_\mt{R}+t_w)-(t_w-\tL)=\tR+\tL$. This leads to the following result for the complexity in this simple epidemic model\footnote{The factor of two in the second term is the same one mentioned in footnote 9 of \cite{Bridges} and is due to the fact that $K/2$ gates act in a unit time step.}
\begin{equation}\label{epbt0}
\mathcal{C}_{\text{epidemic}}(t_\mt{L},t_\mt{R})= K \log \left(1+e^{(t_w-t_\mt{L}-t^*)/\ell}\right)+\frac{K}{2 \ell} \left(t_\mt{L}+t_\mt{R}\right)\,.
\end{equation}
If we further restrict to the symmetric time evolution, we obtain
\begin{equation}\label{casecase234}
\mathcal{C}_{\text{epidemic}}(t)= K \log \left(1+e^{(t_w-\frac{t}{2}-t^*)/\ell}\right)+\frac{K}{ 2\ell}\, t\,.
\end{equation}

Examining this result for a large insertion time $t_w$, we see two regimes. At early times $t \ll 2(t_w-t^*)$, we see that the complexity behaves as
\begin{equation}
\mathcal{C}_{\text{epidemic}}(t) \approx \frac{K}{\ell} (t_w-t^*) + K e^{-(t_w-t^*)/\ell}\,e^{\frac{t}{2\ell}} \,.
\label{epi22}
\end{equation}
Hence there are two contributions, first the constant and second a term which grows exponentially. However the latter growth is suppressed by the exponentially small prefactor $e^{-(t_w-t^*)/\ell}$. Therefore we see that the complexity is approximately constant in this regime and, as we will see below, equal to the complexity of formation. (We will come back to the tiny exponential growth later). This regime (of early times and a simple operator) corresponds to the one in figure \ref{TimeInsertionsShockwave} $(a)$ and indeed, the leading (constant) behaviour above in eq.~\reef{epi22} matches that in eq.~\eqref{boil3} if we identify ${K}/{\ell}\approx 4 M_1$. Of course, the present epidemic model  does not account for the order $\Delta M$ contribution in eq.~\reef{boil3}, while the previous simple circuit model does not account for the small exponential growth in eq.~\reef{epi22}.

At later times $t \gg 2(t_w-t^*)$, we obtain
\begin{equation}
\mathcal{C}_{\text{epidemic}}(t)\approx \frac{K}{2 \ell} t\,,
\label{epi33}
\end{equation}
which restores the linear growth of the unperturbed evolution. This regime corresponds to the sketch in figure \ref{TimeInsertionsShockwave} $(b)$. This result now matches that in eq.~\eqref{boil5} with the previous identification, \ie now the prefactor becomes ${K}/(2\ell)\approx 2 M_1$. Again, eq.~\reef{epi33} does not describe the order $\Delta M$ correction found in the circuit model discussion.

Of course, we can also consider the regime $t_\mt{L}-t_w>0$, which matches the sketch in figure \ref{TimeInsertionsShockwave} $(c)$, in which case there are no cancellations (\ie no switchback) and so the count of necessary gates is simply given by
\begin{equation}
\mathcal{C}_{\text{epidemic}}(t_\mt{L},t_\mt{R})= \frac{K}{2 \ell} (t_\mt{L}-t_w) + \frac{K}{2 \ell} (t_w+t_\mt{R})= \frac{K}{2 \ell}(t_\mt{L}+t_\mt{R})\,.
\end{equation}
Of course, restricting to symmetric time evolution with $\tL=\tR=t/2$, yields $\mathcal{C}_{\text{epidemic}}(t)=\frac{K}{2 \ell} t$, as in eq.~\reef{epi33}. This case matches the result in eq.~\reef{boil6} for the circuit model, up to the order $\Delta M$ correction.

Now we can also set $t=0$ in eq.~\eqref{casecase234} to compare with the complexity of formation, which reads\footnote{This result matches eq.~\reef{csimple32} with the replacement $t\to t_w$. Of course, this is no surprise since with $\tL=0=\tR$, the precursors in question match after this substitution and equating $W=\cOR$.}
\begin{equation}
\Delta\mathcal{C}_{\text{epidemic}}(t_w)= K \log \left(1+e^{(t_w-t^*)/\ell}\right)\,.
\label{epi44}
\end{equation}
Of course, while our notation indicates the complexity of formation associated with the precursor, this quantity cannot predict (the part of) the complexity of formation associated with the state $\left| TFD  \right\rangle$ appearing in the holographic calculations.
In any event, initially with small $t_w$, \ie $t_w\ll t^*$, we obtain $\Delta\mathcal{C}_{\text{epidemic}}(t_w)\approx K e^{(t_w-t^*)/\ell}$, which indicates an exponential growth with $t_w$ but again this term is suppressed by an exponential factor $e^{-t^*/\ell}$. In our discussion of the circuit model, this regime corresponds to the one in figure \ref{TimeInsertionsShockwave} $(b)$ and the result matches that in eq.~\eqref{boil5}. In the second regime with
$t_w\gg t^*$, eq.~\reef{epi44} yields $\Delta\mathcal{C}_{\text{epidemic}}(t_w)\approx K(t_w-t^*)/\ell$ which indicates a linear growth of the complexity of formation with respect to $t_w$ after a delay of duration $t^*$. This regime corresponds to the one in figure \ref{TimeInsertionsShockwave} $(a)$ and our result matches eq.~\eqref{boil3}.

\begin{figure}
\centering
\includegraphics[scale=0.6]{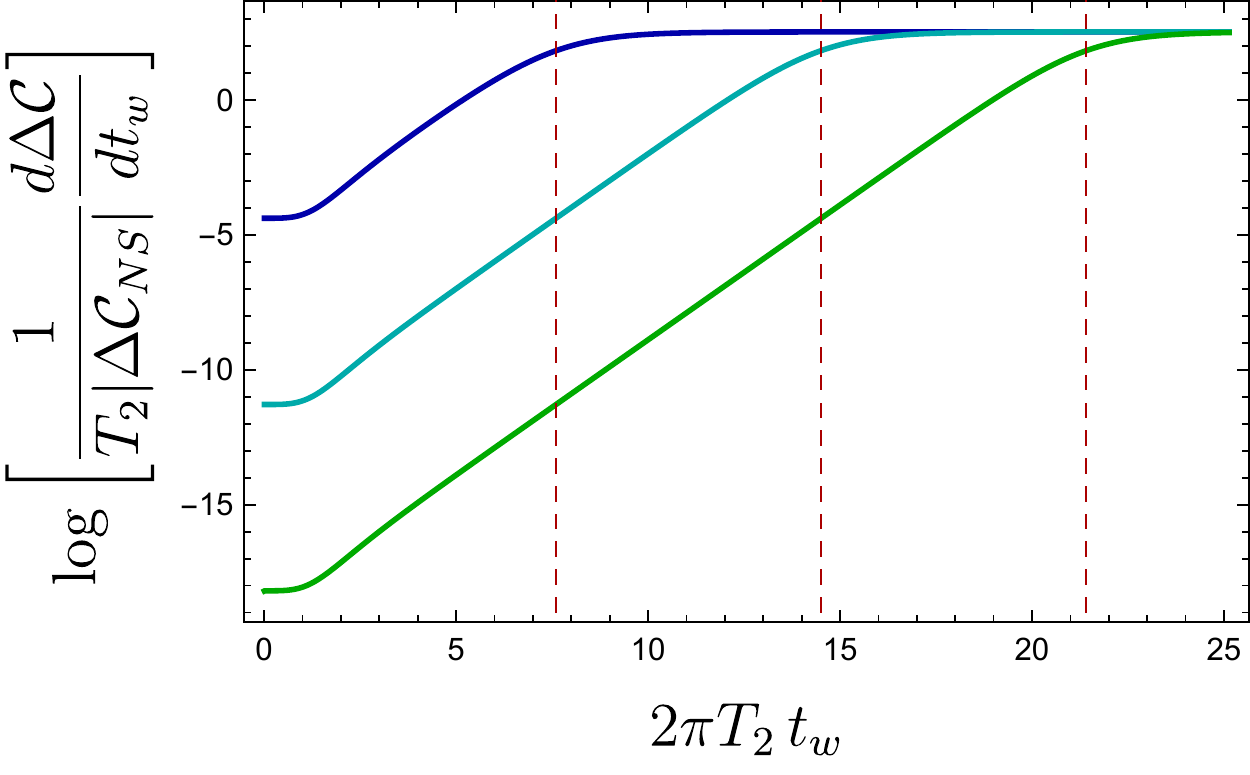}
\includegraphics[scale=0.45]{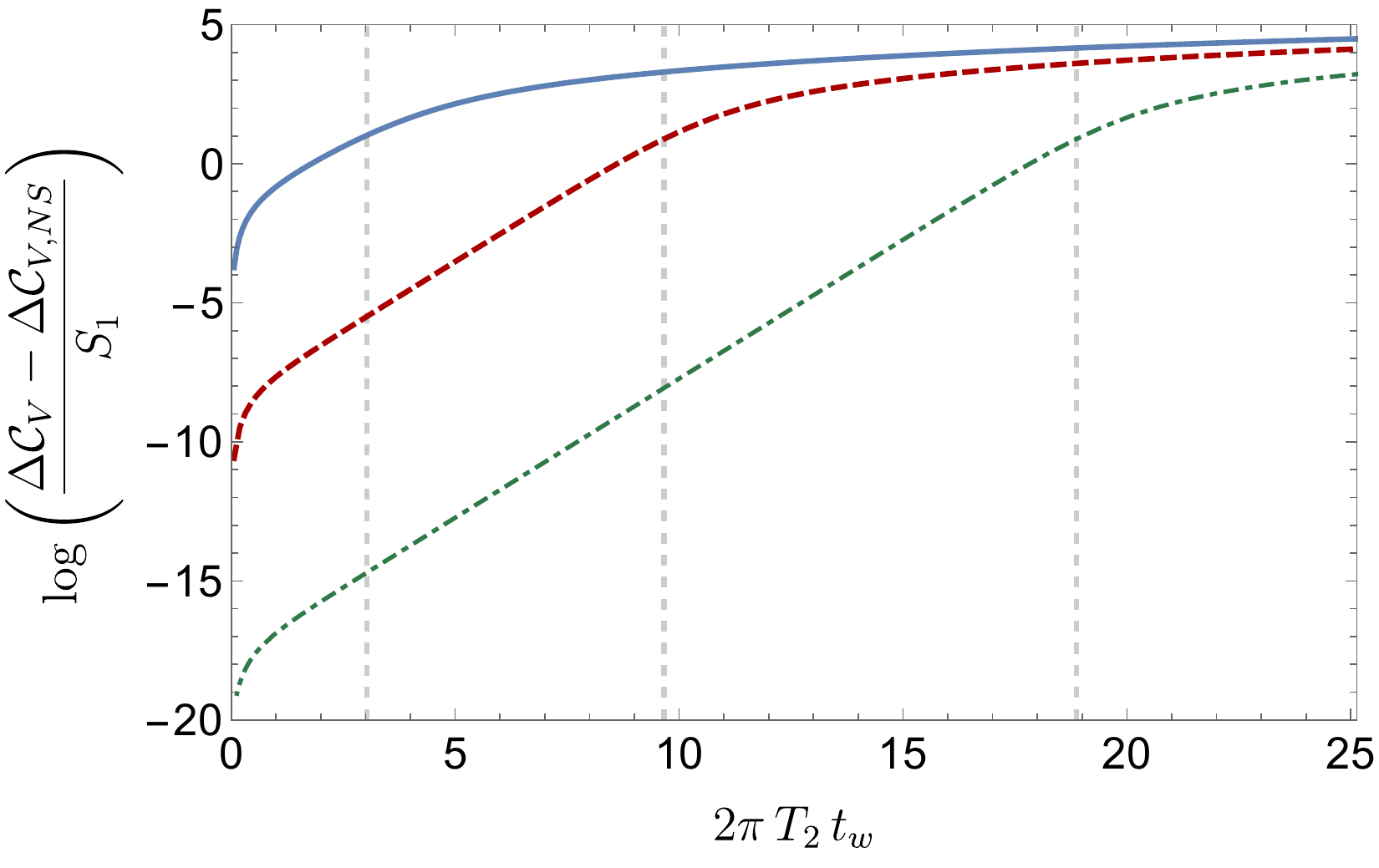}
\caption{Left: The derivative of the complexity of formation for BTZ black holes using the CA conjecture with respect to the insertion time $t_w$, for light shock waves characterized by $w=1+10^{-3}$ (dark blue), $w=1+10^{-6}$ (light blue) and $w=1+10^{-9}$ (green). We observe a period of exponential growth until times of the order of the scrambling time, which then becomes a linear growth at late times where the plot saturates. Right: early exponential dependence of the complexity of formation on the injection time $t_w$ from the CV conjecture with $w=1+10^{-1}$ (blue), $w=1+10^{-4}$ (red dashed) and $w=1+10^{-8}$ (green, dot-dashed). We can read from this plot the correct Lyapunov exponent.}
\label{DerivativeFormBTZ}
\end{figure}

We have seen that the epidemic model is in good agreement with the various different regimes of holographic complexity for light shocks. To obtain a more precise match, it is natural to choose $\ell=1/\lambda_L$ where $\lambda_L=\frac{2 \pi}{\beta}$ is the (quantum) Lyapunov exponent of gravitational systems that saturates the bound on chaos \cite{Maldacena:2015waa}.
We demonstrate in figure \ref{DerivativeFormBTZ} that an exponential growth with this particular exponent is indeed present in our holographic results for BTZ black holes. In addition, our previous identification $K/\ell\approx 4 M_1$ now indicates that $K$ is proportional to the entropy of the system. The scrambling time $t^*$ then becomes approximately the fast scrambling time  $t^*_{\rm scr}\sim \frac{\beta}{2 \pi} \log S$ of black holes \cite{ShenkerStanfordScrambling,Sekino:2008he}.

Let us add that all of the cases considered above correspond to $t_\mt{L}+t_\mt{R}>0$. If instead we take $t_\mt{L}+t_\mt{R}<0$, but keep $\tR+t_w>0$ as in the holographic model, then the right side of the circuit in figure \ref{fig:cmep1} is longer than the left part and the expression for the complexity becomes
\begin{equation}\label{epst0}
\mathcal{C}_{\text{epidemic}}(t_\mt{L},t_\mt{R})= K \log \left(1+e^{(t_\mt{R}+t_w-t^*)/\ell}\right)+\frac{K}{2 \ell}\, |t_\mt{L}+t_\mt{R}|\,.
\end{equation}
Further, let as mention that as discussed around eq.~\eqref{delate} when the energy of the insertion is large the infection will be as fast as a single time step in the system after the insertion which leads to $t_\text{del}\sim 1/T_2$. In the epidemic model this amounts to changing the boundary conditions to eq.~\eqref{difep} and starting with $K_2-K_1$ infected qubits as the initial condition for the circuit after the insertion.

In addition, the authors of \cite{Stanford:2014jda,Diagnosing} proposed that a good approximation for the complexity could be derived by looking at the lengths of geodesics stretching across the Einstein-Rosen bridge of a BTZ black hole\footnote{An infinite (but state independent) constant was subtracted off in order to obtain a finite result \cite{Stanford:2014jda}.}
\beq
{\cal C}_{\text{simple}}(t_\mt{R},t_\mt{L})\equiv \tilde K \,\log \!\left[\cosh \frac{\pi(\tL+\tR)}{\beta} +c \, \exp\!\left[ \frac{\pi}\beta \left(2t_w-2t^*_{\rm scr}+\tR-\tL\right)\right] \right] \, .
\label{hmmm}
\eeq
Above, the normalization constant $\tilde K$ should again reflect the number of degrees of freedom while $c$ is some order one constant.
Note that this expression is very similar to the previous one discussed in the context of the epidemic model. When $t_\mt{L}+t_\mt{R}\gg 0$ we may approximate $\cosh \frac{\pi(\tL+\tR)}{\beta}\approx \frac12\exp \left[\frac{\pi(\tL+\tR)}{\beta}\right]$ which then yields the form \eqref{epbt0} when identifying $\tilde K$ and $K$. On the other hand, with
$t_\mt{L}+t_\mt{R}\ll 0$, we may approximate $\cosh \frac{\pi(\tL+\tR)}{\beta}\approx \frac12\exp \left[-\frac{\pi(\tL+\tR)}{\beta}\right]$ which then exactly takes the form \eqref{epst0}.
Hence, this expression \reef{hmmm} again produces the different behaviours described above.

First, let us consider the symmetric time evolution with $\tL=\tR=t/2$, this expression \reef{hmmm} reduces to
\begin{equation}
{\cal C}_{\text{simple}}(t)= \tilde K \,\log \!\left[\cosh(\pi t/\beta) +c \, \exp\frac{2\pi}\beta (t_w-t^*_{\rm scr}) \right]\,.
\end{equation}
For early shocks (\ie $t_w\gg t^*_{\rm scr}$), the exponential term dominates the argument of the logarithm at early times. Therefore the holographic complexity is essentially constant until we reach the critical time $t\sim 2(t_w-t^*_{\rm scr})$, as described in the discussion around eq.~\eqref{crititi2}  and in figure \ref{LightShockCompEvol}. After this critical time, there is a transition to a linear growth at late times with $d{\cal C}_{\text{simple}}/dt\simeq \pi \tilde K/\beta $. The latter agrees quantitatively with our holographic results in eq.~\eqref{ratesrates} with $\tilde K_\mt{A}= 2 S (d-1)/(d \pi^2)$ for the CA coefficient and $\tilde K_\mt{V}= 8 S/d$ for the CV coefficient, both for planar black holes.

Alternatively, we may set $\tL=\tR=0$ to examine the contribution of eq.~\reef{hmmm} to the complexity of formation. In this case, the above expression simplifies to $\Delta{\cal C}\simeq K^{'} \,\log \!\left[1 +c \, \exp\frac{2\pi}\beta (t_w-t^*_{\rm scr})\right]$. Here the exponential dominates the argument of the logarithm for $t_w>t^*_{\rm scr}$ and in this regime of early shocks, the complexity of formation grows linearly with $\Delta{\cal C}\sim 2 \pi \tilde K (t_w-t^*_{\rm scr})/\beta$, which once again matches the expectation of the complexity of formation results if $\tilde K$ assumes the values discussed above for CA and CV.  We want to emphasize that it is surprising that the simple expression in eq.~\reef{hmmm}, based on geodesics in BTZ, captures so many properties of holographic complexity so well. It would be interesting to better understand this agreement in the future.

While we described the complexity of formation as being constant in the regime $t_w<t^*_{\rm scr}$, the previous discussion indicates an exponential growth with $t_w$, as is characteristic of the epidemic model. Of course, we want to stress that this growth is highly suppressed since the prefactor for this exponential carries a factor of $\exp(-2\pi t^*_{\rm scr}/\beta)$. In figure \ref{DerivativeFormBTZ}, we examine $\Delta\mC$ in this initial regime carefully with a log plot, and we find that there is indeed an exponential growth, even though this is not at all evident in the original plots, \eg the left panel of figure \ref{fig:formationcvcv}. Further, from the slope of the curves in figure \ref{DerivativeFormBTZ}, one can infer the correct Lyapunov exponent $\lambda_L = \frac{2 \pi}{\beta}$ (to a good degree of accuracy).

\subsection*{Firewalls?}

The strong sensitivity of the TFD state to small perturbations injected earlier than the scrambling time was already emphasized in \cite{ShenkerStanfordScrambling}, where it was pointed out that even a few thermal quanta of energy will be enough to completely distort the finely tuned correlations of the TFD state when sent early enough. In addition, as shown in the late time behaviour of the holographic complexity in figure \ref{LightShockCompEvol}, such deviations become indistinguishable at late times. In holography, this was explained by the fact that the energy of the shock wave is exponentially blueshifted as it falls to the event horizon \cite{ShenkerStanfordScrambling}. Of course, the characteristic time scale for this to happen is the scrambling time, and is interpreted as the time it takes for these early perturbations to have been scrambled throughout the system. This blueshift also led the authors of \cite{Cool,EvaporatingFire} to draw connections between such perturbations and firewalls. The point being that the infalling quanta can be viewed as firewall by a (not-too late) infalling observer from the second boundary (\ie the left boundary in our calculations). This also suggests that the appearance of firewalls depends on the system with which the black hole is entangled (e.g., the measurements made on the radiation exiting the black hole). One intriguing possibility is that the growth of complexity can serve as a diagnostic of firewalls \cite{Susskind:2014rva,Diagnosing,ying1}, in particular in the context of the shock wave geometries.
We have already mentioned that the complexity will actually decrease when only $\tL$ is pushed forward while holding $\tR=0$ fixed\footnote{To be more precise, according to \cite{Diagnosing}, using holographic complexity as a diagnostic of firewalls at the left horizon also requires sending $\tR=\infty$ to avoid ambiguities.} as long as $\tL<t_w-t^*$. The suggested interpretation of \cite{Diagnosing,ying1} is that the complexity is decreasing as a function of $\tL$ as long as the shock wave is within a Planck distance from the horizon along the surface and this is precisely a manifestation of a firewall which will be encountered by an observer jumping in from the left side.

\subsection*{Future Directions}

In \cite{Vad1} by examining holographic complexity in one-sided Vaidya spacetimes, we found that the null surface counterterm \reef{counter} was an essential ingredient for the CA proposal \reef{defineCA}. Our results here have reinforced this point. The most dramatic discrepancy was shown in section \ref{CANoCT} where without the counterterm, the complexity of formation did not exhibit the switchback effect for $d=2$. In section \ref{CANoCT} and appendix \ref{app:AppEternalAdS5}, we also found an unusual behaviour for the late time growth rate in the limit of very light shock waves. In particular without the counterterm, the growth rate approached the expected rate found for an eternal black hole, but at some characteristic time \reef{ScramblingDiff} related to the scrambling time, there was a transition to some more rapid growth, as illustrated in figure \ref{LateTimeNoCTT2tw6}. The overall lesson here is the importance of testing various proposals for holographic complexity in dynamical spacetimes, such as the Vaidya geometries \reef{MetricV}.

Some additional topics to explore include generalizing our results to perturbations in charged black hole backgrounds. Another route is to explore localized shocks as in \cite{Roberts:2014isa}, as well as null fluid collapses of finite thickness. In addition, very little is known about higher curvature corrections to properties of complexity \cite{shanming, Cano:2018aqi} in shock wave backgrounds. As we discussed previously in this section, it would be interesting to further investigate the complexity=(spacetime volume) conjecture \reef{defineCSV}, and under which circumstances these proposals diverge from the CA and CV conjectures. Of course, it would also be interesting to better understand the connection between this proposal and the  thermodynamic volume \cite{Couch:2016exn, miny}.

In addition, it would be interesting to investigate to what extent the holographic results can be reproduced by complexity calculations in free field theories. For example, the switchback effect seems to be a very robust feature, which is naturally associated with the geodesic deviation of adjacent trajectories \cite{Brown:2016wib} in negatively curved geometries. Therefore, it would be interesting to investigate the complexity of precursors in a field theory context, \eg using \cite{qft1,qft2} where we have already seen that negatively curved spaces can arise.
The Vaidya geometries studied here have an interpretation in terms of a thermal quench, \eg \cite{quench1,quench2}, where some boundary coupling is rapidly varied at $\tR=-t_w$. Another interesting direction might be to combine the recent discussions of complexity in the thermofield double state \cite{Yang:2017nfn,run,therm0} and in quantum quenches \cite{Alves:2018qfv, prep2} for Gaussian states to study the case of thermal quenches.

Further, it may be interesting to explore the implications of our results in the context of negative energy shock waves. In this case the negative rate of change of $M_2 -M_1$ before the scrambling time should be the main effect. Such negative energy shock waves play an important role in the construction of traversable wormholes \cite{Trav1,Trav2} and it would be interesting to check whether the profile of complexity can serve to diagnose them.

In addition, there has been recent progress on defining complexity for general CFT setups. In particular, it was proposed in \cite{MIyaji:2015mia} that the CV proposal can be related to the quantum information metric. This can be used to find the length of a circuit connecting two ground states whose respective Hamiltonians differ by an insertion of a primary operator. It would be interesting to understand in which cases the quantum information metric could also be used to study the relative complexity of states before and after a global quantum quench. Another proposal \cite{EuclideanComplexity1,EuclideanComplexity2,Bhattacharyya:2018wym} ties the complexity to the minimization of a functional given in terms of a generalized Liouville action. It would be interesting to understand how to generalize this proposal to generic time dependent backgrounds and understand if it can be used to study the time dependence of a state after a quantum quench in order to compare with the holographic results here and in \cite{Vad1}.

\section*{Acknowledgments}
We would like to thank Alice Bernamonti, Adam Brown, Dean Carmi, Lorenzo Di Pietro, Federico Galli, Minyong Guo, Robie Hennigar, Juan Pablo Hernandez, Nick Hunter-Jones, Shan-Ming Ruan, Sotaro Sugishita, Brian Swingle, Beni Yoshida and Ying Zhao for useful comments and discussions. Research at Perimeter Institute is supported by the Government of Canada through the Department of Innovation, Science and Economic Development and by the Province of Ontario through the Ministry of Research, Innovation and Science. HM and RCM thank the Kavli Institute for Theoretical Physics for its hospitality at one stage of this project. At the KITP, this research was supported in part by the National Science Foundation under Grant No. NSF PHY17-48958. SC acknowledges support from an Israeli Women in Science  Fellowship from the Israeli Council of Higher Education. RCM is supported by funding from the Natural Sciences and Engineering Research Council of Canada and from the Simons Foundation through the ``It from Qubit'' collaboration.

\appendix

\section{Counterterm for the Null Boundaries}\label{app:CounterTerm}

As originally discussed in \cite{RobLuis}, the contributions to the gravitational action from the null boundaries give rise to various ambiguities. In particular, $I_\mt{WDW}$ will depend on the parametrization of the null boundaries and so one must choose a `universal' prescription which allows the comparison of this quantity evaluated for arbitrary boundary time slices in arbitrary bulk backgrounds. However, a simple alternative, which was also suggested in \cite{RobLuis}, is to add the following counterterm to the action on the null boundaries,
\beq
I_\mt{ct} =  \frac{1}{8 \pi G_N} \int_{\mathcal{B}'} d \lambda\, d^{d-1}\theta\, \sqrt{\gamma} \ \Theta \log\left({\ctL \Theta}\right) \, ,
\label{counterX}
\eeq
where $\ctL$ is some length scale, and $\Theta$ is the expansion scalar of the null boundary generators, \ie
\beq
\Theta = \partial_{\lambda} \log \sqrt{\gamma} \, .
\label{count3X}
\eeq
The expansion $\Theta$ only depends on the intrinsic geometry of the null boundaries and so this surface term \reef{counterX} plays  no role in producing a well-defined variational principle for the gravitational action \reef{THEEACTION}. However, as shown in \cite{RobLuis}, this counterterm ensures that the action is independent of the parametrization of the null boundaries.
While adding this surface term does not modify the key features of holographic complexity in stationary spacetimes, it was argued in \cite{Vad1} and in the main text here that the counterterm is  essential when applying the CA proposal to dynamical spacetimes.

Here we examine in detail the effect of adding the counterterm \reef{counterX} to the gravitational action for the calculations in section \ref{sec:Shocks}. In particular, we will confirm that the inclusion of this term removes the dependence on the parametrization of the null boundaries, \ie on the normalization of the null normal vectors.
We also examine the role of the scale $\ctL$ appearing in eq.~\reef{counterX} in the UV divergences and in the transient behaviour in the growth of the holographic complexity.

Let us begin by reviewing\footnote{See section 2.3 of \cite{Vad1} for further details.} the computation of the expansion \reef{count3X} on the past null boundary on the right side of the WDW patch, \ie the past boundary which crosses the shock wave in figure \ref{EternalShocktEvol}. Recall that the null normals are actually tangent vectors along the null boundaries, \ie $k^\mu\partial_\mu=\partial_\lambda$, and so $\Theta$ is determined by the normalization of these vectors. For the past boundary extending to the right asymptotic AdS boundary, we can write the null normal \reef{pastrs} as
\beq
k^{p}_{\mu}\, d x^{\mu}  =H(r,v) \left( - d v + \frac{2}{F(r,v)} d r \right)
\label{pastaX}
\eeq
where $F(r,v)$ is the metric function appearing in eq.~\reef{MetricV} and $H(r,v)$ takes the form
\begin{equation}\label{Funcg}
H (r,v) = \alpha\ \mathcal{H} (r - r_s) + \tilde\alpha \, \left(1 - \mathcal{H} (r - r_s)  \right) \, ,
\end{equation}
Here $\mathcal{H}$ denotes the Heaviside function and we leave $\tilde\alpha$ unspecified for now. Further,  we have $dr/d\lambda = H(r,v)$ and hence the null expansion \reef{count3X} becomes
\begin{equation}
\Theta = \frac{H(r,v)}{r^{d-1}}\, \frac{d}{d r} \left( r^{d-1} \right) = \frac{(d-1) H(r,v)}{r} \, . \label{count4X}
\end{equation}

Hence the counterterm contribution \eqref{counterX} for $\mathcal{B}_\past$ can be written as
\begin{equation}\label{count1X}
 I^{(\RN{1})}_{\mt{ct}} = \frac{\Omega_{k, d-1} (d-1)}{8 \pi G_N} \int_{r_\mn}^{r_{\text{max}}} \!\!\! d r \, r^{d-2} \, \log{\left( \frac{ (d-1) \ctL H(r,v)}{r} \right)} \, ,
\end{equation}
where we replaced $d \lambda = d r / H(r,v)$. The upper limit of the radial integral $r_\mx$ is the position of the UV regulator surface. We set the lower limit  $r_\mn=r_m$ where $r_m$ is the position of the intersection of the two past null boundaries, with the understanding that we set $r_m=0$ when these boundaries end on the past singularity.  Hence using eq.~\reef{Funcg}, the integral in eq.~\reef{count1X} yields
\beqa
 I^{(\RN{1})}_{\mt{ct}} &=& \frac{\Omega_{k, d-1}}{8 \pi G_N} \, r_\mx^{d-1} \left[ \log\!{\left( \frac{(d-1)\alpha\ctL }{r_\mx} \right)} +\frac1{d-1}\right] \labell{count02}\\
 &&\qquad - \frac{\Omega_{k, d-1}}{8 \pi G_N} \, r_m^{d-1} \left[ \log\!{\left( \frac{(d-1) \tilde\alpha\ctL}{r_m} \right)} +\frac1{d-1}\right]
+ \frac{\Omega_{k, d-1} }{8 \pi G_N}\, r_s^{d-1} \, \log\!\left( \frac{\tilde\alpha}{\alpha} \right)\,.
\nonumber
\eeqa
Upon substituting $\tilde\alpha$ as given in eq.~\reef{AffinePast}, the above result becomes the expression given in eq.~\reef{count1}.

The future null boundary on the left side of the WDW patch also crosses the shock wave at early times, \ie for $\tL<t_{\mt{L},c2}$. Hence it is straightforward to carry out the above analysis with the corresponding null normal \reef{futurerb} and we find that the counterterm contribution becomes
\beq
 I^{(\RN{2})}_{\mt{ct}} = \frac{\Omega_{k, d-1}}{8 \pi G_N} \, r_\mx^{d-1} \left[ \log\!{\left( \frac{(d-1)\alpha\ctL }{r_\mx} \right)} +\frac1{d-1}\right]
+ \frac{\Omega_{k, d-1} }{8 \pi G_N}\, r_b^{d-1} \, \log\!\left( \frac{\hat\alpha}{\alpha} \right)\,.
\label{count03}
\eeq
Here we have assumed that we are only considering positive times when this boundary ends at $r=0$ (\ie at the future singularity). Further, for $\tL>t_{\mt{L},c2}$, we would drop the second term above since this boundary no longer crosses the shock wave.
Upon substituting $\hat\alpha$ from eq.~\reef{AffineFut}, this result matches that given in eq.~\reef{count2}.

Of course, it is also straightforward to evaluate the counterterm contributions for the null boundaries which are parallel to the trajectory of the null shell --- see figure \ref{EternalShocktEvol}. For the past null boundary extending to the left asymptotic AdS boundary, we find
\beqa
 I^{(\RN{3})}_{\mt{ct}} &=& \frac{\Omega_{k, d-1}}{8 \pi G_N} \, r_\mx^{d-1} \left[ \log\!{\left( \frac{(d-1)\alpha\ctL }{r_\mx} \right)} +\frac1{d-1}\right] \labell{count04}\\
 &&\qquad \qquad - \frac{\Omega_{k, d-1}}{8 \pi G_N} \, r_m^{d-1} \left[ \log\!{\left( \frac{(d-1) \alpha\ctL}{r_m} \right)}
 +\frac1{d-1}\right] \,,
\nonumber
\eeqa
where again, we drop the second term for $\tL<t_{\mt{L},c1}$, \ie in the regime where this boundary ends on the past singularity. Finally, for the future null boundary on the right side of the WDW patch, we have
\beq
 I^{(\RN{4})}_{\mt{ct}} = \frac{\Omega_{k, d-1}}{8 \pi G_N} \, r_\mx^{d-1} \left[ \log\!{\left( \frac{(d-1)\alpha\ctL }{r_\mx} \right)} +\frac1{d-1}\right]\,. \label{count05}
\eeq
Here we have assumed that we are in a regime where this boundary surface ends at the future singularity, as with the left future boundary in eq.~\reef{count03}.

To confirm that the inclusion of the counterterm \reef{counterX} removes the dependence on the parametrization of the null boundaries, we should combine the four counterterm contributions given above with the joint contributions for the corresponding WDW patch. In the main text, we have already evaluated the joint contributions at $r=r_s$, $r_b$ and $r_m$ in eqs.~\reef{Joint_rs},
\reef{Joint_rb} and \reef{Joint_rm}, respectively. Here we are indicating the expressions for these joint contributions before $\tilde\alpha$ and $\hat\alpha$ were fixed.\footnote{Recall that for general values of the normalization constants, $\tilde\alpha$ and $\hat\alpha$, the joint contributions in eqs.~\reef{Joint_rs} and
\reef{Joint_rb} account for the fact that $\kappa$ is nonvanishing as the corresponding null boundaries cross the shock wave \cite{Vad1}.} We must also include the joint contributions arising where the null boundaries intersect the UV regulator surface at $r=r_\mx$,\footnote{There is no contribution from the joints where the null boundaries terminate on the future or past singularities  because the area of these joints at $r=0$ vanishes.} and using the prescription given in \cite{RobLuis,diverg}, we find
\beq
I^\mt{UV}_{\mt{joint}} = -\frac{\Omega_{k, d-1}}{4 \pi G_N} \, r_\mx^{d-1} \left[ \log \frac{\alpha}{\sqrt{f_1(r_\mx)} } +
\log \frac{\alpha}{\sqrt{f_2(r_\mx)} }  \right]\,, \label{jointUV}
\eeq
where the first (second) term corresponds to the contribution from the UV joint at the left (right) asymptotic boundary.

Hence now combining the counterterm and the joint contributions, we find
\beqa
 I^\mt{tot}_{\mt{ct}}+I^\mt{tot}_{\mt{joint}} &=& \frac{\Omega_{k, d-1}}{4 \pi G_N} \, r_\mx^{d-1} \left[ \log\!{\left( \frac{(d-1)^2 \ctL^2  \sqrt{f_1(r_\mx) f_2(r_\mx)} }{r_\mx^2 } \right)} +\frac2{d-1}\right] \labell{finale}\\
 &&\quad\  \ - \frac{\Omega_{k, d-1}}{4 \pi G_N} \, r_m^{d-1} \left[ \log\!{\left( \frac{(d-1) \ctL}{r_m} \,\sqrt{|f_1(r_m)|} \right)} +\frac1{d-1}\right]
 \nonumber\\
&&\quad\ \ \,+ \frac{\Omega_{k, d-1} }{8 \pi G_N}\, r_s^{d-1} \, \log\!\left( \frac{f_1(r_s)}{f_2(r_s)} \right)
- \frac{\Omega_{k, d-1} }{8 \pi G_N}\, r_b^{d-1} \, \log\!\left( \frac{f_1(r_b)}{f_2(r_b)} \right)\,.
\nonumber
\eeqa
Hence we see that the combined result is completely independent of the normalization constants appearing in the null normals, \ie $\alpha$, $\tilde\alpha$ and $\hat\alpha$. Of course, this is simply an explicit verification that introducing the counterterm \reef{counterX} eliminates the dependence of $I_\mt{WDW}$ on the parametrization of the null boundaries \cite{RobLuis}.

At this point, we reiterate that we have left $\tilde\alpha$ and $\hat\alpha$ arbitrary above, rather than fixing them with the conditions, in eqs.~\reef{AffinePast} and \reef{AffineFut}, that the null boundaries are affinely parametrized across the shock wave. Hence we emphasize that the elimination of these normalization constants in eq.~\reef{finale} was independent of any particular choice we might make for the parametrization of the correspond boundaries. We might also note that if we choose $\tilde\alpha=\alpha$ and $\hat\alpha=\alpha$ (\eg as was done in \cite{Moosa,brian,Agon:2018zso}), the counterterm contributions at the two crossing points (\ie $r=r_s$ and $r=r_b$) vanish in eqs.~\reef{count02} and \reef{count03}. There remains a contribution at the meeting point $r=r_m$ coming from eqs.~\reef{count02} and \reef{count04}. However, at late times, these terms make a vanishing contribution to the growth rate and so  with this choice (for $\tilde\alpha$ and $\hat\alpha$), the joint terms alone capture many of the essential features of the time evolution of the complexity growth, \eg as can be seen in comparing the results in \cite{brian} and \cite{Alishahiha:2018tep}.\footnote{Both references study the complexity growth rate in hyperscaling violating geometries, but \cite{brian} does not introduce the counterterm and makes the choice $\tilde\alpha=\alpha=\hat\alpha$, while \cite{Alishahiha:2018tep} uses the counterterm.} It is also straightforward to show in the present context of a shock wave propagating into an eternal black hole background, that the key results for the time evolution of the complexity are reproduced with $\tilde\alpha=\alpha=\hat\alpha$. However, the transient early time behaviour again exhibits some differences from the results in the main text where the counterterm is included.

\subsection*{UV divergences with counterterm}

One of the interesting effects of adding the counterterm \reef{counter} to the action of the WDW patch is that it seems to change the structure of the UV divergences in the corresponding holographic complexity, as first noted in \cite{Simon2}. We would like to review these changes and the leading UV divergences carefully here because it is relevant for the comparison of the holographic complexity with the complexity evaluated in quantum field theories \cite{qft1,qft2}. The latter comparison is considered in more detail in the main text in section \ref{sec:Discussion}.

Of course, the action of the WDW patch diverges because this region of spacetime extends all the way to the asymptotic AdS boundaries, \eg as in figure \ref{EternalShocktEvol}. Hence we
regulate our calculations as usual with a UV regulator surface at
$r=r_\mx$,\footnote{Let us add that there are number of variations of this regulator procedure that one might consider \cite{diverg}. For example, one might: a) choose the null boundaries of the WDW patch to be anchored to the desired time slice on the UV regulator surface; b) choose the null boundaries to be anchored to the time slice on the asymptotic AdS boundary (\ie $r\to\infty$) but terminate the WDW patch at $r=r_\mx$, including the GHY surface term on the small (timelike) portion of the regulator surface that becomes part of the boundary; and c) proceed as in (b) and also include the usual AdS boundary counterterms \cite{Emparan:1999pm}, as well as the GHY surface term, on the portion of the boundary at $r=r_\mx$. These different choices will not change the universal features of the holographic complexity but we note that in fact, the UV divergences will agree for procedures (a) and (c). Further, we are implicitly using procedure (a) in the following, as in \cite{diverg,Simon2}.} \eg see \cite{Emparan:1999pm,deHaro:2000vlm,Skenderis:2002wp} and also the discussion for holographic complexity in \cite{Format,diverg}. Of course, this radius can be expressed in terms of the short-distance cutoff $\delta$ in the boundary theory, \eg for the present Vaidya geometries \reef{MetricV}, we have
\beq
r_\mx = \frac{L^2}\delta\left(1-\frac{k}4\,\frac{\delta^2}{L^2}+\cdots\right)\,.
\label{extra1}
\eeq

Hence the leading UV divergences in the holographic complexity take the form  \cite{diverg}
\beq
\tca^\mt{UV}=\frac{I^\mt{UV}_\mt{grav}}\pi \simeq \frac{L^{d-1}}{4 \pi^2 G_N} \, \frac{{\cal V}(\Sigma)}{\delta^{d-1}}\,\left[ \log\!{\left( \frac{L}{\alpha\,\delta } \right)}-\frac1{d-1}\right]+\cdots \,, \label{totUV1}
\eeq
where ${\cal V}(\Sigma)$ is the total volume of the boundary time slice $\Sigma$,  \eg ${\cal V}(\Sigma)=2\,\Omega_{k, d-1}L^{d-1}$, including both the left and right boundaries, for the constant time slices used in our calculations. The ellipsis indicates the subleading divergences which will involve integrals of geometric curvatures over the boundary time slice. Note that as indicated in eq.~\reef{totUV1}, we are only considering the contributions from eq.~\reef{THEEACTION}\footnote{In particular, only the bulk integral and the joint at the cutoff surface are contributing to these UV divergences.} and so we have adopted the notation of eq.~\reef{laughter} since we are not including the counterterm contribution. An interesting feature of the UV divergences in eq.~\reef{totUV1} is the appearance of the normalization constant $\alpha$ in the logarithmic factor. We might add that this factor is essential for the interpretation of this result as holographic complexity since consistency demands that the sum of the contributions in eq.~\reef{totUV1} must be positive in order for $\tca$ to be positive \cite{diverg}.

However, the counterterm contributions must remove this $\alpha$ dependence in eq.~\reef{totUV1}. Indeed combining the leading contributions from eqs.~(\ref{count02}--\ref{count05}), we find
\beq
\left[\ca^\mt{UV}\right]_\mt{ct} = \frac{L^{d-1}}{4 \pi^2 G_N} \, \frac{{\cal V}(\Sigma)}{\delta^{d-1}}\, \left[ \log\!{\left( \frac{(d-1)\alpha\ctL\delta }{L^2} \right)} +\frac1{d-1}\right]+\cdots\,, \label{totUV2}
\eeq
and then combining these UV contributions with eq.~\reef{totUV1} yields
\beq
\ca^\mt{UV}=\tca^\mt{UV}+\left[\ca^\mt{UV}\right]_\mt{ct} = \frac{L^{d-1}}{4 \pi^2 G_N} \, \frac{{\cal V}(\Sigma)}{\delta^{d-1}}\, \log\!{\left( \frac{(d-1)\ctL }{L} \right)} +\cdots\,. \label{totUV3}
\eeq
We emphasize that the counterterm removes the $\alpha$ dependence  from this leading divergence but also from all of the subleading divergences, as is evident from eq.~\reef{finale}. However, the ambiguity which $\alpha$ introduced in eq.~\reef{totUV3} has been replaced here by the ambiguity in specifying the counterterm scale $\ctL$.\footnote{We note that this ambiguity was implicitly fixed in \cite{Simon2} by setting $\ctL=L$.}

Let us add that the AdS scale $L$ appears in two places in eq.~\reef{totUV3}. The first factor, $L^{d-1}/G_N$, yields the central charge $C_T$ of the boundary theory, \eg see \cite{Buchel:2009sk}. However, the factor of $L$ in the argument of the logarithm must be absorbed by $\ctL$ in order for the final result (which is a quantity in the boundary theory) to be independent of the AdS scale.\footnote{It is straightforward to confirm that this factor of $L$ is the AdS scale, and not the boundary curvature scale, following  \cite{Growth,Cano:2018ckq}.}

\section{Complexity=Action in Higher Dimensions}\label{app:AppEternalAdS5}
In this appendix, we examine higher dimensional examples of a shock wave in an eternal black hole geometry using the CA proposal. In section \ref{sec:Shocks}, we focused on the simple case of $d=2$ in detail since much of the analysis could be carried out analytically. Here, we begin by examining the case of $d=4$ (\ie five bulk dimensions) in detail. There will be some interesting differences when comparing the behaviour of the AdS$_5$ black holes here to the BTZ black holes in section \ref{sec:Shocks}, as the sign of $t_{c2} - t_{c1}$ in eq.~\eqref{diffc} changes depending on the parameters $t_w$ (how early the shock wave is inserted) and $w$ (how heavy the shock wave is). As in section \ref{sec:Shocks}, we examine how the CA proposal is affected when we include or do not include the null surface counterterm \eqref{counter} in the WDW action. We conclude the appendix by presenting some results for the complexity of formation (with and without the counterterm) for general $d$ in the case of planar horizons (\ie $k=0$).

From eqs.~\eqref{energy} and \eqref{effect}, the parameters describing the $d=4$ boundary state dual to the AdS$_5$ black hole are
\begin{align}
&M = \frac{3 \Omega_{k, 3}}{16 \pi G_{N}} \omega^{2}\,, \qquad \, \qquad \, \qquad \omega^{2} = r_h^{2} \left(\frac{r_h^2}{L^2}  + k \right) \, , \nonumber \\
& T = \frac{1}{2 \pi r_h} \left(2  \frac{r_h^2}{L^2} + k \right) \,, \qquad \, \qquad S = \frac{\Omega_{k, 3}}{4 G_N} r_h^{3} \, .  \label{ParametersAdS5}
\end{align}
From eqs.~\eqref{fBH1} and \eqref{fBH2}, the blackening factor becomes
\beqa
{\rm for\ all}\ \vL \ \ \&\ \ \vR< - t_{w} \ :&& \qquad  F(r, v) = f_{1} (r) =  \frac{r^2}{L^2} + k - \frac{\omega_1^{2}}{r^{2}}\,,\labell{fBHd41}\\
\vR> - t_{w}\ :&& \qquad  F(r, v) = f_{2} (r) =  \frac{r^2}{L^2} + k - \frac{\omega_2^{2}}{r^{2}} \, .\labell{fBHd42}
\eeqa
Then following eqs.~\reef{rStar1} and \reef{rStar2}, the tortoise coordinates in the different regions of the black hole geometry become: for all $\vL$ and $\vR< - t_{w}$,
\begin{align}\label{Tortoise1AdS5}
r_1^{*}(r) = &\frac{L^2}{2 \left(2 r_{h, 1}^2+k L^2\right)} \bigg[ 2 \sqrt{r_{h, 1}^2+k L^2}\, \tan ^{-1}\!\left(\frac{r}{\sqrt{r_{h, 1}^2+k L^2}}\right) \nonumber \\
&\qquad-\pi  \sqrt{r_{h, 1}^2+k L^2}+r_{h, 1}\, \log\! \left(\frac{|r-r_{h, 1} |}{r+r_{h, 1}}\right) \bigg] \, ;
\end{align}
and $\vR> - t_{w}$,
\begin{align}\label{Tortoise2AdS5}
r_2^{*}(r) = &\frac{L^2}{2 \left(2 r_{h, 2}^2+k L^2\right)} \bigg[ 2 \sqrt{r_{h, 2}^2+k L^2}\, \tan ^{-1}\!\left(\frac{r}{\sqrt{r_{h, 2}^2+k L^2}}\right) \nonumber \\
&\qquad-\pi  \sqrt{r_{h, 2}^2+k L^2}+r_{h, 2}\, \log\! \left(\frac{|r-r_{h, 2} |}{r+r_{h, 2}}\right) \bigg] \, ;
\end{align}
Recall that the integration constants are chosen here such that $\lim_{r \to \infty} r_{1,2}^{*}(r) = 0$. Using the dimensionless coordinates in eq.~\eqref{eq:wz}, eq.~\reef{rocker}, the ratios of the masses and entropies before and after the shock wave become in $d=4$
\begin{equation}\label{RatioMassd4}
\frac{M_2}{M_1} = w^4 \frac{(1 + k z^2)}{(1 + k z^2 w^2)}
\qquad{\rm and}\qquad
\frac{S_2}{S_1}=w^3 \, .
\end{equation}

\subsection*{Early Time Analysis}

One interesting difference for higher dimensional AdS black holes, with respect to (three-dimensional) BTZ black holes, is that the spacetime singularities `bow' into the Penrose diagram \cite{SingBow}. As a result, when studying complexity=action for such (unperturbed) black holes (\ie with $d\ge3$), there is an initial period during which the WDW patch touches both the future and past singularities and the holographic complexity remains constant, as discussed in \cite{Brown2,Growth}. This geometric property also produces some interesting new features in the early time evolution
of the holographic complexity when we introduce shock waves in these higher dimensional black holes.

In section \ref{sec:Shocks}, we introduced two critical times in the evolution of the CA complexity for $t\ge0$. The critical time $t_{c1}$ in eq.~\eqref{tc1} determines the time when the WDW patch lifts off of the past singularity (\ie when $r_m=0$). Hence for $t > t_{c1}$, $r_m$ is a dynamical variable. The critical time $t_{c2}$ in eq.~\eqref{tc2} determines the time when the point where the shock wave hits the future singularity just moves inside of the WDW patch (\ie when $r_{b}=0$). That is, for $t>t_{c2}$, the (left) future null boundary of the WDW patch does not cross the shock wave and so $r_b$ is a dynamical variable only for $t < t_{c2}$. For the BTZ black hole discussed in section \ref{sec:Shocks}, $t_{c2}$ was always equal to $2 t_w$ and the difference between the critical times $t_{c2} -t_{c1} > 0$ was always positive, which meant that there existed a regime with both $r_m$ and $r_b$ as dynamical variables. For the higher dimensional black holes, this is not always the case, and we will derive the relevant expressions below in this appendix.

Let us begin by evaluating the time derivative of the holographic complexity at $t=0$. Note that from eq.~\eqref{tc2}, the condition that $t_{c2} > 0$ implies that $t_{w} > - 2 r^{*}_{1}(0)$. Therefore, if the shock wave is not sent early enough (\ie the latter inequality is not satisfied), only $r_s$ is a dynamical variable in the time derivative of complexity, which we will call regime $(a)$. If $t_{w} > - 2 r^{*}_{1}(0)$, then we have to consider both $r_b$ as well as $r_s$ as dynamical variables. This was always the case for the BTZ black holes studied in section \ref{sec:Shocks}. Here we denote this regime by $(b)$. The rate of change of the holographic complexity in these two regimes reads
\begin{align}
\frac{d \mathcal{C}_A^{(a)} }{d t} =&  \frac{M_2}{\pi}  - \frac{M_1}{\pi} \frac{f_2(r_s)}{f_1(r_s)}
  - \frac{M_2}{2 \pi} \frac{r_s^{d-2}}{\omega_{2}^{d-2}} f_2 (r_s) \log \frac{f_1 (r_s)}{f_2 (r_s)} \, , \label{Beforetwc} \\
\frac{d \mathcal{C}_A^{(b)} }{d t} =&  \frac{M_2}{\pi} \left( 1 + \frac{f_1(r_b)}{f_2(r_b)} \right) - \frac{M_1}{\pi} \left( 1 + \frac{f_2(r_s)}{f_1(r_s)} \right)  + \, \nonumber \\
& + \frac{M_1}{2 \pi} \frac{r_b^{d-2}}{\omega_{1}^{d-2}} f_1 (r_b) \log \frac{f_2 (r_b)}{f_1 (r_b)} - \frac{M_2}{2 \pi} \frac{r_s^{d-2}}{\omega_{2}^{d-2}} f_2 (r_s) \log \frac{f_1 (r_s)}{f_2 (r_s)} \label{Aftertwc} \, .
\end{align}

\begin{figure}
\centering
\includegraphics[scale=0.6]{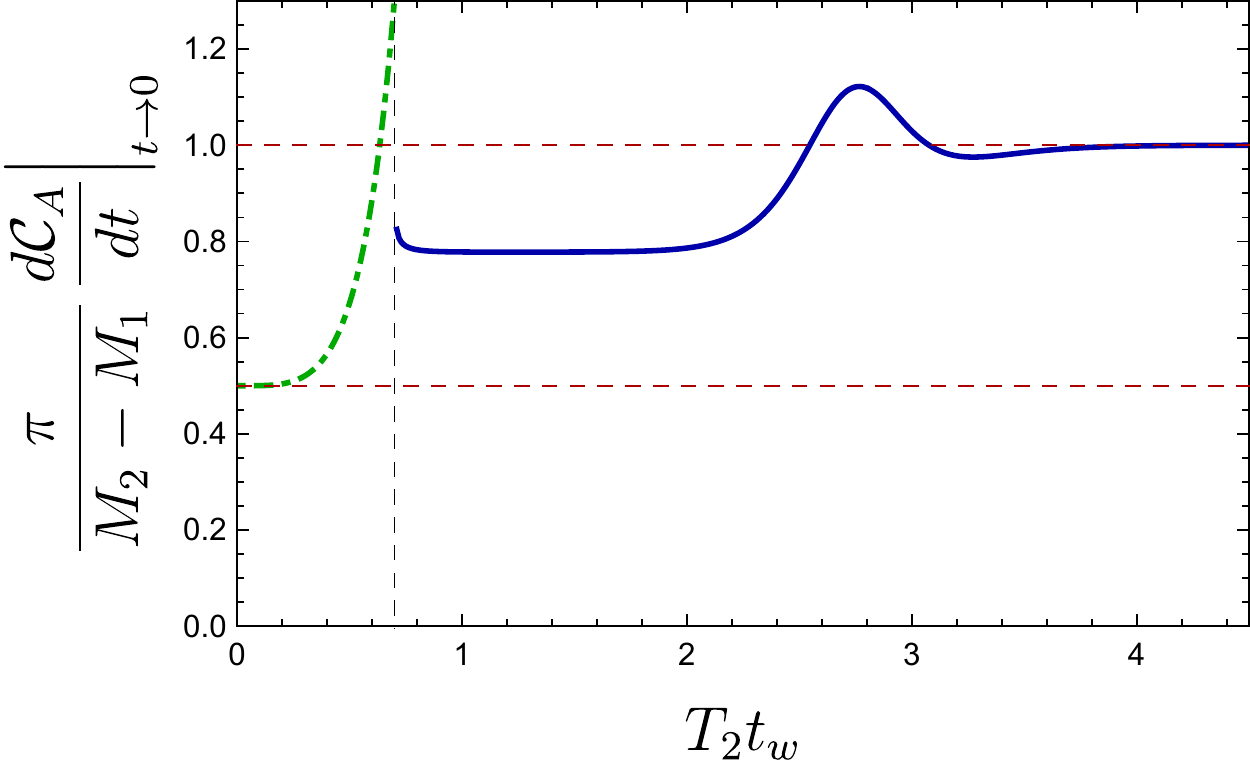}
\includegraphics[scale=0.6]{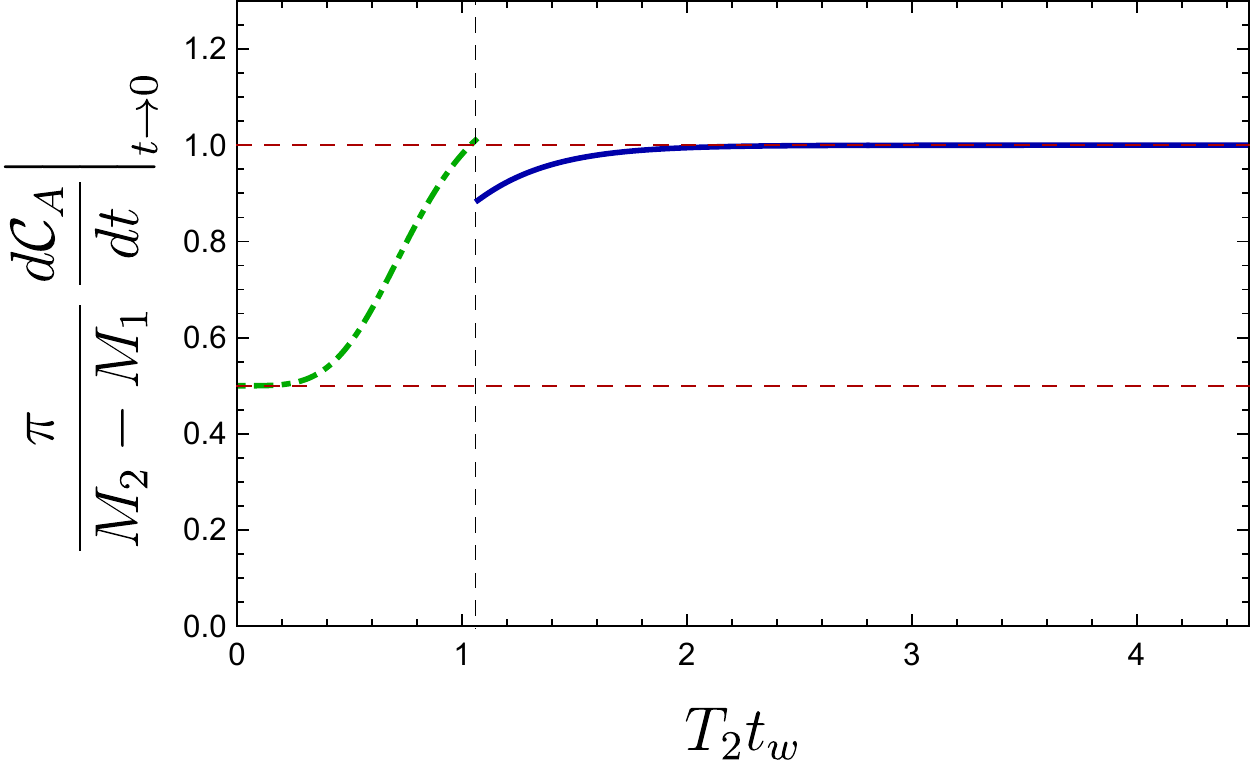}
\caption{The growth rate of the holographic complexity at $t = 0$ for an AdS$_5$ spherical black hole as a function of the perturbation time $t_w$. We choose $z= 1/w$ (as defined in eq.~\eqref{eq:wz}) so that the smaller black hole is at the HP transition. The left and right panels show the results for $w = 1 + 10^{-6}$ and $w=2$, respectively. The vertical dashed line indicates the transition from regimes $(a)$ and $(b)$ in eqs.~ \eqref{Beforetwc} and \eqref{Aftertwc}. The rate of change starts at half of $(M_2 - M_1)/\pi$ (lower dashed red line), as expected from the expansion in eq. \eqref{LimitAdS5Rega}, and for large $t_w$ the rate of $(M_2 - M_1)/\pi$ (indicated by the upper dashed red horizontal line) is reached.}
\label{EarlyShockAdS5Smw}
\end{figure}

Now we consider the rate of complexity growth at $t=0$ in more detail for AdS$_5$ spherical $k=1$ black holes, with $z$ defined in eq.~\eqref{eq:wz} given by $z = 1/w$, which means that the smaller black hole with horizon radius $r_{h,1}$ has the smallest stable horizon radius, right at the Hawking-Page transition. Of course, the overall conclusions are independent of the specific value of $z$, but we will focus on this example for concreteness. In the left panel of figure \ref{EarlyShockAdS5Smw}, we have a very light shock, with $w = 10^{-6}$, and we show the dependence on the perturbation time $t_w$. The vertical dashed line represents the time $t_{w} = - 2 r_{1}^{*}(0)$ (\ie $t_{c2}=0$), which separates between regimes $(a)$ and $(b)$ in eqs.~\eqref{Beforetwc} and \eqref{Aftertwc}. When the shock wave is inserted at very early times (\ie for large values of $t_w$), the initial rate of change becomes the difference of masses $\frac{M_2 - M_1}{\pi}$ (represented by the horizontal dashed red line). The right panel in figure \ref{EarlyShockAdS5Smw} shows the analogous results for a heavier shock wave with $w=2$. For the heavier shocks, the critical time $t_{c2}$ grows in units of $1/T_2$ as $w$ grows, and once again the early growth rate approaches $\frac{M_2 -M_1}{\pi}$ for early enough shock waves.

At $t_w \rightarrow 0$, $r_s$ approaches the AdS boundary. Therefore, if we expand eq.~\eqref{Beforetwc} in inverse powers of $r_s$, we obtain
\begin{equation}\label{LimitAdS5Rega}
\frac{d \mathcal{C}_A^{(a)} }{d t} \bigg|_{t \rightarrow 0} = \frac{M_2 -M_1}{2 \, \pi} + \mathcal{O}\left( \frac{1}{r_{s}^{4}} \right) \, .
\end{equation}
For both spherical and planar black holes, the leading order contribution from eq.~\eqref{LimitAdS5Rega} is simply $(M_2 -M_1)/2$. In contrast, for large $t_w$, it approaches the difference of masses as $r_s$ and $r_b$ approach $r_{h,2}$ and $r_{h,1}$ respectively,
\begin{equation}\label{LimitAdS5Rega}
\frac{d \mathcal{C}_A^{(b)} }{d t} \bigg|_{t \rightarrow 0} = \frac{M_2 -M_1}{\pi} + \mathcal{O}\left( T_2 t_w e^{- T_2 t_w} \right) \, .
\end{equation}

\subsection*{Time evolution in AdS$_5$}

We now turn our attention to the full time evolution in spherical AdS$_{5}$ black hole geometries with shock waves. As the previous early time analysis suggested, there are some interesting differences between the behaviour of higher dimensional black holes and the BTZ case discussed in section \ref{sec:Shocks}. It is still true that for large $t_w$, at first the complexity rate of change is approximately given by $(M_2 - M_1)/\pi$ for a period of time of the order of $2 t_{w}$. This is followed by a transient period, and then the final rate of $(M_2 + M_1)/\pi$ is reached from above, analogously to the unperturbed case in \cite{Growth}. However, there are two possible transient regimes depending on the sign of $t_{c2} - t_{c1}$, which we will analyze next.

As noted above, if the shock wave is sent early enough such that $t_w > -2 r_{1}^{*}(0)$, the dynamics of the growth rate is parametrized by the positions $r_b$ and $r_s$. Now, for light shocks, $t_{c2}- t_{c1}$ is positive, and therefore there is a regime with $r_m$, $r_b$ and $r_s$ contributing to the time derivative, as occurred for the BTZ black hole in section \ref{sec:Shocks}. However, for heavier shocks,  $t_{c2}- t_{c1}$ can be negative. That is, $r_b$ disappears into the future singularity before $r_m$ becomes dynamical. This leads to a different transition between early and late time behaviours. Of course, the dividing line between these different regimes is determined by solving $t_{c2} - t_{c1} = 0$ in eq.~\eqref{diffc}, which yields
\begin{equation}
r^{*}_{1}(r_s) =  2 r^{*}_{1}(0) \, .
\end{equation}
Because generally $r_s$ approaches $r_{h, 2}$ exponentially fast, and we are interested in a regime where $t_{w} > - 2 r_{1}^{*}(0)$, we can approximate the above equation as
\begin{equation}\label{conditionsTrans}
r^{*}_{1}(r_{h, 2}) \approx  2 r^{*}_{1}(0) \, .
\end{equation}
Despite having a simple form, it is still in general a transcendental equation. For AdS$_5$, the above expression can be explicitly written as
\begin{equation} \label{smack}
 \sqrt{k w_c^2 z_c^2+1}\left(\pi +2 \cot ^{-1}\!\left[\frac{\sqrt{kw_c^2 z_c^2+1}}{w_c}\right]\right)
 -2 \coth ^{-1}w_c = 0 \, ,
\end{equation}
where we denote $z_c$ and $w_c$ the parameters at the transition between these regimes. For instance, if we denote $z=1/w_c$ as in the previous discussion, then for $w>w_c$, $t_{c2} - t_{c1} < 0$. For $k=1$ and $z_{c}=1/w_{c}$, we find that $w_c \approx 1.00411$. In order to probe these two regimes, we solve for a very light shock wave and a heavy shock wave.

If $t_{c2} - t_{c1} > 0$, the three regimes to be considered are the same as those discussed in eq.~\eqref{ActionRegimes}. If instead $t_{c2} - t_{c1} < 0$, the time evolution passes through the three following regimes:
\begin{align}
&\ \ \RN{1}\ : -t_{c0}< t < t_{c 2} \, \qquad       \text{$r_b$, $r_s$ exist; $r_m<0$}    \nonumber \\
&\ \RN{2}\ :\ \  t_{c 2} < t < t_{c 1}  \, \qquad       \text{$r_s$ exists;
$r_b,r_m<0$}        \label{ActionRegimesB} \\
& \RN{3}\ :\qquad t > t_{c 1}  \, \qquad  \quad \text{$r_s$, $r_m$ exist; $r_b<0$ }. \nonumber
\end{align}
The rate of change of complexity for regime $\RN{1}$ is again given by eq.~\eqref{RateSymt1}, and for regime $\RN{3}$, by eq.~\eqref{RateSymt3}. For regime $\RN{2}$, it is now given by
\begin{equation}\label{RateSymtNew}
\frac{d \mathcal{C}^{(\RN{2})}_{A}}{d t} = \frac{M_2}{\pi} - \frac{M_1}{\pi} \frac{f_2(r_s)}{f_1(r_s)} - \frac{M_2}{2 \pi} \frac{r_s^{d-2}}{\omega_{2}^{d-2}} f_2(r_s) \log \left[ \frac{f_1(r_s)}{f_2 (r_s)} \right] \, .
\end{equation}
Notice that because generally $r_s$ approaches $r_{h, 2}$ exponentially fast, the rate in the above expression will be approximately constant and equal to $\frac{M_2}{\pi}$.

\begin{figure}
\centering
\includegraphics[scale=0.6]{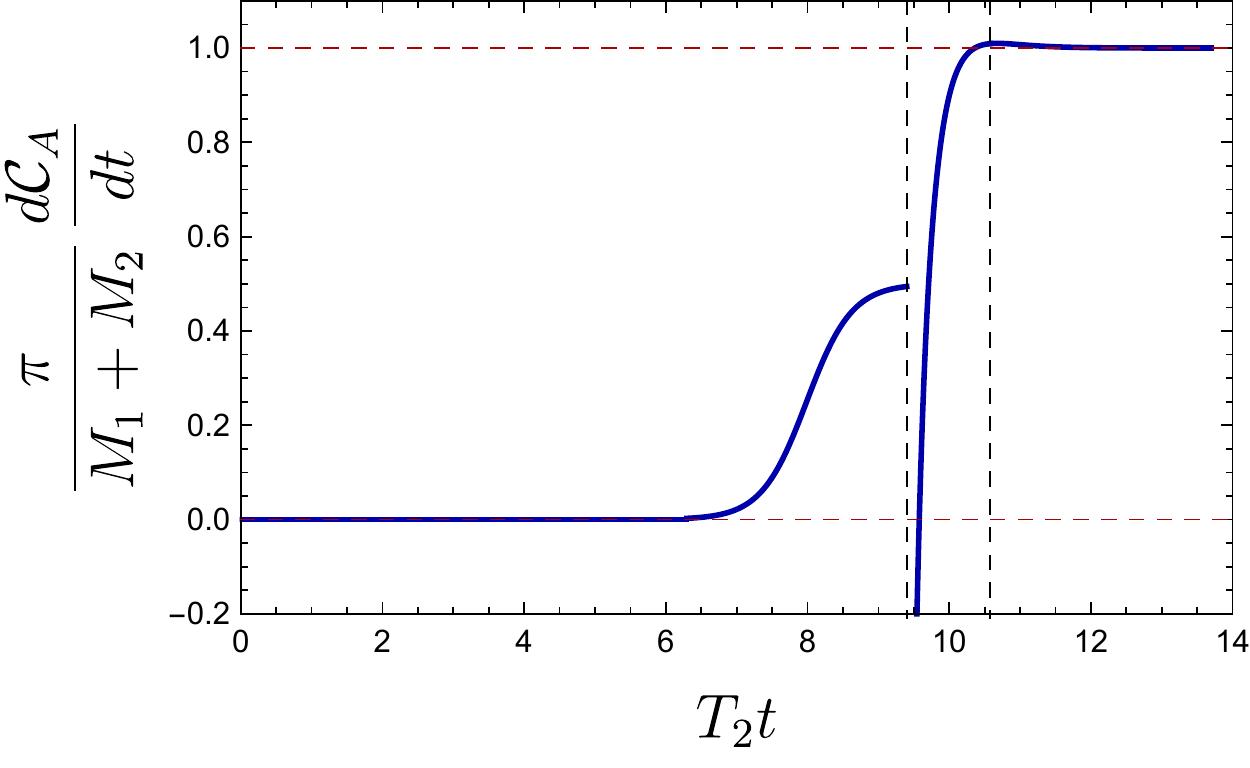}
\includegraphics[scale=0.6]{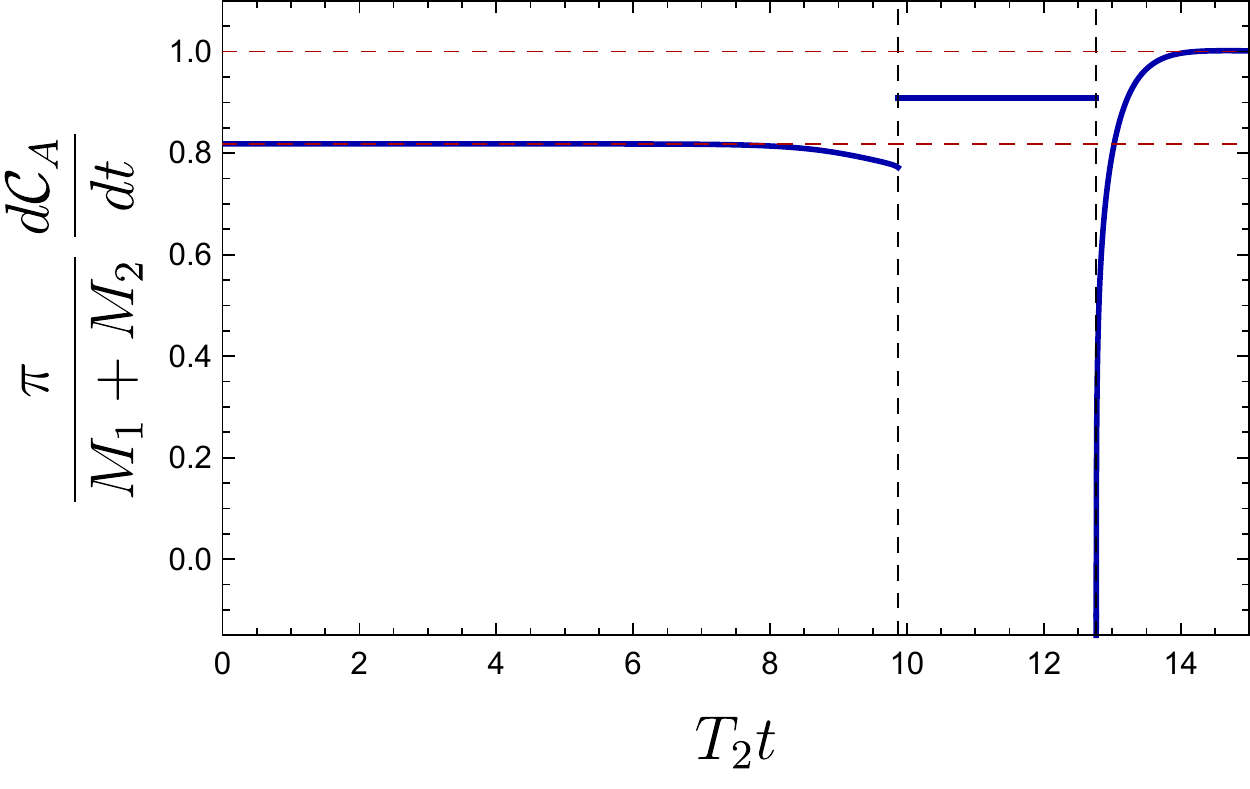}
\caption{Complexity growth rate for AdS$_5$ spherical ($k=1$) black hole with $z=1/w$, $t_w = 6 /T_2$ and $ \ttL = 1$, light shock wave $w=1+10^{-4}$ to the left and heavy shock wave $w=2$ to the right. For the heavy shock wave, $r_b$ disappears into the future singularity before $r_m$ emerges from the past singularity. Around $t=0$, there is a long plateau where the rate is $(M_2-M_1)/\pi$ (indicated by the lower dashed horizontal line). For the heavy shock wave, there is a transient regime ($t_{c2}<t<t_{c1}$) where the rate is approximately constant and given by $M_2/\pi$. In both examples, when $r_m$ emerges from the past singularity (right dashed vertical line), there is a sharp negative peak, then the late time limit of $(M_2+M_1)/\pi$ is approached from above (upper dashed horizontal line). }
\label{AdSSph}
\end{figure}

Figure \ref{AdSSph} shows the evolution of the complexity growth rate in an AdS$_{5}$ spherical black hole geometry, with  $z=1/w$ for early shocks sent at $t_w = 6/ T_2$, with a light energy of $w=1+10^{-4}$ (left) and a heavy one with $w=2$ (right). The lower dashed horizontal line indicates the limit $(M_2-M_1)/\pi$, while the upper dashed line is $(M_2 + M_1)/\pi$, in this normalization. In the heavy shock wave example, for a long time, roughly of the order $2 t_w $, the complexity growth rate is characterized by the difference of masses, then there is a transient regime (\ie $t_{c2}<t<t_{c1}$) with a constant growth rate proportional to $M_2$, as predicted by eq.~\eqref{RateSymtNew}. For the light shock wave we have $t_{c1}<t_{c2}$, which means that the transient behaviour is analogous to the one that was studied for the BTZ black hole. At $t=t_{c1}$ (first vertical line in the left panel and second in the right panel), when $r_m$ emerges from the past singularity, there is a sharp and negative peak in the rate of change. Finally, the late time limit is approached from above, and the rate is proportional to the sum of the masses.

\subsection*{Complexity of Formation}

We now consider the complexity of formation for planar (\ie $k=0$) black holes perturbed by a shock wave in asymptotically AdS$_5$ geometries. In \cite{Format}, we studied the complexity of formation for unperturbed eternal black holes and found that for planar black holes, $\Delta\mC$ is proportional to the horizon entropy (\ie the entanglement entropy between the two copies of the CFT in the thermofield double state).\footnote{At high temperatures,  $\Delta\mC$ could be expressed for the $k=\pm1$ cases as the entropy plus curvature corrections, in an expansion in inverse powers of $LT$, where here $L$ stands for the curvature of the sphere in the boundary theory \cite{Format}.} Using the CA proposal with $d=4$ and $k=0$, the unperturbed result reads
\begin{equation}\label{UnpertForm}
\Delta \mathcal{C}_{A} = \frac{S}{2 \, \pi} \, .
\end{equation}
Since the complexity of formation for stationary planar Schwarzschild-AdS black holes had this simple expression, we will investigate how this quantity behaves in the presence of shock waves.

In order to evaluate the complexity of formation in the Vaidya geometry, we need to examine separately the two regimes, $(a)$ and $(b)$, introduced above. First we consider regime $(a)$, with $r_b<0$ at $t=0$ (and hence there are no contributions from the counterterm that depends on $r_b$ to $\Delta\mC$). In this case, the contributions from the bulk integration are
\begin{align}\label{FormEternalSBulkNotE}
&\Delta I^{(a)}_{\text{bulk}} =  \left( \frac{\Omega_{k, d-1}}{16 \pi \, G_N} \right) \left( - \frac{2 d}{L^2}\right) \bigg[   \int_{r_s}^{r_{max}} d r \, r^{d-1} (- 2 r^{*}_{2}(r))  +  \int_{0}^{r_{max}} d r \, r^{d-1} (- 2 r^{*}_{1}(r)) +   \\
&+ \int_{0}^{r_s} d r \, r^{d-1} \left(t_w + 2 r^{*}_{1}(r_s) - 2 r^{*}_{1}(r) \right)   \bigg] - 2 I_{\text{bulk, Vac}} \nonumber \, .
\end{align}
The Gibbons-Hawking contributions from the future and past singularities are given in eqs.~\eqref{GHFut_Tot_Norway} and \eqref{GHPast_Tot}, respectively, resulting in
\begin{equation}\label{FormEternalSGHNotE}
\Delta I^{(a)}_{GHY} =  \frac{d \, \Omega_{k,d-1}}{16 \pi G_N} \left[ \omega_{1}^{d-2} \left( 2 r_1^{*} (r_s) - 4 r_{1}^{*}(0) \right) - \omega_{2}^{d-2} \left(   2 r_{2}^{*}(r_s)  \right) \right]    \, .
\end{equation}
There are no joint contributions using affine parametrization across the shock wave. With the inclusion of the counterterm to the null boundary that crosses the shock wave at $r_s$ given by eq.~\eqref{count1} with $r_m =0$, we have
\begin{equation}\label{FormEternalSCtNotE}
\Delta I^{(a)}_{\text{ct}} = \frac{\Omega_{k, d-1} }{8 \pi G_N} \left[ r_s^{d-1} \log \left( \frac{f_{1}(r_s)}{f_2(r_s)} \right)   \right] \, .
\end{equation}
Similarly, in the second regime $(b)$, the contribution to the complexity of formation from the counterterm in the future boundary that crosses the shock wave at $r_b$ is given by eq.~\eqref{count2},
\begin{equation}\label{FormEternalSCtb}
\Delta I^{(b)}_{\text{ct}} = \frac{\Omega_{k, d-1} }{8 \pi G_N} \left[ r_b^{d-1} \log \left( \frac{f_{2}(r_b)}{f_1(r_b)} \right)   \right] \, .
\end{equation}

The complexity of formation in regime $(a)$ is then the sum of the above contributions in eqs.~\eqref{FormEternalSBulkNotE}, \eqref{FormEternalSGHNotE} and \eqref{FormEternalSCtNotE},
\begin{equation}\label{CoFSum}
\Delta \mathcal{C}^{(a)}_{A} = \frac{ \Delta I^{(a)}_{\text{bulk}} + \Delta I^{(a)}_{GHY} + \Delta I^{(a)}_{\text{ct}}}{\pi} \, .
\end{equation}
In the second regime $(b)$, which occurs for larger values of $t_w$, the contributions to the complexity of formation are analogous to the expressions arising for the BTZ black hole discussed in section \ref{sec:Shocks}. That is, the result here is simply the sum of eqs.~\eqref{FormEternalSBulk}, \eqref{FormEternalSGH}, \eqref{FormEternalSCtNotE} and \eqref{FormEternalSCtb}, now with $d=4$.

In terms of the dimensionless coordinates \reef{dimx3}, the final result for the complexity of formation for the perturbed planar ($k=0$) AdS$_5$ black holes reads
\begin{align}\label{CoFAdS5Rega}
& \Delta \mathcal{C}_{A}^{(a)} =   \frac{S_1}{4 \pi^2} \bigg[ w^3 \log \left(\frac{x_s+1}{x_s-1}\right)-2 w^3 \tan ^{-1}\left(x_s\right)+2 \tan ^{-1}\left(w x_s\right)+\pi  (w^3+1)  \,  \\
&+ 2 w^3 x_s^3 \log \left(\frac{w^4 x_s^4-1}{w^4 \left(x_s^4-1\right)}\right)+\log \left(\frac{w x_s+1}{w x_s-1}\right)+2 \log \left(\frac{w x_s-1}{w x_s+1}\right) \bigg] \, ,
\nonumber
\end{align}
for regime $(a)$, while for regime $(b)$, the expression becomes
\small
\begin{align}\label{CoFAdS5Regb}
& \Delta \mathcal{C}_{A}^{(b)} =   \frac{S_1}{4 \pi^2 \, w} \bigg[  4 w^4 \tan ^{-1}\left(\frac{x_b}{w}\right)+2 w^4 \tan ^{-1}\left(x_s\right)+2 w \tan ^{-1}\left(w x_s\right)-4 \tan ^{-1}\left(x_s\right)  \,  \nonumber \\
&+2 w x_b^3 \log \left(\frac{x_b^4-w^4}{x_b^4-1}\right)+\left(w^4-2\right) \log \left(\frac{x_s-1}{x_s+1}\right)+w \log \left(\frac{w x_s-1}{w x_s+1}\right)-2 x_b^4 \tan ^{-1}\left(\frac{x_b}{w}\right) \nonumber \\
&+ \left(x_b^4-2 w^4\right) \log \left(\frac{x_b+w}{w-x_b}\right)+2 w \left(x_b^4-2 w^4\right) \tan ^{-1}\left(x_b\right)+\left(2 w^5-w x_b^4\right) \log \left(\frac{x_b+1}{1-x_b}\right)  \nonumber \\
&-\pi  \left((w-1) x_b^4-2 w^5+w^4+w-2\right)-8 w^4 x_s^3 \log (w)+2 w^4 x_s^3 \log \left(\frac{w^4 x_s^4-1}{x_s^4-1}\right) \bigg] \, .
\end{align}
\normalsize
Recall that the cross-over between these two regimes occurs at $t_{w} = - 2 r_{1 }^{*}(0)$ with the tortoise coordinate given in eq.~\reef{Tortoise1AdS5}. Hence this transition occurs when
\beq
t_w = \frac{\pi L^2 \sqrt{r_{h, 1}^2+k L^2}}{ 2 r_{h, 1}^2+k L^2} \, ,
\label{transitab}
\eeq
which for $k=0$, yields $t_w=\pi L^2/(2r_{h,1})=1/(2T_1)$.

\begin{figure}
\centering
\includegraphics[scale=0.6]{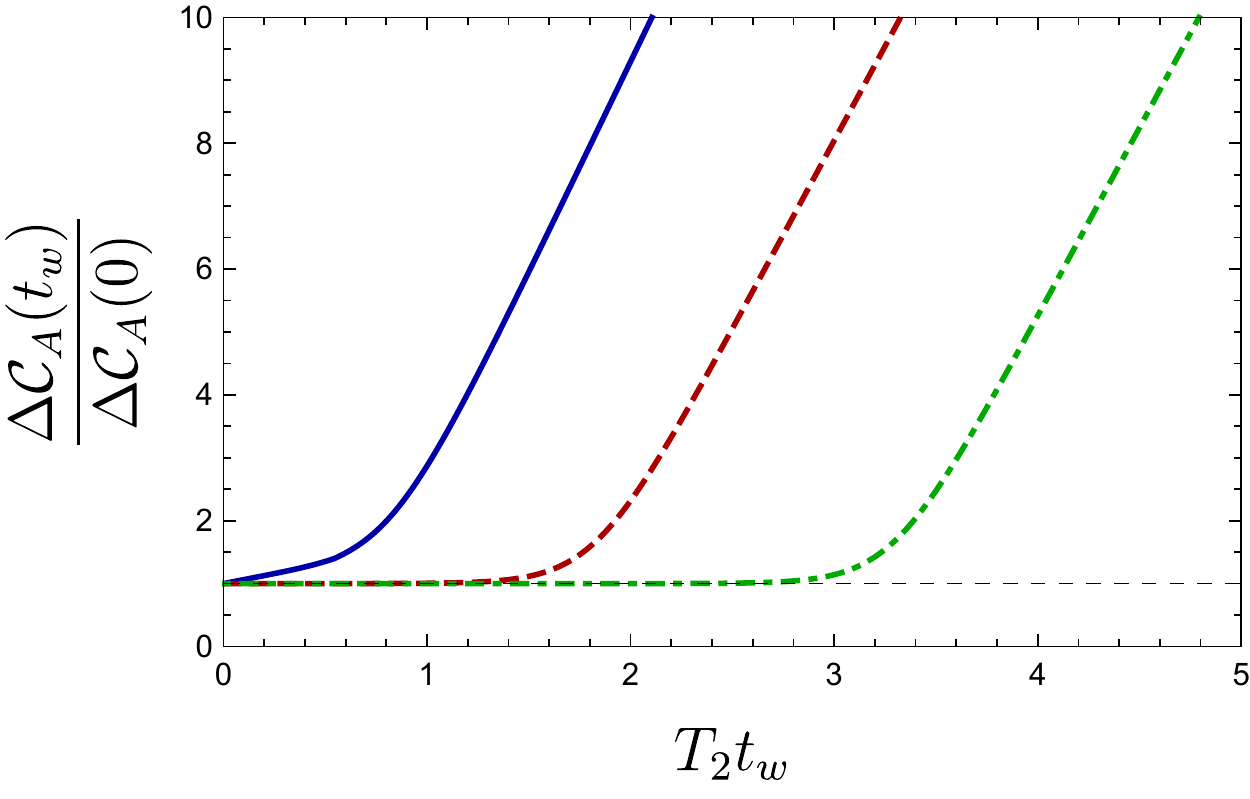}
\includegraphics[scale=0.6]{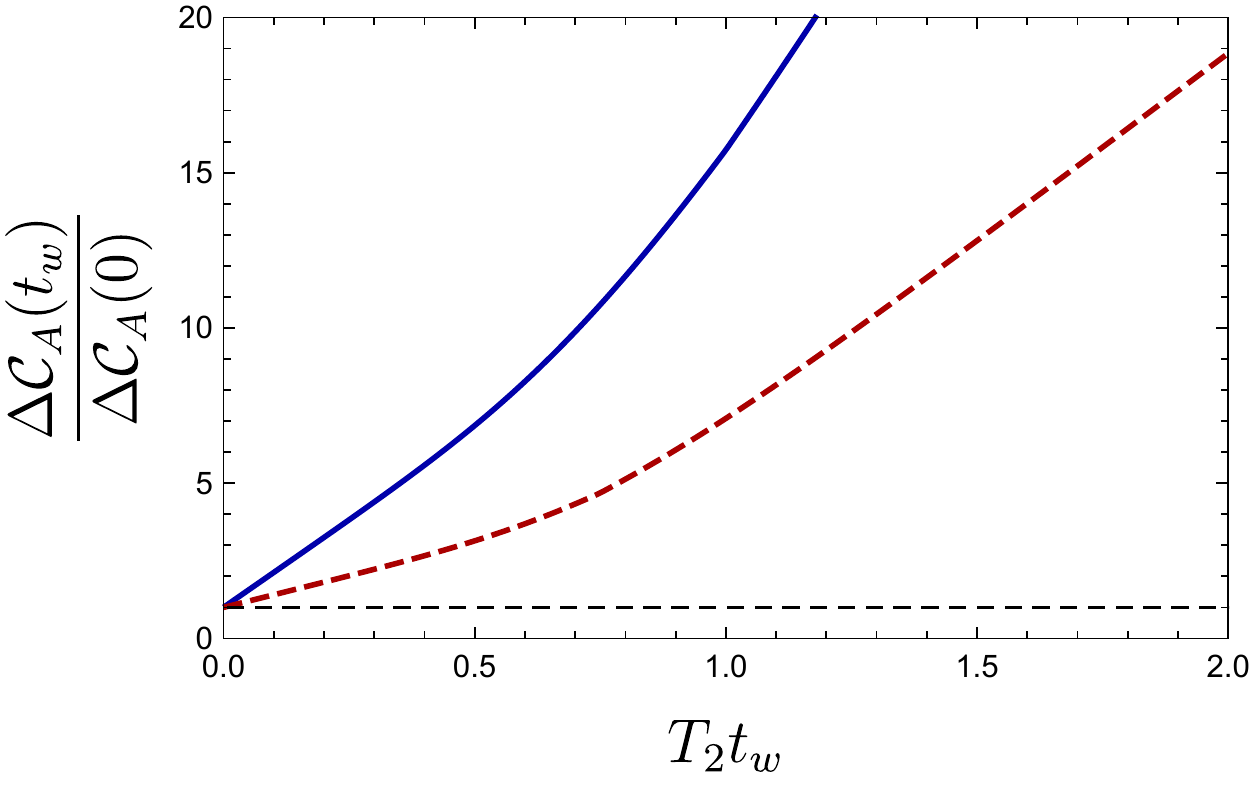}
\caption{The complexity of formation for planar AdS$_5$ black hole (\ie $k=0$) as a function of the shock wave time $t_w$. We normalize by the complexity of formation of the unperturbed black hole in eq.~\eqref{UnpertForm}. In the left panel, we show $\Delta\mC_A(t_w)$ for light shock waves with $w=1+10^{-1}$ (solid blue), $w=1+10^{-4}$ (dashed red) and $w=1+10^{-8}$ (green dot-dashed). For very light shocks, the complexity of formation remains close to the unperturbed value until times of the order of $ t^{*}_{scr}$, then increases approximately linearly with $t_w$. In the right panel, we show $\Delta\mC_A(t_w)$ for heavier shock waves, with $w=2$ (solid blue) and $w = 1.5$ (dashed red). In these cases, the complexity of formation begins growing immediately as $t_w$ moves away from zero.}
\label{FormationAdS5}
\end{figure}

Figure \ref{FormationAdS5} combines these expressions to illustrate the complexity of formation for different temperature ratios $w$ (which for planar black holes is simply $w = \frac{T_2}{T_1}$). First, we investigate the behaviour of light shock waves in the left panel, with $w=1+10^{-1}$ (solid blue), $w=1+10^{-4}$ (dashed red) and $w=1+10^{-8}$ (green dot-dashed). We see a similar overall behaviour to the BTZ case in figure \ref{FormationBTZShock}. The complexity does not change significantly from the unperturbed result until times of the order of the scrambling time $t^{*}_{scr}$, after which it grows approximately linearly with $t_w$. In the right panel of figure \ref{FormationAdS5},  we show $\Delta\mC_A$ for heavier shock waves. In this case, the complexity of formation starts growing immediately from the unperturbed value \reef{UnpertForm}, but again we see that there are two distinct regimes.

To better understand the two regimes shown in figure \ref{FormationAdS5}, we want then to investigate the rate of growth of $\Delta\mC_A(t_w)$ from eqs.~\eqref{CoFAdS5Rega} and \eqref{CoFAdS5Regb} for small and large values of $t_w$, respectively. One can find such slopes by differentiating eqs.~\eqref{CoFAdS5Rega} and \eqref{CoFAdS5Regb} with respect to $x_b$ and $x_s$, and finding the derivative of $x_b$ and $x_s$ with respect to $t_w$ in eq.~\eqref{eq:rwrsrm}. As a consequence, we have the simple expressions for the slopes as
\begin{align}
&\frac{d \, \Delta \mathcal{C}}{d \, t_w} \bigg|_{t_w \rightarrow 0} = \frac{M_2 - M_1}{\pi} \, , \label{SlopeAdS5Small} \\
&\frac{d \, \Delta \mathcal{C}}{d \, t_w} \bigg|_{t_w \rightarrow \infty} = \frac{2( M_2 + M_1) }{\pi} \label{SlopeAdS5Big}  \, .
\end{align}
Hence we see that the initial slope depends on the difference in the masses and hence is essentially zero for the very light shock waves. The expression in eq.~\reef{SlopeAdS5Small} can be compared to that for the BTZ black holes in eq.~\reef{SlopeFormEarly}, which contains an additional term proportional to $\log[M_2/M_1]$. Similarly, eq.~\reef{SlopeAdS5Big} can be compared to the slope implied in eq.~\reef{SlopeFormLate} for $d=2$, and in this case, the two slopes for large values of $t_w$ are identical.

\begin{figure}
\centering
\includegraphics[scale=0.8]{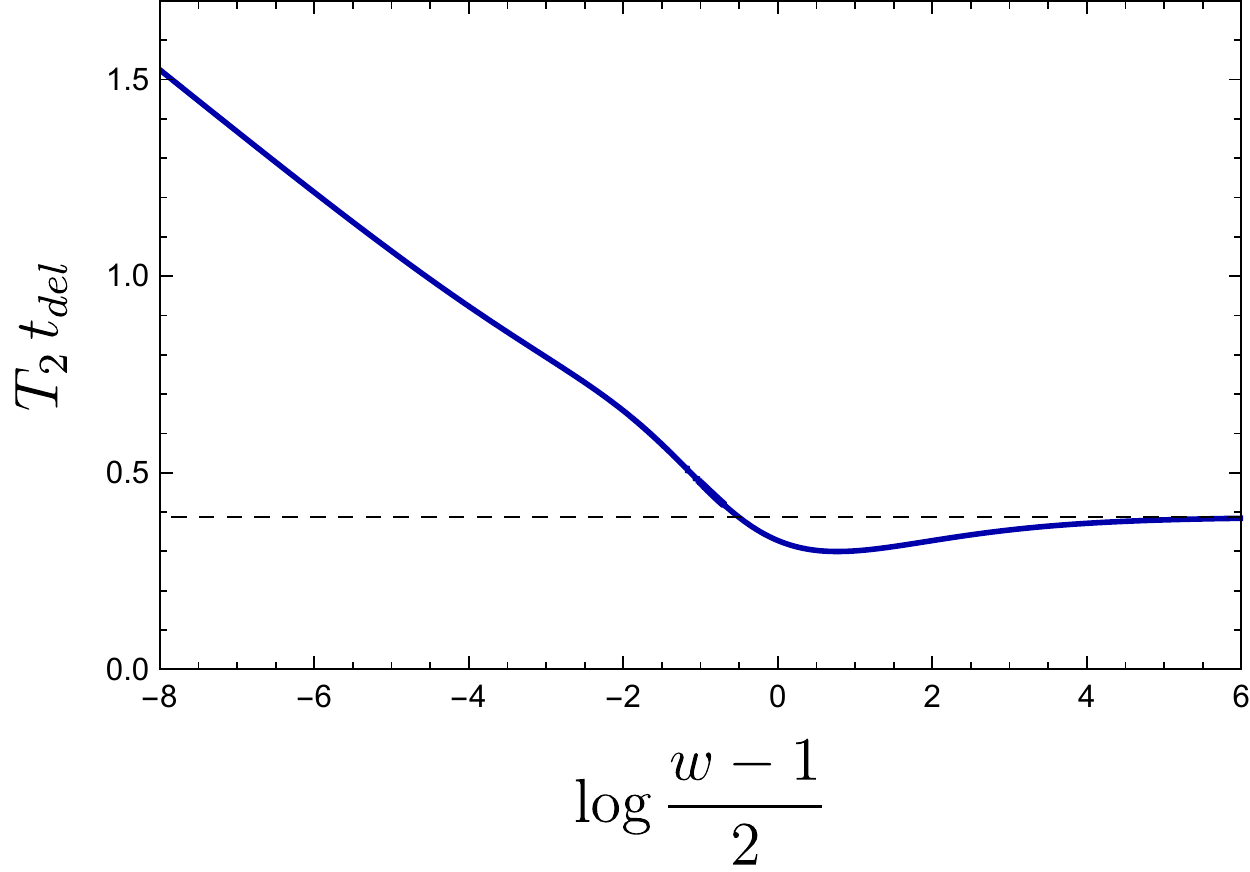}
\caption{The delay time in eq.~\reef{tintAdS5New} as a function of $\log((w-1)/2)$. The left  part of the plot is linear with a slope minus one, as is characteristic of the scrambling time in eq.~\eqref{tintAdS5NewLight}. For heavy shock waves, $t_{\text{del}}$ approaches a constant at large $w$, as given in eq.~\eqref{tintAdS5NewHeavy}. }
\label{twInterceptAdS5New}
\end{figure}

In analogy to eq.~\reef{SlopeFormLate} for $d=2$, we find that the large $t_w$ behaviour of the complexity of formation follows
\begin{equation}
\Delta \mathcal{C}_A = \frac{S_1}{2 \pi} + \frac{2 }{\pi} \left( M_1 + M_2 \right) \left( t_w - t_{\text{del}}\right)+ \mathcal{O} \left(  T_2 t_w e^{-T_2 t_w} \right) \, ,
\end{equation}
with the delay time given by \small
\begin{align}\label{tintAdS5New}
&t_{\text{del}} \equiv \frac{1}{6 \pi T_2 \left(w^4+1\right)} \bigg[  \pi ( w^4-1)-2(2 w^4-1) \cot ^{-1}(w)+4 \pi  w-2 w \tan ^{-1}(w) \nonumber \\
&\qquad+8 w^4 \log (w)-2 w^3(w+1) \log\! \left(\frac{w^4-1}8\right)+(2w^4+w-1) \log\! \left(\frac{w+1}{w-1}\right)  \bigg] \, .
\end{align}
\normalsize
For light shock waves with $w=1+\epsilon$, the delay time agrees with the scrambling time at leading order,\footnote{Combining the expressions in eq.~\reef{ParametersAdS5}, we have $M\propto T^4$, $S\propto T^3$ and $M=\frac34\,S\,T$ for the planar AdS$_5$ black holes (\ie $k=0$). Then we have $4\epsilon \simeq E/M_1$ where $E$ is the energy in the shock wave and so $2/\epsilon=S_1$ if we choose $E=6T_1$. }
\begin{equation}\label{tintAdS5NewLight}
t_{\text{del}} =\frac{1}{2 \pi T_1} \log \left( \frac{2}{\epsilon} \right)+ \frac{1}{4 T_1} + \mathcal{O} \left( \epsilon \, \log \epsilon \right)   = t^{*}_{\text{scr}} + \frac{1}{4 T_1} + \mathcal{O} \left( \epsilon \, \log \epsilon \right) \, ,
\end{equation}
which can be compared to eq.~\reef{newark2}  for the BTZ case.
For heavy shock waves, we have that the delay time approaches a constant proportional to $1/T_2$,
\begin{equation}\label{tintAdS5NewHeavy}
t_{\text{del}} = \frac1{6\pi T_2}\left[
\left(\pi + 6 \log 2\right)  - \frac{8}{w^3}\,\log w+\frac{3 \pi +\frac83+6 \log 2}{w^3} \right]  + \mathcal{O}\left( \frac{1}{T_2 w^4} \right) \, ,
\end{equation}
which can be compared to eq.~\reef{newark} for the BTZ case.
In figure \ref{twInterceptAdS5New} we show how this characteristic time $t_{\text{del}}$ in eq.~\eqref{tintAdS5New} generally behaves as a function of $w$. Overall, this behaviour is very similar to that for the BTZ black hole in figure \ref{twInterceptBTZ}.

\begin{figure}
\centering
\includegraphics[scale=0.6]{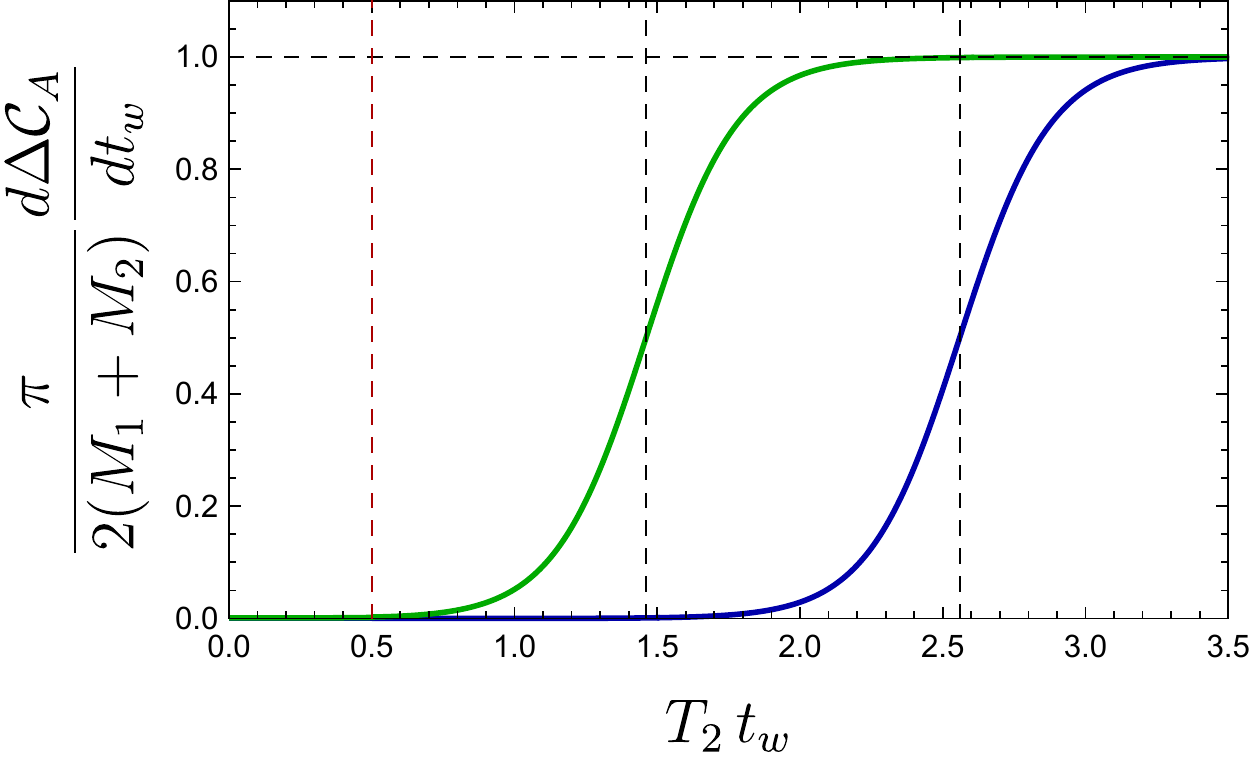}
\includegraphics[scale=0.6]{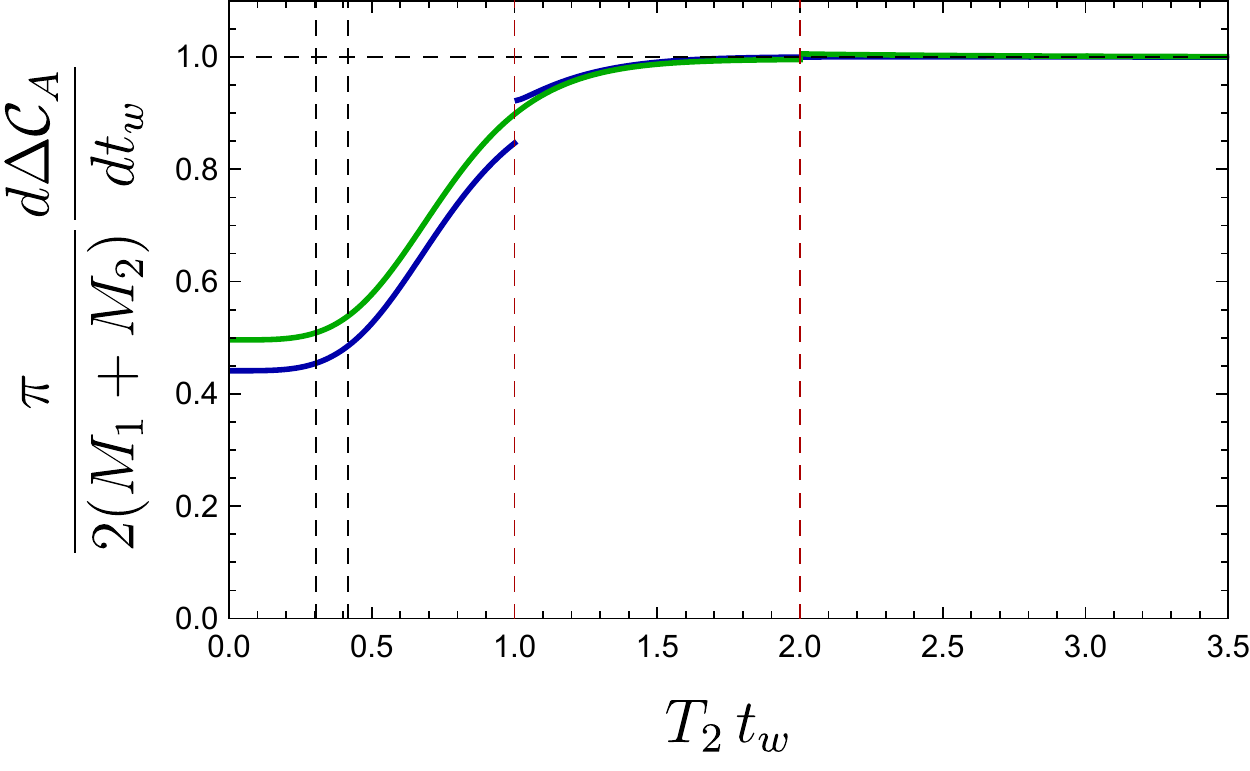}
\caption{The derivative of the complexity of formation with respect to the insertion time $t_w$, for planar AdS$_5$. In the left panel, we evaluate for a light shock waves with $w=1+10^{-6}$ (blue) and $w=1+10^{-3}$ (green), while in the right panel, we have heavier shock waves with $w=2$ (blue) and $w=4$ (green). The results for light shock waves resemble those for the BTZ black hole in figure \ref{FormationBTZShockDer}, with a clear transition between two regimes at the scrambling time, indicated by the vertical dashed black lines. For the heavier shock waves, even though the complexity is a continuous function of $t_w$ as the regime makes a transition between regimes $(a)$ and $(b)$, as in eqs. \eqref{CoFAdS5Rega} and \eqref{CoFAdS5Regb}, the derivative has a non-zero jump proportional to $\log{w}$, as shown in eq. \eqref{DerJump}. The vertical dashed black line denotes the transition from regimes $(a)$ to $(b)$, while the dashed red line stands for the delay time in eq. \eqref{tintAdS5New}. Notice that for heavy shock waves, there is a longer period with a constant derivative, in contrast to the BTZ result in figure \ref{FormationBTZShockDer}. The profile for heavier shock waves is very similar to the $w=4$ example.  }
\label{DerivativesAdS5Formtw}
\end{figure}

We can also calculate the derivative of the complexity of formation with respect to $t_w$ in order to show the transition between shock waves sent before and after the delay time defined in eq.~\eqref{tintAdS5New}. We show these results in figure \ref{DerivativesAdS5Formtw}. For heavy shock waves, we can see that initially the derivative begins as
\begin{equation}
\frac{d \Delta \mathcal{C}_A}{d t_w} = \frac{M_2 -M_1}{\pi} + \mathcal{O} \left( x_s^{-4}  \right) \,.
\end{equation}
Further, there is a more pronounced regime in which this derivative remains constant for small $t_w$, in comparison to the BTZ results in figure \ref{FormationBTZShockDer}. However, this plateau is never very long as it ends before $t_\mt{del}$, which we see from eq.~\reef{tintAdS5NewHeavy} that becomes $T_2 t_\mt{del}\simeq0.39$ for large $w$.
Another notable difference for the heavy shock wave is that the derivative has a discontinuity going from regime $(a)$ to $(b)$, given by eqs.~\eqref{CoFAdS5Rega} and \eqref{CoFAdS5Regb}.\footnote{However, the complexity of formation is continuous, as can be seen from figure \ref{FormationAdS5}.} It is given at leading order by
\begin{equation}\label{DerJump}
\left( \frac{d \Delta \mathcal{C}^{(b)}_A}{d t_w}  - \frac{d \Delta \mathcal{C}^{(a)}_A}{d t_w} \right) \bigg|_{t_w \rightarrow \frac{1}{2 T_1}}  =  \frac{4 ( M_1 +M_2)}{\pi} \frac{\log{w}}{(1+w^4)}  \, .
\end{equation}
As a consequence, the jump for light shock waves is close to zero, making it imperceptible in figure \ref{DerivativesAdS5Formtw}.

\subsection*{Complexity without the counterterm}

We briefly discuss the consequences of not adding eq.~\eqref{counter} to the WDW action for $d=4$. The relevant rates of change of complexity in this regime were calculated and discussed in section \ref{CANoCT}.
For instance, in the planar case with $k=0$ and $d=4$, the rate of change of complexity for very early shock waves in eq.~\eqref{EarlyLimitNoCT} become
\begin{equation}
\frac{d  \mathcal{\tilde C}_A^{(\RN{1})} }{d t}   \bigg{|}_{t_w \rightarrow \infty} = \frac{M_2 - M_1}{3 \pi} + \mathcal{O} \left( T_1 (2 t_w - t) e^{-\pi T_1 (2 t_w -t)} \right) \, .  \label{EarlyLimitBTZNoCT}
\end{equation}
It is proportional to $M_2 - M_1$ and not simply $0$ as for the BTZ black hole. In addition, the late time growth of complexity from eq.~\eqref{LateLimitNoCT} reads
\begin{equation}
\frac{d  \mathcal{\tilde C}_A^{(\RN{3})} }{d t}   \bigg{|}_{t \rightarrow \infty} = \frac{1}{ \pi} \left( M_1 + \frac{M_2}{3} + \frac{5}{6} M_1^{3/4} M_2^{1/4} \right) +  \mathcal{O} \left( T_1 t e^{- \pi T_1 (t - 2 t_w)} \right)\,,
\end{equation}
which does not reproduce the expected late time behaviour with
$d\ca/dt\propto M_1+M_2$. Further, if the shock wave is very light, with $M_1 \approx M_2$, we have
\begin{equation}
\frac{d  \mathcal{\tilde C}_A^{(\RN{3})} }{d t}   \bigg{|}_{M_1 \approx M_2, t \rightarrow \infty} = \frac{13}{6 \pi} \, M_2 \, ,
\end{equation}
which does not recover the expected eternal black hole rate of $2 M_2 /\pi$. The behaviour here is similar to that discussed in the context of the order of limits of light shock wave, without the addition of the counterterm in section \ref{CANoCT}.

Let us also briefly consider the complexity of formation without the inclusion of the counterterm. The only contributions are from the bulk and Gibbons-Hawking-York boundaries in eqs.~\eqref{FormEternalSBulkNotE} and \eqref{FormEternalSGHNotE}
\small
\begin{equation}
\Delta \mathcal{\tilde C}_A^{(a)} = \frac{S_1}{4 \pi^2} \left[ w^3 \log \left(\frac{x_s+1}{x_s-1}\right)-2 w^3 \tan ^{-1}\left(x_s\right)+\log \left(\frac{w x_s-1}{w x_s+1}\right)+2 \tan ^{-1}\left(w x_s\right)+\pi  (w^3+1) \right] \, .
\end{equation}
\normalsize
In the second regime, adding eqs.~\eqref{FormEternalSBulk}, \eqref{FormEternalSGH} and \eqref{FormEternalSCt} with $d=4$,
the complexity of formation reads, \small
\begin{align}
&\Delta \mathcal{\tilde C}_A^{(b)} = \frac{S_1}{4 w \pi^2} \bigg[  -\pi  \left((w-1) x_b^4-2 w^5+w^4+w-2\right)
+\left(w^4-2\right) \log \left(\frac{x_s-1}{x_s+1}\right)  \\
&+w \log \left(\frac{w x_s-1}{w x_s+1}\right) -\left(2 w^4-x_b^4\right) \left(\log \left(\frac{w+x_b}{w-x_b}\right)+w \log \left(\frac{1-x_b}{1+x_b}\right)\right)+ 2 w \tan ^{-1}\left(w x_s\right) \nonumber \\
&+   2 (w^4-2) \tan ^{-1}\left(x_s\right) + 2  \left(2 w^4-x_b^4\right) \left(\tan ^{-1}\left(\frac{x_b}{w}\right)-w\tan ^{-1}\left(x_b\right)\right) \bigg]  \, .\nonumber
\end{align}
\normalsize

\begin{figure}
\centering
\includegraphics[scale=0.6]{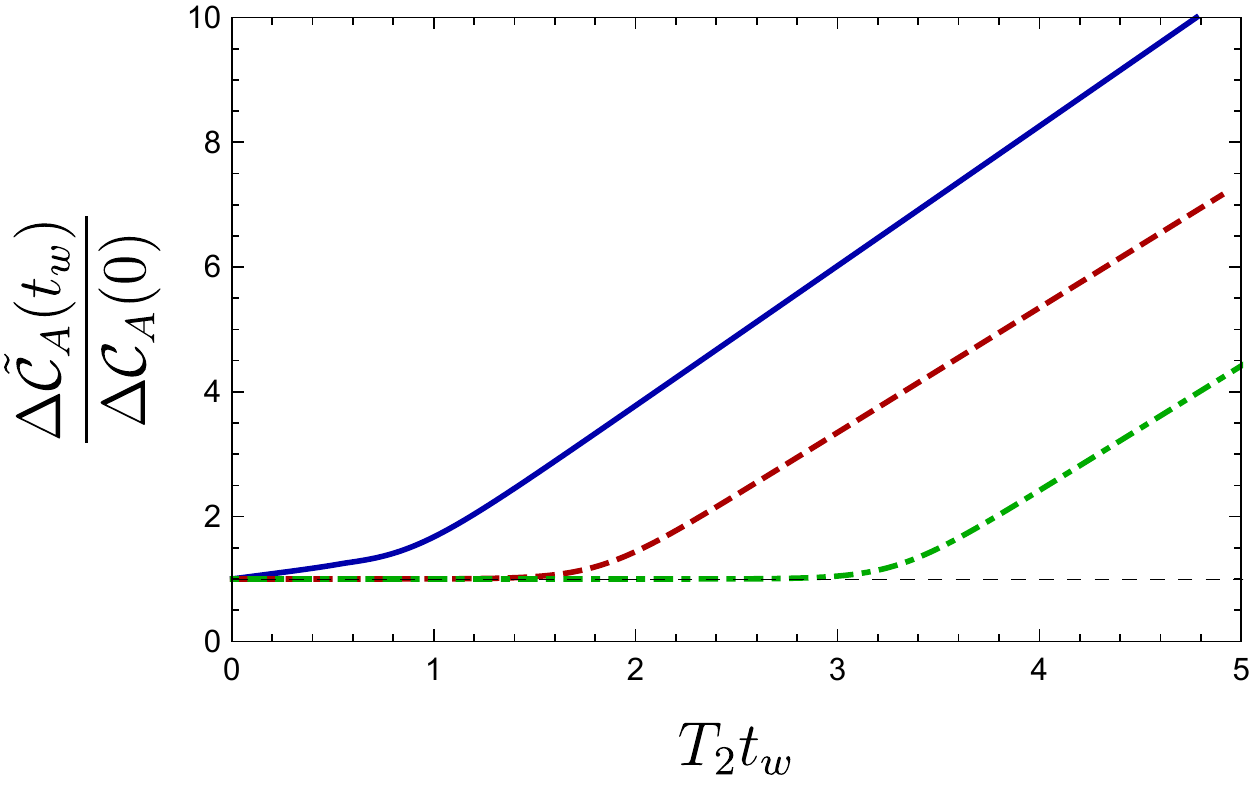}
\includegraphics[scale=0.6]{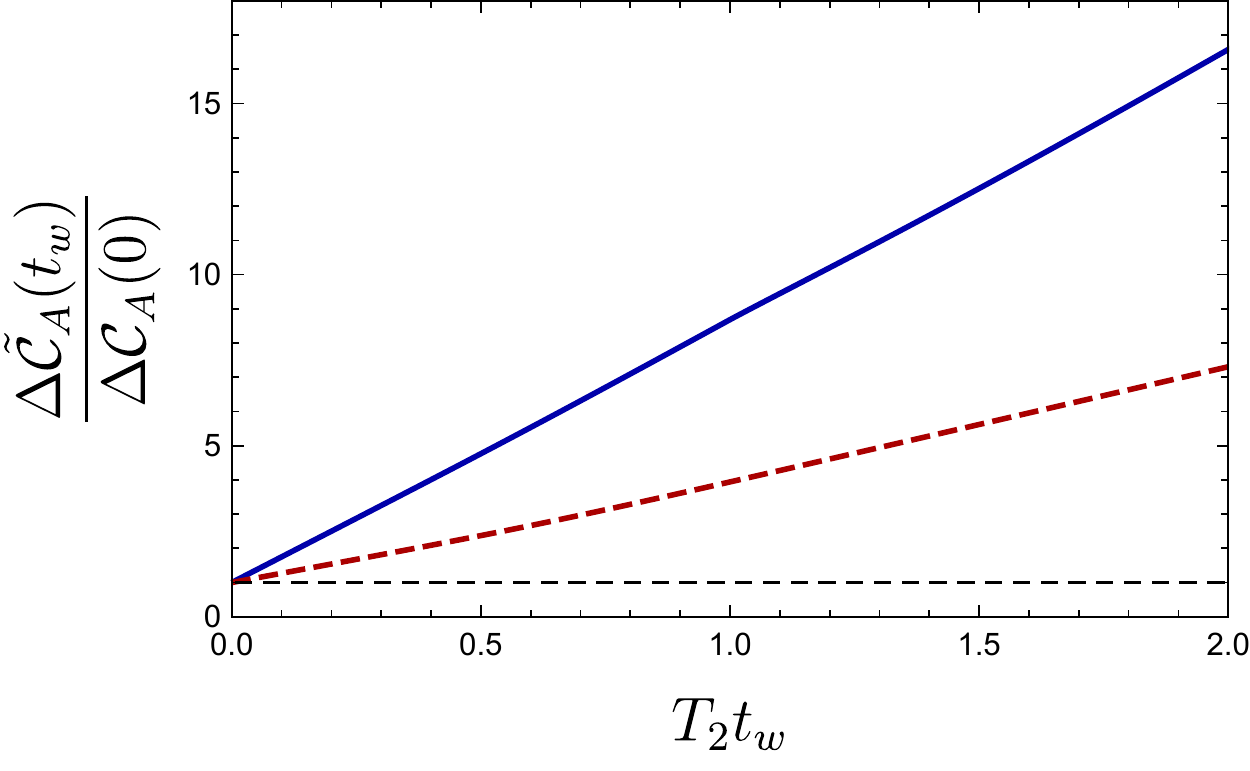}
\caption{The complexity of formation for planar AdS$_5$ black hole as a function of the shock wave time $t_w$, without the inclusion of the counterterm. We normalize by the complexity of formation of the unperturbed black hole in eq.~\eqref{UnpertForm}. In the left panel, we show $\Delta \mathcal{\tilde C}_A(t_w)$ for light shock waves with $w=1+10^{-1}$ (solid blue), $w=1+10^{-4}$ (dashed red) and $w=1+10^{-8}$ (green dot-dashed). For very light shocks, the complexity of formation remains close to the unperturbed value until times of the order of $ t^{*}_{scr}$, then increases approximately linearly with $t_w$. In the right panel, we show $\Delta \mathcal{\tilde C}_A(t_w)$ for heavier shock waves, with $w=2$ (solid blue) and $w = 1.5$ (dashed red). The overall behaviour is similar to the complexity of formation in figure \ref{FormationAdS5}, with the inclusion of the counterterm. Therefore, it contrasts with the BTZ black hole results in section \ref{CANoCT}. }
\label{FormationAdS5NOCT}
\end{figure}

We show the complexity of formation without the addition of counterterm in figure \ref{FormationAdS5NOCT}. In contrast to the complexity of formation of BTZ black holes (\ie $d=2$) in figure \ref{FormationBTZKappaZero}, the overall behaviour of the complexity of formation, up to overall multiplicative constants, is similar to that with the inclusion of the counterterm in figure \ref{FormationAdS5}.

\subsection*{Complexity of Formation for general dimensions (and $k=0$)}

We have shown how the growth rate of complexity behaves in certain regimes in general dimension $d$ in section \ref{sec:Shocks}, with and without the null surface counterterm. It is also possible to derive the dependence of the complexity of formation on $t_w$ for general dimensions if we consider planar horizons, \ie $k=0$. In the case of planar black holes, we can use the tortoise coordinates derived for any dimension $d$ given in \cite{Format}, which we rewrite here with the convention that $r^*$ vanishes at infinity,
\begin{align}
& r_{\text{in} , i}^{*}(r) = \frac{L^2}{r} \left[ \, _2F_1\left(1,-\frac{1}{d};1-\frac{1}{d};\left(\frac{r}{r_{h,i}}\right)^d\right) - 1\right] - \frac{\pi  L^2}{d r_{h,i}}  \cot \left(\frac{\pi }{d}\right) \, , \label{TortoisePlanarGend} \\
& r_{\text{out}, i}^{*}(r)= -
\frac{L^2}{(r^d-r_{h,i}^d)^{1/d}} \  _2F_1\left(\frac{1}{d},\frac{1}{d};1+\frac{1}{d};\frac{r_{h,i}^d}{r_{h,i}^d-r^d}\right)  \nonumber \, .
\end{align}

In the first regime, we have $t_w < - 2 r_{1}^{*}(0)$, which for $k=0$ and general $d$ gives
\begin{equation}
t_w < \frac{1}{2 \cot \left(\frac{\pi }{d}\right)} \frac{1}{T_1} \, .
\end{equation}
We solve eqs.~\eqref{FormEternalSBulkNotE} and \eqref{FormEternalSGHNotE} using the tortoise coordinates given by eq.~\eqref{TortoisePlanarGend}, which has a long but analytic expression,
\small
\begin{align}
& \frac{\pi}{S_1} \, \Delta \mathcal{\tilde C}_A^{(a)}  = \frac{(d-2) \cot \left(\frac{\pi }{d}\right)}{d}  - \frac{d}{\pi  (d-1)} \bigg[ \left( w^d x_s^d -1\right){}^{\frac{d-1}{d}} \, _2F_1\left(\frac{1}{d}-1,\frac{1}{d};1+\frac{1}{d};\frac{1}{1-\left(w x_s\right){}^d}\right) \nonumber \\
&\qquad\qquad - w^{d-1} \left(x_s^d-1\right){}^{\frac{d-1}{d}}   \, _2F_1\left(\frac{1}{d}-1,\frac{1}{d};1+\frac{1}{d};- \frac{1}{x_s^d -1} \right) \bigg]  \label{FormationAnydNOCTa} \\
&\qquad\qquad+\frac{w^{d-1}}{2 \pi } \, \frac{2 x_s^d+d}{\left( x_s^d-1\right){}^{1/d}} \, \,  _2F_1\left(\frac{1}{d},\frac{1}{d};1+\frac{1}{d};- \frac{1}{x_s^d -1}\right) \nonumber \\
&\qquad\qquad
+\frac{1}{2 \pi} \, \frac{2 \left(w x_s\right){}^d-d}{\left(\left(w x_s\right){}^d-1\right){}^{1/d}} \, \, _2F_1\left(\frac{1}{d},\frac{1}{d};\frac{1}{d}+1;\frac{1}{1-\left(w x_s\right){}^d}\right)  \nonumber \, .
\end{align}
\normalsize
Here we have begun with the complexity of formation evaluated without the null surface counterterm \reef{counter}.
If the shock wave is very light, and $t_w \ll t^{*}_{\text{scr}}$, the above expression simply reduces to
\begin{equation}
\Delta \mathcal{\tilde C}_A^{(a)}  = \frac{(d-2) \cot \left(\frac{\pi }{d}\right)}{d \, \pi} \, S_1 , \,\label{oldform}
\end{equation}
which reproduces the complexity of formation for a planar boundary geometry in $d$ dimensions found in \cite{Format}.

The second regime, with $ t_w > - 2 r_{1}^{*}(0)$, consists of solving eqs.~\eqref{FormEternalSBulk} and \eqref{FormEternalSGH} with the general $d$ tortoise coordinates in eq.~\eqref{TortoisePlanarGend}, which results in

\small
\begin{align}
 \frac{\pi}{S_1} \, \Delta \mathcal{\tilde C}_A^{(b)} & =  \frac{\Gamma \left(\frac{1}{d}-1\right)}{\pi  d \Gamma \left(1+\frac{1}{d}\right)} \bigg[ \left(\frac{1}{(w x_s)^d-1}\right)^{\frac{1}{d}-1} \, _2F_1\left(\frac{1}{d}-1,\frac{1}{d};1+\frac{1}{d};- \frac{1}{(w x_s )^d - 1}\right)   \nonumber \\
& - \left(\frac{1}{(w x_s )^d - w^d}\right)^{\frac{1}{d}-1} \, _2F_1\left(\frac{1}{d}-1,\frac{1}{d};1+\frac{1}{d};- \frac{1}{x_s^d- 1}\right) \bigg]    \nonumber \\
&   +  \frac{d w^d - 2 x_b}{2 \pi  x_b}   \bigg[ \, _2F_1\left(1,-\frac{1}{d};\frac{d-1}{d};\left(\frac{x_b}{w}\right){}^d\right)   -  \, _2F_1\left(1,-\frac{1}{d};\frac{d-1}{d};x_b^d\right) \bigg] \label{FormationAnydNOCT}  \\
&+  \frac{ 2 (w x_s)^d - d}{2 \pi  w}   \bigg[ \left( \frac{w^d}{(w x_s)^d -1 } \right)^{1/d}  \, _2F_1\left(\frac{1}{d},\frac{1}{d};1+\frac{1}{d}; - \frac{1}{(w x_s)^d - 1}\right)  \nonumber \\
& -
\left( \frac{1}{x_s^d -1 } \right)^{1/d}  \, _2F_1\left(\frac{1}{d},\frac{1}{d};1+\frac{1}{d}; - \frac{1}{x_s^d - 1}\right) \bigg] + \frac{\cot \left(\frac{\pi }{d}\right)}{2 d w} \left[  2 x^b + (d-4) w - w (2 x_b - d w^d)\right]  \nonumber  \, .
\end{align}
\normalsize

Of course, we can also evaluate the complexity of formation including the counterterm for general dimension and $k=0$. We would simply add the contributions in eq.~\eqref{FormEternalSCtNotE} to eq.~\eqref{FormationAnydNOCTa}
\begin{equation}
\Delta \mathcal{C}_A^{(a)}  = \Delta \mathcal{\tilde C}_A^{(a)}  + \frac{S_1}{2 \pi^2} \left[  w^{d-1} x_s^{d-1} \,  \log \left(\frac{\left(\left(w x_s\right){}^d-1\right)}{ w^d (x_s^d - 1) }\right)    \right] \, \label{FormationAnyda}  ,
\end{equation}
and eqs.~\eqref{FormEternalSCtNotE} and \eqref{FormEternalSCtb} to eq.~\eqref{FormationAnydNOCT}, which results in
\begin{equation}
\Delta \mathcal{C}_A^{(b)}  = \Delta \mathcal{\tilde C}_A^{(b)}  + \frac{S_1}{2 \pi^2} \left[   x_b^{d-1} \log \left(\frac{x_b^d-w^d}{x_b^d-1}\right) +w^{d-1} x_s^{d-1} \,  \log \left(\frac{\left(\left(w x_s\right){}^d-1\right)}{ w^d (x_s^d - 1) }\right)    \right] \, \label{FormationAnyd}  .
\end{equation}
Note that the additional contributions in eq.~\reef{FormationAnyda} do not modify the result in the limit $w\to1$ and $t_w\ll  t^{*}_{\text{scr}}$, \ie we still recover the expected result for the complexity of formation \reef{oldform} without any shock wave perturbation.\footnote{We might note that the original calculations of the complexity of formation in \cite{Format} were made without the counterterm \reef{counter}. }

For large $t_w$, we can simply expand the expressions for $x_s = 1 + A \exp^{-2 \pi T_2 t_w} $ and $x_b =1 - B \exp^{-2 \pi T_1 t_w} $, where $A$ and $B$ are constants that depend on the dimension, but that nonetheless are independent of $t_w$. In addition, we can evaluate the dependence on $t_w$ when $t_w$ is small, which means $r_s$ is close to the boundary. Evaluating the complexity of formation without counterterm in eqs.~\eqref{FormationAnydNOCTa} and \eqref{FormationAnydNOCT}, we find
\begin{equation}
\frac{d \Delta \mathcal{\tilde C}_A}{d t_w}  \bigg{|}_{t_w \rightarrow 0}=  \frac{d-2}{d-1} \,  \frac{ M_2 - M_1}{\pi}   \, , \qquad \quad
 \frac{d \Delta \mathcal{\tilde C}_A}{d t_w}  \bigg{|}_{t_w \rightarrow \infty}=  \frac{d-2}{d-1} \,  \frac{M_1 + M_2}{\pi}  \, . \label{FormtwGendNOCT}
\end{equation}
Also, we can evaluate the complexity of formation with the addition of the counterterm in eqs.~\eqref{FormationAnyda} and \eqref{FormationAnyd}, which in these regimes give us the expected results
\begin{equation}
\frac{d \Delta \mathcal{ C}_A}{d t_w}  \bigg{|}_{t_w \rightarrow 0}=  \frac{(M_2 - M_1)}{\pi}  \, , \qquad \qquad
 \frac{d \Delta \mathcal{ C}_A}{d t_w}  \bigg{|}_{t_w \rightarrow \infty}=  \frac{2 (M_1 + M_2)}{\pi}
 \, , \label{FormtwGend}
\end{equation}
\eg compare to the CA results in eqs.~\reef{boil1} and \reef{boil2}.
Of course, in the limit where $M_2$ is much larger than $M_1$, we should recover the one sided result in \cite{Vad1}. Without the addition of the counterterm, the planar rate of change is just a constant (with the same $d$ dependence), as can be seen in eq.~($3.28$) of \cite{Vad1}. When the counterterm is added, the expressions also agree, with the initial rate of $M_2/\pi$ and the late time rate of $2 M_2/\pi$, as can be seen in eq.~($3.47$) of \cite{Vad1}. In addition, the general $d$ dependence in eq.~\eqref{FormtwGendNOCT} agrees with our analysis of the BTZ black hole in section \ref{CANoCT}, where the complexity of formation without the inclusion of the counterterm saturated to a constant in $d=2$.

\section{Details for Complexity=Volume} \label{app:CVShocksDetails}

In this appendix, we present the detailed derivation behind some of the results presented in section \ref{sec:VolShocks} for our maximal volume surfaces in the shockwave geometries. In particular, we will derive the expressions in eqs.~\eqref{tRshocksF} and \eqref{tLshocksF} for the left and right boundary times, and in eqs.~\eqref{VTshocksF}--\eqref{VtshocksF}  for the maximal volume and its time derivative.
Our derivations are simplified with the following definitions introduced in the main text
\begin{equation}
\tau \left[P,r\right] \equiv \frac{1}{f(r)}-\frac{P}{f(r)\sqrt{f(r) r^{2(d-1)} + P^2}}, \qquad
R \left[P,r\right] \equiv \frac{r^{2(d-1)}}{\sqrt{f(r) r^{2(d-1)} + P^2}}\,,\tag{\ref{tauR}}
\end{equation}
as well as defining
\begin{equation}
\tilde R \left[P,r\right] \equiv \frac{\sqrt{f(r) r^{2(d-1)} + P^2}}{f(r)}-\frac{P}{f(r)}\, .
\end{equation}
In the following, we will add subscripts ($\tau_{1,2}$, $R_{1,2}$) to specify which blackening factor $f_i$ and conserved momentum $P_i$ is being used. These functions obey a number of useful identities which we will use repeatedly throughout the derivation
\begin{subequations}
\begin{alignat}{4}
\label{ID1}
R\left[P,r\right]= &\, \tilde R \left[P,r\right] + P \tau \left[P,r\right] = \tilde R \left[-P,r\right]  -P \tau \left[-P,r\right],
\\
\label{ID2}
\del_P \tilde R\left[P,r\right] = &\, - \tau \left[P,r\right],
\qquad
\del_P \tilde R\left[-P,r\right] =  \tau \left[-P,r\right],
\\
\label{ID3}
\tilde R \left[P,r_t\right]= &\, -\frac{P}{f(r)},
\\
\label{ID4}
\tilde R \left[P,r\right]=&\, \tilde R \left[-P,r\right]-\frac{2 P }{f(r)}= \dot v_+ \left[P,r\right] = - \dot v_- \left[-P,r\right],
\end{alignat}
\end{subequations}
where $r_t$ denotes the turning point (a point in which $\dot r$ vanishes) in the relevant black hole which satisfies (cf. eq.~\eqref{eq_shocks:rturn})
\begin{equation}
P^2 + f(r_t) r_t^{2(d-1)}=0\, .
\end{equation}
In general the time integrals will be given by integrating the coordinates $u$ and $v$ along various parts of the surface according to
\begin{equation}
\begin{split}
\Delta v_{\pm} = \int dv_{\pm} = \int \frac{\dot v_{\pm}}{\dot r_{\pm}} dr = \int \tau \left[\pm P,r\right]  dr
\\
\Delta u_{\pm} = \int du_{\pm} = \int \frac{\dot u_{\pm}}{\dot r_{\pm}} dr = - \int \tau \left[\mp P,r\right]  dr
\end{split}
\end{equation}
and volume integrals will be given by
\begin{equation}
\mathcal{V} = \Omega_{k,d-1} \int \frac{r^{2(d-1)}}{\dot r_{\pm}} dr =  \pm \Omega_{k,d-1} \int R\left[P,r\right] dr.
\end{equation}

\begin{figure*}[t!]
    \centering
    \begin{subfigure}[t]{0.21\textwidth}
        \centering
        \includegraphics[height=1.1in]{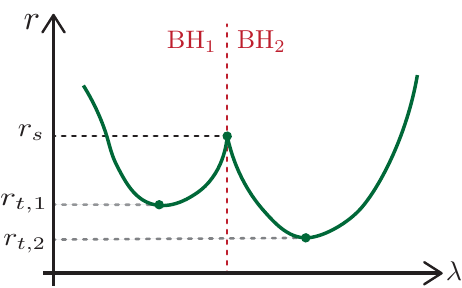}
        \caption{Case A/F}\label{AF}
    \end{subfigure}%
    ~~~~~~~~
    \begin{subfigure}[t]{0.21\textwidth}
        \centering
        \includegraphics[height=1.1in]{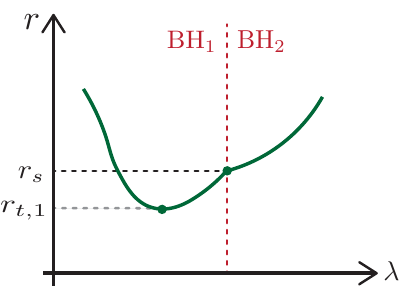}
        \caption{Case B/E}\label{BE}
    \end{subfigure}
    ~~~~
    \begin{subfigure}[t]{0.21\textwidth}
        \centering
        \includegraphics[height=1.1in]{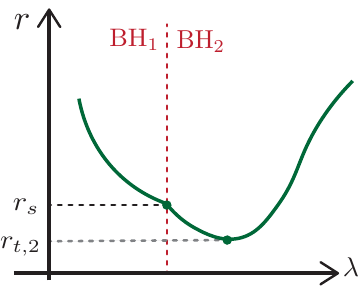}
        \caption{Case C}\label{C}
    \end{subfigure}
    ~~
    \begin{subfigure}[t]{0.21\textwidth}
        \centering
        \includegraphics[height=1.1in]{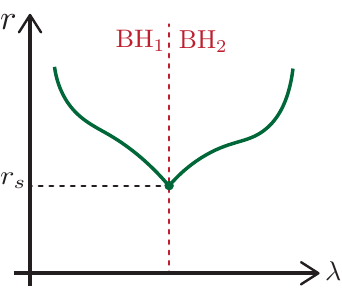}
        \caption{Case D}\label{D}
    \end{subfigure}
    \caption{Different possibilities for the shape of the maximal surface depending on whether it admits a turning point in \BHO and/or in \BHT.}
    \label{surfShapes2}
\end{figure*}

There are six different possibilities to be considered for the shape of our surface:
\begin{enumerate}
\item {\bf Case A} -- The surface passes behind the future horizon in \BHO ($P_1>0$), it admits turning points both in \BHO and in \BHT, see figure \ref{AF}.
\item {\bf Case B} -- The surface passes behind the future horizon in \BHO ($P_1>0$), it admits a turning point in \BHO but not in \BHT, see figure \ref{BE}.
\item {\bf Case C} -- The surface passes behind the future horizon in \BHO ($P_1>0$), it admits a turning point in \BHT but not in \BHO, see figure \ref{C}.
\item {\bf Case D} -- The surface passes behind the future horizon in \BHO ($P_1>0$), it does not admit turning points, see figure \ref{D}.
\item {\bf Case E} -- The surface passes behind the past horizon in \BHO ($P_1<0$), it admits a turning point in \BHO but not in \BHT, see figure \ref{BE}.
\item {\bf Case F} -- The surface passes behind the past horizon in \BHO ($P_1<0$), it admits turning points both in \BHO and in \BHT, see figure \ref{AF}.
\end{enumerate}
We will study these cases one by one and derive in each case expressions for the boundary times and for the maximal volume and its time derivative.

{\bf Case A} -- In this case the surface passes behind the future horizon in \BHO ($P_1>0$), and admits turning points both in \BHO and in \BHT, see figure \ref{AF}. We will describe this surface using the coordinate $u_\mt{L}$ between the left boundary and $r_{t,1}$ and using the coordinate $v_\mt{R}$ everywhere else. To obtain the expression for $t_\mt{L}$ we integrate along the surface in \BHO, first from the left boundary to $r_{t,1}$
\begin{equation}
u_{t,1}-u_\mt{L} = \int_{r_{t,1}}^{\infty} \tau_1[P_1,r] dr,
\end{equation}
then from $r_{t,1}$ to $r_s$
\begin{equation}
v_s-v_{t,1} = \int_{r_{t,1}}^{r_s} \tau_1[P_1,r] dr,
\end{equation}
where we have used $u_{t,1}$ and $v_{t,1}$ to denote the values of $u_\mt{L}$ and $v_\mt{R}$ at the turning point of the surface in \BHO. Summing together and using $v_{t,1}-u_{t,1}=2 r_1^*(r_{t,1})$ we obtain\footnote{To convince oneself that this relation still holds even though  the calculations involve both left and right induced Eddington-Finkelstein coordinates, one may simply translate them to Kruskal-Szekeres coordinates which cover the whole spacetime. Further, we note that this is consistent with the synchronization of the clocks described at the beginning of section \ref{sec:Shocks} since in this coordinate system, it is also easy to show that minimal surfaces going through the bifurcation surface are straight lines of constant (and opposite) $t$ on the two sides.}
\begin{equation}\label{tLcaseA}
t_\mt{L}+v_s-2 r_1^*(r_{t,1}) = \int_{r_{t,1}}^{r_s} \tau_1[P_1,r] dr+ \int_{r_{t,1}}^{\infty} \tau_1[P_1,r] dr .
\end{equation}
For the right boundary time, we integrate along the surface first from $r_s$ to $r_{t,2}$
\begin{equation}
v_{t,2} - v_s = - \int_{r_{t,2}}^{r_s} \tau_2[-P_2,r] dr,
\end{equation}
then from $r_{t,2}$ to the right boundary
\begin{equation}
v_\mt{R} - v_{t,2} =  \int_{r_{t,2}}^{\infty} \tau_2[P_2,r] dr,
\end{equation}
and summing up we obtain
\begin{equation}\label{tRcaseA}
t_\mt{R} - v_s =  - \int_{r_{t,2}}^{r_s} \tau_2[-P_2,r] dr +\int_{r_{t,2}}^{\infty} \tau_2[P_2,r] dr.
\end{equation}
The volume of the maximal surface is given by integrating along the four segments of the surface
\begin{equation}
\frac{1}{\Omega_{k,d-1}}\mathcal{V} =
\int_{r_{t,1}}^{r_s} R_1[P_1,r]dr +\int_{r_{t,1}}^{\infty} R_1[P_1,r]dr+ \int_{r_{t,2}}^{r_s} R_2[P_2,r]dr + \int_{r_{t,2}}^{\infty} R_2[P_2,r]dr.
\end{equation}
We can use the identity \eqref{ID1} as well as the relations \eqref{tLcaseA} and  \eqref{tRcaseA}  for the boundary times, to rewrite it as follows
\begin{align}
\frac{1}{\Omega_{k,d-1}}\mathcal{V}
= & \,
\int_{r_{t,1}}^{r_s} \tilde R_1[P_1,r]dr + \int_{r_{t,1}}^{\infty} \tilde R_1[P_1,r]dr
+ \int_{r_{t,2}}^{r_s} \tilde R_2[-P_2,r]dr+\int_{r_{t,2}}^{\infty} \tilde R_2[P_2,r]dr\nonumber
\\
& \, +P_1(t_\mt{L}+v_s-2 r_1^*(r_{t,1})) +P_2 (t_\mt{R} - v_s).\label{Vol1CaseA}
\end{align}
We note in passing that the fact that $\dot v$ is continuous across the interface implies using \eqref{ID4}
\begin{equation}\label{contCaseA}
\tilde R_1[P_1,r_s]=-\tilde R_2[-P_2,r_s].
\end{equation}
For the time derivative of the volume \eqref{Vol1CaseA}, we obtain
\begin{align}
\frac{1}{\Omega_{k,d-1}}\frac{d \mathcal{V}}{dt}
= &\, \frac{d r_{t,1}}{dt} \left(-2\tilde R_1[P_1,r_{t,1}] - \frac{2P_1}{f_1(r_{t,1})}\right)
+\frac{d r_{t,2}}{dt} \left(-\tilde R_2[-P_2,r_{t,2}]-\tilde R_2[P_2,r_{t,2}]\right)\nonumber
\\
&\,  +\frac{d P_1}{dt} \left(-\int_{r_{t,1}}^{r_s} \tau_1[P_1,r] dr-\int_{r_{t,1}}^{\infty} \tau_1[P_1,r] dr  +(t_\mt{L}+v_s-2 r_1^*(r_{t,1}))\right)\nonumber
\\
&\, +\frac{d P_2}{dt} \left(\int_{r_{t,2}}^{r_s} \tau_2[-P_2,r] dr-\int_{r_{t,2}}^{\infty} \tau_2[P_2,r] dr  +(t_\mt{R} - v_s)\right)\nonumber
\\
&\, +\frac{d r_s}{dt} \left(\tilde R_1[P_1,r_s]+\tilde R_2[-P_2,r_s] \right)
+P_1 \frac{d t_\mt{L}}{d t} + P_2 \frac{d t_\mt{R}}{d t}\label{VolCaseA}
\end{align}
where we have defined $t$ to be a time coordinate specifying the two boundary times via $t_\mt{R}(t)$, $t_\mt{L}(t)$ and used the identity \eqref{ID2} to evaluate some of the derivatives with respect to the momenta.
The various contributions inside the brackets can be shown to vanish using the identity \eqref{ID3}, the expressions for the left and right boundary times \eqref{tLcaseA}-\eqref{tRcaseA} and the continuity relation \eqref{contCaseA}. So we are left with
\begin{equation}
\frac{1}{\Omega_{k,d-1}}\frac{d \mathcal{V}}{dt}
=P_1 \frac{d t_\mt{L}}{d t} + P_2 \frac{d t_\mt{R}}{d t}.
\end{equation}


{\bf Case B} -- In this case the surface passes behind the future horizon in \BHO ($P_1>0$), and admits a turning point in \BHO but not in \BHT, see figure \ref{BE}. We will describe this surface using the coordinate $u_\mt{L}$ between the left boundary and $r_{t,1}$ and using the coordinate $v_\mt{R}$ in all the rest of the surface. The expression for $t_\mt{L}$ is obtained just like for case A and it reads
\begin{equation}\label{tLcaseB}
t_\mt{L}+v_s-2 r_1^*(r_{t,1}) = \int_{r_{t,1}}^{r_s} \tau_1[P_1,r] dr+ \int_{r_{t,1}}^{\infty} \tau_1[P_1,r] dr .
\end{equation}
For the right boundary time, we integrate along the surface from $r_s$ to the right boundary and obtain
\begin{equation}\label{tRcaseB}
t_\mt{R} - v_s =  \int_{r_s}^{\infty} \tau_2[P_2,r] dr.
\end{equation}
The volume of the maximal surface is given by integrating along the three segments of the surface
\begin{equation}
\frac{1}{\Omega_{k,d-1}}\mathcal{V} =
\int_{r_{t,1}}^{r_s} R_1[P_1,r]dr+  \int_{r_{t,1}}^{\infty} R_1[P_1,r]dr + \int_{r_s}^{\infty} R_2[P_2,r]dr .
\end{equation}
We can use the identity \eqref{ID1} as well as the relations \eqref{tLcaseB} and  \eqref{tRcaseB}  for the boundary times, to rewrite it as follows
\begin{equation}\label{Vol1CaseB}
\begin{split}
\frac{1}{\Omega_{k,d-1}}\mathcal{V}
= & \,
\int_{r_{t,1}}^{r_s} \tilde R_1[P_1,r]dr+ \int_{r_{t,1}}^{\infty} \tilde R_1[P_1,r]dr  + \int_{r_s}^{\infty} \tilde R_2[P_2,r]dr
\\
& \, +P_1(t_\mt{L}+v_s-2 r_1^*(r_{t,1})) +P_2 (t_\mt{R} - v_s).
\end{split}
\end{equation}
We note in passing that the fact that $\dot v$ is continuous across the interface implies using eq.~\eqref{ID4}
\begin{equation}\label{contCaseB}
\tilde R_1[P_1,r_s]=\tilde R_2[P_2,r_s].
\end{equation}
For the time derivative of the volume \eqref{Vol1CaseB}, we obtain
\begin{align}
\frac{1}{\Omega_{k,d-1}}\frac{d \mathcal{V}}{dt}
= &\, \frac{d r_{t,1}}{dt} \left(-2\tilde R_1[P_1,r_{t,1}]- \frac{2P_1}{f_1(r_{t,1})}\right)\nonumber
\\
&\,  +\frac{d P_1}{dt} \left(-\int_{r_{t,1}}^{r_s} \tau_1[P_1,r] dr -\int_{r_{t,1}}^{\infty} \tau_1[P_1,r] dr +(t_\mt{L}+v_s-2 r_1^*(r_{t,1}))\right)\nonumber
\\
&\, +\frac{d P_2}{dt} \left(-\int_{r_s}^{\infty} \tau_2[P_2,r] dr  +(t_\mt{R} - v_s)\right)\nonumber
\\
&\, +\frac{d r_s}{dt} \left(\tilde R_1[P_1,r_s]-\tilde R_2[P_2,r_s] \right)
+P_1 \frac{d t_\mt{L}}{d t} + P_2 \frac{d t_\mt{R}}{d t}\label{VolCaseB}
\end{align}
where we have used the identity \eqref{ID2} to evaluate some of the derivatives with respect to the momenta.
All the contributions inside the brackets can be shown to vanish using the identity \eqref{ID3}, the expressions for the left and right boundary times in eqs.~\eqref{tLcaseB} and \eqref{tRcaseB} and the continuity relation \eqref{contCaseB}. So we are left with
\begin{equation}
\frac{1}{\Omega_{k,d-1}}\frac{d \mathcal{V}}{dt}
=P_1 \frac{d t_\mt{L}}{d t} + P_2 \frac{d t_\mt{R}}{d t}.
\end{equation}


{\bf Case C} -- In this case the surface passes behind the future horizon in \BHO ($P_1>0$), and admits a turning point in \BHT but not in \BHO, see figure \ref{C}. We will describe this surface using the coordinate $u_\mt{L}$ between the left boundary and $r_s$ and using the coordinate $v_\mt{R}$ in all the rest of the surface. The expression for $t_\mt{R}$ is obtained just like for case A and it reads
\begin{equation}\label{tRcaseC}
t_\mt{R} - v_s =  - \int_{r_{t,2}}^{r_s} \tau_2[-P_2,r] dr +\int_{r_{t,2}}^{\infty} \tau_2[P_2,r] dr.
\end{equation}
Now $t_\mt{L}$ is obtained by integrating $u_\mt{L}$ from the left boundary to $r_s$
\begin{equation}\label{tLcaseC}
u_s-u_\mt{L} = t_\mt{L}+v_s-2 r_1^*(r_s) = \int_{r_s}^{\infty} \tau_1[P_1,r] dr.
\end{equation}
The volume of the maximal surface is given by integrating along the three segments of the surface
\begin{equation}
\frac{1}{\Omega_{k,d-1}}\mathcal{V} = \int_{r_s}^{\infty} R_1[P_1,r]dr +  \int_{r_{t,2}}^{r_s} R_2[P_2,r]dr+\int_{r_{t,2}}^{\infty} R_2[P_2,r]dr.
\end{equation}
We can use the identity \eqref{ID1} as well as the relations \eqref{tRcaseC}-\eqref{tLcaseC} for the boundary times, to rewrite it as follows
\begin{equation}\label{Vol1CaseC}
\begin{split}
\frac{1}{\Omega_{k,d-1}}\mathcal{V}
= & \, \int_{r_s}^{\infty} \tilde R_1[P_1,r] dr
 + \int_{r_{t,2}}^{r_s} \tilde R_2[-P_2,r]dr
+\int_{r_{t,2}}^{\infty} \tilde R_2[P_2,r]dr
\\
& \, +P_1(t_\mt{L}+v_s-2 r_1^*(r_s)) +P_2 (t_\mt{R} - v_s).
\end{split}
\end{equation}
We note in passing that the fact that $\dot v$ is continuous across the interface implies using eq.~\eqref{ID4}
\begin{equation}\label{contCaseC}
\tilde R_1[-P_1,r_s]=\tilde R_1[P_1,r_s]+\frac{2 P_1}{f_1(r_s)}=\tilde R_2[-P_2,r_s].
\end{equation}
For the time derivative of the volume \eqref{Vol1CaseC}, we obtain
\begin{equation}\label{VolCaseC}
\begin{split}
\frac{1}{\Omega_{k,d-1}}\frac{d \mathcal{V}}{dt}
= &\, \frac{d r_{t,2}}{dt} \left(-\tilde R_2[-P_2,r_{t,2}]-\tilde R_2[P_2,r_{t,2}]\right)
\\
&\,  +\frac{d P_1}{dt} \left(-\int_{r_s}^{\infty} \tau_1[P_1,r] dr  +(t_\mt{L}+v_s-2 r_1^*(r_{t,1}))\right)
\\
&\, +\frac{d P_2}{dt} \left( \int_{r_{t,2}}^{r_s} \tau_2[-P_2,r] dr -\int_{r_{t,2}}^{\infty} \tau_2[P_2,r] dr +(t_\mt{R} - v_s)\right)
\\
&\, +\frac{d r_s}{dt} \left(-\tilde R_1[P_1,r_s]+\tilde R_2[-P_2,r_s] -\frac{2 P_1}{f_1(r_s)}\right)
+P_1 \frac{d t_\mt{L}}{d t} + P_2 \frac{d t_\mt{R}}{d t}
\end{split}
\end{equation}
where we have used again the identity \eqref{ID2} to evaluate some of the derivatives with respect to the momenta.
All the contributions inside the brackets can be shown to vanish using the identity \eqref{ID3}, the expressions for the right and left boundary times \eqref{tRcaseC}-\eqref{tLcaseC} and the continuity relation \eqref{contCaseC}. So we are left with
\begin{equation}
\frac{1}{\Omega_{k,d-1}}\frac{d \mathcal{V}}{dt}
=P_1 \frac{d t_\mt{L}}{d t} + P_2 \frac{d t_\mt{R}}{d t}.
\end{equation}


{\bf Case D} -- In this case the surface passes behind the future horizon in \BHO ($P_1>0$), and admits no turning points, see figure \ref{D}. In fact, our numerical analysis revealed that this case is physically irrelevant since eq.~\eqref{eq:rdotJumpShock} implies that $\dot r_2<\dot r_1$ for planar and spherical black holes for $d\geq 3$, as well as for BTZ black holes, and this is in contradiction with $\dot r_1<0$ and $\dot r_2>0$ implied by this scenario. This case may become relevant when considering hyperbolic black holes with negative mass,  or alternatively in the (unphysical) case of negative energy shocks. We include it here for
completeness.

We will describe the surface using the coordinate $u_\mt{L}$ between the left boundary and $r_s$ and using the coordinate $v_\mt{R}$ between $r_s$ and the left boundary. The expression for $t_\mt{R}$ is obtained just like for case B and it reads
\begin{equation}\label{tRcaseD}
t_\mt{R} - v_s =  \int_{r_s}^{\infty} \tau_2[P_2,r] dr.
\end{equation}
$t_\mt{L}$ is obtained similarly to case C and reads
\begin{equation}\label{tLcaseD}
t_\mt{L}+v_s-2 r_1^*(r_s) = \int_{r_s}^{\infty} \tau_1[P_1,r] dr.
\end{equation}
The volume of the maximal surface is given by integrating along the two segments of the surface
\begin{equation}
\frac{1}{\Omega_{k,d-1}}\mathcal{V} = \int_{r_s}^{\infty} R_1[P_1,r] dr +  \int_{r_s}^{\infty} R_2[P_2,r] dr.
\end{equation}
We can use the identity \eqref{ID1} as well as the relations \eqref{tRcaseD} and \eqref{tLcaseD} for the boundary times, to rewrite it as follows
\begin{equation}\label{Vol1CaseD}
\begin{split}
\frac{1}{\Omega_{k,d-1}}\mathcal{V}
= & \, \int_{r_s}^{\infty} \tilde R_1[P_1,r] dr
+\int_{r_s}^{\infty} \tilde R_2[P_2,r] dr
\\
& \, +P_1(t_\mt{L}+v_s-2 r_1^*(r_s)) +P_2 (t_\mt{R} - v_s).
\end{split}
\end{equation}
We note in passing that the fact that $\dot v$ is continuous across the interface implies using \eqref{ID4}
\begin{equation}\label{contCaseD}
- \tilde R_1[-P_1,r_s]=-\tilde R_1[P_1,r_s]-\frac{2 P_1}{f_1(r_s)}=\tilde R_2[P_2,r_s].
\end{equation}
For the time derivative of the volume \eqref{Vol1CaseD} we obtain
\begin{equation}\label{VolCaseD}
\begin{split}
\frac{1}{\Omega_{k,d-1}}\frac{d \mathcal{V}}{dt}
= &\, \frac{d P_1}{dt} \left(-\int_{r_s}^{\infty} \tau_1[P_1,r] dr  +(t_\mt{L}+v_s-2 r_1^*(r_{t,1}))\right)
\\
&\, +\frac{d P_2}{dt} \left(-\int_{r_{s}}^{\infty} \tau_2[P_2,r] dr  +(t_\mt{R} - v_s)\right)
\\
&\, +\frac{d r_s}{dt} \left(-\tilde R_1[P_1,r_s]-\tilde R_2[P_2,r_s] -\frac{2 P_1}{f_1(r_s)}\right)
+P_1 \frac{d t_\mt{L}}{d t} + P_2 \frac{d t_\mt{R}}{d t}
\end{split}
\end{equation}
where we have used the identity \eqref{ID2} to evaluate some of the derivatives with respect to the momenta.
All the contributions inside the brackets can be shown to vanish using the expressions for the right and left boundary times \eqref{tRcaseD}-\eqref{tLcaseD} and the continuity relation \eqref{contCaseD}. So we are left with
\begin{equation}
\frac{1}{\Omega_{k,d-1}}\frac{d \mathcal{V}}{dt}
=P_1 \frac{d t_\mt{L}}{d t} + P_2 \frac{d t_\mt{R}}{d t}.
\end{equation}


{\bf Case E} -- In this case the surface passes behind the past horizon in \BHO ($P_1<0$), and admits a turning point in \BHO but not in \BHT, see figure \ref{BE}. We will describe this surface using the coordinate $v_\mt{L}$ between the left boundary and $r_{t,1}$, the coordinate $u_\mt{R}$ between $r_{t,1}$ and $r_s$ and the coordinate $v_\mt{R}$ between $r_s$ and the right boundary. The expression for $t_\mt{R}$ is obtained summing the results of the $v_\mt{L}$ and $u_\mt{R}$ integrations in the first two of these segments
\begin{equation}\label{tLcaseE1}
\begin{split}
&v_{t,1}-v_\mt{L} = v_{t,1}+t_\mt{L} = -\int_{r_{t,1}}^{\infty} \tau_1[-P_1,r] dr
\\
&u_s-u_{t,1} = v_s - 2 r_1^*(r_s)-u_{t,1}= - \int_{r_{t,1}}^{r_s} \tau_1[-P_1,r] dr
\end{split}
\end{equation}
which yields
\begin{equation}\label{tLcaseE}
t_\mt{L} + v_s- 2 r_1^*(r_s)+2 r_1^*(r_{t,1}) = -\int_{r_{t,1}}^{r_s} \tau_1[-P_1,r] dr -\int_{r_{t,1}}^{\infty} \tau_1[-P_1,r] dr.
\end{equation}
The right boundary time is obtained just like in case B and reads
\begin{equation}\label{tRcaseE}
t_\mt{R} - v_s =  \int_{r_s}^{\infty} \tau_2[P_2,r] dr.
\end{equation}
The volume of the maximal surface is given by integrating along the three segments of the surface
\begin{equation}
\frac{1}{\Omega_{k,d-1}}\mathcal{V} =\int_{r_{t,1}}^{r_s} R_1[P_1,r] dr+ \int_{r_{t,1}}^{\infty} R_1[P_1,r] dr +  \int_{r_s}^{\infty} R_2[P_2,r] dr.
\end{equation}
We can use the identity \eqref{ID1} as well as the relations \eqref{tLcaseE} and \eqref{tRcaseE} for the boundary times, to rewrite it as follows
\begin{equation}\label{Vol1CaseE}
\begin{split}
\frac{1}{\Omega_{k,d-1}}\mathcal{V}
= & \,
\int_{r_{t,1}}^{r_s} \tilde R_1[-P_1,r] dr
+  \int_{r_{t,1}}^{\infty} \tilde R_1[-P_1,r] dr
+  \int_{r_s}^{\infty} \tilde R_2[P_2,r] dr
\\
& \, +P_1(t_\mt{L}+v_s-2 r_1^*(r_s)+2 r_1^*(r_{t,1})) +P_2 (t_\mt{R} - v_s).
\end{split}
\end{equation}
The fact that $\dot v$ is continuous across the interface implies using \eqref{ID4}
\begin{equation}\label{contCaseE}
\tilde R_1[P_1,r_s]=\tilde R_1[-P_1,r_s]-\frac{2 P_1}{f_1(r_s)}=\tilde R_2[P_2,r_s].
\end{equation}
For the time derivative of the volume \eqref{Vol1CaseE} we obtain
\begin{align}
\frac{1}{\Omega_{k,d-1}} \frac{d \mathcal{V}}{dt}
= &\,
\frac{d r_{t,1}}{dt} \left(-2 \tilde R_1[-P_1,r_{t,1}] + \frac{2 P_1}{f(r_{t,1})}\right)\nonumber
\\
&\,
\frac{d P_1}{dt} \left(\int_{r_{t,1}}^{r_s} \tau_1[-P_1,r] dr +\int_{r_{t,1}}^{\infty} \tau_1[-P_1,r] dr +(t_\mt{L}+v_s-2 r_1^*(r_s)+2 r_1^*(r_{t,1}))\right)\nonumber
\\
&\, +\frac{d P_2}{dt} \left(-\int_{r_s}^{\infty} \tau_2[P_2,r] dr  +(t_\mt{R} - v_s)\right)\nonumber
\\
&\, +\frac{d r_s}{dt} \left(\tilde R_1[-P_1,r_s]-\tilde R_2[P_2,r_s] -\frac{2 P_1}{f_1(r_s)}\right)\label{VolCaseE}
+P_1 \frac{d t_\mt{L}}{d t} + P_2 \frac{d t_\mt{R}}{d t}
\end{align}
where we have used the identity \eqref{ID2} to evaluate some of the derivatives with respect to the momenta.
All the contributions inside the brackets can be shown to vanish using the identity \eqref{ID3}, the expressions for the left and right boundary times \eqref{tLcaseE}-\eqref{tRcaseE} and the continuity relation \eqref{contCaseE}. We are finally left with
\begin{equation}
\frac{1}{\Omega_{k,d-1}}\frac{d \mathcal{V}}{dt}
=P_1 \frac{d t_\mt{L}}{d t} + P_2 \frac{d t_\mt{R}}{d t}.
\end{equation}


{\bf Case F} -- In this case the surface passes behind the past horizon in \BHO ($P_1<0$), and admits turning points in \BHO and \BHT, see figure \ref{AF}. We will describe this surface using the coordinate $v_\mt{L}$ between the left boundary and $r_{t,1}$, $u_\mt{R}$ between $r_{t,1}$ and $r_s$ and $v_\mt{R}$ for the two segments between $r_s$ and the right boundary. The expression for $t_\mt{R}$ is the same as in case E
\begin{equation}\label{tLcaseF}
t_\mt{L} + v_s- 2 r_1^*(r_s)+2 r_1^*(r_{t,1}) = -\int_{r_{t,1}}^{r_s} \tau_1[-P_1,r] dr -\int_{r_{t,1}}^{\infty} \tau_1[-P_1,r] dr.
\end{equation}
The right boundary time is obtained just like in case A and reads
\begin{equation}\label{tRcaseF}
t_\mt{R} - v_s =  -\int_{r_{t,2}}^{r_s} \tau_2[-P_2,r] dr +\int_{r_{t,2}}^{\infty} \tau_2[P_2,r] dr.
\end{equation}
The volume of the maximal surface is given by integrating along the four segments of the surface
\begin{equation}
\frac{1}{\Omega_{k,d-1}}\mathcal{V} =\int_{r_{t,1}}^{r_s} R_1[P_1,r] dr+ \int_{r_{t,1}}^{\infty} R_1[P_1,r] dr +  \int_{r_{t,2}}^{r_s} R_2[P_2,r] dr + \int_{r_{t,2}}^{\infty} R_2[P_2,r] dr.
\end{equation}
We can use the identity \eqref{ID1} as well as the relations \eqref{tLcaseF} and \eqref{tRcaseF} for the boundary times, to rewrite it as follows
\begin{equation}\label{Vol1CaseF}
\begin{split}
\frac{1}{\Omega_{k,d-1}}\mathcal{V}
= & \,
\int_{r_{t,1}}^{r_s} \tilde R_1[-P_1,r] dr
+ \int_{r_{t,1}}^{\infty}\tilde R_1[-P_1,r] dr
+  \int_{r_{t,2}}^{r_s} \tilde R_2[-P_2,r] dr
 + \int_{r_{t,2}}^{\infty} \tilde R_2[P_2,r] dr
\\
& \, +P_1(t_\mt{L}+v_s-2 r_1^*(r_s)+2 r_1^*(r_{t,1})) +P_2 (t_\mt{R} - v_s).
\end{split}
\end{equation}
The fact that $\dot v$ is continuous across the interface implies using \eqref{ID4}
\begin{equation}\label{contCaseF}
\tilde R_1[P_1,r_s]=\tilde R_1[-P_1,r_s]-\frac{2 P_1}{f_1(r_s)}=-\tilde R_2[-P_2,r_s].
\end{equation}
For the time derivative of the volume \eqref{Vol1CaseC} we obtain
\begin{align}
\frac{1}{\Omega_{k,d-1}}\frac{d \mathcal{V}}{dt}
= &\,
\frac{d r_{t,1}}{dt} \left(-2 \tilde R_1[-P_1,r_{t,1}] + \frac{2 P_1}{f(r_{t,1})}\right)\nonumber
\\
&\,
\frac{d r_{t,2}}{dt} \left(- \tilde R_2[-P_2,r_{t,2}] - \tilde R_2[P_2,r_{t,2}]\right)\nonumber
\\
&\,
\frac{d P_1}{dt} \left(\int_{r_{t,1}}^{r_s} \tau_1[-P_1,r] dr +\int_{r_{t,1}}^{\infty} \tau_1[-P_1,r] dr +(t_\mt{L}+v_s-2 r_1^*(r_s)+2 r_1^*(r_{t,1}))\right)\nonumber
\\
&\, +\frac{d P_2}{dt} \left(\int_{r_{t,2}}^{r_s} \tau_2[-P_2,r] dr -\int_{r_{t,2}}^{\infty} \tau_2[P_2,r] dr  +(t_\mt{R} - v_s)\right)\nonumber
\\
&\, +\frac{d r_s}{dt} \left(\tilde R_1[-P_1,r_s]+\tilde R_2[-P_2,r_s] -\frac{2 P_1}{f_1(r_s)}\right)
+P_1 \frac{d t_\mt{L}}{d t} + P_2 \frac{d t_\mt{R}}{d t}\label{VolCaseF}
\end{align}
where we have used the identity \eqref{ID2} to evaluate some of the derivatives with respect to the momenta.
All the contributions inside the brackets can be shown to vanish using the identity \eqref{ID3}, the expressions for the left and right boundary times \eqref{tLcaseF}-\eqref{tRcaseF} and the continuity relation \eqref{contCaseF}. So we are  again left with
\begin{equation}
\frac{1}{\Omega_{k,d-1}}\frac{d \mathcal{V}}{dt}
=P_1 \frac{d t_\mt{L}}{d t} + P_2 \frac{d t_\mt{R}}{d t}.
\end{equation}


{\bf Simpler derivation of volume variation formula:}
The simple form of the volume variation formula,
\begin{equation}
\frac{1}{\Omega_{k,d-1}}\frac{d \mathcal{V}}{dt}
=P_1 \frac{d t_\mt{L}}{d t} + P_2 \frac{d t_\mt{R}}{d t}\,,
\end{equation}
which appears for all of the above cases, is of course not a coincidence. It is a simple consequence of the fact that $\mathcal{V}$ extremizes an `action', \ie the volume functional \reef{actLL1}.
A standard result in classical mechanics states that the change in the on-shell action
when moving between two nearby solutions of the equations of motion is given by boundary terms, as we now review.
Let $x^\mu(s,\lambda)$ be a family of solutions to the equations of motion associated with the action
\begin{equation}
S(s) = \int_{\lambda_\mt{L}}^{\lambda_\mt{R}} d\lambda \, \mathcal{L}(x^\mu, \dot x^{\mu})
\end{equation}
labeled by a parameter $s$ and where we have denoted $\dot x^\mu=\partial_\lambda x^{\mu}$. We also assume that the boundary values of the `time' parameter $\lambda$ are constant for all the solutions in the family.
The variation of the on-shell action with respect to $s$ is then given by
\begin{equation}\label{VarLL1}
\frac{dS}{ds} = \partial_s x^{\mu} k_{\mu} |_{\lambda_\mt{L}} - \partial_s x^{\mu} k_{\mu} |_{\lambda_\mt{R}}
\end{equation}
where the canonical momentum $k_\mu$ is given by $k_\mu \equiv \frac{\partial \mathcal{L}}{\partial \dot x^{\mu}}$.

In our case the onshell action is given by eq.~\eqref{actLL1} and the canonical momentum $k_\mu$ reads\footnote{Note that in the following, we are not using the gauge choice in eq.~\eqref{ParamLL1}. Instead, we will choose parametrizations which ensure that the boundary values of the worldvolume parameter $\lambda$, \ie $\lambda_\mt{L}$ and $\lambda_\mt{R}$, are constant for all the solutions in the family $x^\mu(s,\lambda)$.}
\begin{equation}\label{canmom}
\frac{1}{\Omega_{k,d-1}} k_\mu d x^\mu = P dv + \frac{r^{d-1} \dot v}{\sqrt{-f\dot v^2 +2 \dot v \dot r}} dr = P du -  \frac{r^{d-1} \dot u}{\sqrt{-f\dot u^2 -2 \dot u \dot r}} dr
\end{equation}
where $P$ is the conserved momentum from eq.~\eqref{eq:consEVaid22}. The family of solutions we are interested in is labeled by a parameter $s=t$ which is simply our boundary time parameter. The variation of the solutions with respect to $t$ at the end points of the trajectory is given by
\begin{equation}
\begin{split}
\hspace{-10pt}\frac{\del v_\mt{R}}{\del t}\biggr|_{\lambda_\mt{R}} = \frac{\del u_\mt{R}}{\del t}\biggr|_{\lambda_\mt{R}} = \frac{\del t_\mt{R}}{\del t}, \quad \frac{\del u_\mt{L}}{\del t}\biggr|_{\lambda_\mt{L}}=  \frac{\del v_\mt{L}}{\del t}\biggr|_{\lambda_\mt{L}} = -\frac{\del t_\mt{L}}{\del t}
,\quad \frac{\del r}{\del t}\biggr|_{\lambda_\mt{R}} = \frac{\del r}{\del t}\biggr|_{\lambda_\mt{L}} = 0.
\end{split}
\end{equation}
Substituting into eq.~\eqref{VarLL1} we obtain
\begin{equation}
\frac{1}{\Omega_{k,d-1}}\frac{d \mathcal{V}}{dt}
=P_1 \frac{d t_\mt{L}}{d t} + P_2 \frac{d t_\mt{R}}{d t}.
\end{equation}
Before we conclude this argument let us address two subtle points. First, since the coordinates we are using do not cover the full spacetime, we will need to split our trajectory into different segments using the different coordinate patches. However the new boundary contributions due to this splitting will cancel among themselves, due to the opposite signs of the two contributions in eq.~\eqref{VarLL1} and the fact that the canonical momentum varies smoothly along the trajectory. This is true except for at the exact point in which the shell is being crossed which brings us to the second subtle point. At the point in which the shell is being crossed, the canonical momentum \eqref{canmom} is not smooth, however the different solutions in the family cross the shell at location who vary only by their values of $r$. As a consequence the boundary contributions are proportional there to $k_r \, \delta r / \delta t$. Looking at the equations of motion in the close proximity of the shell $\del_\lambda k_r = \del \mathcal{L}/ \del r$ we conclude by integrating in a small interval around the shell that $k_r$ is continuous across the shell (in our favourite parametrization \eqref{ParamLL1}, this takes the form of the continuity of $\dot v$ when crossing the shell). This is enough to assure the cancelation of the extra boundary contributions at the location of the shell.


{\bf Crossing Between the Past and future regimes:}
the crossing between the two regimes is associated with the case $P_1=0$. In this case the minimal surface in \BHO follows a constant time slice through the bifurcation surface. This constant time slice intersects the shock at
\begin{equation}\label{eq_shocks:tcrit}
 v_s + \tL = r^*_1(r_s)\,.
\end{equation}


Collecting all the above results completes the derivation of eqs. \eqref{tRshocksF}-\eqref{VtshocksF}.

\end{document}